\documentclass[twoside,12pt]{article}
\usepackage{epsfig}
\usepackage{amssymb}
\def\d{{\rm d}}
\def\Tr{{\rm Tr}}

\newcommand{\bqn}{\begin{eqnarray}}
\newcommand{\eqn}{\end{eqnarray}}
\newcommand{\beq}{\begin{equation}}
\newcommand{\eeq}{\end{equation}}
\newcommand{\Hh}{{\textstyle\frac 12}}
\newcommand{\Hf}{{\textstyle\frac 14}}
\newcommand{\mbf}[1]{\mbox{\boldmath$#1$}}

\newcommand{\nrmk}{|\mbox{\boldmath $k$}|}

\def\bk{{\mbox{\boldmath$k$}}}
\def\bq{{\mbox{\boldmath$q$}}}

\def\bp{{\mbox{\boldmath$p$}}}

\def\bP{{\mbox{\boldmath$P$}}}
\def\bPp{{\mbox{\boldmath$P$}^{\prime}}}

\def \dfrac #1#2 {\displaystyle\frac{#1}{#2}}
\def\Upp{\Big[\phi_{\sp}\Big]}
\def\Wpp{\Big[\phi_{\Dp}\Big]}
\def\km{{\cal M}}

\newcommand{\CM}{{\cal M}}

\newcommand{\CO}{{\cal O}}

\newcommand{\gve}{\varepsilon}

\newcommand\el{{\rm e}}

\def\Spp{^3S_1^{++}}
\def\Dpp{^3D_1^{++}}
\def\Smm{^3S_1^{--}}
\def\Dmm{^3D_1^{--}}
\def\sp{^3S_1^{+}}
\def\Dp{^3D_1^{+}}
\def\sm{^3S_1^{-}}
\def\Dm{^3D_1^{-}}
\def\Pte{^3P_1^{e}}
\def\Pto{^3P_1^{o}}
\def\Pe{^1P_1^{e}}
\def\Po{^1P_1^{o}}
\def\Pse{^1P_1^{e}}
\def\Pso{^1P_1^{o}}

\def\Sp{^1S_0^{+}}
\def\Sm{^1S_0^{-}}

\def\PAe{^3P_0^{e}}
\def\PAo{^3P_0^{o}}

\def \ins#1#2#3#4#5#6#7 {
     \begin{figure}[#5]
            \begin{minipage}{#2}
                   \vbox to #3{
                   \hspace*{#4}
                   \special{em:graph {./#1.pcx}}}
                   \caption{#6}
                   \protect\label{#7}
            \end{minipage}\\
      \end{figure}}

\def\eb{E_{\beta}}

\def\spr{{s^{\prime}}}

\def\lmb{{\lambda_{\beta}}}

\def\eb{e_{\beta}}

\def\db{\partial_{\beta^2}}

\def\pp{{\bar p}}

\def\Spp{^3S_1^{+}}
\def\Dpp{^3D_1^{+}}
\def\Smm{^3S_1^{-}}
\def\Dmm{^3D_1^{-}}
\def\Pte{^3P_1^{e}}
\def\Pto{^3P_1^{o}}
\def\Pse{^1P_1^{e}}
\def\Pso{^1P_1^{o}}
\def\Sp{^1S_0^{+}}
\def\Sm{^1S_0^{-}}
\def\Pe{^3P_0^{e}}
\def\Po{^3P_0^{o}}

\def\im{{\rm Im\,}}
\def\re{{\rm Re\,}}

\def\Journal#1#2#3#4{{#1} {#2} (#4) #3 }
\def\Preprint#1#2#3{{#1} {#2} (#3) }
\def\PreNTH#1{{preprint nucl-th/}{#1}}
\def\NCA{{\em Nuovo Cimento} A}

\def\NPA{{\em Nucl. Phys.} A}

\def\PROK{{\em Prog. Theor. Phys. (Kyoto) Suppl.}}
\def\NPB{{\em Nucl. Phys.} B}

\def\PLB{{\em Phys. Lett.} B}

\def\PRL{\em Phys. Rev. Lett.}
\def\PREV{\em Phys. Rev.}
\def\PREP{\em Phys. Rep.}

\def\PRD{{\em Phys. Rev.} D}
\def\PRC{{\em Phys. Rev.} C}

\def\ZPC{{\em Z. Phys.} C}
\def\ZPA{{\em Z. Phys.} A}
\def\ANNP{\em Ann. Phys. (N.Y.)}

\def\PPNP{{\em Progr.~Part.~Nucl.~Phys.}}

\def\INTA{{\em Int. J. Mod. Phys.} A}
\def\PNTH{{\em E-Preprint: nucl-th/}}
\def\ANP{{\em Adv. Nucl. Phys.}}
\def\JPG{{\em J. Phys.} G: {\em Nucl. Phys.}}
\def\PPN{{\em Phys.~Part.~Nucl.}}
\def\SJPN{{\em Sov. J. Part. Nucl.}}
\def\PAN{{\em Phys. Atom. Nucl.}}
\def\SJNP{{\em Sov. J. Nucl. Phys.}}
\def\PRSA{{\em Proc. Roy. Soc.} A}
\def\FBS{{\em Few Body Syst.}}
\def\EPJA{{\em Eur. Phys. J.} A}
\def\EPJC{{\em Eur. Phys. J.} C}

\def\p{\vec p}

\def\q{\vec q}

\def\bl{\mbox{\boldmath $l$}}
\def\eb{E_{\beta}}
\def\d{d{}}
\newcommand{\be}{\begin{equation}}
\newcommand{\ee}{\end{equation}}
\newcommand{\bea}{\begin{eqnarray}}
\newcommand{\eea}{\end{eqnarray}}

\begin{document}

\title{{\bf
Bethe--Salpeter Approach with the Separable Interaction
for the Deuteron.}}
\author{
S.G.\ Bondarenko,$^{1}$ V.V.\ Burov,$^{1,3}$ A.V.\
Molochkov,$^{1,3}$ G.I.\ Smirnov,$^{2}$ H.\ Toki $^{3}$\\
\\
$^{1}$ Bogoliubov Laboratory of Theoretical Physics\\
Joint Institute for Nuclear Research, Dubna, Russia\\
$^{2}$  Laboratory of Particle Physics\\
Joint Institute for Nuclear Research, Dubna, Russia\\
$^{3}$ Research Center for Nuclear Physics\\
Osaka University, Osaka, 567-0047, Japan\\
}
\maketitle
\begin{abstract}
Recent developments of the covariant Bethe--Salpeter (BS) approach
with the use of the separable interaction for the deuteron are
reviewed. It is shown that the BS formalism  allows a covariant
description of various  electromagnetic reactions like the
lepton-deuteron scattering, deuteron electro-disintegration, deep
inelastic scattering (DIS) of leptons on  light nuclei. The
procedure of the construction of the separable nucleon-nucleon
($NN$) interaction  is discussed.
The BS formalism facilitates analysis of the role of the $P$-waves
(negative energy components) in the electromagnetic properties of
the deuteron and its comparison with the nonrelativistic results.
Furthermore the covariant BS approach makes it possible to analyze
DIS of leptons from the deuteron in a model independent way and to
extend the formalism to DIS reactions on the light nuclei.
\end{abstract}
\newpage
\tableofcontents
\newpage

\section{Introduction\label{In}}
The study of electromagnetic properties of light nuclei, $A \leq
4$, facilitates the construction of the theory of strong
interactions and, in particular, the nucleon--nucleon interaction
(see, for example, \cite{brown}). Theoretical  research in this
field is of topical interest which is reflected in recent review
articles~\cite{car98}-\cite{bma3}.
 A large amount of
available experimental data stimulate a further development of
theoretical methods which are often restricted to qualitative
predictions. The forthcoming experiments are expected to provide
high precision data which will allow us to explore  the region  of
large  momentum transfer  in elastic, inelastic and deep inelastic
(DIS) electron-nucleus reactions.

These data will be able to furnish qualitatively new information
about the fine nuclear structure at small distance. This is the
reason why the role of the non-nucleon  degrees of freedom as
mesons, $ \Delta $-isobars, quarks etc. on intermediate and high
energy phenomena is widely discussed (see, for example,
~\cite{baldin77}-\cite{toki00}). Clear understanding and
consistent interpretation of the experimental information is not
possible without the consideration of the relativistic kinematics
of reactions and the dynamics of the interaction. For this reason,
the construction of a covariant  approach and a detailed analysis
of relativistic effects in electro\-mag\-ne\-tic reactions with
light  nuclei become the tasks of the highest priority.

One can identify three steps for the construction of a theoretical
framework of the lepton--nucleus interaction. The first step  is
to introduce  the dynamical degrees of freedom parameterizing
hadronic and nuclear structure. The second one is to construct
from these degrees of freedom
 bound states which are hadrons and nuclei.
The third step is to find a formalism of the interaction of these
degrees of freedom with incident particles which are photons
$\gamma$ and leptons $l$ in the present study.

In some sense this division is artificial, since the three steps
are just different parts of the same problem of dynamics of
interacting fields to be solved in one consistent approach. The
problems are interrelated by the underlying physics in such a way
that  by fixing one of them  one can find solutions for the
others. Nevertheless, such a division helps us to separate the
task within a consistent set of  approximations and to use a
phenomenological technique to set constrains on different parts of
the approach.

The fact that nuclei consist of bound  nucleons introduces a major
problem for theoretical description of  relativistic $l-A$
interactions. The deuteron is naturally the first object in the
row of many other nuclei, and  has received a vast number of
treatments within many different approaches. One finds also that
non-relativistic schemes of calculations are widely employed in
the analysis, which can be justified for a few particular cases.
On the other hand, the consistent consideration of the
relativistic bound states is offered within the Bethe-Salpeter
(BS) formalism~\cite{BS51,MA55},
 which makes it most promising  for the class of the tasks
considered in the present review. What is even more important is
that the BS formalism allows qualitatively a new interpretation of
the physics of the relativistic bound state and should not be
regarded as an alternative scheme only.

The first practical applications of the BS formalism were based on
the three dimensional reduction. The first reduction of the BS
formalism, the so-called quasi-potential equation, has been made
in several works~\cite{brown},\cite{logunov63}-\cite{ks78}. The
main idea behind these approaches is to fix  the relative time of
bound constituents to a  certain value. Since the relative time
(or relative energy in momentum space) is considered as an
unphysical feature of the BS formalism, the different ways of
fixing the relative time should produce equivalent quasipotential
approaches. The  comparison of the different quasipotential
equations was done in review articles~\cite{orden01,gross01}. This
line of realization of the BS formalism allows one to take into
account some of the relativistic effects, but it looses the
general relativistic covariance~\cite{TjonPasc} (see,
papers~\cite{Fl_Tjon:75,Fa_Tjon:86} also). Moreover, a number of
difficulties arises
 which does not allow  one to establish a direct link to
the non-relativistic calculations, for example, the absence of a
non-relativistic reduction of arbitrary kernels, the problem of
the interpretation of the abnormal parity states. This fact
motivates  us to use  the original BS formalism which offers a
consistent covariant description of the interacting particles and
their bound states. The qualitative connection with the
nonrelativistic results can be made on the level of observables.
Such exotic features of the BS formalism as relative time of the
bound constituents and abnormal parity states receive their
nonrelativistic interpretation through the relation to dynamical
degrees of freedom. Analysis of the relative time dependence gives
qualitatively  a new point of view on relativistic bound states.

This review is devoted to the analysis of the three steps
considered above within the Bethe-Salpeter formalism and its
application to the study of the electromagnetic properties of
light nuclei. We emphasize the covariant description of the BS
formalism by taking the separable interaction, which is still a
stage of infancy. In particular, the role of the abnormal parity
states, in not yet confronted with experimental data, though the
necessity is demonstrated in this paper.
  The review is divided
into 6 sections. In section 2, we consider basic properties and
definitions of the Bethe-Salpeter formalism. By using an example
of the $NN$ system we discuss the basic properties of the four
dimensional ($4D$) bound states. In section 3, we investigate the
simplest process for such a study
--- elastic scattering of electrons off the deuteron.
In  section 4,  we  discuss the application of the BS formalism to
the problem of the deuteron electro-disintegration,  which is an
inelastic process with finite momentum transfer. The analysis of
the deep inelastic scattering of leptons on the lightest nuclei
--- $A = 2 \ldots 4$ is presented in section 5.
 The consideration of the DIS reaction in the infinite momentum frame
leads to many simplifications in the analysis and allows us to
understand clearly some of the important physical consequences of
the $4D$ consideration
 and to draw model independent conclusions.
Section 6 is devoted to the summary of  the results of the $4D$ BS
formalism.

\section{Bethe--Salpeter Approach}\label{sec:BS}

\subsection{\em  Formalism}
In study of processes, involving bound states, in local field
theory we have to consider particles which are not asymptotically
free. The standard field theory reduction technique gives a way to
extract all the information about physical states contained in the
matrix elements to a product of field operators (see for example
~\cite{bjorken}). However, this technique is based on the
assumption that the physical states in the matrix elements can be
treated as being asymptotically free, and the interaction is
switched on adiabatically. Therefore, bound states are excluded
from consideration. In order to include bound states, the
reduction formalism  must be supplemented by a procedure allowing
expectation values in such states to be expressed in terms of
vacuum expectation values.

This problem is solved in the nonrelativistic field theory by
introducing an external classical field which allows the bound
state dynamics to be described as a particle motion in a potential
well. As a result, the calculation of the expectation value in a
bound state reduces to the calculation of the expectation value in
a one-particle state, and binding effects are taken into account
by introducing the momentum distribution of this particle. This
simple approach can be used to obtain amplitudes of the lepton
scattering off bound states in the form of convolution. However,
in this case, the role played by relativistic corrections and
off-shell effects remains unclear. In quasipotential approaches,
the solution of the bound state problem reduces to deriving an
analog of the Shr\"odinger equation with a covariant
three-dimensional potential~\cite{gross1}-\cite{Gross}. The
calculation of expectation values in bound states should be
essentially the same as in the nonrelativistic case. However, in
contrast to nonrelativistic approaches, the quasipotential method
allows a qualitative study of the role of the relativistic and the
off-shell effects.

A method of calculating the expectation values of $\rm T$-products
of local operators in bound states was suggested in~\cite{MA55}.
The essence of the method is that the expectation values in bound
states are expressed in terms of the vacuum expectation values of
a ${\rm T}$-product of  local operators and the matrix elements
for the transition between the vacuum and the bound state. We
shall consider the application of this method to processes
involving a bound state of $n$ nucleons.

\subsection{\em  The Bethe--Salpeter Amplitude}\label{formalism:vertex}
A consideration of a scattering process is based on the analysis
of the matrix element of a ${\rm T}$-product of local operators
$\eta_1 (y_1)\dots \eta_k (y_k)$ in the initial $A$ and final
$A^\prime$ bound states. :
\begin{equation}
\langle A^\prime(\alpha^\prime,P^\prime)| {\rm T}(\eta_1(y_1)\dots
\eta_k(y_k))|A(\alpha,P )\rangle. \label{matel}
\end{equation}
$P$ and $P^\prime$ are the initial and final momenta of the bound
state while the $\alpha$ and $\alpha^\prime$ are full sets of
their discrete quantum numbers. The local field operators,
${\eta_i(y_i)}$, are current operators determining the nucleon
interactions with external fields. To express this matrix element
in terms of vacuum expectation values we shall use the fact that
in a definite kinematical region the joint propagation of $n$
interacting nucleons occurs via the formation of a bound state. In
this case the first term of the series expansion of the
$n$-nucleon Green's function in intermediate state has the form:
\begin{eqnarray}
&&\langle 0 | {\rm T}(\Psi (x_1)\dots \Psi (x_n) \overline\Psi
(x^\prime_1))\dots \overline\Psi
(x^\prime_n) | 0 \rangle =\nonumber\\[.2cm] &&=\int \frac{d^3P}{(2\pi)^3}
\sum\limits_{\alpha} \langle 0 | {\rm T}(\Psi (x_1)\dots \Psi
(x_n))|A(\alpha,P)\rangle \times\nonumber\\ &&\times\langle
A(\alpha,P)|{\rm T}(\overline\Psi (x^\prime_1)\dots \overline\Psi
(x^\prime_n))|0\rangle
\theta(min\{|{x_0}_i|\}-max\{|{x^\prime_0}_i|\})+\dots. \label{pole1}
\end{eqnarray}
The functions $\Psi$ and $\overline\Psi$ are nucleon field
operators in the Heinsenberg representation. The $\theta$ function
is arising due to the expansion of the $\rm T$-product in the
matrix elements. This function ensures that the causality
condition is satisfied:
\begin{equation}
min\{{x_0}_i\} > max\{{x^\prime_0}_i\}.\nonumber\\
\end{equation}
The extreme values (minimum and maximum) of the nucleon
coordinates can be determined by introducing the average
coordinate of $n$ fields conjugate to the total momentum $P$,
which in the case of $n$ particles of the same mass has the form,
\begin{equation}
X=\frac{\sum\limits_{i}^{n}x_i}{n}.
\end{equation}
The  nucleon coordinates relative to this point are $\tilde
x_i=x_i-X$. As a result, the maximum and minimum coordinates can
be defined as
\begin{eqnarray}
max\{{x_0}_i\} = X_0 + max\{|\tilde {x_0}_i|\}, \nonumber\\
min\{{x_0}_i\} = X_0 - max\{|\tilde {x_0}_i|\}. \label{minmax}
\end{eqnarray}

Using the integral representation of the $\theta$ function
\begin{equation}
\theta(x_0)=\frac{i}{2\pi}\int\limits_{-\infty}^{+\infty}
\frac{e^{-ip_0x_0}}{p_0+i\delta}dp_0, \label{tint} \end{equation}
we obtain the necessary expression relating the vacuum expectation
value of the nucleon operators and the matrix elements for the
transition from the vacuum to the bound state:
\begin{eqnarray}
&&\langle 0 | {\rm T}(\Psi (x_1)\dots \Psi (x_n)) \overline\Psi
(x^\prime_1)\dots \overline\Psi (x^\prime_n) | 0 \rangle=
\sum_{\alpha}\int
\frac{d^3P}{(2\pi)^3}\frac{dP_0}{2\pi}e^{-i(P_0-E)(X_0-X_0^\prime)}
\times \label{g4poli}\\
&&\times e^{i(P_0-E)(max\{|{\tilde x_{0i}}|\}+max\{|{\tilde
x^\prime_{0i}}|\})} \frac{\langle 0|{\rm T}(\Psi (x_1)\dots \Psi
(x_n))|A(\alpha,P)\rangle \langle A(\alpha,P)| {\rm T}(\Psi
(x^\prime_1)\dots \Psi (x^\prime_n))|0\rangle}{P_0-E+i\delta}.
\nonumber\end{eqnarray}
This expression is inconvenient because of
using
two sets of coordinates simultaneously, $\{x_i\}$ and $\{X,\tilde
x_i\}$. To avoid this, we shall change over to using the second
set of coordinates everywhere in the calculations. Owing to the
translational invariance, the following relation holds for the
fields  $\Psi$:
\begin{eqnarray}
\Psi(x+\tilde x)=e^{i\hat Px}\Psi(\tilde x)e^{-i\hat Px}.
\label{shiftX} \end{eqnarray} Replacing $x_i$ by $X+\tilde x_i$ in
the operators $\Psi$ and $\overline\Psi$ and shifting by $X$,
using transformation~(\ref{shiftX}), we obtain the matrix elements
for the transition from the vacuum to the bound state in the space
of relative nucleon locations: {  \begin{eqnarray}
&&\hspace*{-1.2cm}\langle 0 |{\rm T}(e^{i\hat
PX}\Psi(x_1)e^{-i\hat PX} \dots e^{i\hat PX} \Psi(x_n)e^{-i\hat
PX})|A(\alpha,P)\rangle= \Phi^{A(\alpha, P)}(\tilde x_1,\dots
\tilde x_n)e^{-i(EX_0-{\bf P\cdot X})}, \nonumber\\[.2cm]  &&\hspace*{-1.2cm} \langle A(\alpha,P)|{\rm T}
(e^{i\hat  PX}\overline\Psi(x^\prime_1)e^{-i\hat PX}\dots e^{i\hat
PX}\overline\Psi(x^\prime_n)e^{-i\hat PX})|0 \rangle =
\overline\Phi^{A(\alpha, P)}(\tilde x^\prime_1,\dots \tilde x^
\prime_n) e^{i(EX_0-{\bf P\cdot X})}. \label{chiX0}
\end{eqnarray} } Although the functions $\Phi^{A(\alpha,P)}$ formally
depend on  $n$ variables, only $n-1$ of them are independent,
owing to the equation $\sum\limits_i^n \tilde x_i = 0.$ As a
result of these transformations, Eq.~(\ref{g4poli}) takes the
form:
  {
\begin{eqnarray}
\langle 0 | {\rm T}(\Psi (x_1)\dots \Psi (x_n) \overline\Psi
(x^\prime_1)\dots \overline\Psi (x^\prime_n)) | 0
\rangle=\sum_{\alpha}\hspace*{-.2cm}\int\hspace*{-.2cm}
\frac{d^4P}{(2\pi)^4}e^{-iP(X-X^\prime)} \frac{\Phi^{A(\alpha,
P)}(\tilde x_1,\dots \tilde x_n) \overline\Phi^{A(\alpha,
P)}(\tilde x^\prime_1,\dots \tilde x^\prime_n)}{P_0-E+ i\delta}.
\label{g4pole} \end{eqnarray} } Since the integral with respect to
$P_0$ is determined by the behavior of the integrand near the pole
$P_0=E$, we have omitted the exponential factor
$exp\{i(P_0-E)(max\{|{\tilde x_{0i}}|\}+max\{|{\tilde
x^\prime_{0i}}|\})\}$.

The unknown functions $\Phi^{A(\alpha, P)}$ and
$\overline\Phi^{A(\alpha, P)}$
\begin{eqnarray}
&&\hspace*{-1.2cm}\Phi^{A(\alpha, P)}(x_1,\dots x_n)= \langle 0
|{\rm T}(\Psi(x_1)\dots \Psi(x_n))|A(\alpha,P)\rangle,
\nonumber\\[.2cm]  &&\hspace*{-1.2cm}
\overline\Phi^{A(\alpha, P)}(x^\prime_1,\dots x^ \prime_n)
=\langle A(\alpha,P)|{\rm T}(\overline\Psi(x^\prime_1)\dots
\overline\Psi(x^\prime_n))|0 \rangle, \label{chiX}
\end{eqnarray}
introduced above and entering into the matrix elements for the
transition from the vacuum to the bound state in Eq.~(\ref{chiX})
describe the nuclear state in terms of the degrees of freedom of
the virtual nucleons. These functions are called the
Bethe--Salpeter amplitudes. They give the solution of the
fundamental problem --- the expression of the expectation values
in bound states in terms of the vacuum expectation values.

\subsection{\em  Analysis of the Matrix Elements in the BS Formalism}
In order to explain how the matrix element~(\ref{matel}) is
related to the nucleon Green's functions and the Bethe--Salpeter
amplitudes, we consider the matrix element
\begin{equation}
\langle 0| {\rm T}(\Psi (x_1)\dots \Psi (x_n)\eta_1 (y_1)\dots
\eta_k (y_k) \overline\Psi (x^\prime_1)\dots \overline\Psi
(x^\prime_n)) |0 \rangle \label{me}
\end{equation}
near the singularity of the $n$-nucleon bound state at $P^2=M^2$.
We expand the time-ordered product in the matrix
element~(\ref{matel}) in a product of matrix elements of  $\Psi$,
$\eta$, $\overline\Psi$. In order to do this, we choose the
maximum and minimum zeroth components from the set $\{x_i\}$ in
accordance with (\ref{minmax}) and keeping only terms which
correspond to the first term in~(\ref{pole1}) we write the $\rm
T$-product in the form:
\begin{eqnarray}
&&{\rm T}(\Psi (x_1)\dots \Psi (x_n)\eta_1(y_1)\dots \eta_k(y_k)
\overline\Psi (x^\prime_1)\dots \overline\Psi
(x^\prime_n))=\nonumber\\ &&={\rm T}(\Psi (x_1)\dots \Psi
(x_n)){\rm T}(\eta_1(y_1)\dots
\eta_k(y_k)) {\rm T}(\overline\Psi (x^\prime_1)\dots \overline\Psi (x^\prime_n))\times\nonumber\\
&&\times\theta(X_0-Y_0-max\{|{\tilde x_{0i}}|\}-max\{|{\tilde
y_{0i}}|\}) \theta(Y_0-X^\prime_0-max\{|{\tilde
x^\prime_{0i}}|\}-max\{|{\tilde y_{0i}}|\})+ \nonumber\\ &&+{\rm
T}(\overline\Psi (x^\prime_1)\dots \overline\Psi (x^\prime_n))
{\rm T}(\eta_1(y_1)\dots_k\eta (y_k)) {\rm T}(\Psi (x_1)\dots \Psi
(x_n))\times\nonumber\\ &&\times
\theta(X^\prime_0-Y_0-max\{|{\tilde x^\prime_{0i}}|\}-max\{|{\tilde
y_{0i}}|\}) \theta(Y_0-X_0-max\{|{\tilde x_{0i}}|\}-max\{|{\tilde
y_{0i}}|\}). \nonumber\end{eqnarray}

Inserting a complete set between $\rm T$-products, we rewrite
 Eq.~(\ref{matel}) as a sum over states from the complete set:
\begin{eqnarray}
&&\langle 0 | {\rm T}(\Psi (x_1)\dots \Psi (x_n)\eta_1 (y_1)\dots
\eta_k (y_k) \overline\Psi (x^\prime_1)\dots \overline\Psi (x^\prime_n)) | 0 \rangle =\nonumber\\[.2cm]
&&=\sum_{R,R^\prime} \langle 0 | {\rm T}(\Psi (x_1)\dots \Psi
(x_n))|R\rangle \langle R|{\rm T}(\eta_1 (y_1)\dots
\eta_k (y_k))|R^\prime\rangle\times\nonumber\\[.2cm] &&\times \langle R^\prime|{\rm T}(\overline\Psi (x^\prime_1)\dots
\overline\Psi (x^\prime_n))|0\rangle \theta(X_0-Y_0-max\{{|\tilde x_{0i}}|\}-max\{|{\tilde y_{0i}}|\}) \times\nonumber\\[.2cm]
&&\times \theta(Y_0-X^\prime_0-max\{|{\tilde x^\prime_{0i}}|\}-max\{|{\tilde y_{0i}}|\})+ \nonumber\\[.2cm]
&&+\sum_{R,R^\prime} \langle 0 | {\rm T}(\overline\Psi
(x^\prime_1)\dots
\overline\Psi(x^\prime_n)) |R\rangle \langle R|{\rm T}(\eta_1 (y_1)\dots \eta_k (y_k))|R^\prime\rangle\times\nonumber\\[.2cm]
&&\times\langle R^\prime|{\rm T}(\Psi (x_1)\dots \Psi
(x_n))|0\rangle \theta(X^\prime_0-Y_0-max\{|{\tilde
x^\prime_{0i}}|\}-max\{|{\tilde
y_{0i}}|\}) \times\nonumber\\[.2cm] &&\times\theta(Y_0-X_0-max\{|{\tilde x_{0i}}|\}-max\{|{\tilde y_{0i}}|\})~,
\label{matr1} \end{eqnarray} where $\sum\limits_{R,R^\prime}$
denotes a summation over discrete quantum numbers and an
integration over continuous variables.

Since we are interested in the behavior of Eq.~(\ref{matel}) near
the pole at $P^2=M^2$, it is sufficient to study the contribution
to Eq.~(\ref{matr1}) from the lowest bound state corresponding to
this pole. The bound state corresponds to the first term of the
series~(\ref{matr1}) with $R=A(\alpha,P)$ and
$R^\prime=A^\prime(\alpha^\prime,P^\prime)$. Using the integral
representation (\ref{tint}) for the $\theta$ function, we obtain
the following expression for the matrix element~(\ref{me}): {
\begin{eqnarray} &&\hspace*{-0.8cm}\langle 0| {\rm T}(\Psi
(x_1)\dots \Psi (x_n)\eta_1 (y_1)\dots
 \eta_k (y_k) \overline\Psi (x^\prime_1)\dots \overline\Psi (x^\prime_n)) | 0\rangle=
 \hspace*{-.2cm}\sum\limits_{\alpha, \alpha^\prime}\hspace*{-.2cm}\int\hspace*{-.2cm}\frac{d^3P}{(2\pi)^3}
\frac{d^3P^\prime}{(2\pi)^3} \langle 0 |{\rm T}(\Psi(x_1)\dots \Psi(x_n))|A(\alpha,P)\rangle \nonumber\\
&&\langle A(\alpha,P)|{\rm T}(\eta_1 (y_1)\dots \eta_k
(y_k)|A^\prime(\alpha^\prime,P^\prime))\rangle \langle
A^\prime(\alpha^\prime, P^\prime)|{\rm T}(\overline\Psi
(x^\prime_1)\dots \overline\Psi (x^\prime_n))|0
\rangle\times\label{matr2}\\[.3cm] &&\hspace*{-.5cm}\int
\frac{dP_0}{(2\pi)}\frac{dP_0^\prime}{(2\pi)}
\frac{e^{-i(P_0-E)(X_0-Y_0)}e^{-i(P^\prime_0-E^\prime)(Y_0-X^\prime_0)}
e^{-i(P_0-E)(max\{|{\tilde x_{0i}}|\}-max\{|{\tilde y_{0i}}|\})}
e^{-i(P^\prime_0-E^\prime)(max\{|{\tilde
x^\prime_{0i}}|\}-max\{|{\tilde y_{0i}}|\})}} {
(P_0-E+i\delta)(P_0^\prime-E^\prime+i\delta)}~.
\nonumber\end{eqnarray} } Using Eq.~(\ref{chiX0}) to get over to
the relative variables, we rewrite this expression as
\begin{eqnarray}
&&\langle 0| {\rm T}(\Psi (x_1)\dots \Psi (x_n) \eta_1 (y_1)\dots
\eta_k (y_k) \overline\Psi (x^\prime_1)\dots
\overline\Psi (x^\prime_n)) | 0\rangle=\label{matrX}\\
&&\sum\limits_{\alpha, \alpha^\prime}\int\frac{d^4P}{(2\pi)^4}
\frac{d^4P^\prime}{(2\pi)^4} e^{-i(P_0-E)(max\{|{\tilde
x_{0i}}|\}-max\{|{\tilde y_{0i}}|\})}
e^{-i(P^\prime_0-E^\prime)(max\{|{\tilde
x^\prime_{0i}}|\}-max\{|{\tilde y_{0i}}|\})} \nonumber\\
&&\times \frac{\Phi^{A(\alpha, P)}(\tilde x_1,\dots \tilde x_n)
\langle A(\alpha,P)|{\rm T}(\eta_1 (\tilde y_1)\dots \eta_k
(\tilde y_k))|A^\prime(\alpha^\prime,P^\prime)\rangle
\overline\Phi^{A^{\prime}(\alpha^\prime, P^\prime)}(\tilde
x^\prime_1,\dots \tilde x^\prime_n)} {
(P_0-E+i\delta)(P_0^\prime-E^\prime+i\delta)}e^{-iPX}e^{iP^\prime
X^\prime} e^{-i(P-P^\prime)Y}. \nonumber\end{eqnarray} On the
other hand, owing to the unitarity of the $n$-nucleon Green's
function related to the amplitude (\ref{pole1}) as
\begin{eqnarray}
G_{2n}(x_1\dots x_n, x^\prime_1, \dots x^\prime_n)=\langle 0| {\rm
T}(\Psi (x_1)\dots \Psi (x_n)\overline\Psi (x^\prime_1)\dots
\overline\Psi (x^\prime_n)) | 0\rangle, \label{g2n}
\end{eqnarray}
the matrix element~(\ref{me}) can be rewritten identically as
\begin{eqnarray}
&&\langle 0| {\rm T}(\Psi (x_1)\dots \Psi (x_n)\eta_1 (y_1)\dots
\eta_k (y_k) \overline\Psi (x^\prime_1)\dots \overline\Psi (x^\prime_n)) | 0\rangle = \nonumber\\
&&\int d^4z_1\dots d^4z_n d^4z^\prime_1\dots d^4z^\prime_n
d^4z^{\prime\prime}_1\dots
d^4z^{\prime\prime}_n d^4z^{\prime\prime\prime}_1\dots d^4z^{\prime\prime\prime}_n G_{2n}(x_1\dots x_n, z_1, \dots z_n)\nonumber\\
&&\times G^{-1}_{2n}(z_1\dots z_n, z^\prime_1, \dots z^\prime_n)
\langle 0| {\rm T}(\Psi (z^\prime_1)\dots \Psi (z^\prime_n)\eta_1
(y_1)\dots \eta_k (y_k) \overline\Psi (z^{\prime\prime}_1)\dots
\overline\Psi(z^{\prime\prime}_n))| 0\rangle\nonumber\\ &&\times
G^{-1}_{2n}(z^{\prime\prime}_1, \dots z^{\prime\prime}_n,
z^{\prime\prime\prime}_1, \dots z^{\prime\prime\prime}_n)
G_{2n}(z^{\prime\prime\prime}_1, \dots z^{\prime\prime\prime}_n,
x^\prime_1, \dots x^\prime_n)=\nonumber\\ &&\int d^4z_1\dots
d^4z_n d^4z^{\prime\prime\prime}_1\dots
d^4z^{\prime\prime\prime}_n G_{2n}(x_1\dots x_n, z_1, \dots
z_n)\nonumber\\ &&\times \overline G_{2n+k}(z_1\dots z_n, y_1\dots
y_k, z^{\prime\prime\prime}_1, \dots z^{\prime\prime\prime}_n)
G_{2n}(z^{\prime\prime\prime}_1, \dots z^{\prime\prime\prime}_n,
x^\prime_1, \dots x^\prime_n~), \label{g2nk}\end{eqnarray} where
$\overline G_{2n+k}$ is the truncated Green's function defined as:
\begin{eqnarray}
&&\overline G_{2n+k}(x_1\dots x_n, y_1\dots y_k, x^\prime_1, \dots x^\prime_n)=\nonumber\\
&&=\int d^4z_1\dots d^4z_n d^4z^\prime_1\dots d^4z^\prime_n G^{-1}_{2n}(x_1\dots x_n, z_k, \dots z_n)\times\nonumber\\
&&\times \langle 0| {\rm T}(\Psi (z_1)\dots \Psi (z_n)\eta_1
(y_1)\dots \eta_k (y_k) \overline\Psi (z^\prime_1)\dots
\overline\Psi (z^\prime_n)) | 0\rangle G^{-1}_{2n}(z^{\prime}_1,
\dots z^{\prime}_n, x^\prime_1, \dots x^\prime_n),
\label{barg2nk}\end{eqnarray} where index $n$ is related to the
nuclear operators $\Psi$ and index $k$ is related to the external
field operators $\eta$. Taking into account the behavior of
$n$-nucleon Green's function~(\ref{g4pole}) near the pole at
$P^2=M^2$, we compare this expression with (\ref{matrX}):
\begin{eqnarray}
&&\hspace*{-.8cm}\sum_{\alpha, \alpha^\prime} \int\frac{d^4P}
{(2\pi)^4}\frac{d^4P^\prime}{(2\pi)^4}\int d z_1 \dots dz_n
dz^\prime_1 \dots dz^\prime_n e^{-iP X} e^{iP^\prime
X^\prime}\frac{\Phi^{A(\alpha, P)}(\tilde x_1,\dots
\tilde x_n) \overline{\Phi}^{A(\alpha, P)}(z_1,\dots z_n)}{P_0-E+ i\delta}\nonumber\\[.2cm]
&&\hspace*{-.8cm}\times\overline G_{2n+k}(z_1\dots z_n, \tilde
y_1\dots \tilde y_k, z^\prime_1\dots z^\prime_n)
\frac{\Phi^{A^\prime(\alpha^\prime, P^\prime)}(z^\prime_1,\dots
z^\prime_n) \overline\Phi^{A^\prime(\alpha^\prime, P^\prime)}
(\tilde x^\prime_1,\dots
\tilde x^\prime_n)}{P^\prime_0-E^\prime+ i \delta}=\nonumber\\[.2cm]
&&\hspace*{-.8cm}\sum\limits_{\alpha,
\alpha^\prime}\int\frac{d^4P}{(2\pi)^4}
\frac{d^4P^\prime}{(2\pi)^4} e^{-i(P_0-E)(max\{|{\tilde
x_{0i}}|\}-max\{|{\tilde y_{0i}}|\})}
e^{-i(P^\prime_0-E^\prime)(max\{|{\tilde
x^\prime_{0i}}|\}-max\{|{\tilde y_{0i}}|\})}
e^{-i(P-P^\prime)Y}\nonumber\\[.2cm]
&&\hspace*{-.8cm}\times \frac{\Phi^{A(\alpha, P)}(\tilde x_1,\dots
\tilde x_n) \langle A(\alpha,P)|{\rm T}(\eta_1 (\tilde y_1)\dots
\eta_k (\tilde y_k))|A^\prime(\alpha^\prime,P^\prime)\rangle
\overline\Phi^{A^\prime(\alpha^\prime, P^\prime)}(\tilde
x^\prime_1,\dots \tilde x^\prime_n)} {
(P_0-E+i\delta)(P_0^\prime-E^\prime+i\delta)} e^{-iPX}e^{iP^\prime
X^\prime}. \label{matr3}
\end{eqnarray}

Multiplying both sides of the integrand in~(\ref{matr3}) by
$(P_0-E)(P^\prime_0-E^\prime)$ and passing to the limit  $P_0 \to
E$ and $P^\prime_0 \to E^\prime$, we obtain an expression for
calculating the expectation value in the bound state of the ${\rm
T}$-product of a set of local operators:
\begin{eqnarray}
&&\langle A(\alpha,P)|{\rm T}(\eta (y_1)\dots \eta (
y_k))|A^\prime(\alpha^\prime,P^\prime)\rangle
=\int dz_1 \dots dz_n dz^\prime_1 \dots dz^\prime_n \overline{\Phi}^{A(\alpha, P)}(z_1,\dots z_n)\nonumber\\
&&\overline G_{2n+k}(z_1\dots z_n, y_1\dots y_k, z^\prime_1\dots
z^\prime_n) {\Phi}^{A^\prime(\alpha^\prime,
P^\prime)}(z^\prime_1,\dots z^\prime_n). \label{matrres}
\end{eqnarray}
This expression relates the scattering amplitude of $n$-nucleon
bound state with irreducible Green's function $\overline
G_{2n+k}$, describing scattering on $n$ virtual nucleons, to the
BS amplitudes $\Phi^{A(\alpha, P)}$ and $\overline\Phi^{A(\alpha,
P)}$ describing the nuclear states in terms of the nucleon degrees
of freedom, the equation of which is to be found.

\subsection{\em  The Bethe--Salpeter Equation}

The relation between the BS amplitude $\Phi^{A(\alpha, P)}$ and
the $n$-nucleon Green's function $G_{2n}$ is established in
Eq.~(\ref{g4pole}). Thus, having an equation for $G_{2n}$, we
obtain an equation satisfied by the BS amplitude
$\Phi^{A(\alpha,P)}$. For this task, however, it is insufficient
to know $G_{2n}$ only perturbatively, since the analysis of the
behavior of the Green's function near a bound-state pole requires
summation of the entire perturbation series. Let us therefore
examine what general equations for
 $G_{2n}$ can be obtained without invoking to any perturbation
theory.

The propagation of a free nucleon from a point $x_1$ to a point
$x_2$ is described by the free Green's function $S_{(1)}$
satisfying the equation of the form~\cite{IS}:
\begin{equation}
(i\hat\partial_{x_1} - m)S_{(1)}(x_1,x_2)=\delta(x_1-x_2),
\label{freeeq}\end{equation}
where $m$ is the nucleon mass. In the
case of a nucleon interacting with its own field, a term taking
into account the self-interaction appears on the right-hand side
of the equation for the exact Green's function $G_2$,
\begin{equation}
(i\hat\partial_{x_1} - m)G_2(x_1,x_2)= \delta(x_1-x_2)+ \int
dx^\prime_1\overline G_2(x_1,x^\prime_1) G_2(x^\prime_1,x_2),
\label{inteq}\end{equation} Comparing~(\ref{freeeq}) and
(\ref{inteq}), we see that the function $\overline
G_2(x_1,x^\prime_1)$ satisfies the Dyson equation,
\begin{equation}
\overline G_2(x_1,x_2)=S^{-1}_{(1)}(x_1,x_2)-G^{-1}_2(x_1,x_2),
\label{Dayson}\end{equation}
 i.e., it coincides with the
one-nucleon irreducible self-energy part. The Green's function
$S_{(2)}$ describing the joint propagation of two physical
nucleons which do not interact with each other satisfies an
equation of the form:
\begin{equation}
(i\hat\partial_{x_1} - m^*)\otimes (i\hat\partial_{x_2} - m^*)
S_{(2)}(x_1,x_2,y_1,y_2)= \delta(x_1-y_1)\delta(x_2-y_2),
\label{freeeq2}\end{equation}
 where $m^*=m+G_2(x_1,x_2)$. Inclusion of the interaction between the nucleons leads
 to the appearance of an additional term on the right-hand side:
\begin{eqnarray} & & (i\hat\partial_{x_1} - m^*)\otimes (i\hat\partial_{x_2} - m^*) G_4(x_1,x_2,y_1,y_2)=\nonumber\\
& = & \delta(x_1-y_1)\delta(x_2-y_2) +\int dy^\prime_1
dy^\prime_2\overline G_4(x_1,x_2,y^\prime_1,y^\prime_2)
G_4(y^\prime_1,y^\prime_2,y_1,y_2). \label{inteq2}\end{eqnarray}
Comparing (\ref{freeeq2}) and (\ref{inteq2}), we obtain the
two-particle analog of the Dyson equation, namely, the
inhomogeneous Bethe--Salpeter equation:
\begin{equation} \overline
G_{4}(x_1,x_2,y_1,y_2) = S^{-1}_{(2)}(x_1,x_2,y_1,y_2)-
G^{-1}_{4}(x_1,x_2,y_1,y_2), \label{kernel2}\end{equation} In
analogy with the one-nucleon case, the function describing the
interaction between the nucleons coincides with the irreducible
self-energy part of the two-nucleon system. Generalizing to the
case of  $n$ nucleons, we obtain the equation:
\begin{equation}
\overline G_{2n}(x_1\dots x_n, x^\prime_1\dots x^\prime_n) =
S^{-1}_{(n)}(x_1\dots x_n, x^\prime_1\dots x^\prime_n)-
G^{-1}_{2n}(x_1\dots x_n, x^\prime_1\dots x^\prime_n),
\label{kernel}\end{equation} where the function $S_{(n)}$ is the
direct product of $n$-nucleon propagators:
\begin{equation}
S_{(n)}(x_1\dots x_n, x^\prime_1\dots x^\prime_n)= \langle 0|{\rm
T}(\Psi(x_1) \overline \Psi(x^\prime_1))|0\rangle \otimes \dots
\langle 0|{\rm T}(\Psi(x_n)\overline\Psi(x^\prime_n))|0\rangle.
\end{equation}
Using Eq.~(\ref{kernel}) for $G_{2n}$, we obtain the integral
equation with $\overline G_{2n}$ as the kernel:
\begin{eqnarray}
&&G_{2n}(x_1\dots x_n, x^\prime_1\dots x^\prime_n)=
S_{(n)}(x_1\dots x_n, x^\prime_1\dots x^\prime_n)+ \nonumber\\
&&\int dz_1\dots dz_n dz^\prime_1\dots dz^\prime_n
S_{(n)}(x_1\dots x_n, z_1\dots z_n) \overline G_{2n}(z_1\dots z_n,
z^\prime_1\dots z^\prime_n)\nonumber\\ &&\times
G_{2n}(z^\prime_1\dots z^\prime_n, x^\prime_1\dots x^\prime_n).
\label{BSG}\end{eqnarray} Thus, the exact $n$-nucleon Green's
function is the solution of the integral equation which relates
the two unknown Green's functions  $G_{2n}$ and $\overline G_{2n}$
to each other.

There are several ways of solving this problem:\\[-0.6cm]
\begin{itemize}
  \item dispersion method with Nakanishi integral representation of
perturbation theory;\\[-0.8cm]
  \item separable ansatz for $\overline G_{2n}$;\\[-0.8cm]
  \item perturbative method.
  \end{itemize}

In analogy with the Dyson equation, the Bethe--Salpeter equation
can be studied by using the technique of dispersion relations.
This can be realized by introducing a generalization of the
spectral representation for the exact one-particle Green's
function for the case of $n$ particles --- the Nakanishi integral
representation of perturbation theory~\cite{nakanishi69}. This
approach has been used successfully for solving the
Bethe--Salpeter equation in the case of two scalar
particles~\cite{Kusaka95,Kusaka97}.

On the other hand, Eq.~(\ref{BSG}) offers an excellent possibility
to model the $n$-nucleon Green's function if $\overline G_{2n}$ is
introduced explicitly. Both the separable ansatz and perturbative
methods are related to this strategy.

In the case of a separable form for the kernel of Eq.~(\ref{BSG}),
we write:
\begin{equation}
\overline G_{2n}(z_1\dots z_n, z^\prime_1\dots z^\prime_n)=
\sum^{N}_{i,j}\lambda_{ij}g_i(z_1\dots z_n)g_j(z^\prime_1\dots
z^\prime_n).\label{G2nSep}
\end{equation}
In this case the problem of solving the integral equation is
replaced by the problem of solving a system of linear equations.
This approach has been used successfully to describe a two-nucleon
system~\cite{graz:rel,BBD:2000}. Recently, the combination of the
approaches based on the use of a separable potential and the
spectral representation taking into account the analytic
properties of the two-nucleon Green's function is demonstrated to
serve as the foundation for the construction of a relativistic
separable ansatz for the function $\overline
G_{2n}$~\cite{toki01}.

Most commonly used form of $\overline G_{2n}$ can be obtained by
perturbative methods. Let us consider the iterative solution of
Eq.~(\ref{BSG}). We take the zeroth iteration as
$$G_{2n}^{(0)}(x_1\dots x_n, x^\prime_1\dots x^\prime_n)= S_{(n)}(x_1\dots x_n, x^\prime_1\dots x^\prime_n).$$
Substituting this expression into Eq.~(\ref{BSG}), we obtain the
first iteration:
\begin{eqnarray}
&&G_{2n}^{(1)}(x_1\dots x_n, x^\prime_1\dots x^\prime_n)=
S_{(n)}(x_1\dots x_n, x^\prime_1\dots x^\prime_n)+ \nonumber\\
&&+\int dz_1\dots dz_n dz^\prime_1\dots dz^\prime_n
S_{(n)}(x_1\dots x_n, z_1\dots z_n)
\overline G_{2n}(z_1\dots z_n, z^\prime_1\dots z^\prime_n)\times\nonumber\\
&&\times S_{(n)}(z^\prime_1\dots z^\prime_n, x^\prime_1\dots
x^\prime_n). \label{iter1}\end{eqnarray}

In the course of successive iterations we obtain the expansion of
the exact $n$-nucleon Green's function in an infinite series in
powers of $\overline G_{2n}$:
\begin{eqnarray}
&&G_{2n}(x_1\dots x_n, x^\prime_1\dots x^\prime_n)=
S_{(n)}(x_1\dots x_n, x^\prime_1\dots x^\prime_n)+ \nonumber\\
&&+\sum\limits_{m}^{\infty} \int dz_1\dots dz_n dz^{\prime}_1\dots
dz^{\prime}_n\dots
dz^{(m)}_1\dots dz^{(m)}_n S_{(n)}(x_1\dots x_n, z_1\dots z_n)\times\nonumber\\
&&\times\overline G_{2n}(z_1\dots z_n, z^\prime_1\dots z^\prime_n)
S_{(n)}(z^\prime_1\dots
z^\prime_n, x^\prime_1\dots x^\prime_n) \dots \times\nonumber\\
&&\times\overline G_{2n}(z^{(m-1)}_1\dots z^{(m-1)}_n,
z^{(m)}_1\dots z^{(m)}_n) S_{(n)}(z^{(m)}_1\dots z^{(m)}_n,
x^\prime_1\dots x^\prime_n). \label{iter}\end{eqnarray}

On the other hand, the function  $G_{2n}$  can be expanded in a
perturbation series in a specific nucleon--nucleon interaction
model (for example, the meson-exchange model):
\begin{eqnarray}
&&G_{2n}(x_1\dots x_n, x^\prime_1\dots x^\prime_n)=
\sum_i G^{(i)}_{2n}(x_1\dots x_n, x^\prime_1\dots x^\prime_n),\nonumber\\[.2cm]
&&G^{(0)}_{2n}(x_1\dots x_n, x^\prime_1\dots x^\prime_n)=
S_{(n)}(x_1\dots x_n, x^\prime_1\dots x^\prime_n).
\label{iter2}\end{eqnarray} Comparing ~(\ref{iter})
and~(\ref{iter2}), we obtain:
\begin{eqnarray}
&&\overline G_{2n}(x_1\dots x_n, x^\prime_1\dots x^\prime_n)= \nonumber\\
&&=\sum\limits_{m}^{\infty}\sum\limits_{m_1+m_2=m}\frac{1}{m+1}
\int dz_1\dots dz_n dz^{\prime}_1\dots dz^{\prime}_n {G^{(m_1)}}^{-1}_{2n}(x_1\dots x_n, z_1\dots z_n)\times\nonumber\\
&&\times {G^{(m+1)}}_{2n}(z_1\dots z_n, z^\prime_1\dots
z^\prime_n) {G^{(m_2)}}^{-1}_{2n}(z^\prime_1\dots z^\prime_n,
x^\prime_1\dots x^\prime_n). \label{expansion}\end{eqnarray} In
the meson-exchange model of $NN$ interaction, the first term of
the series ($m=1$) corresponds to the one-boson exchange
approximation in the kernel of~(\ref{BSG}) (the ladder
approximation).

We substitute Eq.~(\ref{g4pole}) into Eq.~(\ref{BSG}), multiply
both sides of the resulting expression by $(P_0-E)$, and take
$P_0\to E$. We obtain:
\begin{eqnarray}
&&{\Phi}^{A(\alpha, P)}(x_1,\dots x_n)= \label{BSchi}\\
&&\hspace*{-1cm}=\int dz_1\dots dz_n dz^\prime_1\dots dz^\prime_n
S_{(n)}(x_1\dots x_n, z_1\dots z_n) \overline G_{2n}(z_1\dots z_n,
z^\prime_1\dots z^\prime_n) {\Phi}^{A(\alpha, P)}(z^\prime_1,\dots
z^\prime_n). \nonumber\end{eqnarray} Thus, the matrix element for
the transition between the vacuum and the $n$-nucleon bound state
satisfies a homogeneous integral equation with kernel~$\overline
G_{2n}$, which is related to the exact $n$-nucleon and one-nucleon
Green's functions by~(\ref{kernel}).

We shall need Eq.~(\ref{BSchi}) for the rest of the calculations.
By means of the Fourier transform, the Bethe--Salpeter amplitude
$\Phi^A$ can be rewritten in momentum space as:
\begin{eqnarray}
\Phi^{A(\alpha)}(p_1\dots p_{n})=\int d^4x_1\dots
d^4x_{n}e^{i\sum_i^{n} p_ix_i} \Phi^{A(\alpha, P)}(x_1\dots
x_{n}), \label{momentum1}\end{eqnarray} where $p_i$ is momentum of
$i$-th nucleon, $\sum_i^n p_i=P$. It is more convenient to use a
set of the variables which includes the total momentum $P$
explicitly. We use the set of the momenta
$\{P,k_i\}_{_{i=1..n-1}}$, where $k_i=p_i-P/n$ is the nucleon
relative momenta. The momentum $k_n$ is not included in the set
because it is not independent, $k_n=-\sum_i^{n-1}k_i$. In terms of
this set the expression~(\ref{momentum1}) can be written as:
\begin{eqnarray}
\Phi^{A(\alpha)}(P,k_1\dots k_{n-1})=\int d^4\tilde x_1\dots
d^4\tilde x_{n}e^{i\sum_i^{n} k_i\tilde x_i} \Phi^{A(\alpha,
P)}(\tilde x_1\dots \tilde x_{n}). \label{momentum2}\end{eqnarray}
 The BS equation in the momentum space takes the
form:
\begin{eqnarray}
&&\Phi^{A(\alpha)}(P,k_1,\dots k_{n-1})= \label{BSgam}\\
&&\hspace*{-1cm}S_{(n)}(P,k_1,\dots k_{n-1}) \int
\frac{d^4k_1}{(2\pi)^4}\dots \frac{d^4k_{n-1}}{(2\pi)^4} \overline
G_{2n}(P, k_1\dots k_{n-1}, k^\prime_1\dots k^\prime_{n-1})
\Phi^{A(\alpha)}(P,k^\prime_1,\dots k^\prime_{n-1}).
\nonumber\end{eqnarray}

Since~(\ref{BSgam}) and~(\ref{BSchi}) are homogeneous integral
equations, the BS amplitude is defined up to some constant. In
order to determine this constant we consider the expectation value
of the baryon current at zero momentum transfer,
\begin{equation}
\langle A(\alpha,P)|J_{\mu}(0)|A(\alpha^\prime,
P)\rangle=inP_\mu\delta^{\alpha,\alpha^\prime}.
\label{norm-cur}\end{equation} Using Eq.~(\ref{matrres}) we obtain
the normalization condition for $\Phi^{A(\alpha)}$:
\begin{eqnarray}
&&\int \frac{d^4k_1}{(2\pi)^4}\dots \frac{d^4k_{n-1}}{(2\pi)^4}
\frac{d^4k^\prime_1}{(2\pi)^4}\dots
\frac{d^4k^\prime_{n-1}}{(2\pi)^4} \overline\Phi^{A(\alpha)} (P,
k_1,\dots k_{n-1}) \times\label{n_chi}\\ &&\times{\overline
G_{2n+1}}_{\mu}(q=0,P, k_1\dots k_{n-1}, k^\prime_1\dots
k^\prime_{n-1})\Phi^{A(\alpha^\prime)} (P, k^\prime_1,\dots
k^\prime_{n-1})=inP_{\mu}\delta^{\alpha,\alpha^\prime}.
\nonumber\end{eqnarray} Using the fact that at zero momentum
transfer the exact truncated photon-$n$-nucleon vertex ${\overline
G_{2n+1}}_{\mu}$ is related to the derivative of the $n$-nucleon
Green's function with respect to the total momentum, {
\begin{eqnarray} {\overline G_{2n+1}}_{\mu}(q=0,P, k_1\dots
k_{n-1}, k^\prime_1\dots k^\prime_{n-1})=-\frac{\partial}{\partial
P_{\mu}} G_{2n}^{-1}(P, k_1\dots k_{n-1}, k^\prime_1\dots
k^\prime_{n-1}),
\end{eqnarray} }
and expressing  $G_{2n}^{-1}$ with the help of Eq.~(\ref{kernel}),
we obtain the normalization condition:
\begin{eqnarray}
&&\int \frac{d^4k_1}{(2\pi)^4}\dots \frac{d^4k_n}{(2\pi)^4}
\frac{d^4k^\prime_1}{(2\pi)^4}\dots
\frac{d^4k^\prime_n}{(2\pi)^4} \overline\Phi^{A(\alpha)} (P, k_1,\dots k_{n-1})\times\nonumber\\
&&\times\left[S^{-1}_{(n)}(P, k_1,\dots k_{n-1})
\left\{\frac{\partial}{\partial {P_{\mu}}} S_{(n)}(P, k_1,\dots
k_{n-1})\right\} S^{-1}_{(n)}(P, k_1,\dots k_{n-1})\right.+
\nonumber\\ &&+\left. \frac{\partial}{\partial {P_{\mu}}}\overline
G_{2n}(P, k_1\dots k_{n-1}, k^\prime_1\dots k^\prime_{n-1})\right]
\Phi^{A(\alpha^\prime)} (P, k^\prime_1,\dots
k^\prime_{n-1})=-inP_{\mu}\delta^{\alpha,\alpha^\prime}.
\label{norm_chi}\end{eqnarray}

We conclude this section by noting the important features of the
BS amplitude:
\begin{itemize}
\item the BS amplitude depends on the zeroth component of the
relative coordinate (relative time)  of the nucleons, which,
according to (\ref{matrres}), is reflected  in the dynamical
observables of the $n$-nucleon bound state. In momentum space this
leads to a dependence on the zeroth component of the nucleon
relative momentum (relative energy). The relative time dependence
is manifested as observable effects in DIS of leptons which will
be discussed in section~\ref{DIS}, \item the analytic properties
of $\Phi^{A(\alpha)}$ in (\ref{g4pole}) are related with
singularities of Green's function $G_{2n}$. This connection can be
used to derive nonperturbatively the kernel of the BS
equation~\cite{toki01}. In  section~\ref{BS:anal} we will consider
this in details. There are poles associated with the external
nucleon propagators, cuts in the relative momenta, and poles
associated with the various bound states formed either by several
or by all  nucleons. The latter ones are isolated in
(\ref{g4pole}) and therefore do not contribute to
$\Phi^{A(\alpha)}$. The poles associated with a bound state of
$n$-nucleons ($n< A$) can be isolated by special
procedure~\cite{myadep,bma4}, discussed in
section~\ref{DIS:Nuclei}, where it is applied to the study of DIS
of leptons off light nuclei. The consideration of these poles
gives the nuclear amplitude in terms of nucleons and lighter
nuclei amplitudes. The first type of singularities can be
explicitly isolated by introducing the BS vertex function:
\begin{eqnarray}
S_{(n)}(P,k_1\dots k_{n-1})\Gamma^{A(\alpha)}(P,k_1\dots k_{n-1})=
\Phi^{A(\alpha)}(P, k_1,\dots k_{n-1}),
\end{eqnarray}
which is widely used in this review.
\end{itemize}

\subsection{\em  Basic Properties of Two--Nucleon BS equation}
We consider now the two-particle case of the BS equation, which
allows one to understand its basic properties in detail. Starting
from the formula~(\ref{BSG}) with $n=2$,
\begin{eqnarray}
&&{G_4}_{\alpha\beta,\gamma\delta}(x_1,x_2,x^{\prime}_1,x^{\prime}_2)=
{S_{(2)}}_{\alpha\beta,\gamma\delta}(x_1,x_2,x^{\prime}_1,x^{\prime}_2)\label{f002}\\
&&\hskip 10mm +i\int \prod\limits_i^4 d^4 w_i
{S_{(2)}}_{\alpha\beta,\sigma\rho}(x_1,x_2,w_1,w_2) {\overline
G_4}_{\sigma\rho,\lambda\omega}(w_1,w_2,w_3,w_4)
{G_4}_{\lambda\omega,\gamma\delta}(w_3,w_4,x^{\prime}_1,x^{\prime}_2),
\nonumber
\end{eqnarray}
where we have introduced explicitly the spinor indices noted by
Greek letters. The repeated spinor indices are assumed to be
summed up. The functions $G_4$ and ${\overline G}_4$ are the exact
and the truncated two-nucleon Green's functions, respectively, and
$S_{(2)}$ is the Green's function of two noninteracting nucleons,
and equals to the direct product of full one-nucleon propagators.
It is a widely used assumption to omit self-energy part in
one-nucleon propagators. In this case
\begin{eqnarray}
{S_{(2)}}_{\alpha\beta,\gamma\delta}(x_1,x_2,x^{\prime}_1,x^{\prime}_2)
= {S^{(0)}_{\alpha\gamma}}(x_1-x^{\prime}_1)
{S^{(0)}_{\beta\delta}}(x_2-x^{\prime}_2). \label{noself}
\end{eqnarray}
To write the BS equation in momentum space, we take Fourier
transform and introduce
\begin{eqnarray}
V(P,k^{\prime},k) = \int d^4 x_{1} d^4 x_{2} d^4 y_{1} d^4 y_{2}
{\overline G_4}(x_1,x_2,y_1,y_2)
\times \hskip 30mm \label{f003}\\
\exp{(iP(\frac{x_{1}+x_{2}}{2}-\frac{y_{1}+y_{2}}{2})+
ik^{\prime}\frac{x_{1}-x_{2}}{2}-ik\frac{y_{1}-y_{2}}{2})},
\nonumber\end{eqnarray} where $P$ is the total momentum,  $k$ and
$k^{\prime}$ are the relative 4-momenta of the two nucleons before
and after the interaction. They are connected with 4-momenta of
first ($p_1$) and second ($p_2$) particles:
\begin{eqnarray}
&& P = p_1+p_2, \qquad k = (p_1-p_2)/2,
\label{moments}\\
&& p_1 = P/2+k, \quad p_2 = P/2-k. \nonumber\end{eqnarray} We
introduce the function $V(P;k^{\prime},k)$ in momentum space for
the kernel ${\overline G_4}(x_1,x_2,x^{\prime}_1,x^{\prime}_2)$ of
the BS equation~(\ref{BSG}). Similar formula could be written for
functions $G_4(x_1,x_2,x^{\prime}_1,x^{\prime}_2)$ and
$S_{(2)}(x_1,x_2,x^{\prime}_1,x^{\prime}_2)$.

Thus, the BS equation for full Green's function of the two-nucleon
system can be written as,
\begin{eqnarray}
&& G_{\alpha\beta,\gamma\delta}(P,k^\prime,k) =\,
S_{\alpha\gamma}(\frac{P}{2}+k) S_{\beta\delta}(\frac{P}{2}-k)
\delta^{(4)}(k^{\prime}-k)
\label{f004}\\
&&\hskip 10mm +i
S_{\alpha\epsilon}(\frac{P}{2}+k^{\prime})S_{\beta\lambda}(\frac{P}{2}-k^{\prime})
\int\,
\frac{d^4k^{\prime\prime}}{(2\pi)^{4}}V_{\epsilon\lambda,\rho\omega}(P,k^{\prime},k^{\prime\prime})
G_{\rho\omega,\gamma\delta}(P,k^{\prime\prime},k),
\nonumber\end{eqnarray} where one-nucleon propagator
$S_{\alpha\beta}(p)$ with assumption~(\ref{noself}) has the form,
\begin{eqnarray}
S_{\alpha\beta}(p) = \left[1/(p \cdot
\gamma-m+i\epsilon)\right]_{\alpha\beta}. \label{f005}
\end{eqnarray}
Here $m$ is the mass of the nucleon,  $p\cdot \gamma$ denotes
$p_\mu \gamma^\mu$, and $\gamma_{\mu}$ are Dirac matrices. The
Greek letters $\alpha\beta$ denote the component in the $4\times4$
matrix.

Introducing two-nucleon $T$-matrix by equation
\begin{equation}
S_{\alpha\sigma}(\frac{P}{2}+k^{\prime})S_{\beta\rho}(\frac{P}{2}-k)
T_{\sigma\rho,\gamma\delta}(P,k^\prime,k)=\int\,
\frac{d^4k^{\prime\prime}}{(2\pi)^{4}}G_{\alpha\beta,\epsilon\lambda}(P,k^\prime,k^{\prime\prime})
V_{\epsilon\lambda,\gamma\delta}(P,k^{\prime\prime},k),
\label{TMatrix}\end{equation} we can write the BS equation for the
$T$-matrix as
\begin{eqnarray}
&& T_{\alpha\beta,\gamma\delta}(P,k^\prime,k) =
V_{\alpha\beta,\gamma\delta}(P,k^\prime,k)
\label{f007}\\
&& \hskip 10mm +i\int\frac{d^4 k^{\prime\prime}}{(2\pi)^4}
V_{\alpha\beta,\epsilon\lambda}(P,k^{\prime},k^{\prime\prime})
S_{\epsilon\eta}(P/2+k^{\prime\prime})
S_{\lambda\rho}(P/2-k^{\prime\prime})
T_{\eta\rho,\gamma\delta}(P,k^{\prime\prime},k). \nonumber
\end{eqnarray}

To solve the BS equation, we should assume some form for the
interaction kernel. Considering a model with exchange particle
(for instance, meson with mass $\mu$) interacting with nucleons we
could formulate the following analytic properties of the
two-nucleon $T$-matrix with all legs on
mass-shell:\label{exchange-properties}
\begin{enumerate}
\item if the two-nucleon system forms a bound state, $T$-matrix
has a simple pole in the total momentum squared ($s=P^2$) at the
point corresponding to the mass of the bound state $s=M^2$; \item
in the region $s>4m^2$, $T$-matrix has so-called unitarity cut
which corresponds to the elastic nucleon-nucleon scattering ($NN
\to NN$); \item in the region $s>(2m+n\mu)^2$, $T$-matrix has cuts
which correspond to the inelastic nucleon-nucleon scattering
resulting in the production of $n$ mesons with mass $\mu$ ($NN \to
NN(n\mu)$); \item in the region $s<(4m^2-(n\mu)^2)$, $T$-matrix
has cuts which correspond to the inelastic nucleon-antinucleon
scattering (in a cross-reaction) with production of $n$ mesons
with mass $\mu$ ($N{\bar N} \to N{\bar N}(n\mu)$) (so-called {\em
left-hand} cuts).
\end{enumerate}
Another choice of the kernel (for instance a widely used separable
form) leads to analytic properties different from the ones
considered here. It will be discussed in section~\ref{BS:Sol}.

The equation for the BS amplitude in momentum space  can be
written by using Eq.~(\ref{BSgam}). We use $A(\alpha)\equiv J\km$
because the total momentum $J$ defines the bound state in the
two-nucleon case.
\begin{eqnarray}
\Phi^{J\km}_{\alpha\beta}(P,k) =
iS_{\alpha\eta}(\frac{P}{2}+k)S_{\beta\rho}(\frac{P}{2}-k) \int
\frac{d^4 k^{\prime\prime}}{(2\pi)^4}
V_{\eta\rho,\epsilon\lambda}(P,k,k^{\prime\prime})
\Phi^{J\km}_{\epsilon\lambda}(P,k^{\prime\prime}), \label{f013}
\end{eqnarray}
and the normalization condition~(\ref{norm_chi}) takes the form:
\begin{eqnarray}
&&\int \frac{d^4k^{\prime}}{(2\pi)^4}\frac{d^4k}{(2\pi)^4}
\overline\Phi^{J\km}(P,k^{\prime}) \left[\phantom{A\over
B}\delta^{(4)}(k-k^{\prime}) S^{-1}(P/2+k)S^{-1}(P/2-k) \right.
\nonumber\\
&&\frac{\partial}{\partial {P_{\mu}}} \left\{
{S}(P/2+k){S}(P/2-k)\right\} {S}^{-1}(P/2+k){S}^{-1}(P/2-k)+
\nonumber\\
&&+\left.\frac{\partial}{\partial {P_{\mu}}}
V(P,k^{\prime},k)\right] \Phi^{J\km^\prime}(P,k) =
-2iP_{\mu}\delta^{\km\km^\prime}. \label{f014}
\end{eqnarray}
If the interaction kernel does not depend on the total momentum
$P$, then Eq.~(\ref{f014}) becomes
\begin{eqnarray}
\int \frac{d^4k}{(2\pi)^4} \overline\Gamma^{J\km}(P,k)
\frac{\partial}{\partial {P_{\mu}}} \left\{
{S}(P/2+k){S}(P/2-k)\right\} \Gamma^{J\km}(P,k) =
-2iP_{\mu}\delta^{\km\km^\prime}, \label{f015}
\end{eqnarray}
where $\Gamma^{J\km}(P,k)$ is the two-nucleon vertex function
defined as
\begin{eqnarray}
S_{\alpha\gamma}(\frac{P}{2}+k)
S_{\beta\delta}(\frac{P}{2}-k)\Gamma_{\gamma\delta}(P,k) =
\Phi_{\alpha\beta}(P,k). \label{f010} \end{eqnarray}

\subsection{\em  Partial--Wave Decomposition of the BS  Amplitude}\label{sec:partial}
In order to solve the BS equation and to calculate the cross
sections of the electromagnetic reactions with a two-nucleon
system, we use the partial wave decomposition of the BS amplitude
separating the radial and the spin-angular parts. The two
representations for the partial-wave decomposition are considered.

\subsubsection{\em Direct Product Representation}\label{DPR}
In the direct product representation, we determine the two
particle spinor basis in laboratory frame as
${U^{\rho_1}_{s_1}}(\bk)\otimes {U^{\rho_2}_{s_2}}(-\bk)$, where
$\mu$ is the spin projection,  $\rho_{1,2}$ is the so called
$\rho$-spin~\cite{kubis}, which distinguishes the positive and
negative energy states.  Both the positive and negative energy
states are necessary in order to prepare a complete set for the
two-particle bound state.  The spinors ${U^{\rho}_{s}}(\bk)$ are
connected with the Dirac free spinors, $u_{s}(\bk)$ and
$v_{s}(\bk)$, as
\begin{eqnarray}
U^{\rho}_{s} (\bk)= \left\{ \begin{array}{ll} u_{s}(\bk), &\rho =
+, \\ v_{-s}(-\bk), &\rho = -. \end{array} \right.
\label{Urho}\end{eqnarray} The Dirac spinors are determined
as~\cite{IS},
\begin{eqnarray}
u_{s}(\bk) = L(\bk) u_{s} (\mbf{0}), \quad v_{s}(\bk) = L(\bk)
v_{s} (\mbf{0}), \label{trans}\end{eqnarray} and the boost
operator for a particle with spin $1/2$ and mass  $m$
is~\cite{lurie}
\begin{eqnarray}
L(\mbf{k})=\frac{m+ k\cdot\gamma
\gamma_0}{\sqrt{2E_{\bk}(m+E_{\bk})}}. \label{lor}
\end{eqnarray}
Here $k=(E_{\bk},\bk)$ is the 4-momentum of a particle on mass
shell, $E_{\bk}=\sqrt{m^2+\bk^2}$ is the energy of the particle.
In laboratory frame the spinors can be written as:
\begin{eqnarray}
u_{s}(\mbf{0}) = \left(\begin{array}{c} \chi_{s}  \\ 0\
\end{array} \right),  \quad v_{s}(\mbf{0})=
\left(\begin{array}{c}         0   \\ \chi_{-s} \ \end{array}
\right), \nonumber
\end{eqnarray}
where $\chi_{s}$ are two-component Pauli spinors. The
normalization conditions are:
\begin{eqnarray}
{\bar u}_s(\bk)\,u_{s^\prime}(\bk) =
\frac{m}{E_{\bk}}\delta_{ss^\prime},\quad\quad {\bar
v}_{s}(\bk)\,v_{s^\prime}(\bk) =
-\frac{m}{E_{\bk}}\delta_{ss^\prime}. \label{spnorm}
\end{eqnarray}

The BS amplitude $\Phi^{J\km}(P,k)$ of the two-particle system
with total angular momenta $J$ and projection $\km$ in laboratory
frame can be written as
\begin{eqnarray}
\Phi^{J\km}(P,k)=\sum \limits_{LS\rho_1\rho_2} \;
\phi_{JLS\rho_1\rho_2}(k_0,\nrmk)\; {\cal
Y}^{JLS\rho_1\rho_2,\km}(\mbf{k}), \label{reldp}
\end{eqnarray}
where $P=(M,{\bf 0})$ is the total momentum,  $k$ is the relative
momentum of the two-particle system ($k=(k_0,\bk), k_0\neq
E_{\bk})$. Here $JLS\rho_1\rho_2$ is combination of quantum
numbers of total angular momentum $J$, orbital momentum $L$, spin
$S$ and $\rho$-spins. We define the spin angular function as
${\cal Y}^{JLS\rho_1\rho_2,\km}(\mbf{k})$:
\begin{eqnarray}
{\cal Y}_{\alpha\beta}^{JLS\rho_1\rho_2,\km}(\mbf{k}) \,=\, i^L
\sum \limits_{s_1 s_2 m_L m_S}\, (L m_L S m_S | J {\km}) \, (\Hh
s_1 \Hh s_2 | S m_S)\, Y_{L m_L}({\hat{\mbf{k}}})
\left({U_{s_1}^{\rho_1}}(\mbf{k})\right)_{\alpha}
\left({U_{s_2}^{\rho_2}}(-\mbf{k})\right)_{\beta},
\label{Ydecomp}
\end{eqnarray}
where $(.|.)$ are Clebsch-Gordan coefficients and
${\hat{\mbf{k}}}={\bk}/{|\mbf{k}|}$. The spin-angular function
${\cal Y}^{JLS\rho_1\rho_2,\km}(\mbf{k})$ is a matrix,
$(1\times4)\otimes(1\times4) = (1\times16)$ in spinor space. The
spinor indices $\alpha\beta$ specify the component of this matrix.

The orthogonalization condition for the spin-angular functions is
\begin{eqnarray}
\int\,\d\Omega_{\bk}\, {{\cal
Y}_{\alpha\beta}^{JLS\rho_1\rho_2,\km}}^{\dag}(\bk)\, {\cal
Y}_{\alpha\beta}^{JL^\prime S^\prime \rho_1^\prime
\rho_2^\prime,{\km}^{\prime}}(\bk)\, =\,
\delta_{LL^\prime}\,\delta_{{\km}
{\km}^{\prime}}\,\delta_{SS^\prime}\,\delta_{\rho_1\rho_1^\prime}\delta_{\rho_2\rho_2^\prime}.
\end{eqnarray}
Where: $d\Omega_{\bk} \equiv d\phi_{\bk}\ d\cos{\theta_{\bk}}$ and
the conjugated spin-angular function can be obtained by
substitution, ${U_{s}}^{\pm}(\mbf{k}) \to {\bar
U_{s}}^{\pm}(\mbf{k})$.

We can write the inverse propagators $[S(P/2+k)]^{-1}$,
$[S(P/2-k)]^{-1}$ in laboratory frame as:
\begin{eqnarray}
[S(P/2+k)]^{-1} &=& P\cdot\gamma/2+k\cdot\gamma-m = \label{propi}\\
&=& \frac{1}{2E_{\bk}} \left[ (p_1\cdot\gamma-m){S_{-}^{(1)}}^{-1}
+ (p_2\cdot\gamma+m){S_{+}^{(1)}}^{-1} \right],
\nonumber\\[0mm]
[S(P/2-k)]^{-1} &=& P\cdot\gamma/2-k\cdot\gamma-m = \nonumber\\
&=& \frac{1}{2E_{\bk}} \left[ (p_2\cdot\gamma-m){S_{-}^{(2)}}^{-1}
+ (p_1\cdot\gamma+m){S_{+}^{(2)}}^{-1} \right], \nonumber
\end{eqnarray}
where   $p_1=(E_{\bk},\bk), p_2=(E_{\bk},-\bk)$, $s=P^2$, and
functions ${S^{i}_{\rho}}$, with $\rho=\pm$ are
\begin{eqnarray}
{S_{\rho}^{(1)}} = 1/(\sqrt{s}/2 + k_0 - \rho E_{\bk}), \quad
{S_{\rho}^{(2)}} = 1/(\sqrt{s}/2 - k_0 - \rho E_{\bk}).
\label{propii}
\end{eqnarray}
Using expressions (\ref{Urho}), (\ref{trans}), (\ref{propi}) and
the Dirac equation, we can write
\begin{eqnarray}
&& [S(P/2+k)]^{-1}\, {U_{\mu}^{\rho}}(\mbf{k}) = \rho\,
{S_{\rho}}^{(1)}\, {U_{\mu}^{\rho}}(-\mbf{k}), \nonumber\\
&& [S(P/2-k)]^{-1}\, {U_{\mu}^{\rho}}(-\mbf{k}) = \rho\,
{S_{\rho}}^{(2)}\, {U_{\mu}^{\rho}}(\mbf{k}) \nonumber
\end{eqnarray}
Here, we can write the expansion of the BS vertex function
$\Gamma^{J\km}_{\alpha\beta}(P,k)$ as
\begin{eqnarray}
\Gamma^{J{\km}}_{\alpha\beta}(P,k) = \sum \limits_{LS\rho_1\rho_2}
\; (-1)^L\;\rho_1\;\rho_2\; g_{JLS\rho_1\rho2} (k_0,\nrmk)\; {\cal
Y}^{JLS\rho_1\rho_2,\km}_{\alpha\beta}(-\mbf{k}). \label{g2p}
\end{eqnarray} Here the radial part $g_{LS\rho_1\rho2}$ of the vertex
function is connected with the radial part $\phi_{LS\rho_1\rho_2}$
of the BS amplitude~(\ref{reldp}) through
\begin{eqnarray}
\phi_{JLS\rho_1\rho_2}(k_0,\nrmk) =
S_{\rho_1}^{(1)}\,S_{\rho_2}^{(2)}\,
g_{JLS\rho_1\rho_2}(k_0,\nrmk). \label{g2pr}
\end{eqnarray}

It is further convenient to introduce symmetrical notation for the
positive and negative energy states. We define states with total
$\rho$-spin $\varrho$: $|\varrho> = \sum_{\rho_1\rho_2}
(\frac{1}{2}\rho_1\frac{1}{2}\rho_2|\varrho\mu_{\varrho})
|\frac{1}{2}\rho_1> \otimes |\frac{1}{2}\rho_2>$, as
\begin{eqnarray}
&& (+) \equiv |11>, \qquad\qquad (-) \equiv |1-1>,
\\
&& (e) \equiv |10> = \frac{1}{\sqrt{2}} ( |\frac{1}{2}
\frac{1}{2}> |\frac{1}{2} -\frac{1}{2}> + |\frac{1}{2}
-\frac{1}{2}> |\frac{1}{2} \frac{1}{2}> ),
\\
&& (o) \equiv |00> = \frac{1}{\sqrt{2}} ( |\frac{1}{2}
\frac{1}{2}> |\frac{1}{2} -\frac{1}{2}> - |\frac{1}{2}
-\frac{1}{2}> |\frac{1}{2} \frac{1}{2}> ).
\end{eqnarray}
Eq.~(\ref{g2pr}) can be written in the following way
\begin{eqnarray}
\phi_{JLS\varrho}(k_0,\nrmk) =
S_{\varrho}(k_0,\nrmk;s)\,g_{JLS\varrho}(k_0,\nrmk), \label{g2pri}
\end{eqnarray}
where $S_{\varrho}(k_0,\nrmk;s)$ is
\begin{eqnarray}
&& S_{+}=(\frac{\sqrt{s}}{2}+k_0-E_{\bk})^{-1}
(\frac{\sqrt{s}}{2}-k_0-E_{\bk})^{-1},
\label{spart}\\
&& S_{-}=(\frac{\sqrt{s}}{2}+k_0+E_{\bk})^{-1}
(\frac{\sqrt{s}}{2}-k_0+E_{\bk})^{-1},
\nonumber\\
&& S_{e}= (\frac{s}{4}-k_0^2-E_{\bk}^2)
((\frac{s}{4}-k_0^2-E_{\bk}^2)^2-4k_0^2E_{\bk}^2)^{-1},
\nonumber\\
&& S_{o}=
(-2k_0E_{\bk})((\frac{s}{4}-k_0^2-E_{\bk}^2)^2-4k_0^2E_{\bk}^2)^{-1}.
\nonumber\end{eqnarray}

\subsubsection{\em  Matrix Representation}\label{MR}
In the matrix representation, we replace the spinor of a second
particle by the transposed spinor and then calculate the direct
product,
\begin{eqnarray}
{U_{s_1}^{\rho_1}}(\mbf{k})\otimes
 {U_{s_2}^{\rho_2}}(-\mbf{k})\,
\longrightarrow \, {U_{s_1}^{\rho_1}}(\mbf{k})\otimes
 {U_{s_2}^{\rho_2}}^{T}(-\mbf{k}).
\label{replace}
\end{eqnarray}
We use matrices $(4\times1)\otimes(1\times4) = (4\times4)$ instead
of matrices $(1\times4)\otimes(1\times4) = (1\times16)$ in spinor
space. Then matrices $(4 \times 4)$ can be expanded by Dirac
$\gamma$-matrices.

The BS amplitude in the rest frame can be written as:
\begin{eqnarray}
\Phi^{J{\km}}(P,k) = \chi^{J{\km}}(P,k)\,U_C =
\sum\limits_{LS\varrho} \phi_{JLS\varrho}
(k_0,\nrmk)\,\Gamma^{JLS\varrho,\km}(\mbf{k})\,U_C, \label{ampmat}
\end{eqnarray}
where $U_C$ is the  charge-conjugate matrix
\begin{equation}\label{Uc}
  U_C = i\gamma_2\gamma_0,
\end{equation}
and the spin-angular part $\Gamma^{JLS\varrho,\km}(\mbf{k})$ has a
structure  similar to ${\cal Y}^{JLS\varrho,\km}(\bk)$, but with
the replacement of (\ref{replace}).

We show here,  as an example, the function
$\Gamma^{^3S_1^{+},\km}(\mbf{k})$, with notation
$^{2S+1}L_J^{\rho}$ for partial states ($JLS\varrho$):
\begin{eqnarray}
&&\sqrt{4 \pi}\,{\Gamma}^{^3S_1^{+},\km}(\, \mbf{k})U_C =\sum
\limits _{s_1 s_2}\, (\Hh s_1 \Hh s_2 |1{\km})\, u_{s_1}( \mbf{k})
u_{s_2}^T(-\mbf{k}) =
\nonumber\\
&&{L}( \mbf{k})\sum \limits _{s_1 s_2} \,(\Hh s_1 \Hh s_2 |1{\km})
\,\Bigl(\begin{array}{c} \chi_{s_1} \\ 0 \end{array} \Bigr)
\,(\begin{array}{cc} \chi_{s_2}^T &0 \end{array} )
\,{L}^T(-\mbf{k}) =
\nonumber\\
&& {L}( \mbf{k}) \,\Biggl( \begin{array}{cc} \sum \limits_{s_1
s_2} (\Hh s_1 \Hh s_2 |1{\km}) \chi_{s_1} \chi^T_{s_2} &0
\\ 0&0 \end{array}\Biggr) \,{L}^T(-\mbf{k}) =
\nonumber\\
&& {L}( \mbf{k}) \,\frac {1+\gamma_0}{2\sqrt{2}} \,\Biggl(
\begin{array}{cc} 0 & -\mbf{\sigma}\cdot \mbf{\xi}_{\km}\\
\mbf{\sigma}\cdot\mbf{\xi}_{\km} &0 \end{array}\Biggr) \,\Biggl(
\begin{array}{cc} 0 & -i \sigma_2\\ -i\sigma_2 &0
\end{array}\Biggr) \,{L}^T(-\mbf{k}) =
\nonumber\\
&& {L}( \mbf{k}) \,\frac {1+\gamma_0}{2\sqrt{2}}
\,(-\mbf{\gamma}\mbf{\xi}_{\km})\,{L}( \mbf{k}) U_C = \nonumber\\
&& \frac{1}{2E_{\bk}(m+E_{\bk})}\,(m+p_1\cdot\gamma)\,
\frac{1+\gamma_0}{2\sqrt{2}}\,
\xi_{\km}\cdot\gamma\,(m-p_2\cdot\gamma)\,U_C~.
\nonumber\end{eqnarray} Here we make use of the relations for
Pauli spinors and Clebsch--Gordan coefficients:
\begin{equation}\label{clebsh}
  \sqrt{2}\sum \limits_{s_1 s_2}
(\Hh s_1 \Hh s_2 |1{\km}) \chi_{s_1} \chi^T_{s_2}=
(\mbf{\sigma}\cdot\mbf{\xi}_{\km}) \,({i \sigma_2}).
\end{equation} $\mbf{\xi}_{\km}$ is a 3-vector of the polarization of
a particle with spin one and the components in the rest frame,
\begin{eqnarray}
\mbf{\xi}_{+1}=(-1,-i,0)/\sqrt{2}, \quad
\mbf{\xi}_{-1}=(1,-i,0)/\sqrt{2}, \quad \mbf{\xi}_{0}=(0,0,1).
\label{vecpol} \end{eqnarray} The polarization 4-vector $\xi_{\km}
= (0,\mbf{\xi}_{\km})$ is determined in the rest frame.

In the general case, we can separate $\rho$-dependence and rewrite
the spin-angular functions ${\Gamma}^{JLS\varrho,\km}(\mbox
{\boldmath$k$}) \equiv {\Gamma}^{JLS\rho_1\rho_2,\km}(\mbf{k})$
as:
\begin{eqnarray}
{\Gamma}^{JLS++,\km}(\mbox {\boldmath$k$}) &=
&\frac{p_1\cdot\gamma + m}{\sqrt{2E_{\bk}(m+E_{\bk})}}\,
\frac{1+\gamma_0}{2}\, {\tilde \Gamma}^{JLS,\km}(\mbox
{\boldmath$k$})\, \frac{p_2\cdot\gamma -
m}{\sqrt{2E_{\bk}(m+E_{\bk})}}, \label{gf}\\
{\Gamma}^{JLS--,\km}(\mbox {\boldmath$k$}) &=
&\frac{p_2\cdot\gamma - m}{\sqrt{2E_{\bk}(m+E_{\bk})}}\,
\frac{-1+\gamma_0}{2}\, {\tilde \Gamma}^{JLS,\km}(\mbox
{\boldmath$p$})\,
\frac{p_1\cdot\gamma + m}{\sqrt{2E_{\bk}(m+E_{\bk})}}, \nonumber\\
{\Gamma}^{JLS+-,\km}(\mbox {\boldmath$k$}) &=
&\frac{p_1\cdot\gamma + m}{\sqrt{2E_{\bk}(m+E_{\bk})}}\,
\frac{1+\gamma_0}{2}\, {\tilde \Gamma}^{JLS,\km}(\mbox
{\boldmath$k$})\, \frac{p_1\cdot\gamma +
m}{\sqrt{2E_{\bk}(m+E_{\bk})}}, \nonumber\\
{\Gamma}^{JLS-+,\km}(\mbox {\boldmath$k$}) &=
&\frac{p_2\cdot\gamma - m}{\sqrt{2E_{\bk}(m+E_{\bk})}}\,
\frac{1-\gamma_0}{2}\, {\tilde \Gamma}^{JLS,\km}(\mbox
{\boldmath$k$})\, \frac{p_2\cdot\gamma -
m}{\sqrt{2E_{\bk}(m+E_{\bk})}}, \nonumber
\end{eqnarray}
The conjugate
functions  can be written in the following form:
\begin{eqnarray}
{\bar {\Gamma}^{JLS\varrho,\km}}(\mbox {\boldmath$k$}) =\gamma_0
\;\left[{\Gamma}^{JLS\varrho,\km} (\mbox
{\boldmath$p$})\right]^{\dagger}\;\gamma_0, \label{conj}
\end{eqnarray}
Then the orthogonalization condition can be presented as
\begin{eqnarray}
\int d \Omega_{\bk}\, \mbox{Tr} \{
{{{\Gamma}}^{JLS\varrho,\km}}^{\dagger}(\mbox {\boldmath$k$})
{\Gamma}^{JL^\prime S^\prime \varrho^\prime,{\km}^{\prime}}
(\mbf{k}) \} = \delta_{{\km} {{\km}^{\prime}}}
\delta_{LL^{\prime}}\delta_{SS^\prime}\delta_{\varrho\varrho^\prime}.
\label{ortm}
\end{eqnarray}
Using the Pauli principle for identical particles
\begin{eqnarray}
\Phi^{J{\km}}(P,k)=-P_{12}\Phi^{J{\km}}(P,k),
\end{eqnarray}
where $P_{12}$ is the permutation operator of two particles, we
can write the BS amplitude $\chi^{_{J\km}}(P,k)$ in the rest frame
as:
\begin{eqnarray}
\chi^{_{J\km}}(P,k)=(-1)^{I+1}U_C\left[\chi^{_{J{\km}}}(P,-k)\right]^T
U_C~, \end{eqnarray} where $I$ is the isospin of the system. These
relations give us the symmetry properties of the radial functions
$\phi_{JLS\varrho}(k_0,\nrmk)$ for the transformations, when we
replace $k_0 \rightarrow -k_0$. Radial functions for different
$LS\varrho$ will be odd or even under this transformation. In
order to have a radial function with the determined symmetry under
the $k_0 \rightarrow -k_0$ transformation.


\paragraph{\boldmath $^1S_0$--Channel.\unboldmath}\label{BS:1s0}
The BS amplitude of the two-nucleon system in the $^1S_0$-channel
has four states: $^1S_0^{+}$, $^1S_0^{-}$, $^3P_0^{e}$,
$^3P_0^{o}$ (or $^3P_0^{+-}$, $^3P_0^{-+}$). For $NN$-scattering
we take $\sqrt{s}=\sqrt{P^2}$. The corresponding spin-angular
parts are shown in table~\ref{tab:1s0} where $p_1=(E_{\bk},\bk)$
and $p_2=(E_{\bk},-\bk)$.

\begin{table}[ht]
{ \caption{\label{tab:1s0}{{\sf Spin-angular parts $\tilde
\Gamma^{JLS\varrho,\km}$ for $~^1S_0$-states}}}
\[ \begin{array}{cc} \hline\hline
0LS&{\sqrt{8\pi} \;\;\tilde \Gamma}^{0LS,0}\\[1ex]
\hline ^1S_0&-\gamma_5\\[1ex]
^3P_0&\nrmk^{-1} (p_1\cdot\gamma-p_2\cdot\gamma) \gamma_5\\
\hline\hline \end{array} \]}
\end{table}

Using the Lorentz invariant expressions for
 $\nrmk$, $k_0$ and $E_{\bk}$:
\begin{eqnarray}
k_0=\frac{(P\cdot k)}{M}, \quad E_{\bk}=\sqrt{\frac {(P\cdot
k)^2}{M^2}-k^2+m^2}, \quad \nrmk= \sqrt{\frac {(P\cdot
k)^2}{M^2}-k^2}, \label{p0pep}\end{eqnarray} we can rewrite the
 expression for the BS amplitude in a covariant form. It  is, of
course, more convenient to use the direct covariant form of the BS
amplitude in the $^1S_0$ channel, written in the $4\times 4$
matrix form. For this purpose, we introduce four Lorentz covariant
functions, $h_i(P\cdot k,k^2)$:
\begin{eqnarray}
\sqrt{4\pi}\; \chi^{00}(P,k) &= &h_1\gamma_5 + h_2\,\frac {1}{m}\,
(p_1\cdot\gamma\gamma_5 +\gamma_5 p_2\cdot\gamma)+
\label{covarj0}\\&& h_3\,(\frac{p_1\cdot\gamma-m}{m}\, \gamma_5
-\gamma_5\, \frac {p_2\cdot\gamma+m}{m}) +h_4\,
\frac{p_1\cdot\gamma-m}{m}\, \gamma_5\,
\frac{p_2\cdot\gamma+m}{m}, \nonumber
\end{eqnarray}
where 4-momentum of the two particles, $p_1$ and  $p_2$, are
determined by~(\ref{moments}), and functions $h_i(P\cdot k,k^2)$
can be expressed via the radial functions $\phi_{JLS\varrho}
(k_0,|{\bf k}|)$ defined in Eq.(\ref{ampmat}):
\begin{eqnarray}
h_1 &= & \sqrt{2}D_1 (\phi_{_{^1S_0^+}}+
\phi_{_{^1S_0^-}})+\sqrt{2}/4
(\phi_{_{^1S_0^+}}-\phi_{_{^1S_0^-}}) -
\mu k_0 \nrmk^{-1} \phi_{_{^3P_0^e}} - 8m\nrmk^{-1}D_0  \phi_{_{^3P_0^o}}, \nonumber\\
h_2 &=&\Hf m \nrmk^{-1}  \phi_{_{^3P_0^e}}, \label{h2pj0}\\ h_3
&=& 8 a_0m^2 (\phi_{_{^1S_0^+}}+\phi_{_{^1S_0^-}}) - \Hh\mu k_0
\nrmk^{-1}\phi_{_{^3P_0^e}}
- 8a_0m\nrmk^{-1}\gve  (m-E_{\bk}) \phi_{_{^3P_0^o}}, \nonumber\\
h_4 &=& -4a_0\sqrt{2}m^2(\phi_{_{^1S_0^+}}+\phi_{_{^1S_0^-}}) +
8a_0m^3\nrmk^{-1}\phi_{_{^3P_0^o}}. \nonumber
\end{eqnarray}
Here
\begin{eqnarray}\label{coeff}
a_0&=&1/(16ME_{\bk}),\quad \gve = 2m+E_{\bk},
\quad \mu=m/M \nonumber\\ D_0&=&a_0(4k_0^2+16m^2-4E_{\bk}^2-M^2),\\
D_1&=&a_0(-M^2/4+k_0^2-3E_{\bk}^2+4m^2).\nonumber
\end{eqnarray}
We note that the functions $h_2$ and  $\phi_{^3P_0^e}$ are odd
under $k_0\rightarrow -k_0$, while other functions are even.

\paragraph{\boldmath $^3S_1-^3D_1$\unboldmath --Channel (Deuteron).}\label{BS:3s1}
The BS amplitude for the deuteron\footnote{Here, in the case of
bound state $M$ denotes the mass of the deuteron, while in the
case  of $np$-pair $M = \sqrt{s}$.} has eight states: $^3S_1^{+}$,
$^3S_1^{-}$,$^3D_1^{+}$,  $^3D_1^{-}$, $^3P_1^{e}$, $^3P_1^{o}$,
$^1P_1^{e}$, $^1P_1^{o}$, (or $^3P_1^{+-}$, $^3P_1^{-+}$,
$^1P_1^{+-}$, $^1P_1^{-+}$), which are numbered as  $1,\dots,8$.
The corresponding spin-angular parts ${\tilde
\Gamma}^{JLS,\km}(\mbox {\boldmath$k$})$ are tabulated in
table~\ref{tab:3s1}.
    \begin{table}[ht]
    { \caption{\label{tab:3s1}{{\sf Spin-angular parts $\tilde
    \Gamma^{1LS,\km}$ for the deuteron}}}
    \[ \begin{array}{cc} \hline\hline 1LS&
    {\sqrt{8\pi}\;\;\tilde \Gamma}^{1LS,\km}\\[1ex]
    \hline
    ^3S_1&{\xi_{\km}\cdot\gamma}\\
    ^3D_1& -\frac{1}{\sqrt{2}} \left[
    {\xi_{\km}\cdot\gamma}+\frac{3}{2}
    ({p_1\cdot\gamma}-{p_2\cdot\gamma})(k\xi_{\km})|\bk|^{-2}\right]\\
    ^3P_1& \sqrt{\frac{3}{2}} \left[ \frac{1}{2}
    {\xi_{\km}\cdot\gamma}({p_1\cdot\gamma}-{p_2\cdot\gamma})
    -(k\xi_{\km}) \right]|\bk|^{-1}\\
    ^1P_1&\sqrt{3} (k\xi_{\km})|\bk|^{-1}\\
    \hline\hline
    \end{array}
    \]
    }
    \end{table}

The BS amplitude has the following covariant matrix form,
\begin{eqnarray}
\sqrt{4\pi}\; \chi^{1{\km}}(P,k) &= &h_1\, \xi_{\km}\cdot\gamma
+h_2\, \frac{k \xi_{\km}}{m} + \label{covar}\\ &&h_3\, \left
(\frac {p_1\cdot\gamma-m}{m}\, \xi_{\km}\cdot\gamma +
\xi_{\km}\cdot\gamma\, \frac{p_2\cdot\gamma+m}{m}\right)+
\nonumber\\
&&h_4\, \left(\frac {p_1\cdot\gamma + p_2\cdot\gamma}{m}\right)
\,\frac {k \xi_{\km}}{m}+ \nonumber\\ && h_5\, \left(\frac
{p_1\cdot\gamma-m}{m}\, \xi_{\km}\cdot\gamma -
\xi_{\km}\cdot\gamma\, \frac{p_2\cdot\gamma+m}{m}\right)+
\nonumber\\
&&h_6\, \left(\frac {p_1\cdot\gamma - p_2\cdot\gamma-2m}{m}\right)
\,\frac {k \xi_{\km}}{m}+ \nonumber \\ &&\frac
{p_1\cdot\gamma-m}{m}\,\left(h_7\, {\xi_{\km}\cdot\gamma} +h_8\,
\frac{k \xi_{\km}}{m} \right )\,\frac{p_2\cdot\gamma+m}{m}.
\nonumber
\end{eqnarray}

Here we have introduced eight covariant functions  $h_i(P\cdot
k,k^2)$, which are connected with the radial functions
$\phi_{JLS\varrho}(k_0,\nrmk)$ in the rest frame via relations:
\begin{eqnarray}
h_1 &= &D^+_1\, (\phi_{_{^3D_1^+}}-\sqrt{2}\phi_{_{^3S_1^+}})
        + D^-_1\, (\phi_{_{^3D_1^-}}-\sqrt{2}\phi_{_{^3S_1^-}})
        + \Hh\sqrt{6}\mu k_0\nrmk^{-1} \,\phi_{_{^3P_1^e}}
        + \sqrt{6}m D_0 \nrmk^{-1} \,\phi_{_{^3P_1^o}},               \nonumber \\
h_2 &= &\sqrt{2}(D^-_2\,\phi_{_{^3S_1^+}} +
D_2^+\,\phi_{_{^3S_1^-}})
        - D_3^+\,\phi_{_{^3D_1^+}}
        - D_3^-\,\phi_{_{^3D_1^-}} \label{h2pj1}\\
      &&-\Hh\sqrt{6} \mu p_0\nrmk^{-1}\,\phi_{_{^3P_1^e}}
        -\sqrt{6} D_4 m\nrmk^{-1} \,\phi_{_{^3P_1^o}}
        +\sqrt{3} m^2 \nrmk^{-1} E_{\bk}^{-1}\,\phi_{_{^1P_1^e}}, \nonumber \\
         h_3 &= &-\Hf\sqrt{6}m \nrmk^{-1}\, \phi_{_{^3P_1^e}}, \nonumber \\
         h_4 &= &8a_1\sqrt{2}mk_0\,(\phi_{_{^3S_1^+}}-\phi_{_{^3S_1^-}})
        +8a_2\gve \,mk_0(\phi_{_{^3D_1^+}}-\phi_{_{^3D_1^-}}) \nonumber\\
        &&-16a_0\sqrt{3}m^2 \nrmk^{-1}\,(k_0\phi_{_{^1P_1^e}}-E_{\bk}\phi_{_{^1P_1^o}}), \nonumber \\
        h_5 &= &16a_0m^2\,[\phi_{_{^3D_1^-}}+\phi_{_{^3D_1^+}}-\sqrt{2}\,(\phi_{_{^3S_1^+}}+\phi_{_{^3S_1^-}})]
        +8a_0\sqrt{6}m \nrmk^{-1} \,[k_0E_{\bk}\,\phi_{_{^3P_1^e}}+\nonumber\\
        &&(2m^2-E_{\bk}^2)\,\phi_{_{^3P_1^o}}],   \nonumber \\
h_6 &= &4a_1\sqrt{2}m[D_6^-\,\phi_{_{^3S_1^+}} + D_6^+\,
\phi_{_{^3S_1^-}}]
        - 4 m^2 \nrmk^{-2} [D_5^+ \,\phi_{_{^3D_1^+}} + D_5^- \,\phi_{_{^3D_1^-}}]  -\nonumber\\
       &&16a_0\sqrt{6}m^2 \nrmk^{-1}\,(m\phi_{_{^3P_1^o}}-M\phi_{_{^1P_1^e}}), \nonumber\\
       h_7 &=& 4a_0 m^2\,[\sqrt{2}(\phi_{_{^3S_1^+}}+\phi_{_{^3S_1^-}})-(\phi_{_{^3D_1^+}}+\phi_{_{^3D_1^-}})]-
        4a_0\sqrt{6}m^3 \nrmk^{-1}\, \phi_{_{^3P_1^o}}, \nonumber \\
h_8 &=& 4a_0m^3 \nrmk^{-2}\, [\sqrt{2}(m-E_{\bk})(\phi_{_{^3S_1^+}}+\phi_{_{^3S_1^-}})-(2E_{\bk}+m)(\phi_{_{^3D_1^+}}+\phi_{_{^3D_1^-}})] +\nonumber\\
        &&4a_0\sqrt{6}m^3 \nrmk^{-1} \,\phi_{_{^3P_1^o}}. \nonumber
\end{eqnarray}
The coefficients in these expressions are:
\begin{eqnarray}
a_1&=&a_0 m/(m+E_{\bk}),\quad a_2=a_1/(m-E_{\bk}),\nonumber\\
  D^\pm_1&=&a_0(4k_0^2+16m^2-M^2-4E_{\bk}^2\pm 4ME_{\bk}),\nonumber\\
  D^\pm_2&=&a_1(16m^2+16mE_{\bk}+4E_{\bk}^2+M^2-4k_0^2\pm 4M\gve ),\nonumber\\
D^\pm_3&=&a_2[-12mE_{\bk}^2+2M^2E_{\bk}-8k_0^2E_{\bk}+16m^3+
mM^2-4mk_0^2+8E_{\bk}^3\nonumber\\
            &&  \pm(16m^2M+ 4mME_{\bk}- 8E_{\bk}^2M)],\nonumber\\ D_4&=&a_0(16m^2-4E_{\bk}^2-k_0^2+m^2),\nonumber\\
D^\pm_5&=&a_0(-2E_{\bk}^2+4m^2+4mE_{\bk}\pm \gve M),\nonumber\\
D_6^\pm&=&a_0(2\gve \pm M),\nonumber
\end{eqnarray}
and $a_0$, $\gve $, $\mu$, $D_0$ are given in Eq.~(\ref{coeff}).
Here functions $h_3$, $h_4$ and $\phi_{_{^3P_1^e}}$,
$\phi_{_{^1P_1^o}}$ are odd and others are even under
$k_0\rightarrow -k_0$.

\subsection{\em  Construction of the Light-Front Wave Function from the BS~Amplitude}\label{sect:lfbs}
Here, we consider the relation between the BS approach~\cite{BS51}
and the light front dynamics (LFD) approach for a two-nucleon
system~\cite{car98}, which is one of possible 3D relativistic
dynamics proposed by Dirac~\cite{dir49}.
In this approach the state vector describing the system is
expanded in Fock components defined on a hypersphere in the
four-dimensional space-time. The LFD approach is intuitively
appealing since it is formally close to the nonrelativistic
description in terms of the Hamiltonian, and state vectors can be
directly interpreted as wave functions.

The equivalence between these two approaches has been a subject of
recent discussions presented in ref.~\cite{car98,equiv,car00} and
references therein. The application of the two approaches for a
two-nucleon system (deuteron) serves to clarify the structure of
the different components of the amplitude. It is also useful in
the context of the three-particle dynamics, where the proper
covariant and/or light-front construction of the nucleon amplitude
in terms of three valence quarks (including spin dependence and
configuration mixing) is presently discussed~\cite{bkw98,karm98}.
Although the relation between the light-front and the
Bethe--Salpeter amplitudes for a two-nucleon system has been
spelled out to some extend in a report~\cite{car98}, we provide
here some useful details.

To proceed, we present different ways of construction of a
complete (covariant) Bethe--Salpeter amplitude, see,
e.g.~\cite{bbbd:elec,bb:cov2}. Beside the spin structure of the
wave function (amplitude), we also present a comparison of the
{\em radial} part of the amplitude on the basis of the Nakanishi
integral representation~\cite{nakanishi69}. The  representation is
well known and elaborated for the scalar case, see,
 e.g.~\cite{Kusaka95,Kusaka97}.
 However, this formulation was not followed up later in the actual applications.
 We employ this integral representation in order to establish a connection between the two
different approaches.

In this context, the direct product representation used in the
rest frame of the two-nucleon system using the $\rho$-spin
notation is close to the nonrelativistic coupling scheme and
provides states of definite angular momentum. To construct the
covariant basis, this form is transformed into a matrix
representation which will then be expressed in terms of the Dirac
matrices. A generalization to arbitrary deuteron momenta finally
leads to the covariant representation of the Bethe--Salpeter
amplitude. This was explained in the previous  section along with
an explicit construction of the deuteron ($J=1$) and the $J=0$
two-nucleon state.

We now compare the BS amplitude  to the covariant LFD form.  The
state vector defining the light-front plane is denoted by
$\omega$, where $\omega=(1,0,0,-1)$ leads to the standard
light-front formulation defined in the frame $t+z=0$. We  start
from the integral that restricts the variation of the arguments of
the Bethe-Salpeter function to those of the light-front plane,

\begin{equation} \label{bs1}I=\int
d^4x_1\ d^4x_2\ \delta (\omega \cdot x_1)\ \delta (\omega\cdot
x_2)\ \Phi^P (x_1,x_2)\exp (il_1\cdot x_1+il_2\cdot x_2)\ ,
\end{equation} where $l_1,l_2$ are the on-shell momenta:
$l_1^2=l_2^2=m^2$,  and
\begin{equation}\Phi^P (x_1,x_2)\equiv
{\Phi}^{A=2(\alpha\equiv JM, P)}(x_1,
x_2)=<0|T(\Psi(x_1)\Psi(x_2))|P>\label{bs2}
 \end{equation}
 is the
Bethe-Salpeter amplitude~(\ref{chiX}).

 To go further we represent first the $\delta $ -functions in
(\ref{bs1}) by the integral form,
\begin{eqnarray}
\delta(\omega\cdot x_1)= \frac{1}{2\pi}\int \exp(-i\omega\cdot
x_1\alpha_1)d\alpha_1,\quad \delta(\omega\cdot x_2)=
\frac{1}{2\pi}\int \exp(-i\omega\cdot x_2\alpha_2)d\alpha_2\ ,
\end{eqnarray}
We introduce the Fourier transform of the Bethe-Salpeter function
$\Phi (P,k)$ as the Fourier transform of $\tilde\Phi^P$ defined
as,
\begin{equation}
\Phi^P (x_1,x_2)=(2\pi )^{-3/2}\exp\left[-iP\cdot
(x_1+x_2)/2\right]\tilde\Phi^P(x)\ ,\quad x=\frac{x_1-x_2}{2}\ .
\end{equation}
The BS function is
\begin{equation}
\label{bs3}\Phi (P,k)=\int \tilde \Phi ^P (x)\exp (ik\cdot x)d^4x\
,
\end{equation}
where $k=(p_1-p_2)/2$, $P=p_1+p_2$, $p_1$ and $p_2$ are usual
off-mass shell four-vectors. Making the change of variables
$\alpha_1+\alpha_2=\tau$, $(\alpha_2-\alpha_1)/2=\beta$, we
obtain:
\begin{eqnarray}
p_1=l_1-\omega\tau/2+\omega\beta\nonumber\\
p_2=l_2-\omega\tau/2-\omega\beta,\nonumber
\end{eqnarray}
and the integral~(\ref{bs1}) takes following form:
\begin{equation}
\label{bs4} I=\sqrt{2\pi}\int_{-\infty }^{+\infty }\delta
^{(4)}(l_1+l_2-P-\omega \tau )d\tau \int_{-\infty }^{+\infty }\Phi
\left(l_1+l_2-\omega\tau,\frac{l_1-l_2}{2}+\omega\beta
\right)d\beta\ .
\end{equation}

On the other hand, the integral (\ref{bs1}) can be expressed in
terms  of the two-body light-front wave function. We assume that
the  light-front plane is the limit of a space-like plane,
therefore  the  operators $\Psi (x_1)$ and $\Psi (x_2)$ commute,
and, hence,  the  symbol of the $\rm T$-product in (\ref{bs2}) can
be omitted. In the considered representation, the Heisenberg
operators $\Psi (x)$ in (\ref{bs2}) are identical on the light
front $\omega\cdot x=0$ to the Schr\"odinger operators (just as in
the ordinary formulation of field  theory the Heisenberg and
Schr\"odinger operators are identical for $t=0$). The
Schr\"odinger operator $\Psi (x)$ (for the spinless case for
simplicity), which for $\omega\cdot x=0$ is the free field
operator, is given in~\cite{car98}:
\begin{eqnarray} \label{bs5}
\Psi(x)& \equiv &\frac{1}{(2\pi)^{3/2}} \int
\tilde{\Psi}(k)\exp(ik\cdot
x)d^4k\nonumber\\
&=& \frac 1{(2\pi )^{3/2}}\int \left[a(\bk ) \exp (-ik\cdot
x)+a^{\dagger}(\bk )\exp (ik\cdot  x)\right]
\frac{d^3k}{\sqrt{2\varepsilon_k}}\ .
\end{eqnarray}
We represent the state vector $|P\rangle $ in (\ref{bs2}) as the
Fock components of the state vector. Since the vacuum  state on
the light front is always ``bare'', the creation operator, applied
to the vacuum state $\langle 0|$ gives zero, and in the operators
$\Psi (x)$  the part containing the annihilation operators only
survives. This cuts out the two-body Fock component in  the  state
vector.  Thus  the integral $I$ can be obtained in the following
form~\cite{car98}:
\begin{equation} \label{bs6}
I=\frac{(2\pi )^{3/2}(\omega\cdot P)}{2(\omega\cdot
l_1)(\omega\cdot l_2) }\int_{-\infty }^{+\infty }\psi
(l_1,l_2,P,\omega \tau )\delta ^{(4)}(l_1+l_2-P-\omega \tau )\
d\tau\ ,
\end{equation}
where $\psi (l_1,l_2,P,\omega \tau )$ is two body wave function
defined on the light front specified by $\omega$. We make a
comparison (\ref{bs4}) and (\ref{bs6}) and find the formal
connection between the LFD wave function and the BS amplitude:
\begin{equation} \label{bs7}
\psi (l_1,l_2,P,\omega \tau )=\frac{(\omega\cdot l_1)(\omega\cdot
l_2)}{\pi (\omega \cdot P)}\int_{-\infty }^{+\infty }\Phi \left(P,
\frac{l_1-l_2}{2}+\beta \omega\right)d\beta\ ,
\end{equation}

 Since $l_1$ and
$l_2$ are on the mass shell, it is possible to use the Dirac
equation after making the replacement of the arguments indicated
in Eq.~(\ref{bs7}). This will be done explicitly for the $J=0$ and
the deuteron channel in the following.

\subsubsection{\em $^1S_0$ -- Channel}\label{BSLF:1s0}
Using the Dirac equations $\bar u(l_1)(\gamma\cdot l_1 -m) =0$,
and $(\gamma\cdot l_2 +m) C u(l_2)^\top=0$, one obtains  the
 light-front wave function from the Bethe--Salpeter amplitude using
Eq.~(\ref{ampmat}), (\ref{bs7})
\begin{equation}
\chi^{_{00}}\rightarrow H^{(0)}_1 \gamma_5 + 2  H^{(1)}_2 \frac
{\beta \gamma\cdot {\omega}}{m\omega\cdot P}\gamma_5~.
\label{1s0lf}
\end{equation}
The functions $H_1(s,x)$ and $H_2(s,x)$, depending now on
$x=\omega\cdot l_1/\omega\cdot P$ and $s=(l_1+l_2)^2=4(q^2+m^2)$,
are obtained from the functions $h_i(P\cdot k,k^2)$ through the
remaining integrals over $\beta$ implied in Eq.~(\ref{bs7}),
\begin{eqnarray}
H^{(0)}_i(s,x)&=& N \int {h_i((1-2x)(s-M^2)+\beta \omega\cdot P,
-s/4+m^2+(2x-1)\beta)\, \omega\cdot P  d\beta} \nonumber\\
&\equiv&N \int {\tilde h_i (s,x,\beta^{\prime})\, d\beta^{\prime}
},\nonumber\\ H^{(k)}_i(s,x)&\equiv&N \int { \tilde
h_i(s,x,\beta^{\prime}) \, (\beta^{\prime})^k d\beta^{\prime} }
\label{eqn:H} \end{eqnarray} where the variable
$\beta^{\prime}=\beta \omega\cdot P$ has been introduced, and
$N=x(1-x)$, $1-x=\omega\cdot l_2/\omega\cdot P$. We would like to
stress that in this case the functions $h_3$ and $h_4$ do not
contribute. Instead of the four structures appearing in the
Bethe-Salpeter wave function, the light front function consists of
only two. Note, that the second term in the parenthesis is defined
by the pure relativistic component of the Bethe-Salpeter
amplitude.

\subsubsection{\em $^3S_1$-$^3D_1$ -- Channel (Deuteron)}\label{BSLF:3s1}
In the deuteron case, starting from the formula Eq.~(\ref{bs7}),
replacing the momenta $p_i$, and applying the Dirac equation we,
arrive at
\begin{eqnarray}
\chi^{_{1M}} &\rightarrow & H^{(0)}_1 \gamma\cdot {\epsilon_M}
+H^{(0)}_2 \frac {k\cdot \epsilon_M}{m} +[H^{(1)}_2+2H^{(1)}_5]
\frac{\omega\cdot \epsilon_M}{m\omega\cdot P}  \nonumber \\
&&+2H^{(1)}_6  \frac{k\cdot\epsilon_M \gamma\cdot {\omega}}{m^2
\omega\cdot P} +2 H^{(1)}_3 \frac { \gamma\cdot {\epsilon_M}
\gamma\cdot  {\omega} -\gamma\cdot  {\omega} \gamma\cdot
{\epsilon_M} }{\omega\cdot P} \nonumber\\&&
+[2H^{(2)}_6+2H^{(2)}_7] \frac { \omega\cdot \epsilon_M
\gamma\cdot {\omega} } {m^2(\omega\cdot P)^2}~, \label{bsd5}
\end{eqnarray}
where $H_i^{(k)}$ are defined in Eq.~(\ref{eqn:H}). In this case
the functions $h_4$ and $h_8$ do not contribute. The expression
$(\gamma\cdot {\epsilon_M}\gamma\cdot {\omega}-\gamma\cdot
{\omega}\gamma\cdot {\epsilon_M})$ at the term $H_5$ given in
Eq.~(\ref{bsd5}) can be transformed to a different one to compare
it directly to the light-front form.  Using in addition the
on-shellness of the momenta $l_1$ and $l_2$, the resulting form is
\begin{eqnarray} \label{bsd7}
\bar u_1(\gamma\cdot {\epsilon_M}\gamma\cdot {\omega} -\gamma\cdot
{\omega}\gamma\cdot {\epsilon_M})C\bar u_2^\top &=&
\frac{4}{s}\bar u_1[-i\gamma_5e_{\mu\nu\rho\gamma}
\epsilon_{\mu}l_{1\nu}l_{2\rho}\omega_{\gamma}\nonumber\\&&
+k\cdot\epsilon_M\;\omega\cdot P-m\;\gamma\cdot {\epsilon_M}
\;\omega\cdot +P \nonumber\\
&&-\frac{1}{2}(s-M^2)(x-\frac{1}{2})\omega\cdot\epsilon_M
\nonumber\\&& +\frac{1}{2}m(s-M^2)\,\frac{\gamma\cdot {\omega}
\;\omega\cdot\epsilon_M} {\omega\cdot P}] C\bar u_2^\top
\end{eqnarray}
The final form of the light-front wave function is then
\begin{eqnarray}
\chi_{_{1M}} &\rightarrow & H_1^{\prime} \gamma\cdot {\epsilon}_M
+H_2^{\prime} \frac {k\cdot \epsilon_M}{m} +H_3^{\prime}
\frac{\omega\cdot \epsilon_M}{m\,\omega\cdot P} +H_4^{\prime}
\frac{k\cdot \epsilon_M\; \gamma\cdot  {\omega}} {m^2 \omega\cdot
P}  \nonumber \\&& +H_5^{\prime} i \gamma_5 e_{\mu \nu \rho
\sigma}\epsilon_{\mu}{l_1}_{\nu} {l_2}_{\rho}{\omega}_{\sigma}
+H_6^{\prime} \frac { \omega\cdot \epsilon_M\; \gamma\cdot
{\omega} }{m^2(\omega\cdot P)^2}~,
\end{eqnarray}
with the functions
\begin{eqnarray}
H_1^{\prime} &=& H^{(0)}_1-\frac{4}{s}2H^{(1)}_3, \nonumber \\
H_2^{\prime} &=& H^{(0)}_2+\frac{4}{s}2H^{(1)}_3, \nonumber \\
H_3^{\prime} &=& [H^{(1)}_2+2H^{(1)}_5] -\frac{ (s-M^2)}{s}(2x-1)
2H^{(1)}_3, \nonumber \\ H_4^{\prime} &=& 2H^{(1)}_6, \nonumber \\
H_5^{\prime} &=& \frac{4}{ms}2H^{(1)}_3, \nonumber \\ H_6^{\prime}
&=& [2H^{(2)}_6+2H^{(2)}_7] +2 \frac{s-M^2}{s}m^2 2H^{(1)}_3~.
\label{eqn:conn}
\end{eqnarray}
Provided the invariant functions $h_i$ are given from a solution
of the BS equation, the above relations allow us to directly
calculate the corresponding light-front components of the wave
functions.

Thus, the projection of the Bethe-Salpeter amplitude to the light
front amplitude reduces the number of independent functions from
eight to six
 for the $^3S_1-^3D_1$ channel, and from four to two for the
$^1S_0$ channel. The reduction takes place because the nucleon
momenta $p_1$ and $p_2$ are on mass shell in the LF formalism. The
result is based on the application of the Dirac equation and the
use of the covariant form. Any other representations (e.g.
spin-angular momentum basis) also lead to a reduction of the
number of amplitudes for the two-nucleon wave function that is
however less transparent. For an early consideration compare, e.g.
ref.~\cite{GT60}.

\subsubsection{\em Integral Representation Method} A deeper insight into
the relation between the Bethe-Salpeter amplitude and the light
front wave function will be provided within the integral
representation proposed by
Nakanishi~\cite{nakanishi69,nakanishi71,nakanishi88}. This method
has recently been fruitfully applied to solve the Bethe-Salpeter
equation both in the ladder approximation and beyond within the
scalar theories~\cite{Kusaka95,Kusaka97}. In this framework the
following ansatz for radial Bethe-Salpeter
amplitudes~(\ref{g2pri}) of orbital momentum $L$ has been
proposed,
\begin{equation}
\phi_{JLS\varrho}(k_0,|\bk|)\equiv \phi_{JLS\varrho}(P\cdot k,k^2)
=\int_0^\infty d\alpha\;\int_{-1}^{+1} dz \;\frac{g_{JLS\varrho}
(\alpha,z)} {(\alpha+\kappa^2-k^2-z\,P\cdot k-i\epsilon)^n},
\label{nak1}
\end{equation}
where $g_{JLS\varrho} (\alpha,z)$ are the densities or weight
functions, $\kappa=m^2-M^2/4$, and the integer $n \geq 2$. The
weight functions $g_{JLS\varrho}(\alpha,z)$ that are continuous in
$\alpha$ vanish at the boundary points $z=\pm 1$.  The form of
Eq.~(\ref{nak1}) opens the possibility to find the Bethe-Salpeter
amplitude in the whole Minkowski space while commonly used
solutions are restricted to the Euclidean space only. In fact, the
densities could be considered as the main object of the
Bethe-Salpeter theory, because knowing them allows one to
calculate all relevant amplitudes.

For the realistic deuteron we need to expand the Nakanishi form to
the spinor case, which has not been done so far. The key point in
this procedure is a proper choice of the
 spin-angular momentum functions as well as the
integration over angles in the Bethe-Salpeter equation.  The
choice of the covariant form of the amplitude allows us to
establish a system of equations for the densities
$g_{ij}(\alpha,z)$, suggesting the following general form for the
radial functions $h_i(P\cdot k,k^2)$ (even in $P\cdot k$)
\begin{eqnarray}
h_i(P\cdot k,k^2)&=&\int_0^\infty d\alpha\;\int_{-1}^{+1} dz\;
\left\{\frac{g_{i1}(\alpha,z)} {(\alpha+\kappa^2-k^2-z\,P\cdot
k)^n}\right.\nonumber\\&& \qquad\qquad\qquad
+\frac{g_{i2}(\alpha,z)\;k^2} {(\alpha+\kappa^2-k^2-z\,P\cdot
k)^{n+1}} \\ \nonumber&& \qquad\qquad\qquad
\left.+\frac{g_{i3}(\alpha,z)\; (P\cdot
k)^2}{(\alpha+\kappa^2-k^2-z\,P\cdot k)^{n+2}}\right\}.
\label{nak2}
\end{eqnarray}
For the functions that are odd in $P\cdot k$ the whole integrand
is multiplied by factor $P\cdot k$.  Although now the number of
densities is larger, the total number of {\em independent}
functions is still eight. The form given in Eq.~(\ref{nak2}) is
valid only for the deuteron case. The continuum amplitudes of the
$^1S_0$ state, e.g., require a different form.

It is a major advantage that we can perform the integration over
$\beta^{\prime}$ in the expressions of Eq.~(\ref{eqn:H}) by using
the integral representation. Substituting the arguments of the
functions $h_i$ into the integral representation Eq.~(\ref{nak2})
leads to a denominator of the form
\begin{equation}
{\cal D}^k(\alpha,z;x,s,\beta')=
(\alpha+\frac{s}{4}(1+(2x+1)z)-\beta^{\prime}
(2x-1+z)-i\epsilon)^k \nonumber
\end{equation}
Using the identity for an analytic function $F(z)$
\begin{equation}
\int_{-1}^{+1} dz\, \int_0^\infty d\beta^{\prime}\, \frac{F(z)}
{{\cal D}^k(\alpha,z;x,s,\beta')}
=\frac{i\pi}{(k-1)}\frac{F(1-2x)} {(\alpha+sx(1-x))^{k-1}}
\end{equation}
we can express the radial amplitudes in terms of Nakanishi
densities. Thus, $H_5(s,x)$ reads
\begin{equation}
H_5(s,x)=\frac{x(1-x)}{s} \int d\alpha \left\{ \frac
{g_{51}(\alpha,1-2x)}{(\alpha+sx(1-x))}+
\frac{g_{52}(\alpha,1-2x)sx(1-x) }{(\alpha+sx(1-x))^2}\right\}
\end{equation}
Note, that the dependence of the amplitude on the light front
argument $x$ is fully determined by the dependence of the density
on the variable $z=1-2x$, which has also been noted in
ref.~\cite{car98} for the Wick-Cutkosky model.

This fully completes the connection between the Bethe-Salpeter
amplitude and the light front form. The evaluation of the
Nakanishi integrals does not lead to cancellations of functions.
Although some functions are cancelled for reasons given above, all
spin-angular momentum functions (or all densities) in principle
contribute to the light-front wave functions.

Once the Bethe-Salpeter amplitudes are given (or the Nakanishi
densities) the light front wave function can  be calculated
explicitly. The Nakanishi spectral densities of the Bethe-Salpeter
amplitudes lead directly to the light-front wave function.

We would like to stress that the two relativistic approaches have
shown qualitatively similar results in the description of the
electro-disintegration near the threshold (see chapter
(\ref{disint})). The functions $f_5$ and $g_2$ (notation of
ref.~\cite{car98}) may be related to the pair current in the
light-front approach whereas the functions $h_5$ and $h_2$ play
this role in the Bethe-Salpeter approach. The results presented
here allow us to specify this relation on a more fundamental
level.

\subsection{\em Solution of the BS Equation}\label{BS:Sol}
In the previous section we have considered the general properties
of the BS amplitude. The different representations for spinor and
angular parts were proposed. The most important part defining the
dynamical structure of the bound state is the radial part which
can be found by solving the BS equation.

In this section, we consider the solution of the BS equation with
a separable interaction (separable {\em ansatz}). The separable
interaction is well known in nonrelativistic approaches. In
particular, this suggestion allows one to reduce the system of
integral equations, in which the Lippmann--Schwinger equation is
transformed, after partial expansion, to a system of linear
equations~\cite{yam,yam1,brown}. The Bethe--Salpeter equation with
separable kernel can be transformed to a system of linear
equations as well.

\subsubsection{\em Separable Interaction}\label{BS:sep}
Let us consider the BS equation for $T$-matrix after partial
decomposition (see subsection~\ref{sec:partial}):
\begin{eqnarray}
&&T_{_{JL^\prime
S^\prime\varrho^\prime,JLS\varrho}}(k_0^{\prime},|\bk^\prime|,k_0,|\mbox{\boldmath
$k$}|;s) = V_{_{JL^\prime S^\prime
\varrho^\prime,JLS\varrho}}(k_0^{\prime},|\bk^\prime|,k_0,|\mbox{\boldmath
$k$}|;s)+
\label{s001}\\
&&\!\!\!\!\!\!\!\!\!\!\!\!\!\!\frac{i}{2\pi^2}\!\int\d
k^{\prime\prime}_0\int {\bk^{\prime\prime}}^2\,\d
|\bk^{\prime\prime}|\!\!\!\!\! \sum\limits_{L^{\prime\prime}
S^{\prime\prime} \varrho^{\prime\prime}}\!\!\!\! V_{_{JL^\prime
S^\prime\varrho^\prime,JL^{\prime\prime} S^{\prime\prime}
\varrho^{\prime\prime}}}(k^\prime_0,|\bk^\prime|,k^{\prime\prime}_0,|\mbox{\boldmath
$k^{\prime\prime}$}|;s)\,
S_{\varrho^{\prime\prime}}(k^{\prime\prime}_0,|\mbox{\boldmath
$k^{\prime\prime}$}|;s)\, T_{_{JL^{\prime\prime} S^{\prime\prime}
\varrho^{\prime\prime},JLS\varrho}}(k^{\prime\prime}_0,|\mbox{\boldmath
$k^{\prime\prime}$}|,k_0,|\bk|;s), \nonumber
\end{eqnarray}
where $S_{\varrho}(k_0,|\bk|;s)$ is given by Eq.~(\ref{spart}).

The separable {\em ansatz} for the interaction is introduced in
the following manner~\cite{graz:rel,rupp3}:
\begin{eqnarray}
V_{_{JL^\prime
S^\prime\varrho^\prime,JLS\varrho}}(k_0^{\prime},|\bk^{\prime}|,k_0,|\bk|;s)=
\sum\limits_{i,j=1}^{N}\, \lambda_{ij}\,g_i^{JL^\prime
S^\prime\varrho^\prime}(k_0^{\prime},|\bk^{\prime}|)\,
g_j^{JLS\varrho}(k_0,|\bk|),\quad \lambda_{ij}=\lambda_{ji},
\label{s002}
\end{eqnarray} where $N$ is a rank of separability, $\lambda_{ij}$
are parameters of the interaction kernel, and
$g_i^{LS\varrho}(k_0,|\bk|)$ are functions which define the
interaction. Then according to Eq.(\ref{s001}) the $T$-matrix can
be expressed in a separable form too. We assume the following
separable form for it
\begin{eqnarray}
T_{_{JL^\prime
S^\prime\varrho^\prime,JLS\varrho}}(k_0^{\prime},|\bk^{\prime}|,k_0,|\bk|;s)
= \sum\limits_{i,j=1}^{N}\, \tau_{ij}(s)\,g_i^{JL^\prime
S^\prime\varrho^\prime}(k_0^{\prime},|\bk^{\prime}|)\,
g_j^{JLS\varrho}(k_0,|\bk|). \label{s003}
\end{eqnarray}
Substituting Eqs.~(\ref{s002}) and~(\ref{s003}) in
Eq.~(\ref{s001}) we can obtain an expression for $\tau(s)$,
\begin{eqnarray}
(\tau^{-1}(s))_{ij} = (\lambda^{-1})_{ij} - H_{ij}(s),
\label{s004} \end{eqnarray}
where $H_{ij}(s)$ is determined by
equation:
\begin{eqnarray}
H_{ik}(s) = \frac{i}{2\pi^2}\sum\limits_{LS\varrho} \int\,\d
k_0\,\int \bk^2\,
\d|\bk|\,S_{\varrho}(k_0,|\bk|;s)\,g_i^{JLS\varrho}(k_0,|\bk|)\,
g_k^{JLS\varrho}(k_0,|\bk|). \label{s005}\end{eqnarray}

The solution for the radial part of the BS amplitude can be
presented in the following form,
\begin{eqnarray}
&& \phi_{_{JLS\varrho}}(k_0,|\bk|) =
\sum_{i,j=1}^{N}\,
S_{\varrho}(k_0,|\bk|;s)\,\lambda_{ij}\,g_{i}^{JLS\varrho}(k_0,|\bk|)\,c_j(s),
\label{s006}
\end{eqnarray}
where the coefficients $c_j(s)$ satisfy the following system
 of linear homogeneous equations:
\begin{eqnarray}
c_i(s)\,-\,\sum\limits_{k,j=1}^{N}\,H_{ik}(s)\,\lambda_{kj}\,c_j(s)\,=\,0.
\label{s007}
\end{eqnarray}

\subsubsection{\em Dispersion Analysis of Nucleon--Nucleon
$T$-matrix}\label{BS:anal} To analyze the analytic properties of
the solution for the $T$-matrix with separable interaction, let us
consider a rank I case. We take only an $S$-state in $^1S_0$- and
$^3S_1-^3D_1$-channels (namely $\Sp$- and $\Spp$-waves). Omitting
all partial waves and separability indices one could write a
solution for $T$-matrix,
\begin{eqnarray}
t(k_0^\prime, |\bk^\prime|, k_0, |\bk|; s) = \tau(s) g(k_0^\prime,
|\bk|) g(k_0, |\bk^\prime|) \label{t05}\end{eqnarray}
with the function $\tau(s)$,
\begin{eqnarray}
\tau(s) = 1/(\lambda^{-1} + h(s)), \label{t06}\end{eqnarray} and
function $h(s)$,
\begin{eqnarray}
h(s) = -\frac{i}{4\pi^3}\, \int\, \d k_0\,\int\, |\bk|^2\, \d
|\bk|\,
\frac{[g(k_0,|\bk|)]^2}{(\sqrt{s}/2-E_{\bk}+i\epsilon)^2-k_0^2}~.
\label{t07}\end{eqnarray} Here we have introduced small letters
for the functions $T$ and $H$ in the simple indexless case.

As a result, the $T$-matrix can be rewritten in the following form:
\begin{eqnarray}
t(k^{\prime}_0, |\bk^{\prime}|, k_0, |\bk|; s) =
\frac{g(k^{\prime}_0, |\bk^{\prime}|) g(k_0, |\bk|)} {\lambda^{-1}
+ h(s)}, \label{t08}\end{eqnarray} and the on-mass-shell
expression is:
\begin{eqnarray}
t(s) = \frac{n(s)}{d(s)} = \frac{[g(0, \pp)]^2}{\lambda^{-1} +
h(s)}. \label{t09}\end{eqnarray} It should be noted that the
Eq.~(\ref{t09}) has the so-called $N/D$-form widely used in the
nonrelativistic $T$-matrix theory~\cite{brown} and some
 methods of relativization of the theory.

We use the following representation for the on-mass-shell
$T$-matrix valid in the region of unitarity:
\begin{eqnarray}
t(s) \equiv t(0,\pp,0,\pp;s) = - \frac{16
\pi}{\sqrt{s}\sqrt{s-4m^2}}\, e^{i\delta(s)}\, \sin{\delta(s)},
\label{t03}\end{eqnarray} with $\pp = \sqrt{s/4-m^2} =
\sqrt{2mT_{lab}}$ and $\delta(s)$ is the phase shift.

Using Eq.~(\ref{t09}), it is easy to relate the $T$-matrix and the
phase shift $\delta(s)$. To achieve this, we assume the imaginary
part of the function $n(s)$ satisfies the following condition:
\begin{eqnarray}
\im{} n(s) = 0. \label{t010}\end{eqnarray} The condition is
related with the specific choice of $g$-functions for the
$NN$-vertex which will be discussed later. Taking into account
Eqs.~(\ref{t09}) and~(\ref{t010}), the phase shift $\delta(s)$ can
be given as
\begin{eqnarray}
{\cot}\, \delta(s) = \frac{\re t(s)}{\im t(s)} = -\,
\frac{\lambda^{-1} + \re h(s)}{\im h(s)}
\label{t011}\end{eqnarray}

To express the low-energy parameters in term of the $T$-matrix
solution, it is suitable to expand the function $h(s)$ in a series
of $\pp$ terms:
\begin{eqnarray}
&& h(s) = h_0 + i\pp h_1 + \pp^2 h_2 + i \pp^3 h_3 + {\cal
O}(\pp^4),
\label{t012a}\\
&& \re h(s) = h_0 + \pp^2 h_2 + {\cal O}(\pp^4),
\label{t012b}\\
&& \im h(s) = \pp (h_1 + \pp^2 h_3 + {\cal O}(\pp^3)).
\label{t012}\end{eqnarray} The low-energy parameters of
$NN$-scattering are introduced by expanding the $T$-matrix into
series of $\pp$-terms following the expression suggested in
ref.~\cite{Bethe:1949}:
\begin{eqnarray}
\pp\, {\cot}\, \delta(s) = - a_0^{-1} + \frac{r_0}{2}\pp^2 + {\cal
O}(\pp^3)
 \label{t03a}\end{eqnarray} and taking into
consideration only the first two terms of the
decomposition~(\ref{t03a}).

Using now the definition~(\ref{t03a}) and
Eqs.~(\ref{t011})-(\ref{t012}) one can find the parameters
 $a_0$ and $r_0$:
\begin{eqnarray}
&& a_0 = \frac{h_1}{\lambda^{-1}+h_0},\\
\label{t013a}
&& r_0 = \frac{2}{h_1}\left[(\lambda^{-1}+h_0)\frac{h_3}{h_1}-h_2\right].
\label{t013b}\end{eqnarray}

At the mass of the bound state squared $M_b^2$, $T$-matrix has a
simple pole at the total momentum squared $s$, and the bound state
condition can be written in the following form:
\begin{eqnarray}
t(k^{\prime}_0, |\bk^{\prime}|, k_0, |\bk|; s) =
\frac{B(k^{\prime}_0, |\bk^{\prime}|, k_0, |\bk|;
s=M_b^2)}{s-M_b^2} + R(k^{\prime}_0, |\bk^{\prime}|, k_0, |\bk|;
s), \label{t03b}\end{eqnarray} where functions $B$ and $R$ are
regular at the point $s=M_b^2$. The bound state energy $E_b$ is
connected to $M_b$ as: $M_b = 2m-E_b$. The bound state
condition~(\ref{t03b}), with the help of Eq.~(\ref{t08}), can be
presented in a form:
\begin{eqnarray}
\lambda^{-1} = -h(s=M_b^2). \label{t014}\end{eqnarray}

Let us consider now the analytic properties of the solution
(namely, function $h$). The simplest choice of the function
$g(k_0,|\bk|)$ is the {\em Yamaguchi} type
function~\cite{yam,yam1}:
\begin{eqnarray}
g(k_0,|\bk|) = (k_0^2-\bk^2-\beta^2+i\epsilon)^{-1}.
\label{t21a}\end{eqnarray} In this case, the function $h$ can be
rewritten as follows:
\begin{eqnarray}
h(s,\beta) = -\,\frac{i}{4\pi^3}\, \db\, \int \d k_0\,\int\,
\bk^2\, \d |\bk|\ \frac{1}{(\sqrt{s}/2-E_\bk+i\epsilon)^2-k_0^2}\
\frac{1}{k_0^2-\eb^2+i\epsilon}~, \label{t21}\end{eqnarray} where
$\db \equiv \partial / \partial \beta^2$ and
$\eb=\sqrt{\bk^2+\beta^2}$. We introduced the second argument
$\beta$ to underline the explicit dependence of the function $h$
on this parameter.

Analyzing Eq.~(\ref{t21}) one can identify four
poles in the complex plane $k_0$, namely:
\begin{eqnarray}
&& k^{(1)}_0(s) = \frac{\sqrt{s}}{2} - E_{\bk} + i\epsilon
\qquad\qquad k^{(2)}_0(s) = - \frac{\sqrt{s}}{2} + E_\bk -
i\epsilon
\label{t22a}\\
&& k^{(3)}_0(s) = - \eb + i\epsilon \qquad\qquad\quad\quad
k^{(4)}_0(s) = \eb - i\epsilon \nonumber\end{eqnarray}

The variation of $s$ results in the move of the poles $k^{(1)}_0$
and $k^{(2)}_0$,  and one confronts the situation where two poles
``pinch'' the real $k_0$ axis. It means the function $h(s)$ has in
this $s$-point the leap and imaginary part. First points in which
this condition is satisfied (branch points) can be found from the
following equations:
\begin{eqnarray}
&& k^{(1)}_0(s) = k^{(2)}_0(s) \qquad \Rightarrow \qquad s_0 =
4m^2,
\label{t23a}\\
&& k^{(1)}_0(s) = k^{(4)}_0(s) \qquad \Rightarrow \qquad s_1 =
4(m+\beta)^2. \label{t23b}\end{eqnarray}

Summarizing the situation, one could say that the function $h(s)$ has two
cuts starting in points $s_0$ and $s_1$ respectively, and
therefore can be written in a dispersion form:
\begin{eqnarray}
&& h(s,\beta) = \int\limits_{4m^2}^{+\infty}
\frac{\rho(s^{\prime},\beta)\,ds^{\prime}}{s^{\prime}-s-i\epsilon},
\label{t24}\\
&& \rho(s^{\prime},\beta) = \theta (t-4m^2)
\rho_{el}(s^{\prime},\beta) + \theta
(t-4(m+\beta)^2)\rho_{in}(s^{\prime},\beta)
\nonumber\end{eqnarray} with two spectral functions $\rho_{el;in}$
({\em el} stands for {\em elastic} and {\em in} --- for {\em
inelastic}) which are connected with the imaginary parts as follows:
\begin{eqnarray}
\rho(s^{\prime},\beta) = \frac{1}{\pi} \im h(s^{\prime},\beta) =
\frac{1}{2\pi i} (h-h^{*}). \label{t25}\end{eqnarray}

Let us note some analytic properties of the obtained solution for
$T$-matrix (see also subsection~\ref{exchange-properties}):
\begin{enumerate}
\item A bound state (deuteron) appears in the $T$-matrix as a
simple pole in total momentum squared  $s$ at the mass of the
bound state squared, $s=M_b^2$; \item in the region $s>4m^2$,
$T$-matrix has the cut corresponding to the elastic $NN$
scattering (Eq. (\ref{t23a}));
\item in the region $s>4(m+\beta)^2$, $T$-matrix has the cut corresponding
to the inelastic $NN$ scattering  (Eq.~(\ref{t23b}));
\item $T$-matrix has no {\em left-hand} cuts but there is a pole
of second order at the point $s=4(m^2-\beta^2)$.
\end{enumerate}

To find spectral functions one should  perform $k_0$-integration
in Eq.~(\ref{t21}) which results in:
\begin{eqnarray}
h(s,\beta) = -\frac{1}{2\pi^2}\,\db\, \int \bk^2\, \d |\bk|\,
\frac{1}{s/4-\sqrt{s}E_{\bk}+m^2-\beta^2+i\epsilon}
\left[\frac{1}{\sqrt{s}-2E_{\bk}+i\epsilon}+\frac{1}{2\eb}\right].
\label{t26}\end{eqnarray} Taking into account the following
symbolic equation:
\begin{eqnarray}
\frac{1}{x-x_0 \pm i\epsilon} = \frac{{\cal P}}{x-x_0} \mp
i\pi\delta(x-x_0) \label{t27}\end{eqnarray}
it is easy to find spectral functions:
\begin{eqnarray}
&& s^{\prime} \ge 4m^2,
\nonumber\\
&& \hskip 10mm
\rho_{el}(s^{\prime},\beta) =
\sqrt{\spr}\sqrt{\spr-4m^2}/(\pi^2(\spr-4m^2+4\beta^2)^2),
\label{d28a}\\
&& \spr \ge 4(m+\beta)^2,
\nonumber\\
&& \hskip 10mm
\rho_{in}(\spr,\beta) =
-(64\beta^6+16\beta^4\spr-192\beta^4m^2-20\beta^2\spr^2
\nonumber\\
&& \hskip 10mm
+192\beta^2m^4-32\beta^2\spr
m^2+16m^4\spr+3\spr^3-64m^6-12m^2\spr^2)/
\nonumber\\
&& \hskip 10mm
(2\pi^2\spr(\spr-4m^2+4\beta^2)^2\sqrt{\spr-4(m+\beta)^2}
\sqrt{\spr-4(m-\beta)^2}).
\label{d28b}
\end{eqnarray}

To perform integration in Eq.~(\ref{t21}) it is suitable to introduce new
variables:
\begin{eqnarray}
&& \lmb = \frac{\beta}{m},\quad\quad\qquad\qquad\qquad
t=\frac{s}{4m^2},
\label{t211a}\\
&& \rho^2 = 1 - \frac{4m^2}{s} = 1 - \frac{1}{t},\qquad
w^2 = \frac{\beta^2}{m^2-\beta^2} = \frac{\lmb^2}{1-\lmb^2},
\label{t211b}\\
&& v^2=\frac{m+\beta}{m-\beta}= \frac{1+\lmb}{1-\lmb}, \qquad
\sigma^2 = - \frac{s - 4(m+\beta)^2}{s - 4(m-\beta)^2}=
-\frac{t-(1+\lmb)^2}{t-(1-\lmb)^2}. \label{t211c}
\end{eqnarray}

If the following conditions are valid:
\begin{eqnarray}
&& m > \beta > 0 \qquad \Rightarrow \qquad 1> \lmb > 0, \\
\label{t212a} && 4(m+\beta)^2 > s >4m^2 \qquad \Rightarrow \qquad
(1+\lmb)^2 > t > 1 \label{t212b}\end{eqnarray}
the parameters are
real and positive:
\begin{eqnarray}
1 > \rho^2 > 0, \qquad\qquad w^2 > 0,
\\
v^2 > 1 > 0, \qquad\qquad \sigma^2 > 0. \nonumber\end{eqnarray}
The condition~(\ref{t212b}) means that the second (inelastic)
imaginary part does not contribute to the function $\im h(s)$ when
phase shifts are calculated in the region $4(m+\beta)^2 > s >4m^2$
and, therefore:
\begin{eqnarray}
\im h(s,\beta) = \im h_{el}(s,\beta) =
\frac{\rho(1-\rho^2)(1+w^2)^2}{4m^2\pi(\rho^2+w^2)^2},\quad\quad
\mbox{ if }\quad 4(m+\beta)^2 > s >4m^2.
\label{t213}\end{eqnarray} Performing integration~(\ref{t24}) one
can obtain the real part of the function $h(s)$ (imaginary is
given by Eq.~(\ref{t213})):
\begin{eqnarray}
&& \re h(s,\beta) = h_{el}(s,\beta) + h_{in}(s,\beta),
\label{t215a}
\\
&& h_{el}(s,\beta) = \label{t215b} \frac{\rho(1-\rho^2)(1+w^2)^2}
{4m^2\pi^2(\rho^2+w^2)^2} \ln{\left|\frac{1-\rho}{1+\rho}\right|}
+\frac{(1-\rho^2)(1+w^2)^2(w^2-\rho^2)}{4m^2\pi^2(\rho^2+w^2)^2w}\arctan{\frac{1}{w}}
+\frac{(1-\rho^2)(1+w^2)}{4m^2\pi^2(\rho^2+w^2)},
\nonumber\\
&& h_{in}(s) = \label{t215c}
\frac{(1+\sigma^2)(1+v^2)^2}{32m^2\pi^2(v^4+\sigma^2)}
\ln{\left|\frac{v^2-1}{v^2+1}\right|}
 -\frac{(1+\sigma^2)(1+v^2)^2}{16m^2\pi^2(v^2-1)(v^2-\sigma^2)}
 \nonumber\\
 &&
\quad
-\frac{(1+\sigma^2)(1+v^2)^2}{16m^2\pi^2(v^2-1)^2v(v^2-\sigma^2)^2}
(v^6+v^4\sigma^2-4v^4+4v^2\sigma^2-v^2-\sigma^2)\arctan{\frac{1}{v}}
 \nonumber\\
 &&
\quad
 +
\frac{(1+\sigma^2)(1+v^2)^3}{16m^2\pi^2\sigma(v^4+\sigma^2)(v^2-\sigma^2)^2\pi^2(v^2-1)^2}
(-v^6+2v^6\sigma^2-3v^4\sigma^2+3\sigma^4v^2-2\sigma^4+\sigma^6)\arctan{\frac{1}{\sigma}}~.
\nonumber\end{eqnarray}

To find also the expressions for the low-energy parameters we
should return to Eqs.~(\ref{t012})-(\ref{t013b}) and expand
function $h(s)$ in a series of $\pp$ terms:
\begin{eqnarray}
h_0(\beta) &=& \frac{(1+w^2)}{4m^2\pi^2w^3}(w+(1+w^2)\arctan{\frac{1}{w}})
\label{t216a}\\
&& + \frac{5+20v^2+5v^8+14v^4+20v^6}{4m^2\pi^2(v^2-1)^3(3v^2+1)}
\sqrt{\frac{3v^2+1}{v^2+3}}\arctan{1/\sqrt{\frac{3v^2+1}{v^2+3}}}
\nonumber\\
&& -\frac{(v^2+1)^4}{4m^2\pi^3v(v^2-1)^3}\arctan{\frac{1}{v}}
-\frac{1}{8m^2\pi^2}\ln{\left|\frac{v^2+1}{v^2-1}\right|}
-\frac{(v^2+1)^2}{4m^2\pi^2(v^2-1)^2},
\nonumber\\
h_2(\beta) &=& -
\frac{(1+w^2)^2}{4m^4\pi^2w^5}(3w+(3+w^2)\arctan{\frac{1}{w}})
\label{t216b}\\
&& -\frac{1}{4m^4\pi^2(v^2-1)^5(v^2+3)(3v^2+1)^2}
(304v^2+1212v^{12}+3790v^8
 \nonumber\\
 &&
+304v^{14}+2704v^{10}+29v^{16}+1212v^4+ 2704v^6+29)
\nonumber\\
&& \times \sqrt{\frac{3v^2+1}{v^2+3}}
\arctan{1/\sqrt{\frac{3v^2+1}{v^2+3}}}
\nonumber\\
&& +
\frac{(v^4+10v^2+1)(v^2+1)^4}{4m^4\pi^2v(v^2-1)^5}\arctan{\frac{1}{v}}
\nonumber\\
&& + \frac{1}{8m^4\pi^2}\ln{\left|\frac{v^2+1}{v^2-1}\right|}
+\frac{11v^8+52v^6+66v^4+52v^2+11)(v^2+1)^2}{8m^4\pi^2(v^2-1)^4(3v^2+1)(v^2+3)},
\nonumber\\
h_1(\beta) &=& \frac{(1+w^2)^2}{4m^3\pi w^4},
\label{t216c}\\
h_3(\beta) &=& -\frac{(1+w^2)^2(4+3w^2)}{8m^5\pi w^6}~.
\label{t216d}\end{eqnarray}

At this moment, we can find internal parameters of the separable
interaction ($\lambda$, $\beta$) to reproduce experimental values
for low-energy parameters $a_{0s}^{exp} = -23.748 \pm 0.010$ fm,
$r_{0s}^{exp} = 2.75 \pm 0.05$ fm for singlet channel ($^1S_0$),
and $a_{0t}^{exp} = 5.424 \pm 0.004$ fm and bound state (deuteron)
energy $E_d^{exp} = 2.224644 \pm 0.000046$ MeV for triplet channel
($^3S_1$). Experimental data are taken from~\cite{Kroll}.

{\bfseries In the case of $^1S_0$-channel} we use
Eq.~(\ref{t013a}) with $a_{0} \equiv a_{0s}^{exp}$ to find
$\lambda$:
\begin{eqnarray}
{\lambda}^{-1} = (a_{0s}^{exp})^{-1} h_1(\beta) - h_0(\beta).
\label{dis01}
\end{eqnarray}
Inserting the above expression into Eq.~(\ref{t013b}) with $r_0
\equiv r_{0s}^{exp}$ we find:
\begin{eqnarray}
r_{0s}^{exp} = \frac{2}{h_1(\beta)} \left[ (a_{0s}^{exp})^{-1}
h_3(\beta) - h_2(\beta)\right]. \label{dis02}
\end{eqnarray}
Solving nonlinear Eq.~(\ref{dis02}) we find the value of $\beta$,
and then using Eq.~(\ref{dis01}) --- the value of $\lambda$.

{\bfseries In the case of $^3S_1$-channel} we obtain $\lambda$
from the bound state condition~(\ref{t014}) with $E_b \equiv
E_d^{exp}$:
\begin{eqnarray}
\lambda^{-1} = - h(s=(M_d^{exp})^2,\beta), \label{dis03}
\end{eqnarray}
where $M_d = 2m-E_d$. Inserting the above expression into
Eq.~(\ref{t013a}) with $a_0 \equiv a_{0t}^{exp}$ we find:
\begin{eqnarray}
a_{0t}^{exp} = \frac{h_1(\beta)}{h_0(\beta) -
h(s=(M^{exp}_d)^2,\beta)}. \label{dis04}
\end{eqnarray}
Solving the nonlinear Eq.~(\ref{dis04}) we can find  $\beta$, and
then using Eq.~(\ref{dis03}) --- the value of $\lambda$.
 As a result we find:
 \begin{eqnarray}
\begin{array}{ll}
\mbox{for $^1S_0$ channel:}
& \lambda = -0.29425404 \mbox{ GeV}^{-2},
\qquad \beta = 0.22412880 \mbox{ GeV},\\
&\\
\mbox{for $^3S_1$ channel:}
& \lambda = -0.79271213 \mbox{ GeV}^{-2},
\qquad \beta = 0.27160579 \mbox{ GeV},\\
\end{array}
\end{eqnarray}

\begin{figure}[ht]
\centerline{
\includegraphics[width=110mm,angle=270]{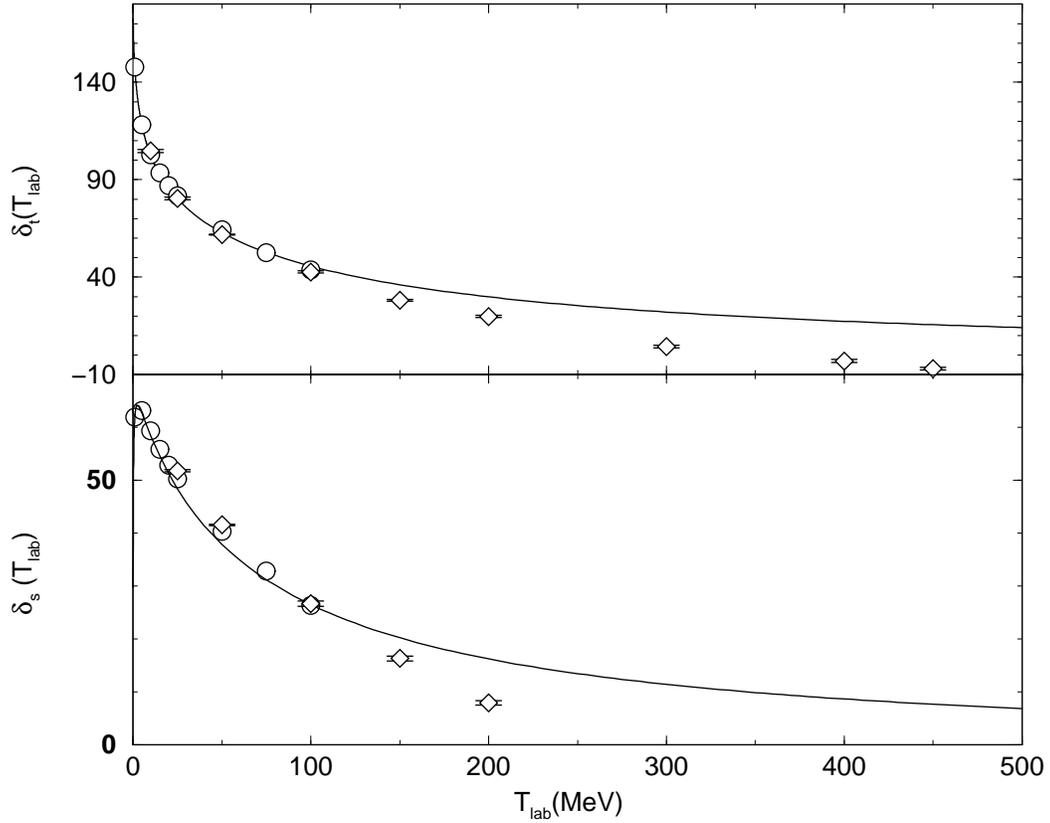}}
\caption{\label{disper-fig1} The  triplet $\delta_t$ (top panel)
and singlet $\delta_s$ (bottom panel) phase shifts calculated with
the separable interaction in Yamaguchi form with rank I.}
\end{figure}

The phase shifts $\delta_s(s)$ and $\delta_t(s)$ calculated with
these parameters are displayed in Fig.~\ref{disper-fig1}.
Experimental data are taken from ref.~\cite{arndt}. As it is seen
from the figure,  the simplest choice of the separable interaction
--- rank I with just two parameters $\lambda$ and $\beta$ ---
is able to provide the low-energy parameters of elastic $NN$
scattering $a_s$ and $r_s$ in singlet channel, and $a_t$ and bound
state (deuteron) energy $E_d$ in triplet channel, with required
accuracy and to reproduce phase shifts up to $T_{lab} \simeq 100$
MeV.

We note in conclusion that the dispersion form of the $T$-matrix
for the elastic $NN$ scattering obtained for the separable
interaction allows to perform analytical calculations and
explicitly connect parameters of the kernel and the
observables~\cite{toki01}. It is interesting to note that the rank
I separable interaction is able to provide the low energy data
even up to $T_{lab}\simeq 100$ MeV and the deuteron properties.

\subsubsection{\em Covariant Graz-II Interaction}\label{BS:Graz}
In the actual calculations for various electromagnetic
observables, we use a more complex separable interaction, namely
rank-III covariant Graz-II kernel, where partial waves with only
the positive energy are taken into account ($^3S_1^+$, $^3D_1^+$).
In this case the functions $g_i$ have the following
form~\cite{graz:rel},
\begin{eqnarray}
&&g_{1}^{^3S_1^+}(k_0,|\bk|)=\frac{1-\gamma_{1}(k_0^2-\bk^2)}
{(k_0^2-\bk^{2}-\beta_{11}^{2})^{2}},
\label{gfactors}\\
&&g_{2}^{^3S_1^+}(k_0,|\bk|)=-\frac{(k_0^2-\bk^2)}
{(k_0^2-\bk^2-\beta_{12}^2)^{2}},
\nonumber \\
&&g_{3}^{^3D_1^+}(k_0,|\bk|)=\frac{(k_0^2-\bk^{2})
(1-\gamma_{2}(k_0^2-\bk^{2}))}
{(k_0^2-\bk^{2}-\beta_{21}^{2})(k_0^2-\bk^{2}-\beta_{22}^{2})^{2}},
\nonumber \\
&&g_{1}^{^3D_1^+}(k_0,|\bk|)=g_{2}^{^3D_1^+}(k_0,|\bk|)=
g_{3}^{^3S_1^+}(k_0,|\bk|)\equiv 0. \nonumber\end{eqnarray} The
parameters of these functions are given in table~\ref{tab:graz}.

\begin{figure}[ht]
\begin{center}
\psfig{figure=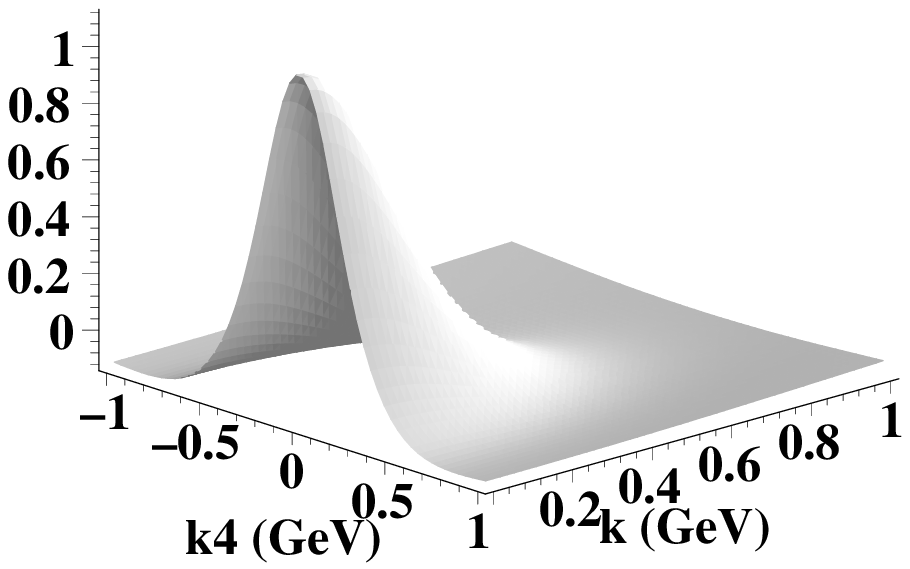,width=120mm} \vbox to 100mm{}
\caption{\label{vert-s}Vertex function $g_{\sp}(k_4,|\bk|)$.}
\end{center}
\end{figure}

\begin{figure}[ht]
\begin{center}
\psfig{figure=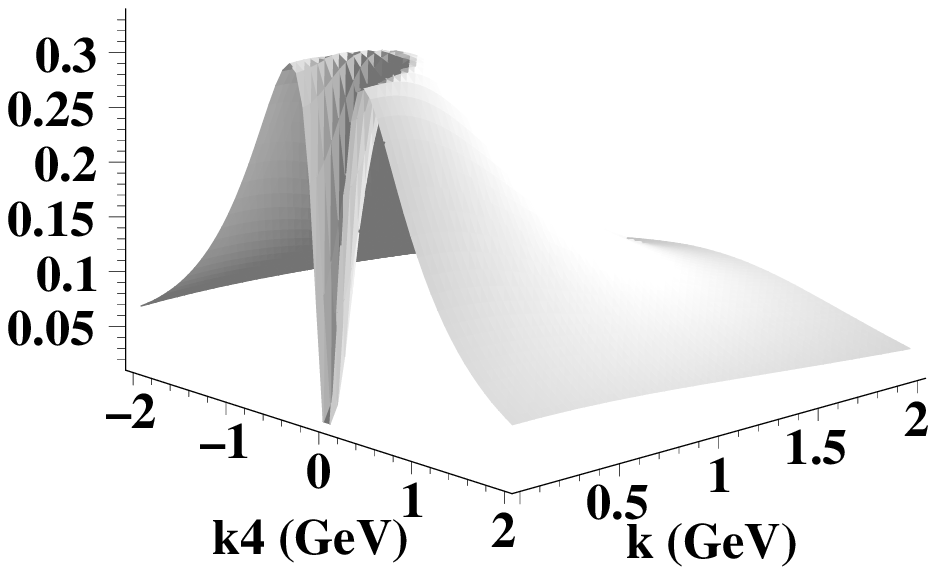,width=120mm} \vbox to 100mm{}
\caption{\label{vert-d} Vertex function $g_{\Dp}(k_4,\nrmk)$.}
\end{center}
\end{figure}

The solution of the BS equation for the vertex function can be
explicitly written as
\begin{eqnarray}
g_{\sp}(k_0,|\bk|) &=& (c_1 \lambda_{11}+c_2 \lambda_{12}+c_3
\lambda_{13})
g_1^{^3D_1^+}(k_0,|\bk|)+ \label{vert-graz}\\
&& (c_1 \lambda_{12}+c_2 \lambda_{22}+c_3 \lambda_{23}) g_2^{\sp}(k_0,|\bk|), \nonumber\\
g_{\Dp}(k_0,|\bk|) &=& (c_1 \lambda_{13}+c_2 \lambda_{23}+c_3
\lambda_{33}) g_3^{\Dp}(k_0,|\bk|), \nonumber\end{eqnarray} where
we take into account that the matrix $\lambda$ is symmetric.
Vertex functions  $g_{\sp}(k_4,|\bk|)$ and $g_{\Dp}(k_4,|\bk|)$ in
Euclidean space ($k_4=ik_0$) are shown in Fig.~\ref{vert-s} and
Fig.~\ref{vert-d}, respectively.

To calculate the phase shifts, we use the following
parametrization for on-mass-shell $T$-matrix:
\begin{eqnarray}
T(s) = -\frac{8}{\sqrt{s(s-4m^2)}} \left( \begin{array}{ll}
\cos{2\epsilon}\ e^{2 i \delta_{\rm S}} - 1 & i\sin{2\epsilon}\
e^{i (\delta_{\rm S}+\delta_{\rm D})} \\ i\sin{2\epsilon}\ e^{i
(\delta_{\rm S}+\delta_{\rm D})} & \cos{2\epsilon}\ e^{2i
\delta_{\rm D}} - 1 \end{array} \right),
\end{eqnarray}
where $\delta_{S}$ ($\delta_{D}$ ) are phase shifts of $^3S_1^{+}$-
($^3D_1^{+}$-) waves and $\epsilon$ is the mixing parameter.

\begin{figure} \hskip 15mm
\psfig{figure=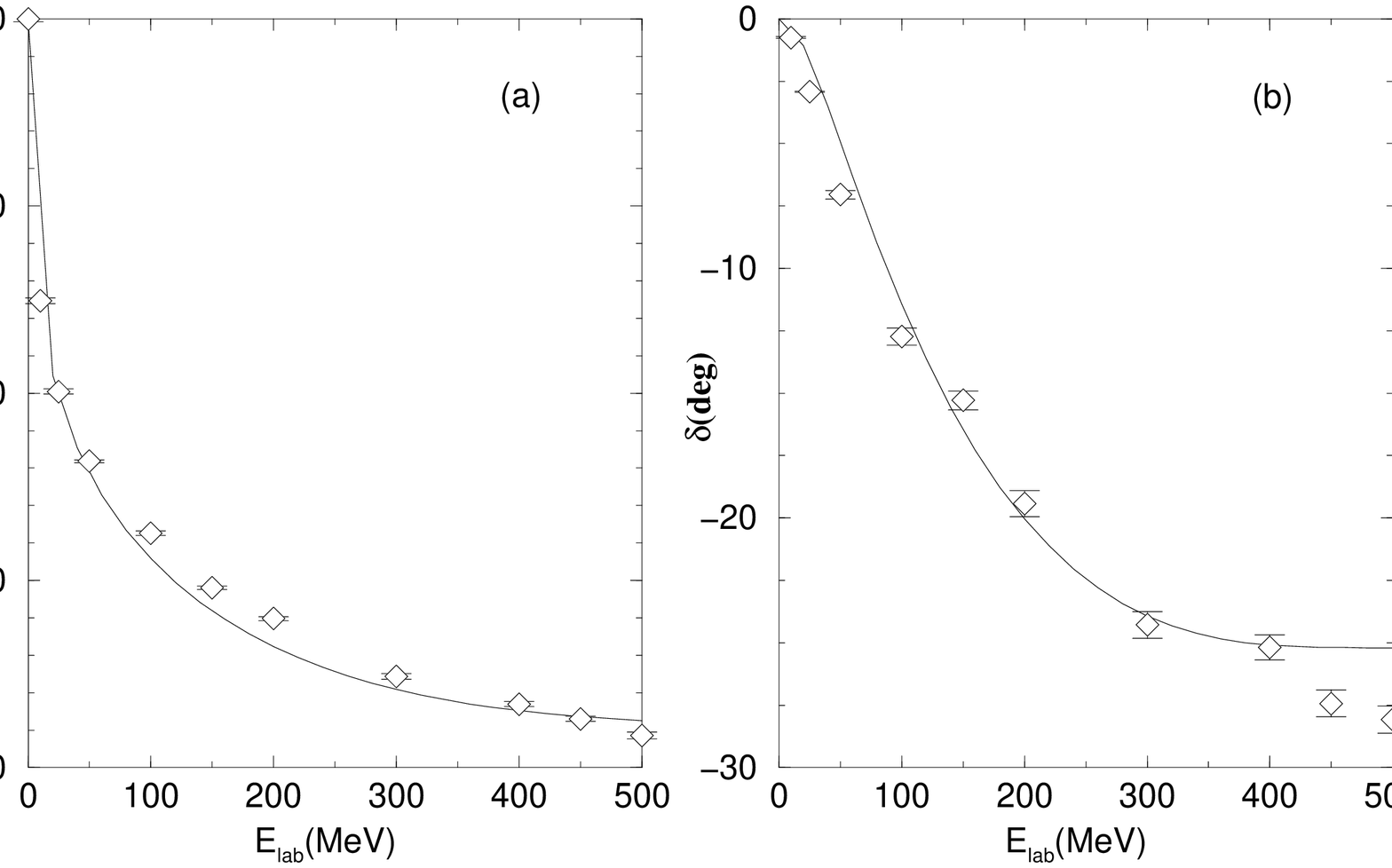,width=0.70\textwidth} \caption{{{(a)
$^3S_1$- and (b) $^3D_1$-phase shifts
 calculated by using  rank III Graz II
kernel. Experimental data are taken from
ref.~\protect\cite{arndt}}}} \label{fig:phase}
\end{figure}

The calculated results  are given in table~\ref{tab:prop} and in
Fig.~\ref{fig:phase}. The Bethe-Salpeter approach with the
separable interaction provides the deuteron properties and also
phase shifts of the nucleon-nucleon scattering in wide energy
region 0--400 MeV in the $S$-$D$ chanel.

\begin{table}[th]
\caption{\label{tab:graz}{{Parameters of the covariant  Graz-II
separable interaction}}} {\[
\begin{tabular}{lrlllrll}
\hline\hline
$\gamma_1$     & 28.69550 &                   & GeV$^{-2}$ &
$\lambda_{11}$ & 2.718930 &$\times$ 10$^{-4}$ & GeV$^{6}$ \\
$\gamma_2$     & 64.9803  &                   & GeV$^{-2}$ &
$\lambda_{12}$ & -7.16735 &$\times$ 10$^{-2}$ & GeV$^{4}$ \\
$\beta_{11}$   & 2.31384  &$\times$ 10$^{-1}$ & GeV        &
$\lambda_{13}$ & -1.51744 &$\times$ $10^{-3}$ & GeV$^{6}$ \\
$\beta_{12}$   & 5.21705  &$\times$ $10^{-1}$ & GeV        &
$\lambda_{22}$ & 16.52393 &                   & GeV$^{2}$ \\
$\beta_{21}$   & 7.94907  &$\times$ $10^{-1}$ & GeV        &
$\lambda_{23}$ & 0.28606  &                   & GeV$^{4}$ \\
$\beta_{22}$   & 1.57512  &$\times$ $10^{-1}$ & GeV        &
$\lambda_{33}$ & 3.48589  &$\times$ $10^{-3}$ & GeV$^{6}$ \\
\hline\hline
\end{tabular}
\]}
\end{table}

\begin{table}[th]
\caption{\label{tab:prop}{{Properties of the deuteron and the low
energy $NN$-scattering parameters in $^3S_1$ chanel with rank III
Graz-II kernel}}}
\[ {\rm
\begin{tabular}{@{}llllllll} \hline\hline & $p_{\rm D}(\%)$ &
$\epsilon_{\rm D}$ & $Q_{\rm D}$ & $\mu_{\rm D}$ & $\rho_{\rm
D/S}$ & $r_0$ (Fm) & $a$ (Fm) \\ &&(MeV)&(Fm$^{-2}$)&($e/2m$)&&&\\
\hline NR Graz II& 4 & 2.225 & 0.2484 & 0.8279 & 0.02408 & 1.7861
& 5.4188 \\ Cov. Graz II& 4.82 & 2.225 & 0.2812 & 0.8522 & 0.0274
& 1.78 & 5.42\\ \hline \multicolumn{2}{@{}l} {Exp.}& 2.2246 &
0.286 & 0.8574 & 0.0263 & 1.759 & 5.424\\ \hline\hline
\end{tabular}} \] \end{table}

\subsection{\em  Separable and One-Meson Exchange Interaction}
\label{sec:ladder} In this section, we  show the meson exchange
interaction (meson-nucleon) in ladder approximation and its
connection with the separable kernel with the Yamaguchi-type
$g$-functions. To achieve this, we introduce simple approximation
for ladder kernel:
\begin{eqnarray}
V(k_0^\prime,|\bk^\prime|;k_0,|\bk|) \to {\tilde
V}(k_0^\prime,|\bk^\prime|;k_0,|\bk|) =
\frac{V(k_0^\prime,|\bk^\prime|;0,0)V(0,0;k_0,|\bk|)}{V(0,0;0,0)}.
\label{ns01}
\end{eqnarray}
Using the expressions of the kernel for scalar meson exchange
($sc$)~\cite{kubis,machl87}:
\begin{eqnarray}
V_{sc}(k_0^\prime,|\bk^\prime|;k_0,|\bk|) =
-\frac{g_{sc}^2}{4\pi}\frac{1}{\pi^2}\frac{1}{4|\bk^\prime| |\bk|
E_{\bk^\prime}E_\bk}
\left[(E_{\bk^\prime}E_\bk+m^2)Q_0(z)-|\bk^\prime|
|\bk|Q_1(z)\right], \label{ns02}
\end{eqnarray}
where $z=(\bk^{\prime 2}+\bk^2-(k^\prime_0-k_0)^2+\mu^2)/(2|\bk^\prime| |\bk|)$, $\mu$
is the meson mass and $Q_{i}(z)$ are the Legendre functions of
second kind, $Q_0(z) = 2^{-1} \ln{(z+1)/(z-1)}$, $Q_1(z) = zQ_0(z)
- 1$. It can be shown that
\begin{eqnarray}
V_{sc}(k_0^\prime,|\bk^\prime|;0,0) = \lim_{k_0\to 0, |\bk|\to 0}
V_{sc}(k_0^\prime,|\bk^\prime|;k_0,|\bk|) = a_{\bk^{\prime}}
\tilde g(k_0^\prime,|\bk^\prime|), \label{ns03}
\end{eqnarray}
where
\begin{eqnarray}
a_{\bk^\prime} =
\frac{g_{sc}^2}{4\pi}\frac{1}{\pi^2}\frac{E_{\bk^\prime}+m}{2E_{\bk^\prime}},
\nonumber\\
\tilde g(k_0^\prime,|\bk^\prime|) = 1/(k_0^{\prime
2}-{\bk^\prime}^2-\mu^2). \label{ns05}
\end{eqnarray}
Expressions for $V_{sc}(0,0;k_0,|\bk|)$ can be obtained from
Eqs.~(\ref{ns03}-\ref{ns05}) by the following substitutions:
$k_0^\prime \to k_0$ and $|\bk^\prime| \to |\bk|$. To make the
connection between parameters, we perform
$|\bk^\prime|/m$-decomposition in the function $a_{\bk^\prime}$ up
to ${\cal O}({\bk^\prime}^2/m^2)$ term:
\begin{eqnarray}
a_{\bk^\prime} = a_\bk = \frac{g_{sc}^2}{4\pi}\frac{1}{\pi^2},
\nonumber\\
V_{sc}(0,0;0,0) =
-\frac{g_{sc}^2}{4\pi}\frac{1}{\pi^2}\frac{1}{\mu^2}. \label{ns06}
\end{eqnarray}

Using the expression~(\ref{ns01}), we find:
\begin{eqnarray}
{\tilde V}_{sc}(k_0^\prime,|\bk^\prime|;k_0,|\bk|) =
-\frac{g_{sc}^2}{4\pi}(\frac{\mu}{\pi})^2 \tilde
g(k_0^\prime,|\bk^\prime|) \tilde g(k_0,|\bk|). \label{ns07}
\end{eqnarray}

By comparing this expression with the separable form of kernel
introduced as
\begin{eqnarray}
v(k_0^\prime, |\bk^\prime|, k_0, |\bk|; s) = \lambda g(k_0^\prime,
|\bk^\prime|) g(k_0, |\bk|). \label{t04}\end{eqnarray} we find the
following relation between parameters:
\begin{eqnarray}
\beta = \mu,\qquad\qquad \lambda_{sc} =
-\frac{g_{sc}^2}{4\pi}(\frac{\mu}{\pi})^2. \label{ns08}
\end{eqnarray}

Equations~(\ref{ns07},\ref{ns08}) are valid also for
vector-meson-exchange kernel ($vc$) with substitution $g_{sc} \to
g_{vc}$ and $\mu$ is  the vector-meson mass. The separable form
and can be derived from the following expression,
\begin{eqnarray}
V_{vc}(k_0^\prime, |\bk^\prime|; k_0, |\bk|) =
-\frac{g_{vc}^2}{4\pi}\frac{1}{\pi^2}\frac{1}{4|\bk^\prime| |\bk|
E_{\bk^\prime} E_{\bk}}
\left[-2(2E_{\bk^\prime}E_{\bk}+m^2)Q_0(z)\right]. \label{ns09}
\end{eqnarray}

For pseudoscalar-meson-exchange, the kernel of interaction has the
form:
\begin{eqnarray}
V_{ps}(k_0^\prime, |\bk^\prime|; k_0, |\bk|) =
-\frac{g_{sc}^2}{4\pi}\frac{1}{\pi^2}\frac{1}{4  |\bk||\bk^\prime|
E_{\bk^\prime} E_{\bk}} \left[-(E_{\bk^\prime}
E_{\bk}-m^2)Q_0(z)+|\bk^\prime| |\bk| Q_1(z)\right], \label{ns10}
\end{eqnarray}
and for $|\bk^\prime|,|\bk| \to 0$ we can write:
\begin{eqnarray}
{\tilde V}_{ps}(k_0^\prime, |\bk^\prime|; k_0, |\bk|) \sim
{\bk^\prime}^2 \bk^2 \tilde g(k_0^\prime,|\bk^\prime|) \tilde
g(k_0,|\bk|). \label{ns11}
\end{eqnarray}
In the latter case ${\tilde V}_{ps}$ tends to zero for
$|\bk^\prime|,|\bk| \to 0$, and it is impossible to find a
relation between parameters similar to Eq.~(\ref{ns08}).

To illustrate the  relations between parameters we used
Eq.~(\ref{ns08}) and calculated values $\lambda_{\mu}$
corresponding to parameters $\mu$ and $g^2/4\pi$ for scalar- and
vector-exchange-mesons from ref.~\cite{machl87}. Results are given
in table~\ref{ns12}.

\begin{table}[htbp]
\caption{\label{ns12} Connection between parameters of two
kernels.}
\begin{center}
\begin{tabular}{lllll}
\hline\hline
$J^P$ & & $\mu$ (GeV) & $g^2/4\pi$ & $\lambda$ (GeV$^{2}$) \\
\hline
$0^+$ & $NN\delta$ & 0.983 & 0.64 & -.06265954891 \\
$0^+$ & $NN\sigma^{\prime}$ & 0.550 & 7.07 & -.2166930823 \\
$1^-$ & $NN\rho$ & 0.769 & 0.43 & -.02576448048 \\
$1^-$ & $NN\omega$ & 0.7826 & 10.6 & -.6577877886\\
\hline\hline
\end{tabular}
\end{center}
\end{table}


\section{Elastic Electron Deuteron Scattering\label{Elastic}}

The previous section was devoted to the analysis of the basic
objects and methods of the BS formalism. It was shown how
different dynamical processes involving bound states of particles
can be included into the field-theoretical consideration. In this
section, we consider the application of these methods to elastic
electron scattering off the simplest bound system
---  the deuteron. We will highlight most important consequences
which follow from the relativistic nature of the bound state.

Our interest in the electron-deuteron scattering is connected
first of all with recent experimental studies of the deuteron
electromagnetic forms factors in the region of high transfer
momenta (see, for example,~\cite{Alexa:1999}-\cite{Bosted:1990}),
where the relativistic effects {\em a priori} play essential role,
and with recent tensor polarization
data~\cite{Garcon:1994}-\cite{Abbott:2000:EPJ}.

Traditional nonrelativistic  methods are based on the impulse
approximation with allowance for relativistic corrections such as
the meson-exchange currents (MEC) and retardation effects. In a
number of investigations (see for
example~\cite{BDS:1992},\cite{gari:huga}-\cite{plessas} and
references therein) it has been shown that the correct account of
these effects is necessary to explain the experimental results.
Generally, the deuteron elastic form factors are known to be
sensitive to the choice of the strong nucleon form factor and to
the MEC models. On the other hand, the recent relativistic
investigations~\cite{karmanov}-\cite{bb:magnet} show that some of
the meson-exchange currents (in particular, the pair current) are
automatically included in the  relativistic impulse approximation.

Beside the relativistic effects, it is important to study the
contribution of the nucleon form factor to the deuteron elastic
form factors and to its polarization properties. There exists a
number of theoretical and phenomenological models of the nucleon
form factors. One finds largest differences in the results of
evaluation of the electric neutron form factor $G_{\rm E}^{\rm
n}(q^2)$ as well as of the ratio $G_{\rm E}/G_{\rm M}$ for the
proton at $Q^2 = -q^2 >$ 1~(GeV/c)$^2$. A further progress is
related with the appropriate choice of the polarization
observables, which would allow the consistent analysis of the
structure of the bound nucleon. In this section we apply the BS
approach to the analysis of elastic electron deuteron scattering
including such topics as elastic form factors and polarization
tensor of the deuteron.

\subsection{\em Relativistic Kinematics\label{RK}}
\begin{figure}[ht]
\begin{tabular}{cc}
\begin{minipage}[t]{0.5\linewidth}
\hskip 10mm
\includegraphics[width=70mm]{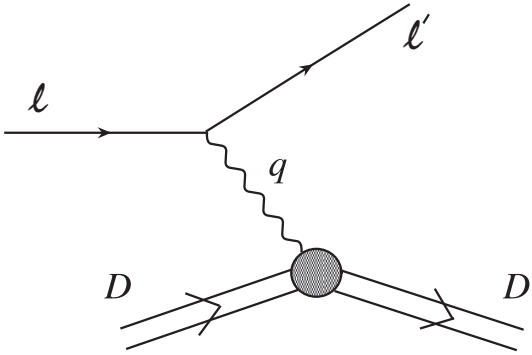}
\end{minipage}
&
\begin{minipage}{0.4\linewidth}
\vskip -60mm \caption{\label{elastic:OPA} Electron-deuteron
elastic scattering in the one-photon approximation.}
\end{minipage}
\end{tabular}
\end{figure}

The differential cross section for unpolarized elastic
electron--deuteron scattering in the one-photon-exchange
approximation (Fig.~\ref{elastic:OPA}) is expressed in terms of
the Mott cross section and deuteron structure functions $A(q^2)$
and $B(q^2)$ (the electron mass is neglected):
\begin{eqnarray}
\frac{d\sigma}{d\Omega_{\el}^{\prime}} =
\Bigl(\frac{d\sigma}{d\Omega_{\el}^{\prime}}\Bigr)_{\rm Mott}
\Bigl[A(q^2)+B(q^2)\tan^2{\frac{\theta_{\el}}{2}}\Bigr],
\label{cross}\end{eqnarray}
\begin{eqnarray}
\Bigl(\frac{d\sigma}{d\Omega_{\el}^{\prime}}\Bigr)_{\rm Mott}=
\frac{\alpha^2
\cos^2{\theta_{\el}/2}}{4E_{\el}^2(1+2E_{\el}/M\sin^4{\theta_{\el}/2})},
\end{eqnarray}
where $\theta_{\el}$ is the electron scattering angle, $M$ is the
deuteron mass, $E_e$ is the incident electron energy, and
\begin{eqnarray}
&&A(q^2)=F_{{\rm C}}^2(q^2)+\frac{8}{9}\eta^2F_{\rm Q}^2(q^2)+
\frac{2}{3}\eta F_{\rm M}^2(q^2),
\nonumber\\
&&B(q^2)=\frac{4}{3}\eta(1+\eta)F_{\rm M}^2(q^2),
\label{structf}\end{eqnarray}
 where $\eta=-{q^2}/{4M^2} = Q^2/4M^2$. The
electric $F_{\rm C}(q^2)$, the quadrupole $F_{\rm Q}(q^2)$ and the
magnetic $F_{\rm M}(q^2)$ form factors are normalized as
\begin{eqnarray}
F_{\rm C}(0)=1, \quad F_{\rm Q}(0)=M^2 \, Q_{\rm D}, \quad F_{\rm
  M}(0)=\mu_{\rm D}\frac{M}{m}
\label{normf}\end{eqnarray} where $m$ is  the nucleon mass,
$Q_{\rm D}$ and $\mu_{\rm D}$ are quadrupole and magnetic moments
of the deuteron, respectively. The tensor polarization components
of the final deuteron are expressed through the deuteron form
factors as follows:
\begin{eqnarray}
&& T_{20}\ \bigl[A+B\tan^2{\frac{\theta_{\el}}{2}}\bigr]=
-\frac{1}{\sqrt{2}}\bigl[\frac{8}{3}\eta F_{\rm C}F_{\rm Q}+
\frac{8}{9}\eta^2F_{\rm
  Q}^2+\frac{1}{3}\eta(1+2(1+\eta)\tan^2{\frac{\theta_{\el}}{2}})F_{\rm M}^2\bigr],
\nonumber\\
&& T_{21}\ \bigl[A+B\tan^2{\frac{\theta_{\el}}{2}}\bigr]=
\frac{2}{\sqrt{3}}\eta(\eta+\eta^2\sin^2{\frac{\theta_{\el}}{2}})^{1/2}
F_{\rm M} F_{\rm Q} \sec{\frac{\theta_{\el}}{2}},
\label{tensormom}\\
&& T_{22}\ \bigl[A+B\tan^2{\frac{\theta_{\el}}{2}}\bigr]=
-\frac{1}{2\sqrt{3}}\eta F_{\rm M}^2. \nonumber\end{eqnarray}

Equation~(\ref{cross}) can be obtained by using the standard
technique~\cite{bjorken} from the following amplitude of the
process
\begin{eqnarray}
M_{\rm fi}=i e^2{\bar u_{m^{\prime}}(l^{\prime})}
\gamma^{\mu}u_{m}(l)\;\frac{1}{q^2} \;\langle D^{\prime}{\cal
M}^{\prime} | J_{\mu} | D {\cal M} \rangle,
\label{eqn:M}\end{eqnarray} where $u_{m}(l)$ denotes the free
electron spinor with 4-momentum $l$ and spin projection $m$, and
$q=l-l^{\prime}=P^{\prime}-P$ is the 4-momentum transfer,
$P(P^{\prime})$ is the initial (final) deuteron momentum; $|D{\cal
M}\rangle$ is the deuteron state with total angular momenta
projection ${\cal M}$, and $J_{\mu}$ is the electromagnetic
current operator.

The deuteron current matrix element is usually parameterized in
the following way (due to $P$- and $T$-parity conservation and
gauge invariance):
\begin{eqnarray}
\langle D^\prime {\cal M}^\prime |J_\mu|D{\cal M}\rangle= &-&e\;
\xi^*_{\alpha\;{\cal M}^\prime}(P^{\prime})\;\xi_{\beta\;{\cal
M}}(P)\; \Biggl[ (P^\prime+P)_{\mu}
\Bigl(g^{\alpha\beta}F_1(q^2)-\frac{q^\alpha
q^\beta}{2M^2}F_2(q^2)\Bigr)
\nonumber\\
&-& (q^{\alpha} g^\beta_\mu - q^{\beta} g^\alpha_\mu)G_1(q^2)
\Biggr], \label{deuteronc}\end{eqnarray} where $\xi_{\cal M}(P)$
and $\xi^{*}_{{\cal M}^\prime}(P^\prime)$ are the polarization
4-vectors of the initial and final deuteron, respectively. Form
factors $F_{1,2}(q^2)$, $G_1(q^2)$ are related to $F_{\rm
C}(q^2)$, $F_{\rm Q}(q^2)$ and $F_{\rm M}(q^2)$ by the equations
\begin{eqnarray}
F_{\rm C} = F_1 + \frac{2}{3}\eta \bigl[F_1 + (1+\eta)F_2 - G_1
\bigr], \quad F_{\rm Q} = F_1 + (1+\eta)F_2 - G_1, \quad F_{\rm M}
= G_1. \label{ffconnect}\end{eqnarray} The normalization condition
for the deuteron current matrix element has the form (in contrast
to~(\ref{norm-cur}) $i$ is included into the matrix element):
\begin{eqnarray}
\lim_{q^2\to 0} \langle D^\prime {\cal M}^\prime |J_\mu|D{\cal
M}\rangle = 2eP_{\mu}\; \delta_{{\cal M}{\cal M}^{\prime}}~.
\nonumber\end{eqnarray}

To calculate the deuteron form factors one should use the
particular system of reference. In the laboratory frame the
4-vectors have the following form (the $Z$--axis is along the
photon momentum):
\begin{eqnarray}
&&P=(M,{\bf 0}), \quad P^{\prime}=(M(1+2\eta),0,0,2 M
\sqrt{\eta}\sqrt{1+\eta}),
\nonumber\\
&&q=(2M\eta,0,0,2M\sqrt{\eta}\sqrt{1+\eta}),
\label{vlab}\end{eqnarray}
\begin{eqnarray}
&&\xi_{{\cal M}=+1}(P) = \xi_{{\cal M}=+1}(P^{\prime})=
-\frac{1}{\sqrt{2}} (0,1,i,0),
\nonumber\\
&&\xi_{{\cal M}=-1}(P) = \xi_{{\cal M}=-1}(P^{\prime})=
\frac{1}{\sqrt{2}} (0,1,-i,0), \nonumber\\
&&\xi_{{\cal M}=0}(P) = (0,0,0,1),\quad \xi_{{\cal
M}=0}(P^{\prime})= (2\sqrt{\eta}\sqrt{1+\eta},0,0,1+2\eta).
\label{xibreit}\end{eqnarray}

Using expressions (\ref{vlab},\ref{xibreit}) and the
parameterization in the form of Eq.~(\ref{deuteronc}) one obtains:
\begin{eqnarray}
&&\langle {\cal M}^{\prime} | J_{0} | {\cal M} \rangle =\; 2Me\;
(1+\eta)\; \Bigl\{ F_1 \delta_{{\cal M} {\cal M}^{\prime}} + 2\eta
\bigl[F_1 + (1+\eta)F_2 - G_1 \bigr] \delta_{{\cal M}^{\prime} 0}
\delta_{{\cal M} 0}\Bigr\}, \nonumber\\
&&\langle {\cal M}^{\prime} | J_{x} | {\cal M} \rangle =\;
\frac{2Me}{\sqrt{2}}\; \sqrt{\eta}\;\sqrt{1+\eta}\; G_1\; \Bigl\{
\delta_{{\cal M}^{\prime} {\cal M}+1} - \delta_{{\cal M}^{\prime}
{\cal M}-1} \Bigr\}. \label{emcmatel}\end{eqnarray}

To calculate the deuteron form factors, one should know three
matrix elements with different total angular momentum projections
and current components.

\subsection{{\em Gauge Invariance and Gauge Independence in the Bethe--Salpeter Approach}}\label{sect:gauge}

   It is well  known that the principle of gauge invariance imposes
stringent constraints on the amplitudes of electromagnetic
interactions with bound systems. In the first order of
perturbation theory in the charge $e$, this principle leads to the
continuity equation for the electromagnetic-current-density
operator and to the Ward-Takahashi (WT) identity for the
five-point Green's function. We will consider the WT identity for
the five-point Green's function and its implications for the
Mandelstam current, which determines the amplitude of electron
scattering on deuterons in the Bethe — Salpeter formalism.

   The continuity equations for the bare (Noether) current and
effective currents (for example, the Mandelstam current or
conserved currents in nonrelativistic quantum mechanics) are not
sufficient for ensuring the gauge invariance of the amplitudes of
electromagnetic transitions ~\cite{Bentz}-\cite{Zu_Tjon:81}.
 Moreover, both
initial and final states must correspond to the current used in
the analysis.

   As a rule, the conserved deuteron electromagnetic current
includes two-particle contributions associated with meson-exchange
currents or interaction currents. At the same time, it was shown
in \cite{graz:rel,Zu_Tjon:80,Zu_Tjon:81}, that the amplitude of elastic $e{\rm
D}$ scattering can be gauge invariant in the relativistic impulse
approximation, which is based on the concept of the one-particle
scattering mechanism. This result was obtained in the BS formalism
with one boson exchange (OBE) potentials  and with separable
interactions. At first glance, it is at odds with the common point
of view on the problem of gauge invariance \cite{Kazes}.

   Our objective here is to study in detail the conditions under
which the gauge-invariant description of elastic electron
scattering can be achieved in other models for constituent
interaction. In addition, we discuss a certain extension of the WT
identity \cite{Naus:92,Friar:92} for an arbitrary system of
charged particles.

   For a two-fermion system, the Mandelstam current can be
represented as
\begin{eqnarray}\label{gauge1}
J_\mu = J^{(1)}_\mu + J^{(2)}_\mu ,
\end{eqnarray}
where $J^{(1)}_\mu \equiv J^{RIA}_\mu$ and $J^{(2)}_\mu$, are,
respectively, the one-particle, and two-particle contributions
that satisfy the relations~\cite{Bentz}
\begin{eqnarray}\label{gauge2}
i q^\mu && \hspace*{-.9cm}J^{(RIA)}_\mu (k^\prime,k; P^\prime, P)
= e_1 \delta (k^\prime -k-q/2) \left[S^{(1)} (P/2+k)^{-1}-S^{(1)}
(P^\prime /2+k^\prime)^{-1}
\right]S^{(2)} (P/2-k)^{-1}+\nonumber\\
&+&e_2 \delta (k^\prime-k+q/2) \left[S^{(2)} (P/2-k)^{-1}-S^{(2)}
(P^\prime /2-k^\prime)^{-1} \right]S^{(1)} (P/2+k)^{-1},\\[.2cm]
i q^\mu && \hspace*{-.9cm}J^{(2)}_\mu (k^\prime ,k; P^\prime, P) =
e_1 V(P,k^\prime-q/2,k)-
V(P^\prime,k^\prime,k+q/2;) e_1 +\nonumber\\
&+& e_2 V(P, k^\prime +q/2,k)-V(P^\prime, k^\prime ,k-q/2) e_2.
\label{gauge3}\end{eqnarray} Here, $k$ and $P$ ($k^\prime$ and
$P^\prime$)  are the relative and total 4-momenta in the initial
(final) states, respectively; $q = (\omega, \q)$ is the 4-momentum
transfer; $P^\prime= P+q$; $S^{(1,2)}(k)$ is the dressed nucleon
propagator; and $V(P,k^\prime, k)$ is the kernel of the BS
equation. This kernel depends on the relative momenta $k$ and
$k^\prime$ and on the total momentum $P$. In the isospin
formalism, we have
\begin{eqnarray}\label{gauge4}
e_i=|e|\frac{1+\tau_z(i)}{2}
\end{eqnarray}
where $e$ is an electron charge, and $\tau_z(i)$ are the Pauli
matrices (i = 1, 2).

   It is worth noting that equations (\ref{gauge2}) and (\ref{gauge3}) do not define
the current completely. They only impose certain constraints on
the longitudinal component of the current.

  The amplitude of elastic $e{\rm D}$ scattering can be represented in
the form (see (\ref{eqn:M}))
\begin{equation}\label{gauge5}
  M^J_{M^\prime M}= \epsilon^\mu \int \frac{d^4k}{(2\pi)^4}
  \frac{d^4k^\prime}{(2\pi)^4}\bar \Phi^{JM^\prime}
  (P^\prime,k^\prime)J_\mu (k^\prime,k,P^\prime, P)\Phi^{JM}(P,k)\equiv \epsilon^\mu ({\cal M}^J_{M^\prime M})_\mu,
\end{equation}
where $\Phi^{JM}(P,k)$ ($\Phi^{JM^\prime}
(P^\prime,k^\prime)$)~(\ref{reldp}) is the BS amplitude, which
describes the initial (final) state, and $\epsilon^\mu$ is the
virtual-photon polarization vector. The gauge-independence
condition
\begin{equation}\label{gauge6}
    q^\mu ({\cal M}^J_{M^\prime M})_\mu=q^\mu [({\cal M}^J_{M^\prime M})_\mu ^{RIA}+({\cal M}^J_{M^\prime M})_\mu ^{(2)}]=0
\end{equation}
is met if the current satisfies identities (\ref{gauge2}) and
(\ref{gauge3}) and if the amplitudes $\Phi^{JM}(P,k)$ and
$\Phi^{JM^\prime} (P^\prime,k^\prime)$ satisfy the BS equation
with the same kernel~(\ref{f013}).

  In the relativistic impulse approximation illustrated in  Fig.~\ref{elastic:RIA}, the
deuteron matrix element ${\cal M}^{RIA}$, defined by the
 Eq.(\ref{gauge6}), for on-shell $\gamma NN$ vertex satisfies the gauge-independence
 condition~\cite{gauge}
\begin{equation}\label{gauge11}
        q^\mu ({\cal M}^J_{M^\prime M})_\mu ^{RIA}=0.
\end{equation}
It implies that
\begin{equation}\label{gauge12}
        q^\mu ({\cal M}^J_{M^\prime M})_\mu ^{(2)}=\int \frac{d^4k}{(2\pi)^4}
  \frac{d^4k^\prime}{(2\pi)^4}\bar \Phi^{JM^\prime}
  (P^\prime,k^\prime)J^{(2)}_\mu (k^\prime,k,P^\prime,
  P)\Phi^{JM}(P,k)=0.
\end{equation}

It turns out that, for certain models of interaction, this
relation holds. Let us isolate isospin structure in $V$. We have
\begin{equation}\label{gauge13}
    V(P,k^\prime,k) = \sum\limits_{T=0,1}\Pi_T V_T(P,k^\prime,k),
\end{equation}
where $\Pi_T$ is the projection operator onto the state with total
isospin $T$, and $V_T$ is the corresponding component of
interaction. In the ladder approximation,
\begin{equation}\label{gauge14}
    V_T(P,k^\prime,k)={\cal V}_T(k^\prime-k),
\end{equation}
 we can verify that
the condition (\ref{gauge12}) is satisfied for any function ${\cal
V}_T(k^\prime-k)$. This result is independent of the isospin value
in  the initial or final state, although only one ${\cal
V}_0(k^\prime-k)$ component contributes to (\ref{gauge12}) in the
case  of $eD$-scattering. We note that all OBE interactions can be
represented in the form (\ref{gauge14}) and that the corresponding
amplitudes are gauge independent, in accord with the result
reported in~\cite{Zu_Tjon:80,Zu_Tjon:81}.

   As the next example, we consider the interaction described by a
separable potential~(\ref{s001}). The authors of
\cite{graz:rel,Rupp:88} proved that the amplitude in the impulse
approximation is a gauge-independent quantity in this case. Their
proof is based on the transformation properties of the quantities
in Eq.(\ref{gauge12}) under boosts.

   However, this result can be obtained in a simpler way by
calculating the contraction in (\ref{gauge6}) and by considering
that, for interaction of the form~(\ref{s001}), the BS amplitude
in (\ref{gauge12})
 is independent of the total momentum $P$ (see~\cite{gauge}).

  Thus, we can conclude that, in the ladder and separable
approximations for the kernel of the BS equation, the amplitude of
elastic scattering in the impulse approximation is gauge
independent. The common feature of the kernels in this section is
that they are independent of the total momentum of the pair.

   These conclusions do not mean that other gauge-independent
contribution can always be disregarded (for example, the
contribution of the exchange currents). Indeed, the results
reported in \cite{IBG,Anikin}, where the $q\bar q$ system with
separable interaction was studied, revealed that two-particle
currents exert a noticeable effect on the pion charge form factor,
especially at high momentum transfers. We can state with
confidence that, for a certain model of $NN$ interaction,
two-particle and more complicated electromagnetic currents must be
taken into account in a consistent manner even in calculating the
elastic nuclear form factors. However, there is no universal
recipe for constructing these currents for a given interaction of
particles in a bound system.

\subsection{\em Relativistic Impulse Approximation}\label{BS:eD-RIA}
\begin{figure}[ht]
\begin{tabular}{cc}
\begin{minipage}[t]{0.5\linewidth}
\hskip 10mm
\includegraphics[width=0.6\linewidth]{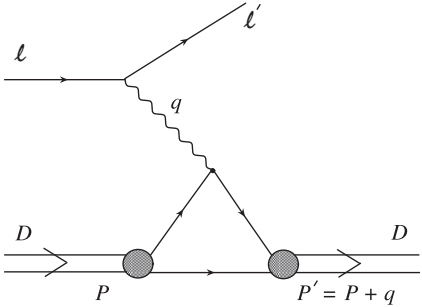}
\end{minipage}
&
\begin{minipage}{0.4\linewidth}
\vskip -40mm \caption{\label{elastic:RIA} Relativistic impulse
approximation for the elastic electron deuteron scattering.}
\end{minipage}
\end{tabular}
\end{figure}

In the relativistic impulse approximation illustrated in
Fig.~\ref{elastic:RIA}, the deuteron current matrix element
can be written as
\begin{eqnarray}
\langle D^{\prime}{\cal M}^{\prime} | J^{RIA}_{\mu} | D {\cal M}
\rangle =ie \int \frac{d^4k}{(2\pi)^4} \Tr \biggl\{ {\bar
\chi^{_{1{\cal M}^{\prime}}}}(P^{\prime},k^{\prime}) \Gamma^{\rm
(S)}_{\mu}(q) \chi^{_{1{\cal
M}}}(P,k)({P\cdot\gamma}/{2}-{k\cdot\gamma}+m) \biggr\},
\label{fff}\end{eqnarray} where $\chi^{_{1{\cal M}}}(P,k)$ is the
BS amplitude of the deuteron, $P^{\prime}=P+q$ and
$k^{\prime}=k+q/2$. The vertex of $\gamma NN$ interaction,
\begin{eqnarray}
\Gamma_{\mu}^{\rm (S)}(q)=\gamma_{\mu} F_1^{\rm (S)}(q^2)
-\frac{\gamma_{\mu} {q\cdot\gamma} - {q\cdot\gamma}
\gamma_{\mu}}{4m} F_2^{\rm (S)}(q^2), \label{nngs}\end{eqnarray}
is chosen to be the form factor on mass shell. The isoscalar form
factors of the nucleon $F_{1,2}^{\rm (S)}(q^2) = (F_{1,2}^{\rm
(p)}(q^2) + F_{1,2}^{\rm
  (n)}(q^2))/2$, which
appeared due to the summation of  two nucleons, are normalized as
$F_1^{\rm (S)}(0) = 1/2$ and $F_2^{\rm (S)}(0) =
(\varkappa_p+\varkappa_n)/2$ with $\varkappa_p = \mu_p - 1$ and
$\varkappa_n = \mu_n$ being anomalous parts of the proton
($\mu_p$) and neutron ($\mu_n$) magnetic moments, respectively.

It should be noted that the choice of the $\gamma NN$ vertex in a
form~(\ref{nngs}) and disregard  the current interaction (CI)
(two-nucleon currents), in general, breaks the gauge independence
of the reaction. Nevertheless, as it was shown in~\cite{gauge},
there exist two cases when RIA and CI are gauge independent
separately. The first case corresponds to the separable
interaction with no dependence on the total momentum in the radial
functions,  and the second one to the one-meson exchange kernel.

First, the trace was taken in Eq.~(\ref{fff}). The covariant form
for the BS amplitude~(\ref{covar}) was used. After taking the
trace, the scalar products of 4-momenta ($P,k,q$) and the deuteron
polarization 4-vectors ($\xi_{\cal M},\xi^{*}_{{\cal
M}^{\prime}}$) with definite spin projections were inserted. Then,
using Eqs.~(\ref{h2pj1}), functions $h_i$ were expressed in terms
of functions $\phi_{JLS\varrho}$ (see section~\ref{BS:3s1}). All
scalar products were evaluated in the laboratory frame.

The resulting expressions for the deuteron current matrix element
can be written as
\begin{eqnarray}
&& \langle D^{\prime}{\cal M}^{\prime} | J^{RIA}_{\mu} | D {\cal
M} \rangle ={\cal I}_{1\;\mu}^{{\cal M}^{\prime}{\cal
M}}(q^2)\;F_1^{\rm (S)}(q^2)+ {\cal I}_{2\;\mu}^{{\cal
M}^{\prime}{\cal M}}(q^2)\;F_2^{\rm (S)}(q^2),
\label{fffd}\\
&& {\cal I}_{1,2\;\mu}^{{\cal M}^{\prime}{\cal M}}(q^2)=ie \int d
k_0\;|\bk|^2\;d|\bk|\;d(\cos{\theta})
\hspace*{-.5cm}\sum_{L^{\prime}S^\prime\varrho^\prime,LS\varrho}\hspace*{-.5cm}\phi_{_{1L^{\prime}S^\prime\varrho}}(k_0^{\prime},|\bk^{\prime}|)
\phi_{_{1LS\varrho}}(k_0,|\bk|)\;I^{L^{\prime},L}_{1,2\;{\cal
M^{\prime}}{\cal M}\;\mu} (k_0,|\bk|,\cos{\theta},q^2),
\nonumber\end{eqnarray} where the function
$I^{L^{\prime},L}_{1,2\;{\cal M^{\prime}}{\cal M}\;\mu}
(k_0,|\bk|,\cos{\theta},q^2)$ is a result of the trace calculation
and the substitution of the scalar products into Eq.~(\ref{fff}).

In Eq.~(\ref{fffd}), the radial part of the BS amplitude for the
final deuteron
$\phi_{1L^{\prime}S^\prime\varrho^\prime}(k_0^{\prime},|\bk^{\prime}|)$
depends on the components of the 4-vector $k^{\prime}$ calculated
in the rest frame ($rf$). The vectors in the rest frame and in the
laboratory frame are related via Lorentz transformation:
\begin{eqnarray}
P^{\prime}_{lab} = {\cal L} P^{\prime}_{rf} = {\cal L} (M,{\bf
0}), \quad k^{\prime}_{lab} = {\cal L} k^{\prime}_{rf},
\label{transfv}\end{eqnarray} where the Lorentz transformation
matrix ${\cal L}$ is of the form:
\begin{eqnarray}
{\cal L} = \left(
\begin{array}{cccc}
1+2\eta & 0 & 0 & 2\sqrt{\eta}\sqrt{1+\eta} \\
0 & 1 & 0 & 0 \\
0 & 0 & 1 & 0 \\
2\sqrt{\eta}\sqrt{1+\eta} & 0 & 0 & 1+2\eta
\end{array}
\right) \label{mtrtransf}\end{eqnarray}

To simplify the notations, the components  of the 4-vector
$k^{\prime}_{rf}$ are denoted by $k^{\prime} \equiv
k^{\prime}_{rf} =
(k_0^{\prime},k_x^{\prime},k_y^{\prime},k_z^{\prime})$, and
$|{\bk}^{\prime}| = |{\bk}^{\prime}_{rf}| =\sqrt{k_x^{\prime\,2} +
k_y^{\prime\,2} + k_z^{\prime\,2}}$. Using relations (\ref{vlab}),
(\ref{transfv}) and (\ref{mtrtransf}) one finds
\begin{eqnarray}
k_0^{\prime} &=& (1+2\eta)k_0 - 2\sqrt{\eta}\sqrt{1+\eta}k_z -
M\eta,
\nonumber\\
k_x^{\prime} &=& k_x, \quad k_y^{\prime} = k_y,\nonumber \\
k_z^{\prime} &=& (1+2\eta)k_z - 2\sqrt{\eta}\sqrt{1+\eta}k_0 +
M\sqrt{\eta}\sqrt{1+\eta}, \label{pprimecm}\end{eqnarray} where
$k_0, k_x, k_y, k_z$ are the components of the 4-vector $k$ in the
deuteron rest frame.

\subsection{\em Numerical Results}\label{BS:eD-res}
The calculations were performed with a covariant separable kernel
of $NN$-interaction Graz II~(for details, see
section~\ref{BS:Graz}). Considering only $\Spp$ and $\Dpp$ states
one writes (see Eq.~(\ref{g2pri}))
$$\phi_{_{1LS+}}(k_0,|\bk|) = S_+(k_0,|\bk|;s)g_{_{1LS+}}(k_0,|\bk|),$$
where $g_{_{1LS+}}$ is the radial part of the vertex function.
Thus, the Bethe-Salpeter amplitude involves singularities in the
$k_0$ plane, which are infinitesimally close to the real axis.
Some of the singularities arise from the propagator, while the
others come from the radial part of the vertex function --- in
other words, from the functions $g_i^{1LS+}$ defined in
Eqs.~(\ref{gfactors}).

For the initial deuteron, the singularities do not depend on $q^2$
(or $\eta$) and always remain in the same quadrant:
\begin{eqnarray}
&& \mbox{for the propagator} \qquad\quad k_0=\pm M/2 \mp
\sqrt{\bk^2+m^2} \pm i\epsilon,
\nonumber\\
&& \mbox{for the functions $g_i^{1LS+}$} \quad\quad k_0=\pm
\sqrt{\bk^2+\beta_i^2}\mp i\epsilon. \nonumber\end{eqnarray}

The situation changes for the final deuteron. Due to the boost of
the arguments~(\ref{pprimecm}) of the amplitude, the singularities
depend on $q^2$ (or $\eta$) and can go across the imaginary axis
and appear in another quadrant (mobile singularities). The
positions of the singularities are the following:
\begin{eqnarray}
&& \mbox{for the propagator}\hskip 120mm
\nonumber\\
&& \hskip 5mm
k_0=-(1+4\eta)M\pm\sqrt{\bk^2+m^2+4\sqrt{\eta(1+\eta)}M|\bk|\cos{\theta}
+4\eta^2(1+\eta)^2M^2}\pm i\epsilon,
\nonumber\\
&& \mbox{for the functions $g_i^{(L)}$}\hskip 100mm
\nonumber\\
&& \hskip 5mm k_0=-\eta
M\pm\sqrt{\bk^2+\beta_i^2+2\sqrt{\eta(1+\eta)}M|\bk|\cos{\theta}
+\eta^2(1+\eta)^2M^2}\mp i\epsilon. \nonumber\end{eqnarray}

The mobility of the singularities does not affect the calculations
if the Cauchy theorem is applied. But for the Wick rotation
procedure, this means that the additional contributions (the
residues at these mobile singularities) should be taken into
account. The minimal value of $-q^2$ for which the imaginary axis
is traversed is: for propagator $-q^2=M(2m-M)\approx 4.17\times
10^{-3}$ (GeV/c)$^2$, for $g_i^{1LS+}$ functions
$-q^2=4M\beta_i\approx 1.17$ (GeV$/c)^2$. The contributions of the
residues from the functions $g_i^{1LS+}$ are negligible (about
$1\%$) in the region $-q^2 < 2$ (GeV/c)$^{2}$,
 and are getting larger with growing
momentum transfer. But the contribution of the residue from the
propagator is very large and can modify the curves significantly
even in the region $-q^2 < 2$ (GeV/c)$^{2}$.

\begin{figure}
\begin{center}
\epsfxsize=90mm \epsfbox{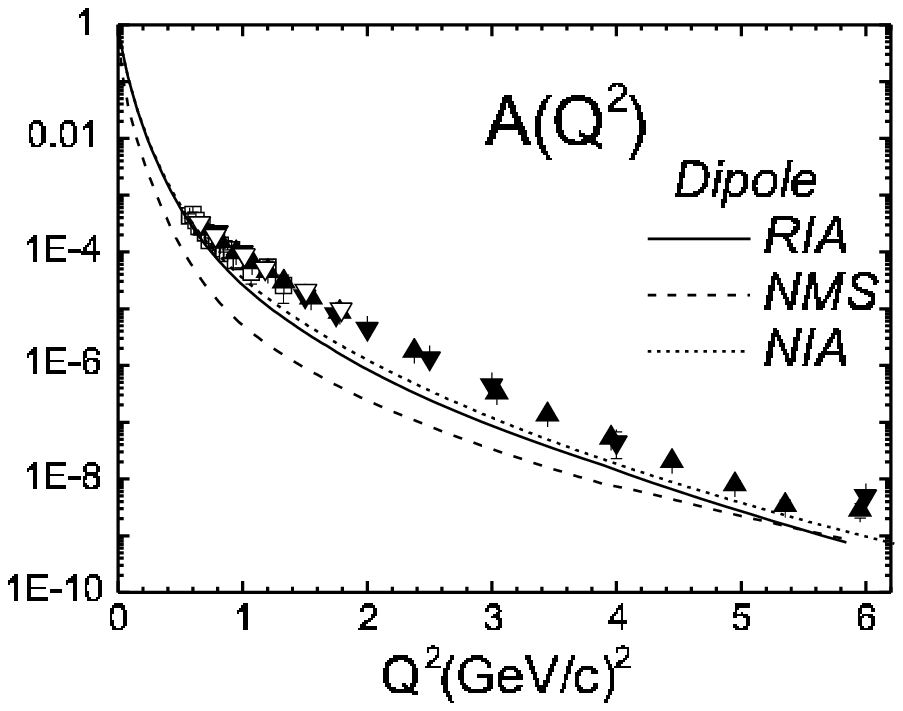} \epsfxsize=90mm
\epsfbox{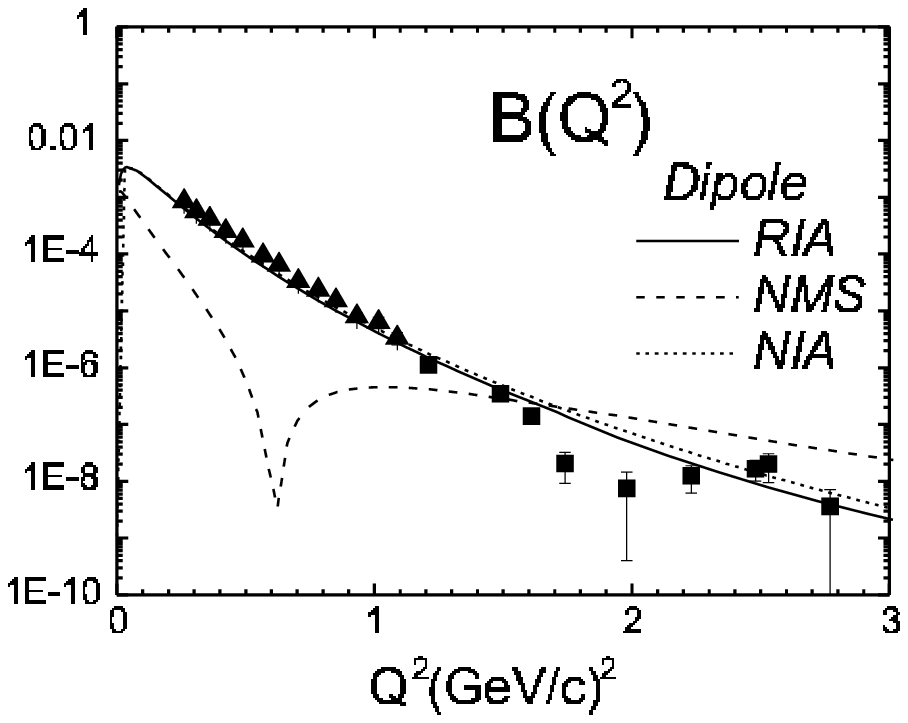} \caption{\label{el-fig34}(a) Structure
function $A(q^2)$. Long dashes represent the calculation without
the contribution of mobile singularities (no mobile singularities
- NMS). The solid curve shows the full relativistic impulse
approximation calculation. Short dashes correspond to the
nonrelativistic impulse approximation calculation (nonrelativistic
Graz II potential). Experimental data are taken
from~\protect\cite{data:A1}-\protect\cite{arnold:A},\protect\cite{Alexa:1999,Abbott:1999}.
(b) Structure function $B(q^2)$. Notations for the curves are
identical to those of Fig.~\protect\ref{el-fig34}a. Experimental
data are taken
from~\protect\cite{data:A1,data:B2,data:B3,arnold:B,Bosted:1990}.}
\end{center}
\end{figure}

The contribution of the residue from the propagator is shown in
Fig.~\ref{el-fig34} for the functions $A(q^2)$ and $B(q^2)$. The
contribution is substantially large both for the function $A(q^2)$
and the function $B(q^2)$ (for the function $B(q^2)$, this
contribution fills the minimum, which does not exist in the
experimental data). This result can be considered as a specific
relativistic effect caused by the Lorentz transformation in the
arguments of the Bethe-Salpeter amplitude (vertex functions and
propagator).
The comparison of the relativistic (RIA) and the nonrelativistic
(NRIA) calculations with the separable kernel of the
$NN$-interaction Graz~II is shown in Fig. \ref{el-fig78}, also.
One can see that the difference between the RIA and the NRIA
calculations rather small. However we must keep in mind that the
RIA calculations does not include the contribution of the
$P$-waves to the deuteron (further development of the RIA
calculations must take into account this contribution) and the
NRIA calculations does not take into account the mesonic exchange
currents. Only after that we can draw the conclusion about the
full contribution of the relativistic effects to the structure
functions of the deuteron.

Yet another interesting result of the investigations is the
dependence of the deuteron form factors on the nucleon form
factors --- in particular, on the neutron electric form factor
$G_{\rm E}^{\rm n}(q^2)$. The electric and the magnetic form
factors of nucleons ($G_{\rm E}(q^2)$ and $G_{\rm M}(q^2)$,
respectively) are related to the Dirac and Pauli form factors
($F_1(q^2)$ and $F_2(q^2)$, respectively) by the equations
\begin{eqnarray}
G_{\rm E}(q^2) &= &F_1(q^2) + \frac{q^2}{4m^2} F_2(q^2),
\nonumber\\
G_{\rm M}(q^2) &= &F_1(q^2) + F_2(q^2).
\label{eqn:sachs}\end{eqnarray}

Three sets of the nucleon form factors are used in the
calculations. The first set is so-called {\em dipole fit}
\begin{eqnarray}
&& G_{\rm M}^{\rm p}(q^2)=(1+\kappa_{\rm p})G_{\rm E}^{\rm
p}(q^2), \qquad
G_{\rm M}^{\rm n}=\kappa_{\rm n} G_{\rm E}^{\rm p}(q^2), \qquad
G_{\rm E}^{\rm n}(q^2)=0,
\nonumber\\
&& G_{\rm E}^{\rm p}(q^2)=1/(1-q^2/0.71{\mbox ({\rm
GeV}/c)}^2)^{2}, \label{dipole}\end{eqnarray} where $\kappa_{\rm
p}=1.7928$, $\kappa_{\rm n}=-1.9130$ are the anomalous magnetic
moments of the nucleons. The second set was suggested by the {\em
vector meson dominance model} (VMDM)~\cite{VMDM}
\begin{eqnarray}
&& F_1^{(\rm S)}(t) = \left[
\frac{m_{\omega}^2}{m_{\omega}^2-q^2}\gamma_{\omega}
+(1-\gamma_{\omega}) \right] F_{1L},
\label{vmdm}\\
&& F_2^{(\rm S)}(t) = \left[ \frac{m_{\omega}^2}{m_{\omega}^2-q^2}
\kappa_{\omega}\gamma_{\omega}
+(1+\kappa_p+\kappa_n-\kappa_{\omega}\gamma_{\omega}) \right]
F_{2L},
\nonumber\\
&& F_{1L} = \frac{\lambda_1^2}{\lambda_1^2+{\hat q}^2}
\frac{\lambda_2^2}{\lambda_2^2+{\hat q}^2}, \qquad F_{2L} =
\frac{\lambda_1^2}{\lambda_1^2+{\hat q}^2}
\left(\frac{\lambda_2^2}{\lambda_2^2+{\hat q}^2}\right)^2,
\nonumber\\ && {\hat q}^2\, =\, -q^2\,
\frac{\ln{(\lambda_2^2-q^2)/\lambda_{3}^2}}
{\ln{\lambda_2^2/\lambda_{3}^2}}, \nonumber\end{eqnarray} where
$\lambda_1 = 0.795$ (GeV/c), $\lambda_2 = 2.27$ (GeV/c),
$\lambda_{3} = 0.29$ (GeV/c), $\kappa_{\omega} = 0.163$,
$\gamma_{\omega} = 0.411$, $m_{\omega} = 0.784$ (GeV/c). The third
set is that from {\em relativistic harmonic oscillator model}
(RHOM)~\cite{RHOM}
\begin{eqnarray}
&&G_{\rm E}^{\rm p}=I^{(3)}(q^2),
\label{rhom}\\
&&G_{\rm E}^{\rm n}=-\frac{q^2}{2m^2}I^{(3)}(q^2),
\nonumber\\
&&\frac{G_{\rm M}^{\rm p}(q^2)}{1+\kappa_p} = \frac{G_{\rm M}^{\rm
n}(q^2)}{\kappa_n} = I^{(3)}(q^2),
\nonumber\\
&&I^{(3)}(q^2) = \frac{1}{(1-q^2/2m^2)^2}\, {\rm
exp}\left(\frac{1}{2\alpha_3}\ \frac{q^2}{1-q^2/2m^2}\right),
\qquad\qquad \alpha_3 = 0.42 ({\rm GeV}/c)^2.
\nonumber\end{eqnarray} The first model assumes that the neutron
electric form factor is equal to zero. Two other models lead to a
nonzero $G_{\rm E}^{\rm n}$.

\begin{figure}
\begin{center}
\epsfxsize=90mm \epsfbox{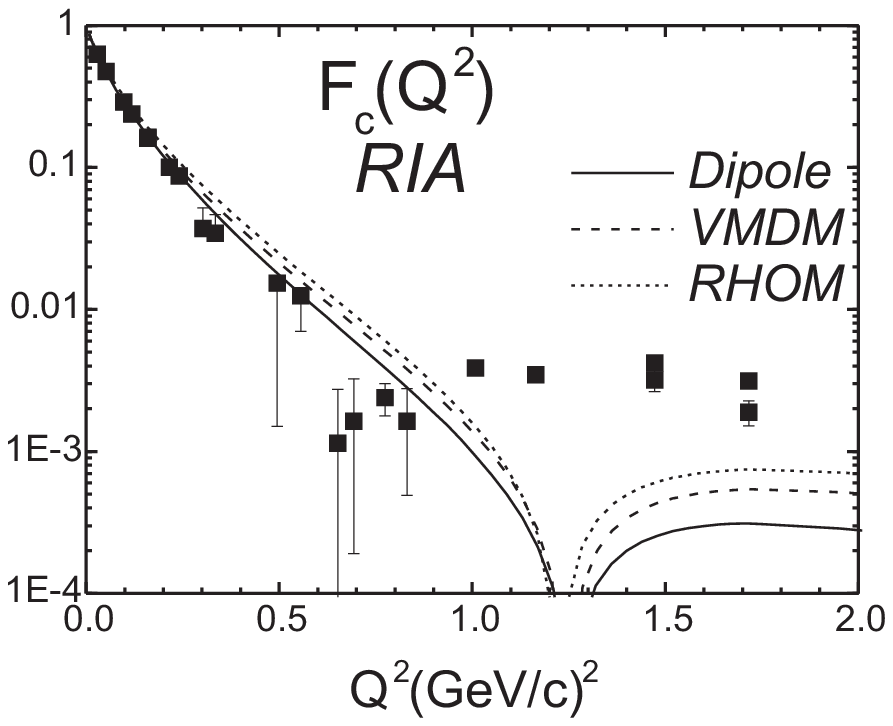} \epsfxsize=90mm
\epsfbox{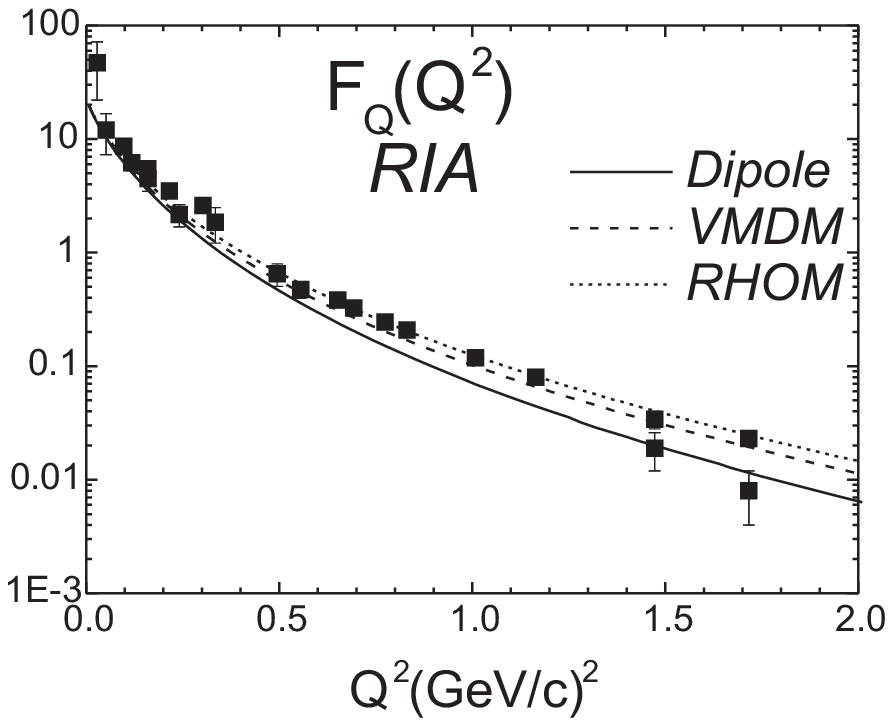} \caption{\label{el-fig56}(a) Charge form
factor $F_{\rm C}(q^2)$. Long and short dashes represent
calculations with the VMDM and RHOM nucleon form factors,
respectively. The solid curve corresponds to the dipole fit. The
experimental data are taken from
ref.~\protect\cite{Abbott:2000:EPJ}. (b) Quadrupole form factor
$F_{\rm Q}(q^2)$. Notations for the curves are identical to those
of Fig.~\protect\ref{el-fig56}a, and the experimental data
originate from the same source.}
\end{center}
\end{figure}

\begin{figure}
\begin{center}
\epsfxsize=90mm \epsfbox{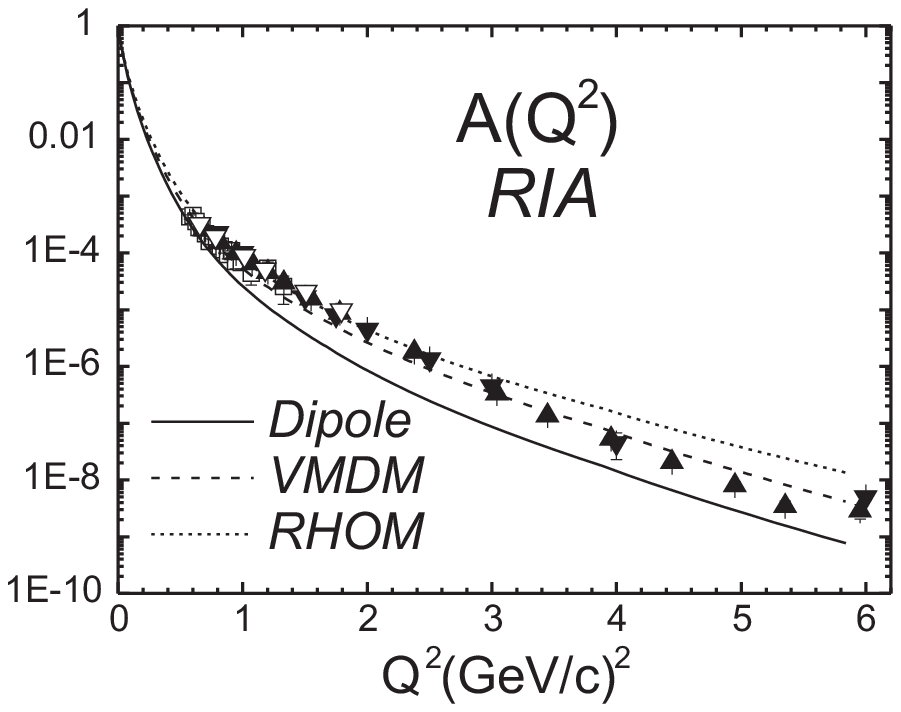} \epsfxsize=90mm
\epsfbox{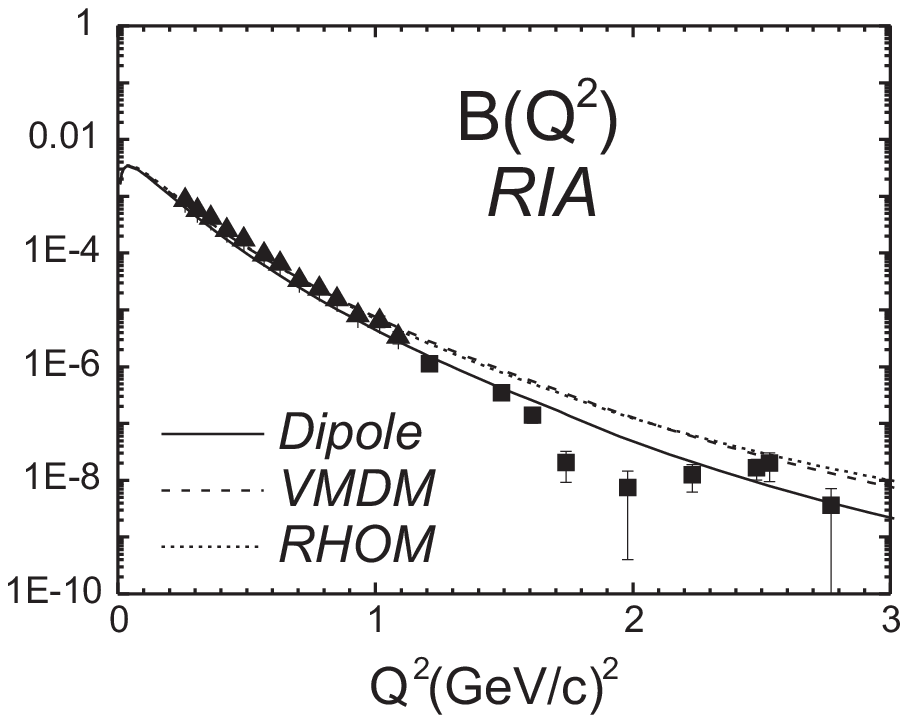} \caption{\label{el-fig78}(a) Deuteron
structure function $A(q^2)$. Notations for the curves are
identical to those of Fig.~\protect\ref{el-fig56}a. Experimental
data originate from the same source as in
Fig.~\protect\ref{el-fig34}a. (b) Deuteron structure function
$B(q^2)$. Notations for the curves are identical to those of
Fig.~\protect\ref{el-fig56}b. The experimental data originate from
the same source as in Fig.~\protect\ref{el-fig34}b.}
\end{center}
\end{figure}

\begin{figure}
\begin{center}
\epsfxsize=90mm \epsfbox{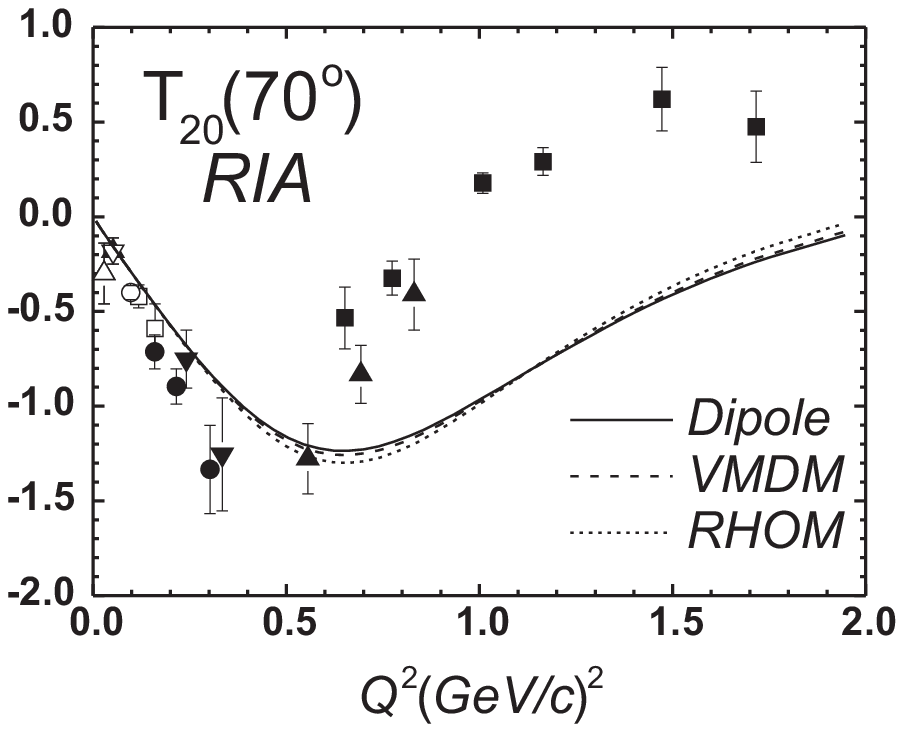} \epsfxsize=90mm
\epsfbox{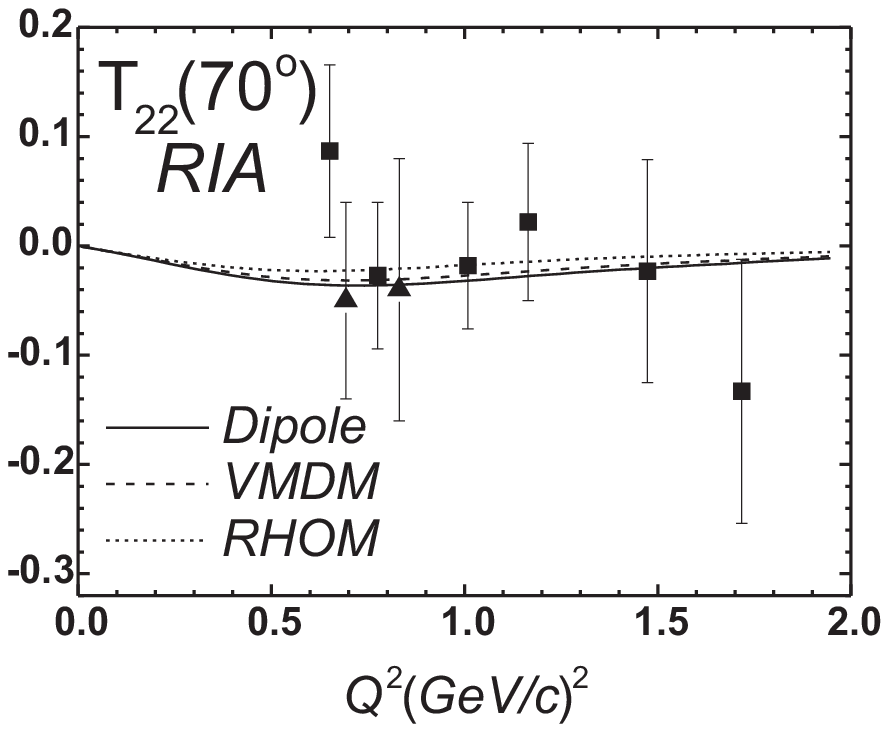}
\end{center}
\caption{\label{el-fig911}(a) Tensor polarization component
$T_{20}(q^2)$ calculated at $\theta_e = 70^{\circ}$. Notations for
the curves are identical to those of Fig.~\protect\ref{el-fig56}a.
Experimental data are taken from
ref.~\protect\cite{Abbott:2000:EPJ} (values from
~\protect\cite{Garcon:1994,Abbott:2000:PRL},\protect\cite{Dmitriev:1985}-\protect\cite{Bouwhuis:1999}
were recalculated at $\theta_e = 70^{\circ}$). (b) Tensor
polarization components $T_{22}(q^2)$ calculated at $\theta_e =
70^{\circ}$. Notations for the curves are identical to those of
Fig.~\protect\ref{el-fig56}a. Experimental data are taken from
ref.~\protect\cite{Garcon:1994,Abbott:2000:PRL}.}
\end{figure}

\begin{figure}
\begin{center}
\epsfxsize=90mm \epsfbox{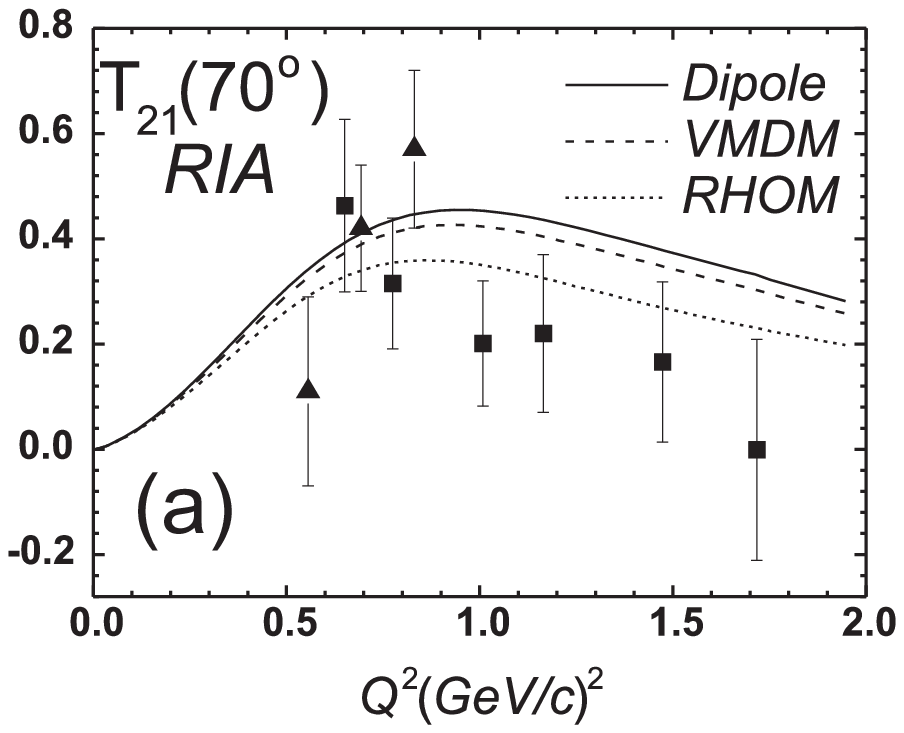} \epsfxsize=90mm
\epsfbox{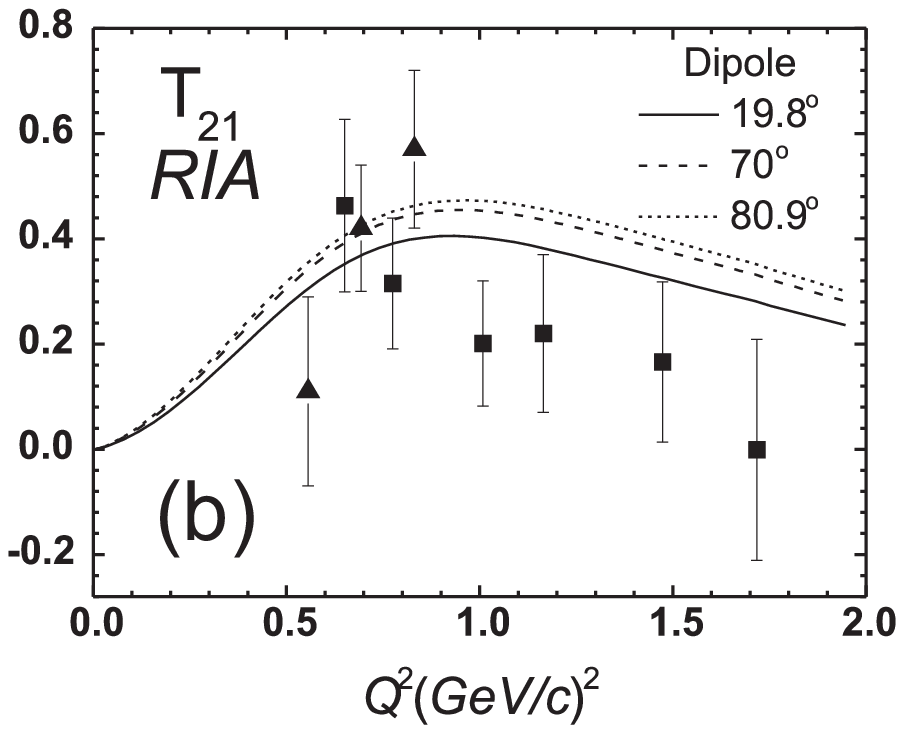}
\\
\vskip 2cm \epsfxsize=90mm \epsfbox{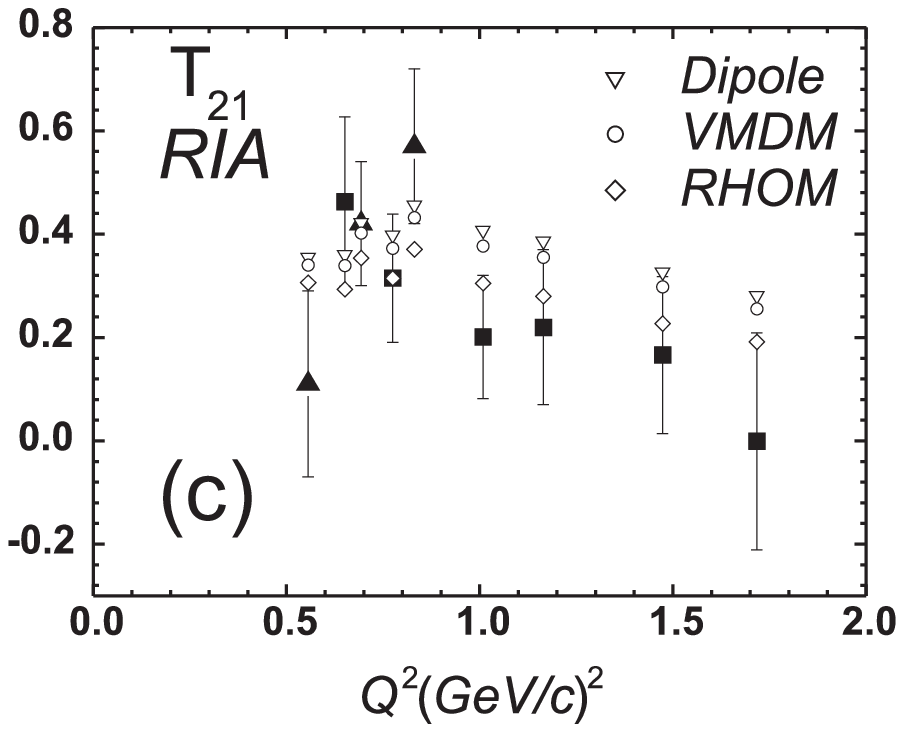}
\end{center}
\caption{\label{el-fig911a} (a) Tensor polarization component
$T_{21}(q^2)$ calculated at $\theta_e = 70^{\circ}$. Notations
 for the curves are
identical to those of Fig.~\protect\ref{el-fig56}a. Experimental
data originate from the same source as in
Fig.~\protect\ref{el-fig56}b. (b) Tensor polarization component
$T_{21}(q^2)$ calculated at different angles
$\theta_{\el}=19.8^{\circ}, 70^{\circ}, 80.9^{\circ}$ (solid
curve, long and short dashes, respectively) with dipole nucleon
form factor. The data points are from the same source as in
Fig.~\protect\ref{el-fig911a}b. (c) Tensor polarization component
$T_{21}(q^2)$ calculated with dipole fit ($\nabla$), VMDM
($\circ$) and RHOM ($\diamond$) at experimental electron
scattering angle values. The  data points originate from the same
source as in Fig.~\protect\ref{el-fig911}b.}
\end{figure}

Figure~\ref{el-fig56} shows the charge and quadrupole form factors
($F_{\rm C}(q^2)$ and $F_{\rm Q}(q^2)$, respectively). The zero of
the form factor $F_{\rm C}(q^2)$ is in the range of $-q^2 = 1.2 -
1.35$ (GeV/$c$)$^{2}$, but experimental data yield $-q^2 = 0.69 -
0.83$ (GeV/$c$)$^{2}$~\cite{Garcon:1994}-\cite{Abbott:2000:EPJ}.
This dip comes from  the specific choice  of the separable Graz II
kernel (in the calculations with non-relativistic Graz II
potential, the zero of $F_{\rm C}(q^2)$ is shifted too). The
nucleon form factors do not shift the zero in the  form factor
$F_{\rm C}(q^2)$. The nucleon form factors with the nonzero
electric from factor for the neutron (VMDM and RHOM) are more
suitable for the description of  the experimental data on the
quadrupole form factor $F_{\rm Q}(q^2)$.
We stress that the contriburion of the $P$-waves (the negative
energy states of the BS amplitude) for the deuteron must shift the
dip of the $F_C(Q^2)$ and must give better agreement with the
experimental data analogously the NRIA calculations with the
taking into account of the mesonic exchange
currents~\cite{BDS:1992}.
The structure functions $A(q^2)$ and $B(q^2)$ are shown in
Fig.~\ref{el-fig78}.
The RIA calculations of the structure functions of the deuteron
show the strong dependence on choice the form factors of the
nucleon. After taking into account the contribution of the
$P$-waves for the deuteron we can make more exact conclusion about
the choice of the  nucleon form factors.
Figures~\ref{el-fig911} and~\ref{el-fig911a}a show the tensor
polarization components $T_{20}(q^2), T_{21}(q^2)$ and
$T_{22}(q^2)$ for the final deuteron calculated with $\theta_e =
70^{\circ}$. It is seen that $T_{20}(q^2)$ and $T_{22}(q^2)$ have
very weak dependence on the nucleon form factors, but for
$T_{21}(q^2)$ it is more pronounced. The $T_{21}(q^2)$ calculated
with three different electron scattering angles $\theta_e =
19.8^{\circ},\ 70^{\circ},\ 80.9^{\circ}$ is shown in the
Fig.~\ref{el-fig911a}b. The change in the electron scattering
angle affects sizably the component $T_{21}(q^2)$.
Figure~\ref{el-fig911a}c shows the result of calculation of the
component $T_{21}(q^2)$ with different nucleon form factors at
electron scattering angles fixed in experiment. This result can be
used to choose between the models for the nucleon form factors.
Unfortunately, large uncertainties in experimental data prevent
from choosing between the sets, and future measurements of the
component $T_{21}(q^2)$ can be very useful for the analysis.

Note that the calculated $T_{20}(q^2)$ differs from the
experimental data in the region $-q^2 > 0.6$ (GeV/c)$^{2}$. This
fact, apparently, could be explained by several reasons. It is
necessary to improve the description of the zero of the charge
form factor $F_{\rm C}(q^2)$ by changing the separable kernel of
$NN$-interaction and taking into account the negative energy
states of the Bethe-Salpeter amplitude for the deuteron. It is
also important to investigate the contribution of the two-body
electromagnetic current. Information about the effect of these
factors provide a powerful tool in the study of the on- and
off-shell behavior of the nucleon form factors in elastic electron
deuteron scattering.

\subsection{\em Electromagnetic Moments of the Deuteron}\label{em-moms}
In the  previous section, the  electromagnetic form factors of the
deuteron were discussed. The relativistic effects
 in the form factors  were analyzed in detail.
The sensitivity of the polarization observables to the models for
the nucleon form factors was discussed too.

Now we consider static electromagnetic characteristics of the
deuteron --- magnetic and quadrupole moments --- in details. As it
was shown in ref.~\cite{Honzawa:92}, the relativistic effects in
BS approach, such as effects of relativistic kinematics,
retardation effects or negative energy states, play essential role
in evaluation of static moments. Our main task is to obtain full
expressions for the moments in RIA and compare results with
nonrelativistic ones.

\subsubsection{\em Definitions}
We define the magnetic moment, $\mu_{\rm D}$, and the quadrupole,
$Q_{\rm D}$, moment of the deuteron from the normalization
condition~(\ref{normf}) of the for form factors
\begin{eqnarray}
\mu_{\rm D} = \frac{m}{M} \lim_{q \to 0} F_{\rm M}(q^2),\quad
Q_{\rm D} = \frac{1}{M^2} \lim_{q \to 0} F_{\rm Q}(q^2).
\label{momdef}\end{eqnarray} Analytical  expressions for the
moments can be obtained from the
 deuteron form factors. We use the Breit system which is defined as
\begin{eqnarray}
P_0=P_0^\prime=E_{\rm D}=\sqrt{M^2+\bP^2},\quad \bP =
-\frac{\bq}2,\quad {\bP}^\prime=\frac{\bq}2.
\end{eqnarray}
Taking the vector ${\bq}$ in the  $Z$-direction we  can write the
4-vectors as
\begin{eqnarray}
&&P=M(\sqrt{1+\eta},0,0,-\sqrt{\eta}), \quad P^{\prime}=M
(\sqrt{1+\eta},0,0,\sqrt{\eta}),
\label{vbreit}\\
&&q=(0,0,0,2M\sqrt{\eta}),
\nonumber\\
&&\xi_{{\cal M}=+1}(P) = \xi_{{\cal M}=+1}(P^{\prime})=
-\frac{1}{\sqrt{2}} (0,1,i,0),
\label{xbreit}\\
&&\xi_{{\cal M}=-1}(P) = \xi_{{\cal M}=-1}(P^{\prime})=
\frac{1}{\sqrt{2}} (0,1,-i,0),
\nonumber\\
&&\xi_{{\cal M}=0}(P) = (-\sqrt{\eta},0,0,\sqrt{1+\eta}),\quad
\xi_{{\cal M}=0}(P^{\prime})= (\sqrt{\eta},0,0,\sqrt{1+\eta}).
\nonumber\end{eqnarray} Taking into account the relations  between
the form factors $F_1$, $F_2$, $G_1$ and $F_{\rm C}$, $F_{\rm M}$,
$F_{\rm Q}$~(\ref{ffconnect}) and using the
formulas~(\ref{vbreit}-\ref{xbreit}) and the deuteron current
parametrization~(\ref{deuteronc}), we obtain
\begin{eqnarray}
&&F_{\rm M} = \frac{e}{M\sqrt{2}} \frac{\langle P^{\prime} {\cal
M}^\prime=+1|J_x|P {\cal M}=0\rangle} {\sqrt{\eta}\sqrt{1+\eta}},
\label{ffbreit}\\
&&F_{\rm Q} = \frac{e}{2M} \frac{\langle P^{\prime} {\cal
M}^\prime=0 |J_0|P {\cal M}=0\rangle- \langle P^{\prime} {\cal
M}^\prime = +1 |J_0|P {\cal M}=+1\rangle} {2\eta\sqrt{1+\eta}}.
\label{ffbreit1}
\end{eqnarray}
With the help of the definitions~(\ref{momdef}) and the last two
equations, the moments $\mu_{\rm D}$ and $Q_{\rm D}$ can be
written as
\begin{eqnarray}
&&\mu_{\rm D}=\frac{m}{M^2\sqrt{2}} \lim_{\eta\to 0} \frac{\langle
P^{\prime} {\cal M}^\prime=+1|J_x|P {\cal M}=0\rangle}
{\sqrt{\eta}\sqrt{1+\eta}},
\label{magnetic}\\
&& \label{quadrupole1} Q_{\rm D}=\frac{1}{2M^3}\lim_{\eta\to 0}
\frac{\langle{P^{\prime} \cal M}^\prime=0 |J_0|P {\cal
M}=0\rangle- \langle P^{\prime} {\cal M}^\prime = +1 |J_0|P {\cal
M}=+1\rangle} {2\eta\sqrt{1+\eta}}.
\end{eqnarray}

These formulas are essential in the calculation of the deuteron
moments. To find the analytical expressions we refer to the RIA
deuteron current matrix element~(\ref{fff}). It should be noted
that in contrast to the BS amplitude in the rest frame ($rf$), the
moments are calculated in the  Breit system. Therefore, the BS
amplitudes and the propagators should be transformed (boosted)
from the rest frame to the Breit frame. These quantities  have the
form ($a\cdot\gamma\equiv a_{\mu}\gamma^{\mu}$)
\begin{eqnarray}
&&\chi^{J{\cal M}}(P,k)=\Lambda({\cal L}) \chi^{J{\cal
M}}(P_{rf},{\cal L}^{-1}k) \Lambda^{-1}({\cal L}),
\label{l1}\\
&&\bar\chi^{J{\cal M^{\prime}}}(P^\prime,k^\prime)=
\Lambda^{-1}({\cal L})\bar\chi^{J{\cal M}^{\prime}}(P_{rf},{\cal
L}k^\prime) \Lambda({\cal L}),
\label{l2}\\
&&\Lambda^{-1} ({\cal L})
(P\cdot\gamma/2-k\cdot\gamma+m)\Lambda({\cal L}) =
(P_{rf}\cdot\gamma/2-{\cal L}^{-1}k\cdot\gamma+m),\hskip 10mm
\label{l3}
\end{eqnarray}
where $\Lambda$ is the Lorenz transformation of the BS amplitude,
\begin{eqnarray}
\Lambda({\cal L})=\frac{M+ P\cdot\gamma\gamma_0}{\sqrt{2M(E_{\rm
D}+M)}},
\label{lorlam}
\end{eqnarray}
and  $P={\cal L}P_{rf}$, $P^\prime={\cal L}^{-1}P_{rf}$, $p={\cal
L}p_{rf}$, $p^\prime={\cal L}^{-1}p^{\prime}_{rf}$. The matrix of
the Lorenz transformation is written as (compare with
Eq.~(\ref{mtrtransf}))
\begin{eqnarray}
{\cal L} = \left(
\begin{array}{cccc}
\sqrt{1+\eta} & 0 & 0 & -\sqrt{\eta} \\
0 & 1 & 0 & 0 \\
0 & 0 & 1 & 0 \\
-\sqrt{\eta} & 0 & 0 & \sqrt{1+\eta}
\end{array}
\right). \label{mtrtransf1}
\end{eqnarray}
Applying expressions~(\ref{l1}-\ref{l3}) to Eq.~(\ref{fff}) one
finds
\begin{eqnarray}
\langle P^\prime {\cal M}^\prime |J^{RIA}_\mu |P {\cal M}\rangle =
i\,e\int \frac{\d^4 k}{(2\pi)^4}\, {\mbox Tr} \left\{
\bar\chi^{J{\cal M}^\prime}(P_{rf},k^\prime)
\tilde\Gamma_\mu(q)\chi^{J{\cal M}}(P_{rf},k) \right.
\nonumber\\
\left. \times (P_{rf}\cdot\gamma/2-k\cdot\gamma+m)
[\Lambda^{-1}({\cal L})]^2\right\}, \label{cur}
\end{eqnarray}
where $k^\prime$ $\equiv k^\prime_{rf}$ is connected with $k$
$\equiv k_{rf}$ and $q$ as (components are given
by~(\ref{pprimecm}))
\begin{eqnarray}
k^{\prime} = {\cal L}^{2} k + \frac{1}{2} {\cal L}q. \label{ltr}
\end{eqnarray}
The boosted  $\gamma N N$ vertex has the form:
\begin{eqnarray}
\tilde\Gamma_\mu(q)=\Lambda({\cal
L})\Gamma_\mu(q)\Lambda^{-1}({\cal L}). \label{boostvertex}
\end{eqnarray}

To find the expressions for the moments we should expand
Eq.~(\ref{cur}) in $\eta$ up to the order $\sqrt{\eta}$ for the
magnetic moment and the order $\eta$ for the quadrupole moment,
and then take  the limit $\eta \to 0$ following
Eqs.~(\ref{magnetic},\ref{quadrupole1}). As it follows from
Eq.~(\ref{cur}), there are objects which give term of order $\eta$
($\sqrt{\eta}$), namely: (i) the BS amplitude of the final
deuteron, including vertex function itself and propagator
$\bar\chi^{J{\cal M}^\prime}(P_{rf},k^\prime)$; (ii) boosted
$\gamma NN$-vertex $\tilde\Gamma_\mu(q)$ ; (iii) transformation
operator of the BS amplitudes $[\Lambda^{-1}({\cal L})]^2$.

Analyzing these factors we could note the following peculiarities
of the current matrix element, which come from the covariant and
the relativistic nature of the formalism: (i) effects connected
with the dependence of the BS amplitude on $k_0 \neq
\sqrt{\bk^2+m^2}$; (ii) effects connected with Lorenz
transformations of the BS amplitude arguments~(\ref{mtrtransf1});
(iii) effects connected with Lorenz transformations of the BS
amplitude itself~(\ref{lorlam}); (iv) effects connected with
negative energy states and transitions between $\sp$-, $\Dp$- and
$^1P_1^{(e),(o)}$-, $^3P_1^{(o),(e)}$-waves.

We note here that $\Lambda({\cal L})$ is written in the limiting
case, $\eta \to 0$ as
\begin{eqnarray}
\Lambda({\cal L}) = 1 - \frac{\sqrt{\eta}}{2}\gamma_3\gamma_0 +
\frac{\eta}{8} +{\cal O}(\eta^{3/2}). \label{lamb1}
\end{eqnarray}

\subsubsection{\em Magnetic Moment of the Deuteron}\label{magnet}
From Eqs.~(\ref{magnetic}), (\ref{cur}) and (\ref{lamb1}) we
obtain the following result for the deuteron magnetic moment:
\begin{eqnarray}
&&\mu_D=\mu_{+} + \mu_{1-} + \mu_{2-} + \mu_{3-}, \label{mmain}
\end{eqnarray}
where the matrix elements corresponding to the transitions between
states with positive energies are denoted by subscript $+$, while
the subscript $-$ means that at least one negative energy state is
included in the matrix element. These terms $\mu$`s have the form
\begin{eqnarray}
&&\mu_+=(\mu_p+\mu_n)(P_{\sp}+P_{\Dp}) -
\frac{3}{2}(\mu_p+\mu_n-\frac{1}{2})P_{\Dp} + R_{+},
\label{muplus}\\
&&\mu_{1-}=\frac{1}{2}(\mu_p+\mu_n)(P_{\Pte}+P_{\Pto})+
\frac{1}{4}(P_{\Pte}+P_{\Pto}) + \label{mu1minus}
\frac{1}{2}(P_{\Pse}+P_{\Pso}) + R_{1-},
\label{mumin1}\\
&&\mu_{2-}=-(\mu_p+\mu_n)P_{\sm}+P_{\sm}+ \label{mu2minus}
\frac{1}{2}(\mu_p+\mu_n)P_{\Dm} - \frac{5}{4}P_{\Dm}+R_{2-},
\label{mumin2}\\
&&\mu_{3-}= \sum_{LS\varrho,L^\prime S^\prime\varrho^\prime}
C^{1LS\varrho,1L^\prime S^\prime\varrho^\prime}, \label{mumin3}
\end{eqnarray}
where $P_{1LS\varrho}$ are the pseudo-probabilities of the
corresponding states defined as
\begin{eqnarray}
&& \sum_{LS\varrho} {P_{1LS\varrho}} = 1,
\nonumber\\
&& P_{1LS\varrho} = i\int \frac{\d k_0 \bk^2 \d
|\bk|}{4M(2\pi)^4}\, \omega_{\varrho}\,
|S_{\varrho}(k_0,|\bk|)\,g_{1LS\varrho}(k_0,|\bk|)|^2,
\\
&& \omega_{+} = 2E_{\bk}-M,\quad \omega_{-} = -2E_\bk-M,\quad
\omega_{e} = \omega_{o} = -M,\quad E_\bk=\sqrt{\bk^2+m^2}.
\label{pseudoprob}
\end{eqnarray}
In Eqs.~(\ref{muplus}-\ref{mumin3}), the diagonal terms are given
explicitly while nondiagonal ones are included into $\mu_{3-}$ and
$R$'s, which are defined as
\begin{eqnarray}
R_{+} &=& -\frac{1}{3}(\mu_p+\mu_n-1+\frac{2m}{M}) H_1^{\sp} -
\frac{m}{M} H_2^{\sp} - \frac{m}{M} H_3^{\sp} - (1-\frac{2m}{M})
P_{\sp}
\label{Rplus}\\[1mm]
&& - \frac{1}{6}(\mu_p+\mu_n-1-\frac{4m}{M}) H_1^{\Dp} -
\frac{m}{M}H_2^{\Dp} - \frac{m}{M} H_3^{\Dp} - \frac{1}{4}
(1-\frac{2m}{M}) P_{\Dp}
\nonumber\\
&& + \frac{\sqrt{2}}{3}(\mu_p+\mu_n-1-\frac{m}{M}) H_1^{\sp,\Dp},
\nonumber\\[1mm]
R_{1-} &=& - \frac{1}{2}(1-\frac{2m}{M}) (\mu_p+\mu_n+\frac{1}{2})
(P_{\Pte} + P_{\Pto}) - \frac{1}{2}(1-\frac{2m}{M}) (P_{\Pe} +
P_{\Po})
\label{R1minus}\\[1mm]
&&
+\frac{2}{5}(2H_4^{\Pto,\Pte}-H_8^{\Pto,\Pte})+\frac{1}{5}H_9^{\Pto,\Pte}-
\frac{2}{5}H_{10}^{\Pto,\Pte}+\frac{2}{5}(H_4^{\Pse,\Pso}+2H_8^{\Pse,\Pso})
\nonumber\\[1mm]
&&-\frac{2}{5}H_9^{\Pse,\Pso}+\frac{4}{5}H_{10}^{\Pse,\Pso}+
\sqrt{2} (\mu_p+\mu_n-1+\frac{4m^2}{M^2})H_5^{\Pte,\Pso} + +
\sqrt{2} (\mu_p+\mu_n-1)H_5^{\Pto,\Pse}
\nonumber\\[1mm]
&& - \frac{\sqrt{2}}{2}H_6^{\Pto,\Pse} - 2\sqrt{2}
H_7^{\Pto,\Pse},
\nonumber\\[1mm]
R_{2-} &=& - \frac{1}{3}(\mu_p+\mu_n-1-\frac{2m}{M}) H_1^{\sm} -
\frac{m}{M} H_2^{\sm} + \frac{m}{M} H_3^{\sm} -
\frac{1}{6}(\mu_p+\mu_n-1+\frac{4m}{M}) H_1^{\Dm}
\label{R2minus}\\[1mm]
&& - \frac{m}{M} H_2^{\Dm} + \frac{m}{M} H_3^{\Dm} + \frac{3}{4}
(1-\frac{2m}{M}) P_{\Dm} + \frac{\sqrt{2}}{3}
(\mu_p+\mu_n-1+\frac{m}{M}) H_1^{\sm,\Dm}. \nonumber\end{eqnarray}
Functions $C^{1LS\varrho,1L^\prime S^\prime\varrho^\prime}$ and
$H_i^{1L^\prime S^\prime\varrho^\prime,1LS\varrho}$ are given in
ref.~\cite{kkb}. Rewriting equation for $\mu_+$ we find
\begin{eqnarray}
\mu_+=\mu_{NR} + \Delta \mu_+, \label{mmainplus}
\end{eqnarray}
where the equation
\begin{eqnarray}
\mu_{NR} =
(\mu_p+\mu_n)-\frac{3}{2}(\mu_p+\mu_n-\frac{1}{2})P_{\Dp}~,
\nonumber\end{eqnarray} gives the nonrelativistic formula for the
deuteron magnetic moment, while the expression
\begin{eqnarray}
\Delta \mu_+=R_+ - (\mu_p+\mu_n)
(P_{\sm}+P_{\Dm}+P_{\Pse}+P_{\Pso}+P_{\Pte}+P_{\Pto})
\label{mnorm}\end{eqnarray} represents the relativistic
correction. The final expression for the deuteron magnetic moment
is
\begin{eqnarray}
&& \mu_D=\mu_{NR} + \Delta \mu~,
\label{mmmain}\\[1mm]
&& \Delta \mu=R_{+} + \Delta \mu_{-} + \mu_{3-}~,
\nonumber\\[1mm]
&& \Delta \mu_{-} =
-(\mu_p+\mu_n)\left[\frac{1}{2}(P_{\Pte}+P_{\Pto})+
(P_{\Pse}+P_{\Pso})+2P_{\sm}+\frac{1}{2}P_{\Dm}\right]
\nonumber\\[1mm]
&& \hskip 10mm +\frac{1}{4}(P_{^3P_1^{e}}+P_{\Pto}) +
\frac{1}{2}(P_{^1P_1^{e}}+P_{^1P_1^{o}}) + P_{\sm} -
\frac{5}{4}P_{\Dm} + R_{1-} + R_{2-}~.
\label{deltamom}\end{eqnarray} We introduce the nonrelativistic
value for the magnetic moment of the deuteron $\mu_{NR}$ and group
all the relativistic corrections into $\Delta\mu$ term.

\subsubsection{\em Isoscalar Pair Current}\label{sec:isopair}
We want to understand the  physical meaning of the $P$-waves in
the BS amplitude. To this end, we consider nonrelativistic
reduction of the expressions of  the magnetic moment. Starting
from Eq.~(\ref{mmmain}) we perform $k_0$-integration in functions
$H_i^{1LS\varrho,1L^\prime S^\prime\varrho^\prime}$, $R_i$ and
$C^{1LS\varrho,1L^\prime S^\prime\varrho^\prime}$. Using the
Cauchy theorem, we take into account only the pole of the positive
energy propagator component $S_+(k_0,|\bk|;s)$ (see
Eq.~(\ref{spart})), namely $k_0={\bar
k}_0={M}/{2}-E_\bk+i\epsilon$. We disregard the contributions from
the singularities of the other parts of propagator and vertex
functions.

We now introduce the nonrelativistic analogs of the Bethe-Salpeter
vertex functions for $\sp$- and $\Dp$-waves, corresponding to the
positive nucleon energy,
\begin{eqnarray}
\frac{g_{\sp}({\bar k_0},|\bk|)}{M-2E_\bk} \rightarrow -\alpha_1\,
u(|\bk|), \quad\quad \frac{g_{\Dp}({\bar k_0},|\bk|)}{M-2E_\bk}
\rightarrow -\alpha_1\, w(|\bk|), \label{r2nc}
\end{eqnarray}
with $\alpha_1=4\pi\sqrt{2M}$, and the normalization condition for
the nonrelativistic wave functions
\begin{eqnarray}
\int \bk^2 d |\bk|\, (u^2(|\bk|)+w^2(|\bk|)) = \frac{\pi}{2}.
\end{eqnarray}

Performing the integration over  $k_0$ and introducing
nonrelativistic analogs of the functions $g_{\sp,\Dp}$, we
represent the terms appearing in expression
\begin{eqnarray}
\mu_{\rm D} = \mu_{\rm D}^{S^+, D^+} + \mu_{\rm D}^{S^+,P} +
\mu_{\rm D}^{D^+,P}. \label{m4ain}
\end{eqnarray}
The first term ($p_{\rm D} =2/\pi \int {\bk}^2 \d{|\bk|} w^2$ is
the probability of $D$-wave)
\begin{eqnarray}
\mu_{\rm D}^{S^+,D^+}\,=\,\mu_{\rm D}^{NIA}\,
&=&\frac{\alpha_1^2}{8M(2\pi)^3}\,\int\,{\bk}^2\,d{|\bk|}\, \Bigl[
4\,(\mu_p+\mu_n)\,u^2({|\bk|}) - (2(\mu_p+\mu_n)-3)\,w^2({|\bk|})
\Bigr]
\nonumber\\
&=&(\mu_p+\mu_n)-\frac{3}{2}\,(\mu_p+\mu_n-\frac{1}{2})\,p_{\rm
D}, \label{nia1}
\end{eqnarray}
reproduces nonrelativistic value for the magnetic moment in
impulse approximation, while terms
\begin{eqnarray}
\mu_{\rm D}^{S^+,P}\,&=&\,\frac{\alpha_1}{8M(2\pi)^3}\,\bigl( S_1
+ S_2 + S_3 \bigr),
\label{nia2}\\
\mu_D^{D^+,P}\,&=&\,\frac{\alpha_1}{8M(2\pi)^3}\,\bigl( D_0 + D_1
+ D_2 + D_3 + D_4 + D_5 +D_6 \bigr), \label{nia3}\end{eqnarray}
correspond to matrix elements between $\sp-$, $\Dp$- and
$P$-states. The functions $S_i$ and $D_i$ are given as
\begin{eqnarray}
&& S_1\,=\,\frac{\sqrt{3}}{6m^2}\,\int\,|\bk|^3\,d
|\bk|\,\Bigl(g_{\Pte}({\bar k_0},|\bk|)- g_{\Pto}({\bar
k_0},|\bk|)\Bigr)\,u(|\bk|),
\label{exp101a}\\
&&
S_2\,=\,(\mu_p+\mu_n-1)\,\frac{\sqrt{3}}{3m^2}\,\int\,|\bk|^3\,d
|\bk|\, \Bigl(g_{\Pte}({\bar k_0},|\bk|)-g_{\Pto}({\bar
k_0},|\bk|)\Bigr)\,u(|\bk|),
\nonumber\\
&& S_3\,=\,-\frac{\sqrt{6}}{3}\,\int\,\bk^2\,d |\bk|\,
\Bigl(g_{\Pse}({\bar k_0},|\bk|)-g_{\Pso}({\bar
k_0},|\bk|)\Bigr)\,u^{\prime}(|\bk|),
\nonumber\\
&& D_0\,=\,-(\mu_p+\mu_n-1)\frac{\sqrt{6}}{6m^2}\,\int\,|\bk|^3\,d
|\bk|\, \Bigl(g_{\Pte}({\bar k_0},|\bk|)-g_{\Pto}({\bar
k_0},|\bk|)\Bigr)\,w(|\bk|),
\label{exp101b}\\
&& D_1\,=\,\frac{3\sqrt{6}}{10}\,\int\,|\bk|\,d |\bk|\,
\Bigl(g_{\Pte}({\bar k_0},|\bk|)-g_{\Pto}({\bar
k_0},|\bk|)\Bigr)\,w(|\bk|),
\nonumber\\
&& D_2\,=\,\frac{\sqrt{6}}{15m^2}\,\int\,|\bk|^3\,d |\bk|\,
\Bigl(g_{\Pte}({\bar k_0},|\bk|)-g_{\Pto}({\bar
k_0},|\bk|)\Bigr)\,w(|\bk|),
\nonumber\\
&& D_3\,=\,\frac{\sqrt{6}}{10}\,\int\,\bk^2\,d |\bk|\,
\Bigl(g_{\Pte}({\bar k_0},|\bk|)-g_{\Pto}({\bar
k_0},|\bk|)\Bigr)\,w^{\prime}(|\bk|),
\nonumber\\
&& D_4\,=\,\frac{2\sqrt{3}}{5}\,\int\,|\bk|\,d |\bk|\,
\Bigl(g_{\Pse}({\bar k_0},|\bk|)-g_{\Pso}({\bar
k_0},|\bk|)\Bigr)\,w(|\bk|),
\nonumber\\
&& D_5\,=\,\frac{\sqrt{3}}{5m^2}\,\int\,|\bk|^3\,d |\bk|\,
\Bigl(g_{\Pse}({\bar k_0},|\bk|)-g_{\Pso}({\bar
k_0},|\bk|)\Bigr)\,w(|\bk|),
\nonumber\\
&& D_6\,=\,\frac{2\sqrt{3}}{15}\,\int\,|\bk|^2\,d |\bk|\,
\Bigl(g_{\Pse}({\bar k_0},|\bk|)-g_{\Pso}({\bar
k_0},|\bk|)\Bigr)\,w^{\prime}(|\bk|), \nonumber
\end{eqnarray}
where $u^{\prime}(|\bk|) \equiv \partial u(|\bk|)/\partial |\bk|$,
$w^{\prime}(|\bk|) \equiv \partial w(|\bk|)/\partial |\bk|$.
In the formulas shown  above, we perform the expansion in $k/m$ to
terms of order $(|\bk|/m)^2$. It should be noted that the matrix
elements in~(\ref{mmmain}) which correspond to transitions between
$\sm$- and $\Dm$-states, $\sm$-, $\Dm$- and $P$-states vanish as
the result of integration with respect to $k_0$ upon performing
the expansion in $|\bk|/m$.

The term $\mu_D^{S^+,D^+}$ coincides with the contribution of
nonrelativistic impulse approximation. To connect terms
$\mu_D^{S^+, P}$ and $\mu_D^{D^+, P}$ with the nonrelativistic
ones we introduce an one-iteration approximation.

\paragraph{One-Iteration Approximation.}
\label{sec:oneit}

The Bethe--Salpeter equation for the deuteron is commonly solved
iteratively. After a partial-wave decomposition, the BS equation
reduces to the set of integral equations for the vertex functions
$g_{JLS\varrho}(k_0,|\bk|)$. They are related to the partial
amplitudes $\phi_{JLS\varrho}(k_0,|\bk|)$ which are used here for
simplicity on the right-hand side of the following equation
\begin{eqnarray}
g_{_{JLS\varrho}}(k_0,|\bk|)\!\!=\!\!\sum_{\mu}\frac{g_{\mu
NN}^2}{4\pi}\frac{-i}{\pi^2}
\!\!\int\limits_{-\infty}^{+\infty}\!\!\!dk^\prime_0\!\!\!\int\limits_{0}^{+\infty}\!\!\!
\frac{1}{4E_{\bk^\prime}E_{\bk}}\frac{|\bk^\prime|}{|\bk|}d|\bk^\prime|\!\!\sum_{L^\prime
S^\prime\varrho^\prime} V^{(\mu)}_{_{JLS\varrho, JL^\prime
S^\prime\varrho^\prime}}(k_0,|\bk|;k^\prime_0,|\bk^\prime|)
\phi_{_{JL^\prime
S^\prime\varrho^\prime}}(k^\prime_0,|\bk^\prime|), \label{bseq1}
\end{eqnarray}
where the index $\mu$ labels the type of exchanged meson, $g_{\mu
NN}$ is the coupling constant and $V^{(\mu)}_{JLS\varrho,
JL^\prime S^\prime\varrho^\prime}$ is the matrix element for the
interaction kernel between the states $JLS\varrho$ and $JL^\prime
S^\prime\varrho^\prime$. To get fast convergence to the solution
of this equation one needs a good educated guess for the initial
vertex function $g_{JLS\varrho}(k_0,|\bk|)$ and respectively
$\phi_{JL^\prime
S^\prime\varrho^\prime}(k^\prime_0,|\bk^\prime|)$. The solution of
the equivalent nonrelativistic problem may be helpful for this
task. After several iterations one usually gets correct solution.

For a zero-order approximation, it is convenient to relate the
vertex functions for the $\sp$- and $\Dp$-states to the
corresponding $S$- and $D$-wave components ($u(|\bk|)$ and
$w(|\bk|)$, respectively) of the nonrelativistic deuteron wave
function as
\begin{eqnarray}
&& g^{(0)}_{\sp}(k_0,|\bk|) = -\alpha_1\, (M-2E_\bk)\,u(|\bk|),
\label{eqn:psi1}\\
&& g^{(0)}_{\Dp}(k_0,|\bk|) = -\alpha_1\, (M-2E_\bk)\,w(|\bk|),
\label{eqn:psi3}
\end{eqnarray}
and to set the vertex functions for all the remaining states to
zero:
\begin{eqnarray}
g^{(0)}_{JLS\varrho}(k_0,|\bk|) = 0,\qquad JLS\varrho \neq \sp,
\Dp. \label{eqn:psi4}
\end{eqnarray}

Note that we need the vertex functions at a fixed value of the
relative energy $k_0 = {\bar k}_0 = M/2-E_\bk$ (see
Eqs.~(\ref{exp101a},\ref{exp101b})). We substitute
relations~(\ref{eqn:psi1}-\ref{eqn:psi4}) into the right-hand side
of Eq.~(\ref{bseq1}) and perform integration with respect to
$k_0$. As before, we do this with the aid of Cauchy theorem,
taking into account only pole of the propagator, which corresponds
to positive nucleon energy. This yields
\begin{eqnarray}
g_{_{JLS\varrho}}({\bar k_0},|\bk|)\!\! =\!\! \frac{\alpha_1}{2\pi
m^2} \sum_{\mu} \frac{g_{\mu NN}^2}{4\pi}\!\!
\int\limits_{0}^{+\infty}
\frac{\bk^\prime}{\bk}d|\bk^\prime|\left( \tilde
V^{(\mu)}_{_{JLS\varrho, \sp}}(|\bk|,|\bk^\prime|)
u(|\bk^\prime|)\!\! +\!\! \tilde V^{(\mu)}_{_{JLS\varrho,
\Dp}}(|\bk|,|\bk^\prime|) w(|\bk^\prime|) \right), \nonumber\\
JLS\varrho = \Pte..\Pso, \label{bseq2}
\end{eqnarray}
where we introduced the functions $\tilde V^{(\mu)}_{JLS\varrho,
(\sp,\Dp)}(|\bk|,|\bk^\prime|)$
that can be obtained from the functions\\
$V^{(\mu)}_{JLS\varrho,(\sp,\Dp)}(k_0,|\bk|;k^\prime_0,|\bk^\prime|)$
by setting $k_0=\bar k_0$ and $k^\prime_0=\bar k^\prime_0$ and by
expanding $E_\bk$ and $E_{\bk^\prime}$ in powers of $|\bk|/m$ to
second-order terms.

We consider now the  $\pi$-meson exchange kernel in the ladder
approximation and calculate in this case the functions
$g_{\Pte..\Pso}$ in first iteration. The matrix elements of the
interaction kernel in this case are given by (index superscript
$(\mu)$ is omitted),
\begin{eqnarray}
&& V_{\Pte\, {\sp}}(k_0,|\bk|;k^\prime_0,|\bk^\prime|) = V_{\Pte\,
{\Dp}}(k_0,{|\bk|};k^\prime_0,{|\bk^\prime|}) = 0,
\label{vfa}\\
&& V_{\Pto\, {\sp}}(k_0,{|\bk|};k^\prime_0,{|\bk^\prime}) =
\label{vfb}
\frac{2}{\sqrt3}\left(\frac{1}{3}\,{|\bk|}\,(E_{\bk^\prime}-m)\,Q_2(z)
+m\,{|\bk^\prime|}\,Q_1(z)-\frac{1}{3}\,{|\bk|}\,(E_{\bk}+2m)Q_0(z)\right),
\nonumber\\
&& V_{\Pto\, {\Dp}}(k_0,{|\bk|};k^\prime_0,{|\bk^\prime|}) =
\label{vfc} \frac{\sqrt{2}}{\sqrt{3}}\left(\frac{1}{3}\,{|\bk|}\,
(2E_{\bk^\prime}+m)\,Q_2(z)-m\,{|\bk^\prime|}\,Q_1(z)-\frac{2}{3}\,{|\bk|}\,(E_{\bk^\prime}-m)Q_0(z)\right),
\nonumber\\
&& V_{\Pse\, {\sp}}(k_0,{|\bk|};k^\prime_0,{|\bk^\prime|}) =
\label{vfd}
-\frac{\sqrt{2}}{\sqrt{3}}\left(\frac{2}{3}\,{|\bk|}\,(E_{\bk^\prime}-m)Q_2(z)
+m\,{|\bk^\prime|}\,Q_1(z)-\frac{1}{3}\,{|\bk|}\,(2E_{\bk^\prime}+m)Q_0(z)\right),
\nonumber\\
&& V_{\Pse\, {\Dp}}(k_0,{|\bk|};k^\prime_0,{|\bk^\prime|}) =
\label{vfe}
\frac{2}{\sqrt{3}}\left(\frac{1}{3}\,{|\bk|}\,(E_{\bk^\prime}+2m)Q_2(z)
-m\,{|\bk|}\,Q_1(z)-\frac{1}{3}\,{|\bk|}\,(E_{\bk^\prime}-m)Q_0(z)\right),
\nonumber\\
&& V_{\Pso\, {\sp}}(k_0,{|\bk|};k^\prime_0,{|\bk^\prime|}) =
V_{\Pso\, {\Dp}}(k_0,{|\bk|};k^\prime_0,{|\bk^\prime|}) = 0,
\nonumber \label{vff}\end{eqnarray} where $Q_\ell(z)$ are the
 Legendre functions of the second kind and $z=(\bk^2+\bk^{\prime
2}+\mu_{\pi}^2$ - $(k^\prime_0-k_0)^2)/2|\bk||\bk^\prime|$ with
$\mu_{\pi}$ being the $\pi$-meson mass. In the above expressions
we have omitted the isospin factor which is $-3$ for the deuteron.
The functions $\tilde V_{11S\varrho,1LS+}(|\bk|,|\bk^\prime|)$
then become
\begin{eqnarray}
&& {\tilde V}_{\Pte\, \sp}({|\bk|},{|\bk^\prime|}) = {\tilde
V}_{\Pte\, \Dp}({|\bk|},{\bk^\prime|}) = 0,
\label{pot1}\\
&& {\tilde V}_{\Pto\, \sp}({|\bk|},{|\bk^\prime|}) =
\frac{2m}{\sqrt{3}}({|\bk^\prime|}\,Q_1(\tilde
z)-{|\bk|}\,Q_0(\tilde z)),
\label{pot2}\\
&& {\tilde V}_{\Pto\, \Dp}({|\bk|},{|\bk^\prime|}) =
\frac{\sqrt{2}m}{\sqrt{3}}({|\bk|}\,Q_2(\tilde
z)-{|\bk^\prime|}\,Q_1(\tilde z)),
\label{pot3}\\
&& {\tilde V}_{\Pse\, \sp}({|\bk|},{|\bk^\prime|}) =
-\frac{1}{\sqrt{2}}\, {\tilde V}_{\Pto S}({|\bk|},{|\bk^\prime|}),
\label{pot4}\\
&& {\tilde V}_{\Pse\, \Dp}({|\bk|},{|\bk^\prime|}) =
\sqrt{2}{\tilde V}_{\Pto, \Dp}({|\bk|},{|\bk^\prime|}),
\label{pot5}\\
&& {\tilde V}_{\Pso, \sp}({|\bk|},{|\bk^\prime|}) = {\tilde
V}_{\Pso, \Dp}({|\bk|},{|\bk^\prime|}) = 0,
\label{pot6}\end{eqnarray} with $\tilde z={(\bk^2+\bk^{\prime
2}+\mu_{\pi}^2)/2|\bk||\bk^\prime|}$.

To perform the integration over  ${|\bk|}$ in Eq.~(\ref{bseq2}),
we
 introduce functions $u(r)$ and $w(r)$ in the configuration space, which are
analogs of functions $u({|\bk|})$ and $w({|\bk|})$ ($w_0\equiv u$,
$w_2\equiv w$, $\ell = 0,2$),
\begin{equation}
w_\ell({|\bk|})=\int\limits_{0}^{+\infty}\,r\,d r\,
w_\ell(r)\,j_\ell({|\bk|}r),\quad\quad \frac{w_\ell(r)}{r} =
\frac{2}{\pi}
\int\limits_{0}^{+\infty}\,{\bk}^2\,d{|\bk|}\,w_\ell({|\bk|})\,j_\ell({|\bk|}r),
\label{nr2k}
\end{equation}
where $j_\ell(x)$ are the spherical Bessel functions. Using
formulas~(\ref{bseq2}),~(\ref{pot1})-(\ref{pot6}) and
Eq.~(\ref{nr2k}), and recalling that the Legendre functions admit
the integral representation,
$$Q_\ell(\tilde z)=\,2{|\bk|}{|\bk^\prime|}\,\int\limits_{0}^{+\infty}\,
e^{-\mu_{\pi}r}\,r\,d r\,j_\ell({|\bk^\prime|}r)\,
j_l({|\bk|}r),$$ we obtain
\begin{eqnarray}
&&g_{\Pte}({\bar k_0},{|\bk|})=0,
\label{find}\\
&&g_{\Pto}({\bar k_0},{|\bk|})= \label{finda}
-(-3)\,\frac{\alpha_1}{\sqrt{3}m}\, \frac{g_{\pi NN}^2}{4\pi}\,
\int\limits_{0}^{+\infty} \,d
r\,\frac{e^{-\mu_{\pi}r}}{r}\,(1+\mu_{\pi}r)
\bigl(u(r)+\frac{1}{\sqrt{2}}w(r)\bigr)\,j_1({|\bk|}r),
\\
&&g_{\Pse}({\bar k_0},{|\bk|})= \label{findb}
-(-3)\,\frac{\alpha_1}{\sqrt{3}m}\, \frac{g_{\pi NN}^2}{4\pi}\,
\int\limits_{0}^{+\infty}\, d
r\,\frac{e^{-\mu_{\pi}r}}{r}\,(1+\mu_{\pi}r)
\bigl(-\frac{1}{\sqrt{2}}u(r)+w(r)\bigr)\,j_1({|\bk|}r),
\\
&&g_{\Pso}({\bar k_0},{|\bk|})=0, \label{find0}\end{eqnarray}
where the isospin factor ($-3$) has been taken into account
explicitly.

Summarizing the above analysis, we can say that, by using the
one-iteration approximation and by specifying the zeroth-order
approximation for the vertex functions with the aid of the
relations~(\ref{eqn:psi1}-\ref{eqn:psi4}), we related the $P$-wave
components of the Bethe-Salpeter vertex function to the
nonrelativistic deuteron wave function. Note that the resulting
relations are proportional to ${g_{\pi NN}^2}/{4\pi}$.

\paragraph{Pair Current}
We should note how to relate the $P$-wave components with the
magnetic moments of the deuteron. Let us now make use of the
expressions obtained above. Substituting
relations~(\ref{find}-\ref{find0}) into
Eqs.~(\ref{exp101a}-\ref{exp101b}), we perform an expansion in
terms of $g^2_{\pi NN}/4\pi$. As a result we find:
\begin{eqnarray}
&&\mu_{\rm D}^{S^+,P}\,=\,N_{\alpha_1}\,\Biggl(
-(\mu_p+\mu_n)\,\frac{\sqrt{3}}{3m^2}\,S_A
+\frac{\sqrt{3}}{6m^2}\,S_A+\frac{\sqrt{6}}{3}\,S_B\Biggr),
\label{exp102}\\
&&\mu_{\rm D}^{D^+,P}\,= \label{exp102a}
N_{\alpha_1}\,\Biggl( (\mu_p+\mu_n)\,\frac{\sqrt{6}}{6m^2}\,D_A
-\frac{7\sqrt{6}}{30m^2}\,D_A -\frac{\sqrt{6}}{10}\,D_B
+\frac{2\sqrt{3}}{15}\,D_C +\frac{\sqrt{3}}{5m^2}\,D_D\Biggr),
\nonumber\end{eqnarray} where
$N_{\alpha_1}={\alpha_1}/{8M(2\pi)^3}$. We have also introduced
the integrals
\begin{eqnarray}
&& S_A\,=\,\frac{\pi}{2}\,\int\,\frac{d r}{r^2}\,
g_{\Pto}(r)\,[u(r)-ru^{\prime}(r)],
\label{exp103}\\
&& S_B\,=\,\frac{\pi}{2}\,\int\,d r\,g_{\Pse}(r)\,[u(r)],
\nonumber\\
&& D_A\,=\,\frac{\pi}{2}\,\int\,\frac{d r}{r^2}\,
g_{\Pto}(r)\,[2w(r)+rw^{\prime}(r)],
\label{exp103a}\\
&& D_B\,=\,\frac{\pi}{2}\,\int\,d r\,g_{\Pto}(r)\,[w(r)],
\nonumber\\
&& D_C\,=\,\,\frac{\pi}{2}\,\int\,d r\,g_{\Pse}(r)\,[w(r)],
\nonumber\\
&& D_D\,=\,\frac{\pi}{2}\,\int\,\frac{d r}{r^2}\,
g_{\Pse}(r)\,[2w(r)+rw^{\prime}(r)] \nonumber
\end{eqnarray}
and functions
\begin{eqnarray}
&& g_{\Pto}(r)\,=\, \sqrt{3}\,\frac{\alpha_1}{m}\,\frac{g_{\pi
NN}^2}{4\pi}\, \frac{e^{-\mu_{\pi}r}}{r}\,
(1+\mu_{\pi}r)\,\bigl(u(r)+\frac{1}{\sqrt{2}}w(r)\bigr),
\label{exp104}\\
&& g_{\Pse}(r)\,=\, \sqrt{3}\,\frac{\alpha_1}{m}\,\frac{g_{\pi
NN}^2}{4\pi}\, \frac{e^{-\mu_{\pi}r}}{r}\,
(1+\mu_{\pi}r)\,\bigl(-\frac{1}{\sqrt{2}}u(r)+w(r)\bigr).
\label{exp104a}\end{eqnarray}

It is convenient to rearrange the above expressions in such a way
as to isolate explicitly the component corresponding to the
$\pi$-meson pair current. For this purpose we single out three
terms:
\begin{eqnarray}
\mu_{\rm D}^{(1)} &=& \frac{\alpha_1}{8M(2\pi)^3}\,
\Bigl(-(\mu_p+\mu_n)\,\frac{\sqrt{3}}{3m^2}\,S_A
+(\mu_p+\mu_n)\,\frac{\sqrt{6}}{6m^2}\,D_A\Bigr)\, =
\label{exp105}\\
&& (\mu_p+\mu_n)\,\frac{g_{\pi NN}^2}{4\pi}\,\frac{1}{4m^3}
\int\,d r\,\frac{e^{-\mu_{\pi}r}}{r^2}\,(1+\mu_{\pi}r)\,
\Bigl[uu^{\prime}-\frac{u^2}{r}+
\frac{1}{\sqrt{2}}((uw)^{\prime}+\frac{uw}{r})
+\frac{1}{2}(ww^{\prime}+\frac{2w^2}{r})\Bigr],
\nonumber\end{eqnarray}
\begin{eqnarray}
\mu_{\rm D}^{(2)} &=& \frac{\alpha_1}{8M(2\pi)^3}\,
\Bigl(\frac{\sqrt{3}}{6m^2}\,S_A
-\frac{7\sqrt{6}}{30m^2}\,D_A+\frac{\sqrt{3}}{5m^2}\,D_D \Bigr)\,
=
\label{exp106}\\
&& \frac{g_{\pi NN}^2}{4\pi}\,\frac{1}{8m^3}\, \int\,d
r\,\frac{e^{-\mu_{\pi}r}}{r^2}\,(1+\mu_{\pi}r)
\Bigl[ -uu^{\prime} + \frac{u^2}{r} - \frac{1}{\sqrt{2}}
(u^{\prime}w+4uw^{\prime}+ \frac{7uw}{r}) -
\frac{1}{5}(ww^{\prime}+\frac{2w^2}{r})  \Bigr], \nonumber
\end{eqnarray}
\begin{eqnarray}
\mu_{\rm D}^{(3)} &=& \frac{\alpha_1}{8M(2\pi)^3}\,
\Bigl(\frac{\sqrt{6}}{3}\,S_B-\frac{\sqrt{6}}{10}\,D_B
+\frac{2\sqrt{3}}{15}\,D_C \Bigr)\,=
\label{exp106a}\\
&&-\frac{g^2_{\pi NN}}{4\pi}\,\frac{1}{4m}\, \int\,d
r\,\frac{e^{-\mu_{\pi}r}}{r}\,(1+\mu_{\pi}r)\, \Bigl[ u^2 -
\frac{1}{\sqrt{2}} uw-\frac{1}{10}w^2 \Bigr].
\nonumber\end{eqnarray} The first term is proportional to
$(\mu_p+\mu_n)$ and coincides with  the $\pi$-meson pair current
of the deuteron magnetic moment. The second and third terms have
no analogs in the nonrelativistic expressions.

The final expression for the deuteron magnetic moment has the
form,
\begin{eqnarray}
\mu_D=\mu_D^{NIA}+\mu_D^{PC}+\Delta\mu_D, \quad\quad \mu_D^{PC} =
\mu_D^{(1)}, \quad\quad \Delta\mu_D=\mu_D^{(2)}+\mu_D^{(3)},
\label{result}
\end{eqnarray}
where $\mu_D^{NIA}$ is the result obtained in the nonrelativistic
impulse approximation (see Eq.~(\ref{nia1})). $\mu_D^{PC}$ is the
contribution of the $\pi$-meson pair current, and $\Delta\mu_D$ is
the additional relativistic contribution to the magnetic moment.

\subsubsection{\em Quadrupole Moment of the Deuteron}\label{quad_mom}
We  derive now the expression for the quadrupole moment of the
deuteron. Using Eqs.~(\ref{quadrupole1}), (\ref{cur}) and
(\ref{lamb1}) we get
\begin{eqnarray}
Q_{\rm D}=\sum\limits_{a,a^\prime}\sum\limits_{\rho,\rho^\prime}
\langle {a^\prime}^{\rho^\prime}|\hat Q|a^\rho\rangle =
\sum\limits_{a,a^\prime}\sum\limits_{\rho,\rho^\prime} \left[
\langle {a^\prime}^{\rho^\prime}|\hat Q_{C}|a^\rho\rangle +\langle
{a^\prime}^{\rho^\prime}|\hat Q_{C}^{LB}|a^\rho\rangle \right.
\label{quadmat} \left. +\langle {a^\prime}^{\rho^\prime}|\hat
Q_{M}|a^\rho\rangle +\langle {a^\prime}^{\rho^\prime}|\hat
Q_{M}^{LB}|a^\rho \rangle\right]. \nonumber\end{eqnarray} Here we
label partial-waves of the deuteron BS amplitude by $a,a^{\prime}$
with $\rho$-spin $\rho, \rho^{\prime}$. The subscripts $C$ and $M$
mean the corresponding contributions of charge and magnetic parts
of the $\gamma NN$ vertex~(\ref{nngs}). The terms with the index
$LB$ indicate the contributions  due to the factor
$([\Lambda({\cal L})^{-1}]^2-1)$ in the matrix
element~(\ref{cur}), which come from the Lorentz boost.

The matrix elements of the operator $\hat Q$ in Eq.(\ref{quadmat}) schematically
represent the combinations, defined by Eq.(\ref{quadrupole1}), of the
following matrix elements ($P\equiv P_{rf} =
P^{\prime}_{rf}$, $k\equiv k_{rf}$, $k^{\prime}\equiv
k^{\prime}_{rf}$, ${\cal M} = 0,1$):
\begin{eqnarray}\label{QC}
&& \langle P {\cal M}|J_{0}^{RIA}|P {\cal M} \rangle_{C} =
\frac{e}{2M}\int\frac{\d^4 k}{ i (2\pi)^4}\, {\mbox Tr}\left\{
\bar\chi_{{\cal M}}(P, k^\prime) \gamma_0\chi_{\cal M}(P,k)
(P\cdot\gamma/2-k\cdot\gamma+m)\right\},
\\
\label{QCLB} && \langle P {\cal M}|J_{0}^{RIA}|P {\cal M}
\rangle_{C}^{LB} =
\\
&&\hskip 40mm \sqrt{\eta}\frac{e}{2M}\int\frac{ d^4 k}{ i
(2\pi)^4}\, {\mbox Tr} \left\{ \bar\chi_{{\cal M}}(P,k^\prime)
\gamma_0\chi_{\cal M}(P,k)
(P\cdot\gamma/2-k\cdot\gamma+m)\gamma_0\gamma_3\right\},
\nonumber\\
\label{QM} && \langle P {\cal M}|J_{0}^{RIA}|P {\cal M}
\rangle_{M} =
\\
&&\hskip 10mm -\frac{e}{2M}\frac{\varkappa}{4m}\int\frac{ d^4k}{ i
(2\pi)^4}\, {\mbox Tr} \left\{ \bar\chi_{{\cal M}}(P,k^\prime)
\Lambda({\cal L}) (\gamma_0{q\cdot\gamma}-{q\cdot\gamma}\gamma_0)
\Lambda({\cal L}) \chi_{\cal M}(P,k)
(P\cdot\gamma/2-k\cdot\gamma+m)\right\}, \nonumber
\end{eqnarray}
\begin{eqnarray}\label{QMLB}
&& \langle P {\cal M}|J_{0}^{RIA}|P {\cal M} \rangle_{M}^{LB} =
\\
&&\hskip 12mm
-\frac{e}{2M}\frac{\varkappa\sqrt{\eta}}{4m}\int\frac{ d^4 k}{ i
(2\pi)^4}\,{\mbox Tr} \left\{ \bar\chi_{{\cal M}}(P,k^\prime)
(\gamma_0{q\cdot\gamma}-{q\cdot\gamma}\gamma_0) \chi_{\cal
M}(P,k)(P\cdot\gamma/2-k\cdot\gamma+m)\gamma_0\gamma_3\right\},
\nonumber
\end{eqnarray}
with $\varkappa = \varkappa_p+\varkappa_n$.

We manipulate the above expressions to find the quadrupole moment,
where the main contribution to the quadrupole moment comes from
the transition between the positive energy components $\sp$ and
$\Dp$ of the BS amplitude,
\begin{eqnarray}
&&Q_{\rm D}^C=\sum\limits_{a,a^\prime=S,D} \langle
{a^\prime}^+|\hat Q_{C}|a^+\rangle
=Q^{(+,+)}_{\bk}+Q^{(+,+)}_{k_0}, \label{twoterms}
\end{eqnarray}
The explicit expressions for $Q^{(+,+)}_{\bk}$ and
$Q^{(+,+)}_{k_0}$ are given in Appendix~A. Performing integration
in Eq.~(\ref{upwp}) with respect to $k_0$ and taking into account
only the positive energy nucleon pole in the propagators we obtain
the following expressions,
\begin{eqnarray}
Q_{\rm D}=-\frac{1}{20} \int\!\frac{ d^3 \bk}{(2\pi)^3} \left\{
\sqrt{8} \left[\bk^2\frac{d u(|\bk|)}{d|\bk|} \frac{d
w(|\bk|)}{d|\bk|}+3|\bk| w(|\bk|) \frac{d u(|\bk|)}{d|\bk|}\right]
+\bk^2\left(\frac{d w(|\bk|)}{d|\bk|}\right)^2
+6\left(w(|\bk|)\right)^2\right\}. \label{QNRmom}\end{eqnarray} In
the last equation we have introduced the expansion over $|\bk|/m$
up to terms of the second order and used substitutions by
following Eq.~(\ref{r2nc}). We then find  that Eq.~(\ref{QNRmom})
coincides with the nonrelativistic expression.

Second term $Q_{k_0}^{(+,+)}$ in Eq.~(\ref{twoterms}) and matrix
element from Lorentz transformation~(see (\ref{QCLB})) have purely
relativistic nature and are responsible for relativistic
contributions to the quadrupole moment of the deuteron. For
instance, for positive energy states
 after integration of the Eq.~(\ref{LB1}) in parts we get
\begin{eqnarray}
&&Q^{(++)}_{LB}=\frac{e}{2M} \int\!\frac{ d k_0 \bk^2 d |\bk|}
{i(2\pi)^4}
(E_{\bk}-\frac{M}{2}+k_0)(1-\frac{2k_0}{M})\frac{2}{5M}
\frac{1}{E_{\bk}} \left\{
\frac{2E_{\bk}^2-mE_{\bk}-m^2}{E^2_{\bk}} \left[ \sqrt{2}\Upp\Wpp
\right.\right.
\nonumber\\
&& \left.\left. +\frac{1}{2}\Wpp^2\right]\right\} +\frac{e}{2M}
\int\!\frac{ d k_0\bk^2 d|\bk|}{ i (2\pi)^4} \frac{1}{5}
\frac{\bk^2}{M^2E_{\bk}}
\left(1-\frac{M}{E_{\bk}}\right)\label{LB4}
\left\{ \sqrt{2}\Upp\Wpp+\frac12\Wpp^2\right\}.\nonumber
\end{eqnarray}
The terms are of the order $Q^{(++)}_{LB}\approx
\langle\frac{\bk^2}{M^2}\rangle Q_{\bk}^{(+,+)}$, and they vanish
in the nonrelativistic limit.

\subsubsection{{\em Numerical Estimates}}\label{estimates}
In this section we presented analytical calculations of the
magnetic and the quadrupole moments of the deuteron. We showed
that both moments in one-iteration approximation have three terms.
The first is pure the NRIA contribution, the second is related to
pair-mesonic current and the third is pure relativistic which has
no analogy with the NRIA calculations. To perform numerical the RIA
calculations of the both moments of the deuteron first of all we
must take into account the contribution of the $P$-waves in the deuteron. It will
be done in future. Now we estimate the contribution of the RIA
calculations~\cite{Zu_Tjon:80,Zu_Tjon:81,np9601040} into the magnetic moment of
the deuteron.

There are eight states in the deuteron channel (instead of two in
the non-relativistic case), viz. $\Spp$, $\Dpp$, $\Smm$, $\Dmm$,
$\Pte$, $\Pto$, $\Pe$, $\Po$ (see section~\ref{BS:3s1}). The
normalization condition for this functions can be written as: \bqn
P_{+}+P_{-}=1,\qquad\qquad
P_+ &=& P_{^3S_1^{++}} + P_{^3D_1^{++}}, \nonumber \\
P_- &=& P_{^3S_1^{--}} + P_{^3D_1^{--}} + P_{^3P_1^{e}} +
P_{^1P_1^{o}} + P_{^3P_1^{e}} + P_{^1P_1^{e}}, \eqn introducing
pseudo-probabilities $P_\alpha$ that are negative for the states
$\Smm$, $\Dmm$, $\Pte$, $\Pto$, $\Pe$, $\Po$, and positive for
$\Spp$, $\Dpp$ \cite{Zu_Tjon:80,Zu_Tjon:81}. The calculation with realistic
vertex functions give the following values:
\begin{center}
\begin{tabular}{|c|c|c|c|c|c|}
 \hline  $^{2S+1}L_J^{\rho}$ &$\Dpp$ & $\Dmm$         & $\Pte +\Pto$        & $\Pe +\Po$& \\
\hline [P\%]& $4.8$ & $-6\cdot 10^{-4}$  & $-0.88 \cdot 10^{-2}$&
$-2.5 \cdot
10^{-2}$&\protect{\cite{Zu_Tjon:80}}\\[0ex]
\hline\end{tabular}
\end{center}
It is obvious that the main contribution to the normalization is
due to the states with positive energies, and the contribution of
the $P$--states is larger than that of the negative energies
states by at least one order of magnitude.

Now we calculate the relativistic corrections $\mu^{PC}_D$ and
$\Delta \mu_D$ to $\mu_D$~(\ref{result}).
Thus~\cite{np9601040,PPD},
\begin{eqnarray}\label{mag:mom}
\mu_p = 2.792847337;\quad \mu_n = -1.9130427;\quad \quad \mu^{exp}_D = 0.857406;
\quad \mu^{NRIA}_D &=& 0.852458703; \nonumber\\
 \mu^{PC}_D/\mu^{NRIA}_D = 0.0022;\quad\quad\quad  \Delta \mu_D/\mu^{NRIA}_D  = -0.00058;\quad\quad\quad \mu_D &=& 0.8538396861.
\end{eqnarray}
We have shown that  the expression
for the magnetic moment in the Bethe--Salpeter approach
can be written in a form closer to non-relativistic calculations. The
additional terms in  equation (\ref{result}) can be considered as
relativistic corrections to the non-relativistic formula.

The non-relativistic value reflects
only the $D$-state probability. Whereas in the relativistic corrections
$P$-states  play  the dominant role and improve the agreement with the experimental data.

The magnitude  of the corrections can be compared
with the contributions of mesonic
exchange currents to the magnetic moment as extracted from ref. \cite{BDS:1992}.
The main contribution is due to the pair term, which leads to
$\Delta \mu /\mu_{NR}=0.21-0.22\%$
for different forms of the Bonn potential.
The same size of this  correction as compared to (\ref{mag:mom})
may be considered as an indication that both corrections
are of  the same physical origin.

\section{Deuteron Electrodisintegration}\label{disint}



We examine here in particular the nonrelativistic reduction of the
Bethe--Salpeter approach. We do not attempt to solve this problem
in a general way, but merely consider a particular process, namely
the electro-disintegration of the deuteron, in order to
investigate some relativistic corrections. We consider the
threshold region only, where a single transition amplitude to the
$^1S_0$ final state dominates (see, e.g., ref.~\cite{are94}),
which offers some technical simplifications. The extension to
other partial waves  is straightforward but tedious and beyond the
scope of the present discussion.

In fact, the deuteron disintegration reaction near threshold
energy region~\cite{edexp} invariably attracts attention from both
the theoretical and the experimental sides. It has been an
excellent process  to examine non-nucleonic degrees of freedom and
relativistic effects.  It is well known that IA alone, which
considers nucleons as non-relativistic objects, fails to describe
the double differential cross section at momentum transfer squared
$-q^2 > 9$ fm$^{-2}$. The experimental data do not show the deep
minimum present in the results of calculations~\cite{edtheor}. The
minimum can be filled if meson-exchange currents are calculated.
The contributions of $\pi$-, $\rho$-currents, and $\Delta
\Delta$-configurations further provide  a satisfactory agreement
with data~\cite{mathiot,suskov}.  The question of a consistent inclusion
of all relativistic corrections (at least for the pion sector) has
recently been addressed in ref.~\cite{ritz96} and supports the
above statement. Nevertheless, some conceptional problems of the
theory still remain open questions. Among them is the problem of
gauge invariance and the choice of the nucleon form factor to be
used in the exchange-current
contributions~\cite{f1ge}-\cite{shebeko}.

By now it became apparent that, some relativistic methods
reproduce such corrections as meson-exchange currents in the
impulse approximation. It has been shown that in the framework of
the light-cone approach, the so-called extra components $f_5$ of
the deuteron wave function, and $g_2$ of the $^1S_0$-wave function
introduced in ref.~\cite{karmanov} give an expression in the
nonrelativistic limit, that equals to the contribution of the pair
current~\cite{karmanov}. In this context, the first iteration
assumption that will be explained below for solving the dynamical
equation is substantial. Another important result is the
calculation of the static electromagnetic properties within the
Bethe--Salpeter approach. We find, that the contribution of the
$P$-states to the magnetic moment of the deuteron is numerically
close to the contribution of mesonic currents and agrees in
sign~\cite{BDS:1992,kkb}. It has already been noticed earlier, by
considering covariant reductions of the BS equation to the three
dimensional equation, that negative energy components in the wave
function are responsible for pair-current-type contributions, see,
e.g., ref.~\cite{fewbody1997}.

\subsection{\em Relativistic Kinematics}

\begin{figure}[ht]
\begin{tabular}{cc}
\begin{minipage}[t]{0.5\linewidth}
\hskip 10mm
\includegraphics[width=70mm]{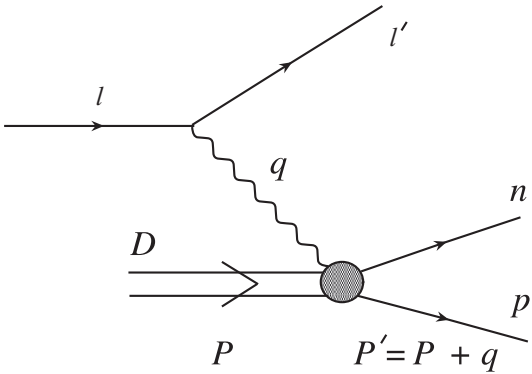}
\end{minipage}
&
\begin{minipage}{0.4\linewidth}
\vskip -60mm \caption{\label{ednp:OPA} Deuteron
electro-disintegration reaction, $eD\to e^\prime(np)$, in the
one--photon approximation. }
\end{minipage}
\end{tabular}
\end{figure}

The amplitude of the electrodisintegration of the deuteron,
$M_{fi}$, in the one-photon-exchange approximation has the form
(Fig.~\ref{ednp:OPA}),
\begin{eqnarray}
M_{fi}\;=\;ie^2\;{\bar u(l^{\prime},s_e^{\prime})}\;
\gamma^{\mu}u(l,s_e)\;\frac{1}{q^2} \;\langle np | J_{\mu} | D
{\cal M} \rangle, \label{eqn:M1}
\end{eqnarray}
where $u(l,s_e)$ is a spinor of the electron with 4-momentum $l$
and electron spin 4-vector $s_e$, $q=l-l^{\prime}$ is 4-momentum
transfer. the matrix element of hadronic current $\langle np |
J_{\mu} | D {\cal M} \rangle$ represents a transition from the
deuteron state $| D{\cal  M} \rangle$ with 4-momentum $P$ and
total spin projection $\cal M$ to the final state of a $np$-pair
with 4-momentum $P^\prime=P+q$, and $J_\mu$ is the electromagnetic
current operator.

Considering the final $np$-pair being near the threshold,  we can
assume $^1S_0$-state for the $np$-pair as a basic state, and
because of the Lorentz covariance, transformation properties under
parity and time reversal as well as current conservation, the
general structure of the transition matrix element for the
$1^+\rightarrow 0^+$ transition current can be written as
\begin{eqnarray}
\langle np (^1S_0) | J_{\mu} | D {\cal M} \rangle =
i\epsilon_{\mu\alpha\beta\gamma}\,\xi^{\alpha}_{\cal M}\,
q^{\beta}\,P^{\gamma}\,V(s,q^2), \label{formc} \end{eqnarray}
where $\xi^{\alpha}_{\cal M}$ is the deuteron polarization
4-vector, $V(s,q^2)$ is a Lorentz invariant function. Due to the
asymmetric tensor, $\epsilon_{\mu\alpha\beta\gamma}$, the matrix
element is gauge invariant,
$$q^{\mu} \langle np (^1S_0) | J_{\mu} | D {\cal M} \rangle=0.$$

Using Eq.~(\ref{eqn:M1}) we obtain the differential cross-section
in terms of the leptonic and hadronic tensors, $L_{\mu\nu}$ and
$W^{\mu\nu}$, respectively.
\begin{eqnarray}
\frac{d^{2}\sigma}{dE_e^{\prime}d\Omega_e^{\prime}}\; =\;
\frac{\alpha^2}{q^4}\; \frac{|{\bl}^{\prime}|}{|{\bl}|}\;
L_{\mu\nu}\; W^{\mu\nu},
\end{eqnarray}
with $\alpha=e^2/4\pi$. The leptonic tensor, $L_{\mu\nu}$, is
\begin{equation}
L_{\mu\nu} = 2 (l_{\mu}l^{\prime}_{\nu}+l^{\prime}_{\mu}l_{\nu})+
q^2 g_{\mu\nu}
+2im_{e}{\epsilon}_{\mu\nu\alpha\beta}q^{\alpha}s_{e}^{\beta},
\end{equation}
where $s_{e}$ is the electron spin 4-vector. The hadronic tensor,
$W_{\mu\nu}$, has the form
\begin{eqnarray}
\label{W:def} W^{\mu\nu} = \langle np (^1S_0) | J^{\mu} | D {\cal
M} \rangle \langle D {\cal M} | J^{\dagger\nu} | np (^1S_0)\rangle
\frac{(2\pi)^3}{2M}\, \int \delta(P+q-p_p-p_n)\,
\frac{d^3{\bp}_p}{2E_{\bp_p}(2\pi)^3}\,
\frac{d^3{\bp}_n}{2E_{\bp_n}(2\pi)^3},
\end{eqnarray}
where ${\bp}_p$ (${\bp}_n$) is the 3-momentum of the proton
(neutron), and $E_{\bp_p} = \sqrt{{\bp}_p^2+m^2}$ ($E_{\bp_n} =
\sqrt{{\bp}_n^2+m^2}$) is the energy of the proton (neutron). With
the help of Eq.~(\ref{formc}) the hadronic tensor can be written
as
\begin{equation}
W^{\mu\nu} = R\;G^{\mu\alpha}\;\rho_{\alpha\beta}\;G^{*\nu\beta}\;
\Bigl|V(s,q^2)\Bigr|^2, \label{eqn:W2}
\end{equation}
where $R$ is a kinematic factor ($s=(P+q)^2$),
\begin{equation}
R=\frac{1}{8\pi^2}\frac{1}{2M} \frac{|\bk^{*}|}{\sqrt{s}}, \qquad
|\bk^{*}|=\sqrt{\frac{s}{4}-m^2},
\end{equation}
and $\rho_{\alpha\beta}$ is the density matrix of the deuteron
\begin{eqnarray}
\rho_{\alpha\beta} = \frac{1}{3}(-g_{\alpha\beta}+
\frac{P_{\alpha}P_{\beta}}{M^2}) +\frac{1}{2M}i
{\epsilon}_{\alpha\beta\gamma\delta} P^{\gamma} s_{D}^{\delta}&-&
\label{dend}
\bigl[\frac{1}{2} \bigl( {(W_{\lambda_1})}_{\alpha\rho}
{(W_{\lambda_2})}^{\rho}_{~\beta}+ {(W_{\lambda_2})}_{\alpha\rho}
{(W_{\lambda_1})}^{\rho}_{~\beta} \bigr)
\\
&+& \frac{2}{3}
(-g_{\lambda_1\lambda_2}+\frac{P_{\lambda_1}P_{\lambda_2}}{M^2})
(-g_{\alpha\beta}+\frac{P_{\alpha}P_{\beta}}{M^2}) \bigr]
p_{D}^{\lambda_1 \lambda_2}, \nonumber\end{eqnarray} where
${(W_{\lambda})}_{\alpha\beta}=
i\epsilon_{\alpha\beta\gamma\lambda}P^{\gamma}/M$, $s_{D}$ is the
spin 4-vector and $p_{D}$ is the alignment tensor of the deuteron.
Using the explicit form of the deuteron density
matrix~(\ref{dend}), the hadronic tensor becomes (the electron
mass is neglected)
\begin{eqnarray}
W^{(u)}_{\mu\nu}&=&\frac{1}{3}\;R\;
\bigl[g_{\mu\nu}(q^2M^2-(Pq)^2)
+(P_{\mu}q_{\nu}+q_{\mu}P_{\nu})(Pq)- \label{tensora}
P_{\mu}P_{\nu}q^2-q_{\mu}q_{\nu}M^2\bigr]\;\Bigl|V(s,q^2)\Bigr|^2,
\\
W^{(v)}_{\mu\nu}&=&\frac{1}{2}\;R\;M\;(s_Dq)\; i
{\epsilon}_{\mu\nu\alpha\beta}\;q^{\alpha}\;P^{\beta}\;
\Bigl|V(s,q^2)\Bigr|^2,
\label{tensor}\\
W^{(t)}_{\mu\nu}&=&R\;\Bigl[\frac{1}{2}[\epsilon_{\mu\lambda_2\alpha\beta}
\epsilon_{\lambda_2\nu\gamma\delta}+\epsilon_{\mu\lambda_2\alpha\beta}
\epsilon_{\lambda_1\nu\gamma\delta}]P^{\alpha}P^{\gamma}q^{\beta}q^{\delta}+
\label{tensorb}
\frac{1}{3}\bigl(-g_{\lambda_1\lambda_2}+\frac{P_{\lambda_1}P_{\lambda_2}}
{M^2}\bigr)
\\
&& \times \bigl[g_{\mu\nu}(q^2M^2-(Pq)^2)+
(P_{\mu}q_{\nu}+q_{\mu}P_{\nu})(Pq)-P_{\mu}P_{\nu}q^2
-q_{\mu}q_{\nu}M^2\bigr]\;p_D^{\lambda_1\lambda_2}
\Bigr]\;\Bigl|V(s,q^2)\Bigr|^2. \nonumber\end{eqnarray} The
superscripts ${(u,v,t)}$ denote unpolarized, vector-polarized and
tensor-polarized parts, respectively.

For the case of unpolarized electrons and deuterons the
differential cross section can be written as
\begin{eqnarray}
\left(\frac{d^2\sigma}{dE^{\prime}_e d
\Omega^{\prime}_e}\right)_{unpol}= \Bigl(\frac {d \sigma}{d
\Omega}\Bigr)_M
\frac{M\,|\bk^{*}|\,m}{12\pi^2}\;
[(E_e+E_e^{\prime})^2-2E_eE_e^{\prime}\cos^2{\frac{\theta_e}{2}}]\;
\tan^2{\frac{\theta_e}{2}}\; \Bigl|V(s,q^2)\Bigr|^2,
\label{main}\end{eqnarray} where
\begin{equation}
\Bigl(\frac {d \sigma}{d \Omega}\Bigr)_M = \frac{\alpha^2\,
\cos^2{\frac{\theta_e}{2}}}{4\,E^2\,\sin^4{\frac{\theta_e}{2}}}
\end{equation}
is the Mott cross-section. We use the following normalization
conditions,
$$<D\bPp \CM^{\prime}|D\bP \CM>=2E_D\,(2\pi)^3\,\delta_{\CM \CM^{\prime}}
\delta(\bPp-\bP)$$
$$<N\bp^\prime,\mu^\prime |N \bp,\mu> =
2E_{\bp}\,(2\pi)^3\,\delta_{\mu
\mu^{\prime}}\delta(\bp^\prime-\bp).$$

\subsection{\em Asymmetries for the Polarized Deuteron}

With the general form of the  hadronic tensor~$W^{\mu\nu}$ in
Eqs.~(\ref{tensora})-(\ref{tensorb}) we can calculate various
polarized deuteron asymmetries,
\begin{eqnarray}
A=\frac {d\sigma(\uparrow,D)-d\sigma(\downarrow,D)}
{d\sigma(\uparrow,D)+d\sigma(\downarrow,D)}, \label{assym}
\end{eqnarray}
where $d\sigma$ is the differential cross section,
$\uparrow(\downarrow)$ is the  helicity $\lambda_e=+1(-1)$ of the
incident electron and $D$ the polarization state of the
vector-polarized deuteron. We take  the momentum of the incident
electron  directed along the $Z$-axis. The 4-vectors $l$ and
$l^{\prime}$ then take the form,
\begin{eqnarray}
l=(E_e,0,0,E_e), \quad\quad l^{\prime}=
(E_e^{\prime}-E_e^{\prime}\sin{\theta_e},0,E_e^{\prime}\cos{\theta_e}),
\end{eqnarray}
where $\theta_e$ is the electron scattering angle, and
$\mbox{\boldmath ${l}$}^{\prime}$ is in the $XZ$-plane defined by
the incident and the outgoing electrons.

We consider, first the case of the {\em vector} polarized
deuterons. If the deuteron is polarized {\em parallel} to the
$Z$-axis, then vector polarization asymmetry is
\begin{eqnarray}
A_{\parallel} &=& \frac{3}{2}\ \kappa \frac{(E_e+E_e^{\prime})
(E_e-E_e^{\prime}\cos{\theta_e})}{(E_e+E_e^{\prime})^2
-2E_eE_e^{\prime}\cos^{2}{\theta_e/2}}~,
\label{A:parallel}\end{eqnarray} where $\kappa$ is the degree of
the polarization of the deuteron. For the asymmetry, the
dependence on $V(s,q^2)$ disappears. For the case of the backward
scattering ($\theta_e=180^{\circ}$) Eq.~(\ref{A:parallel}) gives
\begin{equation}
A_{\parallel} = \frac{3}{2}\ \kappa.
\end{equation}
If the deuteron is polarized {\em parallel} to  $X$-axis, the
vector asymmetry is
\begin{eqnarray}
A_{\perp} = \frac{3}{2}\ \kappa\ \frac{(E_e+E_e^{\prime})
E_e^{\prime}\sin{\theta_e}}
{(E_e+E_e^{\prime})^2-2E_eE_e^{\prime}\cos^{2}{\theta_e/2}}.
\label{A:perp}
\end{eqnarray}

Generalizing formulas for the asymmetries to the arbitrary
direction of the polarization $(\vartheta,\varphi)$, we write
\begin{eqnarray}
A(\vartheta,\varphi) = \frac{3}{2}\ \kappa
\frac{(E_e+E_e^{\prime})(E_e^{\prime}\sin{\theta_e}
\sin{\vartheta}\cos{\varphi}+
(E_e-E_e^{\prime}\cos{\theta_e})\cos{\vartheta})}
{(E_e+E_e^{\prime})^2-2E_eE_e^{\prime}\cos^{2}{\theta_e/2}}.
\label{A:arb}\end{eqnarray}



We consider then  the case of the {\em tensor} polarization of the
initial deuteron. If the initial deuteron is only aligned due to
the $p_{D\,zz}$ component, then
\begin{eqnarray}
d\sigma(p_{D\,zz})=d\sigma[1 + A_{zz} p_{D\,zz}],
\label{ratio}\end{eqnarray}
\begin{eqnarray}
A_{zz}=\frac{4E_e^2+E_e^{\prime\;2}
+4E_eE_e^{\prime}-4E_eE_e^{\prime}\cos{\theta_e}+
3E_e^{\prime\;2}\cos{2\theta_e}}
{4((E_e+E_e^{\prime})^2-2E_eE_e^{\prime}\cos^2{\theta_e/2})},
\nonumber\end{eqnarray} where $A_{zz}$ is the tensor analyzing
power. For the backward scattering ($\theta_e=180^{\circ}$ the
analyzing power is $A_{zz} = 1$.

\subsection{\em Relativistic Impulse Approximation}

\begin{figure}[ht]
\begin{tabular}{cc}
\begin{minipage}[t]{0.5\linewidth}
\hskip 10mm
\includegraphics[width=60mm]{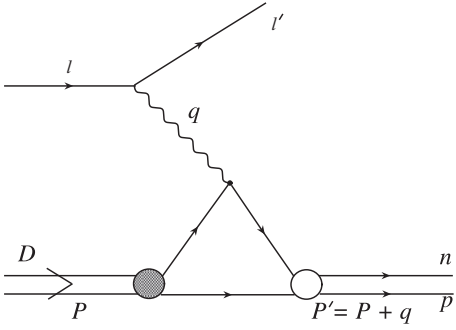}
\end{minipage}
&
\begin{minipage}{0.4\linewidth}
\vskip -50mm \caption{\label{ednp:RIA} Deuteron
electro-disintegration
 in the relativistic impulse approximation. }
\end{minipage}
\end{tabular}
\end{figure}

We evaluate the electromagnetic current matrix element $\langle np
(^1S_0) |J_{\mu}| D {\cal M} \rangle$  in the relativistic impulse
approximation (see diagram in Fig.~\ref{ednp:RIA}). Within the
framework of the BS approach, using the Mandelstam procedure of
constructing of the electromagnetic current
operator~\cite{MA55,Shebeko:1991}, it can be written as (compare
with Eq.~(\ref{fff}))
\begin{eqnarray}
\langle np (^1S_0) | J^{RIA}_{\mu} | D {\cal M} \rangle =i \int
\d^4k\, \Tr \biggl\{ {\bar \chi^{00}}(P^\prime,k^\prime)
\Gamma^{\rm (V)}_{\mu}(q) \chi_{_{\cal M}}(P,k)
({P\cdot\gamma}/{2}-{k\cdot\gamma}+m) \biggr\},
\label{fff1}\end{eqnarray} where $\chi^{00}(P^\prime,k^\prime)$ is
the BS amplitude of the $^1S_0$ state of the $np$-system,
$\chi_{_{J{\cal M}}}(P,k)$ is the BS amplitude of the deuteron,
and $k^\prime=k+q/2$. The vertex of the $\gamma NN$ interaction is
chosen to be on mass shell,
\begin{eqnarray}
\Gamma_{\mu}^{\rm (V)}(q)=\gamma_{\mu} F_1^{\rm (V)}(q^2)
-\frac{\gamma_{\mu} {q\cdot\gamma} - {q\cdot\gamma}
\gamma_{\mu}}{4m} F_2^{\rm (V)}(q^2). \nonumber\end{eqnarray} The
isovector form factors of the nucleon $F_{1,2}^{\rm (V)}(q^2) =
(F_{1,2}^{\rm (p)}(q^2) - F_{1,2}^{\rm (n)}(q^2))/2$, appear
 due to summation of the two nucleons. They are normalized as $F_1^{\rm
(V)}(0) = 1/2$ and $F_2^{\rm (V)}(0) =
(\varkappa_p-\varkappa_n)/2$.

In order to extract $V(s,q^2)$ from the general
expansion~(\ref{formc}) for the $1^+\rightarrow 0^+$ transition,
we first take the trace. According to the Lorentz structure of
Eq.~(\ref{fff1}), seven integrals have to be evaluated, which are
expressed through Lorentz scalar quantities abbreviated by
$(\dots)$ (the index ${\cal M}$ for the polarization vector $\xi$
is suppressed for simplicity).
\begin{eqnarray}
{\cal I}_1&=&i{\epsilon_{\mu\alpha\beta\gamma}}\;
\xi^{\alpha}q^{\beta}P^{\gamma} \  \int\ d^4k\ (...),
\nonumber\\
{\cal I}_2&=&i{\epsilon_{\mu\alpha\beta\gamma}}\;
\xi^{\alpha}P^{\beta} \  \int\ d^4k\ (...)\ k^{\gamma},
\nonumber\\
{\cal I}_3&=&i{\epsilon_{\mu\alpha\beta\gamma}}\;
\xi^{\alpha}q^{\beta} \  \int\ d^4k\ (...)\ k^{\gamma},
\nonumber\\
{\cal I}_4&=&i{\epsilon_{\mu\alpha\beta\gamma}}\;
\xi^{\delta}q^{\beta}P^{\gamma} \  \int\ d^4k\ (...)\
k^{\alpha}k_{\delta},
\nonumber\\
{\cal I}_5&=&i{\epsilon_{\alpha\beta\gamma\delta}}\;
\xi^{\alpha}q^{\beta}P^{\gamma} \  \int\ d^4k\ (...)\
k^{\delta}k_{\mu},
\nonumber\\
{\cal I}_6&=&i{\epsilon_{\alpha\beta\gamma\delta}}\;
\xi^{\alpha}q^{\beta}P^{\gamma}P_{\mu} \  \int\ d^4k\ (...)\
k^{\delta},
\nonumber\\
{\cal I}_7&=&i{\epsilon_{\alpha\beta\gamma\delta}}\;
\xi^{\alpha}q^{\beta}P^{\gamma}q_{\mu} \int\ d^4k\ (...)\
k^{\delta}. \label{structd}
\end{eqnarray}
From inspection of Eq.~(\ref {structd}) it becomes clear that
beside ${\cal  I}_1$, which needs no further consideration, two
generic types of integrals have to be evaluated, namely,
\begin{eqnarray}
&& \int\ d^4k\ (...)\ k_{\alpha}=C_1q_{\alpha}+C_2P_{\alpha}~,
\\
&& \int\ d^4k\ (...)\ k_{\alpha}k_{\beta}= D_1 M^2 g_{\alpha\beta}
+ D_2 q_{\alpha} q_{\beta} + D_3 (q_{\alpha} P_{\beta} +
P_{\alpha} q_{\beta}) + D_4 P_{\alpha} P_{\beta}~,
\nonumber\end{eqnarray} where we have already indicated that after
integration the expressions should depend on the external
4-momenta only, i.e. the transferred momentum $q$ and the total
deuteron momentum $P$. Due to the antisymmetric tensor, the number
of terms is reduced substantially, and all the terms with $D_i$,
$i>1$ vanish. So, beside ${\cal I}_1$, it is necessary to evaluate
the following integrals:
\begin{eqnarray}
{\cal I}_2&=&i{\epsilon_{\mu\alpha\beta\gamma}}\;
\xi^{\alpha}P^{\beta} q^{\gamma} \  \int\ d^4k\ (...)\ c_1(k,q,P),
\nonumber\\
{\cal I}_3&=&i{\epsilon_{\mu\alpha\beta\gamma}}\;
\xi^{\alpha}q^{\beta} P^{\gamma} \  \int\ d^4k\ (...)\ c_2(k,q,P),
\nonumber\\
{\cal I}_4&=&i{\epsilon_{\mu\alpha\beta\gamma}}\;
\xi^{\alpha}q^{\beta}P^{\gamma} \ M^2 \int\ d^4k\ (...)\ d(k,q,P),
\nonumber\\
{\cal I}_5&=&i{\epsilon_{\alpha\beta\gamma\mu}}\;
\xi^{\alpha}q^{\beta}P^{\gamma} \ M^2 \int\ d^4k\ (...)\ d(k,q,P),
\end{eqnarray}
where the functions $c_i=c_i(k,q,P)$ and $d=d(k,q,P)$ are given by
\begin{eqnarray}
c_1(k,q,P) &= &\frac{M^2(kq)-(Pk)(Pq)}{M^2q^2-(Pq)^2},
\nonumber\\
c_2(k,q,P) &=  &\frac{(Pk)q^2-(Pq)(kq)}{M^2q^2-(Pq)^2},\label{eqn:C}\\
d(k,q,P) &= &\frac{(Pk)^2q^2+M^2(kq)^2+(Pq)^2k^2-M^2q^2k^2
-2(Pq)(Pk)(kq)}{2M^2(M^2q^2-(Pq)^2)}. \label{eqn:D}
\end{eqnarray}
The integration and the comparison with the structure of the
transition matrix element given in  Eq.(\ref{formc}) finally leads
to the expression for $V(s,q^2)$,
\begin{equation}
V(s,q^2) = V_1(s,q^2)F_1^{\rm (V)}(q^2) + V_2(s,q^2)F_2^{\rm
(V)}(q^2). \label{eqn:V}
\end{equation}
The expressions for $V_{1,2}(s,q^2)$ are
\begin{eqnarray}
\label{v1v2} V_1(s,q^2) &=& \int d^4k
\left(V_1^{(1)}(s,q^2,k)+V_2^{(1)}(s,q^2,k)
+V_3^{(1)}(s,q^2,k)+V_4^{(1)}(s,q^2,k)\right),\\
V_2(s,q^2) &= &\int d^4k
\left(V_1^{(2)}(s,q^2,k)+V_2^{(2)}(s,q^2,k)
+V_3^{(2)}(s,q^2,k)+V_4^{(2)}(s,q^2,k)\right). \nonumber
\end{eqnarray} The explicit expressions for the functions
$V^{(1,2)}_i$ are given in Appendix~B.

\subsection{\em Isovector Pair Current}\label{sec:nonrel}
We calculate now the function $V_{1,2}(s,q^2)$ (Eq.~(\ref{v1v2}))
from the underlying dynamics in the Bethe--Salpeter approach. We
have two representations of the BS amplitude for the concrete
calculations: 1) the covariant representation and 2) the
partial-wave representation. Implementation of the {\em covariant
representation} for the initial and the final states allow us to
get rid of the functions $V_{1,2}(s,q^2)$ from the proper current
matrix elements.  However, in order to study the relation to the
nonrelativistic expressions, it is more suitable to work out the
formulas for $V_{1,2}(s,q^2)$  in terms of a {\em partial-wave
representation} of the vertex functions. Hence, the expressions
for $V_{1,2}(s,q^2)$ can be written as
\begin{equation}
V_{1,2}(s,q^2)= i \int d^4k \sum_{JLS\varrho,J^\prime L^\prime
S^\prime\varrho^\prime}\ g_{JLS\varrho}(k^\prime_0,|\bk^\prime|)\
{\cal O}_{JLS\varrho,J^\prime L^\prime S^\prime\varrho^\prime}\
g_{J^\prime L^\prime S^\prime\varrho^\prime}(k_0,|\bk|),
\label{a2}
\end{equation}
where the partial vertex functions
$g_{JSL\varrho}(k^\prime_0,|\bk^\prime|)$ are connected to the
partial amplitudes $\phi_{_{LS\varrho}}(k_0,|\bk|)$ via simple
relations~(\ref{g2pri}). These vertex functions represent the
states $^1S_0^{+}$...$^3P_0^{o}$ of the $np$-pair, and the states
$^3S_1^{+}$...$^1P_1^{o}$ of the deuteron (see
paragraph~\ref{BS:3s1}).  In general, the lengthy functions ${\cal
O}_{JSL\varrho, J^\prime L^\prime S^\prime \varrho^\prime}$ depend
on the Lorentz scalars and the explicit expressions are omitted
here. We will specify them below after having introduced
appropriate approximations. The expressions for $k^\prime_0$ and
$|\bk^\prime|$ are given in a formally covariant way (compare with
Eq.~(\ref{p0pep}))
\begin{equation}
k^\prime_0 = \frac{(P+q,k+q/2)}{\sqrt{s}},\
|\bk^\prime|=\left(\frac{(P+q,k+q/2)^2}{s}-(k+q/2)^2\right)^{1/2}.
\label{a3}
\end{equation}

We note that through the use of the Bethe--Salpeter vertex
functions the denominators of ${\cal O}$ appearing in
Eq.~(\ref{a2}) contain products of $(M/2 \pm k_0 \pm E_\bk \pm
i\epsilon)$, $(M/2 \pm k_0 \mp E_\bk \mp i\epsilon)$, $(\sqrt{s}/2
\pm k^\prime_0 \pm E_{\bk^\prime} \pm i\epsilon)$ and $(\sqrt{s}/2
\pm k^\prime_0 \mp E_{\bk^\prime} \mp i\epsilon)$, that stem from
the nucleon propagators (see Eq.~(\ref{spart})). We evaluate the
integrals in the laboratory frame (deuteron at rest). At the
threshold, because of the small deuteron binding energy, it is
possible to utilize the static approximation that preserves the
analytic structure of Eq.~(\ref{a2}) through the following
equations,
\begin{equation}
k^\prime_0 = k_0, \quad |\bk^\prime| = | \bk + \frac{\bq}{2} | =
\left(\bk^2 + \frac{\bq^2}{4} + |\bk| |\bq |{x}\right)^{1/2}.
\label{eqn:static}\end{equation} Here and in the following we use
$\hat{\bf a}={\bf a}/|{\bf a}|$,
\begin{eqnarray}
|\bq|&=&((M^2+s-q^2)^2-4M^2s)^{1/2 }/(2M),\
{x}=\hat\bk\cdot\hat\bq, \nonumber\\ q&=&(\omega,\bq),\
\omega=(s-M^2-q^2)/(2M). \label{qaoth} \end{eqnarray}

We illustrate the approximation by looking closer at the
expressions involving the nucleon positive energy states. The full
denominator of the integrand in Eq.~(\ref{a2}) leads to a
complicated pole structure and reads explicitly
\begin{eqnarray}
\left(\frac{s}{4} -(k_0+\frac {\omega}{2})^2+\bk^{\prime 2}-\sqrt{s}
\left(\frac
{|\bq|^2}{s}(k_0+\frac{\omega}{2})^2+E_{\bk^\prime}^2\right)^{1/2}
-i\epsilon\right)
\left(\frac{M}{2}+k_0-E_{\bk}+i\epsilon\right).\nonumber
\label{eq8}
\end{eqnarray}
It can be simplified by using Eq.~(\ref{eqn:static}):
\begin{eqnarray}
\left(\frac {\sqrt{s}}{2}+k_0-E_{\bk}+i\epsilon\right) \left(\frac
{\sqrt{s}}{2}-k_0-E_{\bk}+i\epsilon\right) \left(\frac
{M}{2}+k_0-E_{\bk}+i\epsilon\right).
\end{eqnarray}
Thus, the static approximation means in particular that $\omega=0$
(no retardation) and the Lorentz boost transformation of the
$np$-pair vertex function is neglected.  We can go  beyond the
static approximation  by expanding the full expression in terms of
$\omega/M $ and $|\bq|^2/s$  that leads to additive corrections.

The integration over $k_0$ can now be performed using the Cauchy
theorem, namely by choosing a proper integration contour and
specifying the corresponding poles, e.g., closing the upper half
plane leads to poles for $k_0$ at $\bar
k^\prime_0=\sqrt{s}/2-E_{\bk^\prime}$. The vertex functions are
then evaluated at $(\bar k^\prime_0, k)$.  Since in the reaction
under the consideration $s\approx 4m^2 \approx M^2$, we expand the
vertex functions near $\bar k_0=M/2-E_{\bk}$ for $g_1$ and $G_1$,
respectively, that allows us to derive analytical expressions in
the one-iteration approximation. The analogous procedure holds for
other partial wave  vertex functions. With this choice, one  of
the nucleons in the deuteron is taken on shell.

We now perform the $k_0$ integration for the interaction in
Eq.~(\ref{a2}). The angular integration is simplified by taking
$\bq$ along the $Z$-axis, and  by replacing $x$  in
Eq.~(\ref{qaoth}) as follows:
\begin{equation}
x=\frac{{\bk^{\prime}}^2-\bk^2-|\bq|^2/4}{|\bk||\bq|},\quad \d x =
\frac{2|\bk|^{\prime}}{|\bk||\bq|}\,\d |\bk|^{\prime}, \label{ttt}
\end{equation}
Finally, to go to the nonrelativistic limit we introduce a
$|\bk|/m$-expansion excluding terms of the order $\CO (\bk^2/m^2)$
(i.e. $E_\bk=m+\CO (\bk^2/m^2)=E_{\bk^\prime}$). The resulting
structure functions ${\tilde V}_{1,2}(s,q^2)$ are then given by
\begin{eqnarray}
&&\tilde V_1(s,q^2) = \frac{\pi}{mM}\frac{1}{|\bq|}
\int\limits_{0}^{+\infty}|\bk|\d |\bk|
\hspace*{-.6cm}\int\limits_{\big||\bk|-|\bq|/2\big|}^{|\bk|+|\bq|/2}\hspace*{-.6cm}|\bk^{\prime}|\d
|\bk^{\prime}|\, \Biggl\{ \frac{g_{\Sp}({\bar
k_0^{\prime}},|\bk^{\prime}|)}
{\sqrt{s}-2E_{\bk^{\prime}}+i\epsilon} \biggl(\frac{g_{\sp}({\bar
k_0},|\bk|)}{M-2E_\bk}- \frac{1}{\sqrt{2}}\frac{g_{\Dp}({\bar
k_0},|\bk|)}{M-2E_\bk} P_2(x)\biggr)
\nonumber\\
&& -\frac{\sqrt{3}}{2}\,\frac{g_{\Sp}({\bar
k_0^{\prime}},|\bk^{\prime}|)}
{\sqrt{s}-2E_{\bk^{\prime}}+i\epsilon}\, \biggl( g_{\Pte}({\bar
k_0},|\bk|)-g_{\Pto}({\bar k_0},|\bk|) \biggr)\,\frac{x}{|\bq|}
-\frac{\sqrt{2}}{4} \biggl( -g_{\PAe}({\bar
k_0^{\prime}},|\bk^{\prime}|)+g_{\PAo}({\bar
  k_0^{\prime}},|\bk^{\prime}|)
\biggr)
\nonumber\\
&& \times\biggl( \frac{g_{\sp}({\bar k_0},|\bk|)}{M-2E_\bk}\,
\frac{|\bq|+2|\bk|x}{|\bq||\bk^{\prime}|} -\frac{1}{\sqrt{2}}\,
\frac{g_{\Dp}({\bar k_0},|\bk|)}{M-2E_\bk}\,
\frac{|\bq|P_2(x)+2|\bk|x}{|\bq||\bk^{\prime}|} \biggr) \Biggr\},
\label{v1nred}\\[.6cm]
&&\tilde V_2(s,q^2) = \frac{\pi}{mM}\frac{1}{|\bq|}
\int\limits_{0}^{+\infty}|\bk|\d |\bk|
\hspace*{-.6cm}\int\limits_{\big||\bk|-|\bq|/2\big|}^{|\bk|+|\bq|/2}\hspace*{-.6cm}|\bk|^{\prime}\d
|\bk|^{\prime} \Biggl\{ \frac{g_{\Sp}({\bar
k_0^{\prime}},|\bk^{\prime}|)}
{\sqrt{s}-2E_{\bk^{\prime}}+i\epsilon} \biggl(\frac{g_{\sp}({\bar
k_0},|\bk|)}{M-2E_\bk} -\frac{1}{\sqrt{2}}\frac{g_{\Dp}({\bar
k_0},|\bk|)}{M-2E_\bk} P_2(x)\biggr)
\nonumber\\
&&+\frac{\sqrt{3}}{4}\, \frac{|\bq|^2}{4m^2}\, \frac{g_{\sp}({\bar
k_0^{\prime}},|\bk^{\prime}|)}
{\sqrt{s}-2E_{\bk^{\prime}}+i\epsilon}\, \biggl( g_{\Pte}({\bar
k_0},|\bk|)-g_{\Pto}({\bar k_0},|\bk|) \biggr)\,\frac{x}{|\bq|}
\nonumber\\ && + \frac{3\sqrt{2}}{16}\; \frac{|\bq|^2}{4m^2}\,
\biggl( -g_{\PAe}({\bar k_0^{\prime}},|\bk^{\prime}|)+
g_{\PAo}({\bar k_0^{\prime}},|\bk^{\prime}|) \biggr) \nonumber\\
&&\times\biggl( \frac{g_{\sp}({\bar k_0},|\bk|)}{M-2E_\bk}\,
\frac{|\bq|+2|\bk|x}{|\bq||\bk^{\prime}|}
-\frac{1}{\sqrt{2}}\,\frac{g_{\Dp}({\bar k_0},|\bk|)}{M-2E_\bk}\,
\frac{|\bq|\,P_2(x)+2|\bk|x}{|\bq||\bk^{\prime}|} \biggr)
\Biggr\}, \label{v2nred}
\end{eqnarray}
where ${\bar k_0^{\prime}}\equiv {\bar
p}_0=\sqrt{s}/2-E_{\bk^{\prime}}$, and ${\bar k_0}=M/2-E_\bk$, and
$P_2(x)=(3x^2-1)/2$ is the Legendre polynomial. The functions
$g_{\sm}$, $g_{\Dm}$, $g_{\Pse}$, $g_{\Pso}$, $g_{\Sm}$ disappear
in the above expressions after $k_0$ integration and because of
the $|\bk|/m$ expansion.  Within this approximation, we are left
with $(+)$ to $(+)$ and $(+)$ to $(e,o)$ transitions only. All
other matrix elements, such as $(-)$ to all and $(e,o)$ to $(e,o)$
vanish. We now examine the expressions for ${\tilde
V}_{1,2}(s,q^2)$  more closely. To recover the nonrelativistic
result, we neglect the vertex functions $g_{\Pte}$, $g_{\Pto}$,
$g_{\Pe}$, $g_{\Po}$ that correspond to the negative $\rho$-spin
components (i.e. do not exist in the nonrelativistic scheme).  If
we replace the functions $g_{\sp}$ and $g_{\Dpp}$ by the
nonrelativistic $S$ and $D$ wave functions of the deuteron using
Eqs.~(\ref{r2nc}) and $g_{\Sp}$ by the $^1S_0$ continuum wave
function of the $np$-pair in the following way
\begin{eqnarray}
\frac{g_{\Sp}({\bar k_0^{\prime}},|\bk|^{\prime})}
{\sqrt{s}-2E_{\bk^{\prime}}+i\varepsilon} &\rightarrow &
-\alpha_2\, u_0(|\bk^{\prime}|),\quad\quad
\alpha_2=\frac{1}{\sqrt{4\pi}}\frac{1}{2\pi}, \label{r2n}
\end{eqnarray}
and insert them for the respective vertex functions in
Eqs.~(\ref{v1nred},\ref{v2nred}) we obtain
\begin{equation}
V^{(0)}(s,q^2) = \frac{\alpha_1\alpha_2\pi}{mM}\frac{1}{|\bq|}\,
G_M^{(V)}(q^2)\, \int\limits_{0}^{+\infty}\,|\bk|\,d|\bk|\,
\int\limits_{\big||\bk|-|\bq|/2\big|}^{|\bk|+|\bq|/2}\,|\bk^{\prime}|\,d|\bk^{\prime}|\,
u_0(|\bk^{\prime}|) \biggl(
u(|\bk|)-\frac{1}{\sqrt{2}}\,w(|\bk|)\,P_2(x) \biggr)~,
\label{vnrnP}
\end{equation} where we have introduced the magnetic isovector
form factor $G_{\rm M}^{\rm (V)}=F_1^{\rm (V)}+F_2^{\rm (V)}$. In
configuration space, the respective integral is found using the
following transformations
for the scattering state (for the deuteron states see
Eq.~(\ref{nr2k}))
\begin{eqnarray}
u_0(|\bk|) =
\frac{2}{\pi}\,\int\limits_{0}^{+\infty}\,r\,dr\,u_0(r)\,j_0(|\bk|r),\quad\quad
\frac{u_0(r)}{r} =
\int\limits_{0}^{+\infty}\,\bk^2\,d|\bk|\,u_0(|\bk|)\,j_0(|\bk|r).
\nonumber\end{eqnarray} The resulting expression is
\begin{equation}
V^{(0)}(s,q^2)=\frac{\alpha_1\alpha_2\pi}{mM}\, G_M^{(V)}(q^2)\,
\int\limits_{0}^{+\infty}\,dr\, u_0(r) \biggl(
u(r)\,j_0(|\bq|r/2)-\frac{1}{\sqrt{2}}\,w(r)\,j_2(|\bq|r/2)
\biggr). \label{vnrn} \end{equation} This result reflects the
so-called nonrelativistic impulse approximation and represents the
lowest order nonrelativistic expansion of the transition form
factors given in Eqs.~(\ref{v1nred},\ref{v2nred}).

\paragraph{One-Iteration Approximation.}\label{sec:oneit0}
Since the one-iteration procedure for the deuteron channel was
discussed in detail in section~\ref{sec:oneit},  we consider here
the $^1S_0$-channel of the $np$ pair.

The inhomogeneous Bethe--Salpeter equation  for the amplitudes in
the $^1S_0$-channel reads
\begin{eqnarray}
\phi_{JLS\varrho}(k_0,|\bk|)=\phi_{JLS\varrho}^{(0)}(k_0,|\bk|)
&+&\sum_{\mu}\frac{g_{\mu NN}^2}{4\pi}\,\frac{-i}{\pi^2}
\int\limits_{-\infty}^{+\infty}\,dk_0\,\int\limits_{0}^{+\infty}\,
\frac{1}{4E_{\bk^\prime}
E_\bk}\,\frac{|\bk^\prime|}{|\bk|}\,d|\bk^\prime|\,
\nonumber\\
&&\!\!\!\!\!\!\!\!\!\!\!\!\!\!\! \times S_{\varrho}(k_0,|\bk|;s)
\sum_{L^\prime S^\prime \varrho^\prime}
V^{(\mu)}_{JLS\varrho,JL^\prime S^\prime
\varrho^\prime}(k_0,|\bk|;k^\prime_0,|\bk^\prime|)\phi_{JL^\prime
S^\prime \varrho^\prime}(k^\prime_0,|\bk^\prime|),
\end{eqnarray}
where $\mu$ is the exchange meson and $\phi_{JLS\varrho}^{(0)}$
denotes the plane-wave function,
\begin{eqnarray}
\phi_{JLS\varrho}^{(0)}(k_0,|\bk|)=\frac{1}{\sqrt{4\pi}}\;\delta_{JLS\varrho,
\Sp}\; \delta(k_0)\;\frac{1}{\bk^2}\;\delta(|\bk|-|\bk^*|),
\end{eqnarray}
and $k^*$ is the on-energy-shell momentum given by
$|{\bk}^*|=\sqrt{s/4-m^2}$.

For the subsequent discussion it is more convenient to split the
system of equations as follows:
\begin{eqnarray}
&&\phi_{\Sp}(k_0,|\bk|)=\phi_{\Sp}^{(0)}(k_0,|\bk|)
+\sum_{\mu}\frac{g_{\mu NN}^2}{4\pi}\,\frac{-i}{\pi^2}
\int\limits_{-\infty}^{+\infty}\,dk^\prime_0\,\int\limits_{0}^{+\infty}\,
\frac{1}{4E_{\bk^\prime}E_{\bk}}\,\frac{|\bk^\prime|}{|\bk|}\,d|\bk^\prime|\,
\nonumber\\
&&\times \sum_{L^\prime S^\prime\varrho^\prime} V^{(\mu)}_{\Sp,
JL^\prime
S^\prime\varrho^\prime}(k_0,|\bk|;k^\prime_0,|\bk^\prime|)
\frac{\phi_{_{JL^\prime
S^\prime\varrho^\prime}}(k^\prime_0,|\bk^\prime|)}{(\sqrt{s}/2+k_0-E_\bk+i\epsilon)
(\sqrt{s}/2-k_0-E_\bk+i\epsilon)},\\
&&g_{\tilde J\tilde L\tilde S\tilde\varrho}(k_0,|\bk|)=\sum_{\mu}
\frac{g_{\mu NN}^2}{4\pi}\,\frac{-i}{\pi^2}
\int\limits_{-\infty}^{+\infty}dk^\prime_0\int\limits_{0}^{+\infty}
\frac{1}{4E_{\bk^\prime}E_{\bk}}\frac{|\bk^\prime|}{|\bk|}d|\bk|
\nonumber\\
&&\hspace*{4cm}\times \sum_{L^\prime S^\prime\varrho^\prime}
V^{(\mu)}_{\tilde J\tilde L\tilde S\tilde\varrho,\tilde J L^\prime
S^\prime\varrho^\prime}(k_0,|\bk|;k^\prime_0,|\bk^\prime|)\phi_{_{\tilde
J L^\prime S^\prime \varrho^\prime}}(k^\prime_0,|\bk^\prime|),
\label{1s0system}
\\
&& \hskip 120mm \tilde J\tilde L\tilde S\tilde\varrho\neq  \Sp.
\nonumber\end{eqnarray}
As in the case of the deuteron channel, we consider the
one-iteration approximation. To this end, we chose a similar
expression for the zero approximation which is helpful in finding
relation with the nonrelativistic solution (compare
with~(\ref{eqn:psi1}-\ref{eqn:psi4})),
\begin{eqnarray}
\phi_{\Sp}(k_0,|\bk|) = \frac{-\alpha_2\,
(\sqrt{s}-2E_{\bk})\,u_0(|\bk|)}{(\sqrt{s}/2+k_0-E_{\bk}+i\epsilon)
(\sqrt{s}/2-k_0-E_{\bk}+i\epsilon)}, \label{eqn:ph1}\end{eqnarray}
\begin{eqnarray}
\phi_{LS\varrho}(k_0,|\bk|) = 0, \qquad LS\varrho \neq \Sp~.
\label{eqn:ph2}\end{eqnarray} Here $u_0(|\bk|)$ is the
nonrelativistic continuum wave function in $^1S_0$ channel given
by
\begin{eqnarray}
u_0(|\bk|)=\frac{1}{\bk^2}\;\delta(|\bk|-|\bk^*|)
+\frac{m\;t(|\bk|,|\bk^{*}|;E_{\bk^{*}})}{{\bk^{*}}^{2}-\bk^2+i\epsilon},
\end{eqnarray}
and $t(|\bk|,|\bk^{*}|;E_{\bk^{*}})$ is the nonrelativistic
half-off-shell $t$-matrix for the $^1S_0$ channel normalized
through  the condition
\begin{eqnarray}
t(E_{\bk^*}) \equiv
t(|\bk^*|,|\bk^*|;E_{\bk^*})=-\frac{2}{\pi}\;\frac{1}{m\;|\bk^*|}\;
\sin{\delta_0}\;e^{i\delta_0},
\end{eqnarray}
where $\delta_0$ is the phase shift, and $E_{\bk^*}={\bk^*}^2/m$.

Analogously to the deuteron case, we finally arrive at the
one-iteration solution for the amplitudes in the $^1S_0$ channel,
\begin{eqnarray}
g_{\Pe}({\bar k_0},|\bk^{\prime}|)&=&
-(+1)\,\frac{\alpha_2\sqrt{2}}{\pi m}\,\frac{g_{\pi NN}^2}{4\pi}\,
\int\limits_{0}^{+\infty}\,dr\,\frac{e^{-\mu_{\pi}r}}{r}\,(1+\mu_{\pi}r)
u_0(r)\,j_1(|\bk^{\prime}|r),
\label{finp}\\
g_{\Po}({\bar k_0},|\bk^{\prime}|)&=&0, \label{finp1}
\end{eqnarray}
where $(+1)$ is the isospin factor.

We have shown in this section that  proper choice of zero
approximation wave function (i.e. the nonrelativistic one) allows
one to obtain
 additional partial amplitudes  through the
Bethe--Salpeter equation after one iteration only. They are
connected to the interaction kernel and in electromagnetic
processes give rise to the so-called pair-current correction as is
shown below.

\paragraph{Pair Current.}
\label{sec:emc} We are now in the position to turn to the first
order corrections to $V^{(0)}(s,q^2)$ given in Eq.~(\ref{vnrn}).
To this end, we expand the expressions given in
Eqs.~(\ref{v1nred},\ref{v2nred}) into a power series of $g_{\pi
NN}^2/4\pi$ to extract the pionic contribution only. Also we
consider the $P$-states contribution only, i.e. components with
one negative $\rho$-spin. The resulting expression for the
transition form factor $V(s,q^2)$ will be denoted by
$V^{(\pi)}(s,q^2)$.  Substituting Eqs.~(\ref{find},\ref{finda}) as
well as Eqs.~(\ref{finp},\ref{finp1}) into Eqs.~(\ref{v1nred}) and
(\ref{v2nred}), and using the replacements of Eq.~(\ref{r2nc}) and
Eq.~(\ref{r2n}) we obtain
\begin{eqnarray}
V^{(\pi)}(s,q^2) &=& \frac{\alpha_1\alpha_2\pi}{2m^2M|\bq|}\;
\frac{g_{\pi NN}^2}{4\pi} \int\limits_{0}^{+\infty}\,\d r
\,\frac{e^{-\mu_{\pi}r}}{r}\,(1+\mu_{\pi}r)\,
\int\limits_{0}^{+\infty}\,|\bk|\,d|\bk|\,
\int\limits_{\big||\bk|-|\bq|/2\big|}^{|\bk|+|\bq|/2}\,|\bk^{\prime}|\,\d
|\bk^{\prime}|
\\
&& \times\Biggl\{
(-3)\,\left(F_1^{(V)}(q^2)-\frac{1}{2}\frac{|\bq|^2}{4m^2}\,F_2^{(V)}(q^2)\right)
\,u_0(|\bk^{\prime}|)
\biggl(u(r)+\frac{1}{\sqrt{2}}w(r)\biggr)\,j_1(|\bk|r)\,\frac{x}{|\bq|}
\nonumber\\
&&+
(+1)\,\left(F_1^{(V)}(q^2)-\frac{3}{4}\frac{|\bq|^2}{4m^2}\,F_2^{(V)}(q^2)\right)
\frac{1}{\pi} u_0(r)\,j_1(|\bk^{\prime}|r)\,
\nonumber\\
&&\times\biggl(
u(|\bk|)\,\frac{|\bq|+2|\bk|x}{|\bq||\bk^{\prime}|}
-\frac{1}{\sqrt{2}}\,w(|\bk|)\,\frac{|\bq|P_2(x)+2|\bk|x}{|\bq||\bk^{\prime}|}
\biggr)\Biggr\}. \nonumber\end{eqnarray} The
$k^{\prime}$-integration can be solved analytically which yields
\begin{eqnarray}
\label{ve:final} V^{(\pi)}(s,q^2) &=&
\frac{\alpha_1\alpha_2\pi}{2m^2M|\bq|}\; \frac{g_{\pi NN}^2}{4\pi}
\int\limits_{0}^{+\infty}\,\d r\,
\frac{e^{-\mu_{\pi}r}}{r^2}\,(1+\mu_{\pi}r)\,
\\
&&\times\Biggl\{ 3\left(F_1^{\rm
(V)}(q^2)-\frac{1}{2}\frac{|\bq|^2}{4m^2}\,F_2^{(V)}(q^2)\right)
u_0(r)\,\biggl(u(r)+\frac{1}{\sqrt{2}}w(r)\biggr)\,j_1(|\bq|r/2)
\nonumber\\
&&+\left(F_1^{\rm
(V)}(q^2)-\frac{3}{4}\frac{|\bq|^2}{4m^2}\,F_2^{(V)}(q^2)\right)
 u_0(r)\,\biggl(u(r)+\frac{1}{\sqrt{2}}w(r)\biggr)\,j_1(|\bq|r/2)
\Biggr\}
\nonumber\\
&=&\frac{2\alpha_1\alpha_2\pi}{m^2M|\bq|} H(q^2)\frac{g_{\pi
NN}^2}{4\pi}\!\! \int\limits_{0}^{+\infty} \d
r\frac{e^{-\mu_{\pi}r}}{r^2}(1+\mu_{\pi}r)
u_0(r)\biggl(u(r)+\frac{1}{\sqrt{2}}w(r)\biggr)j_1(|\bq|r/2).
\nonumber\end{eqnarray} Here we have introduced the function
\begin{eqnarray}
H(q^2) = F_1^{\rm (V)}(q^2)-\frac{9}{16}\frac{|\bq|^2}{4m^2}\,
F_2^{\rm (V)}(q^2). \label{gamma}
\end{eqnarray}

This first order contribution in $g_{\pi NN}^2$ supplements the
lowest order relativistic expansion as given in Eq.~(\ref{vnrn}).
We then arrive at the following expression for the transition form
factor:
\begin{eqnarray}
V(s,q^2) = \frac{\alpha_1\alpha_2\pi}{mM} \Biggl\{G_{\rm M}^{\rm
(V)}(q^2)\, \int\limits_{0}^{+\infty}\,\d
r\,u_0(r)\,\biggl(u(r)\,j_0(|\bq|r/2)-
\frac{1}{\sqrt{2}}w(r)\,j_2(|\bq|r/2)\biggr)
\nonumber\\
+H(q^2)\frac{2}{m|\bq|}\frac{g_{\pi NN}^2}{4\pi}\!\!
\int\limits_{0}^{+\infty}\!\!\d r\!
\frac{e^{-\mu_{\pi}r}}{r^2}(1+\mu_{\pi}r)
u_0(r)\biggl(u(r)+\frac{1}{\sqrt{2}}w(r)\biggr)j_1(|\bq|r/2)\Biggr\}~.
\label{vnr:final}\nonumber\end{eqnarray}
The lowest order in the $\pi NN$ coupling constant $g_{\pi NN}^2$
leads to an additional contribution after iterating the $P$-wave
channel once. Comparing this result to the one achieved within the
nonrelativistic scheme that introduces meson-exchange currents, we
find  that the first term coincides analytically with the
nonrelativistic impulse approximation contribution, and the second
one with the $\pi$-pair-current contribution.

\paragraph{Nonrelativistic impulse approximation and pair current contribution.}
The differential cross section for the $D \rightarrow {}^{1}S_{0}$
- transition has the form (nonrelativistic case)
\begin{equation}
\frac{d^{2}\sigma}{d\Omega d\omega} =
\frac{16}{3}\alpha^{2}\frac{l^{\prime 2}}{\bq^2} \frac{|\bk^*|
m}{t^{2}}\,\sin^{2}\frac{\theta}{2}\, ((l + l^\prime)^{2} -
2ll^\prime
\cos^{2}\frac{\theta}{2}) \\
|\langle{}^{1}S_{0}\parallel T_{1}^{\rm Mag} \parallel D
\rangle|^{2}, \label{NR}\end{equation} where  $q=(\omega,{\bq})$
is the momentum transfer, $t=-q^2$.  The momentum $\bk^*$ is
related to the relative energy $E_{\bk^*}$ of the $np$ system as
given before $E_{\bk^*} = \bk^{*\,2}/m$, and the relation between
kinematical quantities is given by $|\bq| = \sqrt{ ((2m +
E_{\bk^*})^{2} - M^{2} + t)^{2}/4M^{2} + t}$ and $l^\prime = (-
\omega + \sqrt{\omega^{2} + t/\sin^{2}\theta/2})/2$, $\omega =
E_{e} - E^\prime_{e}$.

In the general case, the current matrix element is a sum of the
nonrelativistic impulse approximation contribution and the
contributions from the meson-exchange current and the
retardation-currents. Our concern is only with $\pi$-meson
pair-current part, and hence we find for $\langle{}^{1}S_{0}
\parallel T_{1}^{\rm Mag} \parallel D\rangle$ the following
expression:
\begin{eqnarray}
\langle{}^{1}S_{0}  \parallel T_{1}^{\rm Mag}
\parallel D\rangle&=& \langle{}^{1}S_{0}  \parallel T_{1,ia}^{\rm
Mag} \parallel D\rangle + \langle{}^{1}S_{0}  \parallel T_{1,\pi
c}^{\rm Mag} \parallel D\rangle~,
 \label{mal} \end{eqnarray}
where we introduced $T_{1,ia}^{\rm Mag}$, which reflects the
impulse approximation operator and $T_{1,\pi c}^{\rm Mag}$, which
is the $\pi$-meson pair (contact) operator.

\paragraph{Results.}\label{sec:results}
We have shown that the nonrelativistic reduction of the
Bethe--Salpeter approach utilizing the one-iteration approximation
leads to the results that exhibit the same {\em analytical
structure} as the nonrelativistic result plus pair-current
corrections. Some details differ as will be discussed below. One
may now use the ``exact'' nonrelativistic wave functions to
calculate the different contributions to the cross sections. This
is done here for an illustration.

\begin{figure}[htb]
\begin{tabular}{cc}
\begin{minipage}{0.45\textwidth}
\begin{center}
\psfig{figure=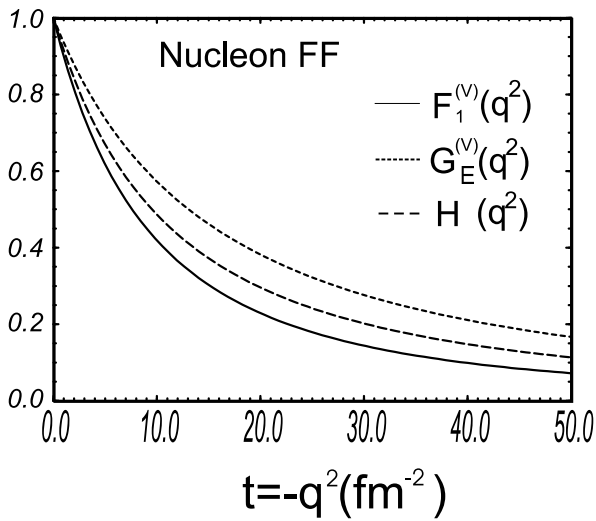,width=\textwidth}
\end{center}
\end{minipage}
&
\begin{minipage}{0.45\textwidth}
\caption{\label{fig:ffl} Nucleon electromagnetic form factors
which enter the calculations for the nonrelativistic pair current
contribution as discussed in the text.  Solid line displays
$F_1^{\rm (V)} (q^2)$, dotted line --- $G_{\rm E}^{\rm (V)}
(q^2)$, and dashed line --- $H (q^2)$.}
\end{minipage}
\end{tabular}
\end{figure}

The formula for the cross section is given in Eq.~(\ref{NR}). The
dominant M1 transition matrix element (i.e. $D\rightarrow
np(^1S_0)$) of the multipole decomposition given there directly
corresponds to the nonrelativistic reduction of Eq.~(\ref{formc})
given above. The contribution of the nonrelativistic impulse
approximation given, e.g. in ref.~\cite{mathiot} coincides with
the formula given in Eq.~(\ref{vnrn}) if $\alpha_1$ and $\alpha_2$
are chosen as in Eq.~(\ref{r2nc}) and (\ref{r2n}). The analytical
structure of the nonrelativistic pair-current contribution equals
to that of the $P$-state contribution derived from the
Bethe--Salpeter approach given in Eq.~(\ref{ve:final}).  We note
that the nonrelativistic pionic pair-current contribution given in
ref.~\cite{mathiot} depends on the nucleon form factor $F_1^{\rm
(V)}(q^2)$. Sometimes, the electric nucleon Sachs form factor
$G^{(\rm V)}_{\rm E}(q^2)= F^{\rm (V)}_1(q^2) +
\frac{q^2}{4m^2}F^{\rm (V)}_2(q^2)$ is  used (see
ref.~\cite{suskov}). The Bethe--Salpeter approach yields a
different dependence on the nucleon form factor that is also
consistent with current conservation and is given by the function
$H(q^2)$ defined in Eq.~(\ref{gamma}).  As an illustration of the
different behavior we display the form factors $F_1^{\rm
(V)}(q^2)$ (solid line), $G_{\rm E}^{\rm (V)}(q^2)$ (short-dashed
line), and the function $H(q^2)$ (long-dashed line) in
Fig.~\ref{fig:ffl}.

As a parametrization for the nucleon form factors we use the one
given in ref.~\cite{hoehler}. The form factor $F_1(q^2)$ is larger
than the form factor $G_{\rm E}(q^2)$ and the function $H(q^2)$
and, therefore, the respective pionic pair current is expected to
be larger than the one using the other two form factors. The
function $H(q^2)$ being in between the two others, the respective
contribution of the $P$-states in Bethe--Salpeter approach is
different from the nonrelativistic calculations which normally use
either $F_1^{\rm (V)}(q^2)$ or $G_{\rm E}^{\rm (V)}(q^2)$.

To investigate the influence of the nucleon form factors more
closely we calculate the impulse approximation and pionic
pair-current contributions to the differential cross section. The
calculation is performed with the Paris $NN$
potential~\cite{paris-wf} at $E_{np}=1.5$ MeV and
$\theta=155^{\circ}$.

\begin{figure}[htb]
\begin{tabular}{cc}
\begin{minipage}{0.5\textwidth}
\begin{center}
\psfig{figure=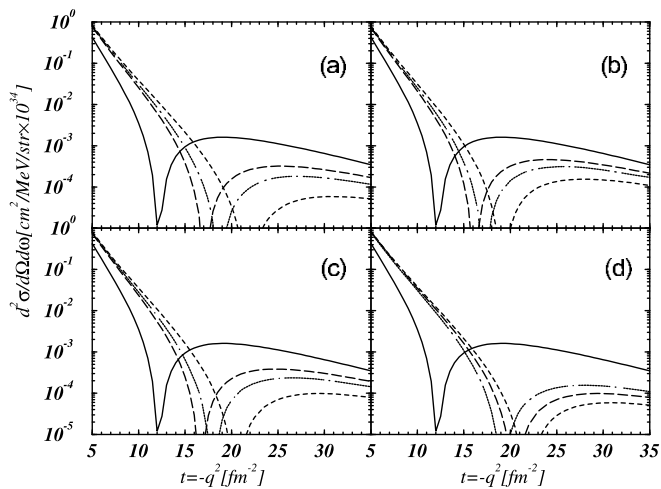,width=\textwidth}
\end{center}
\end{minipage}
&
\begin{minipage}{0.45\textwidth}
\caption{\label{fig:iapc} Impulse approximation (ia, solid lines)
and $\pi$-meson pair-current contributions (pc) to the
differential cross section: (a)-(c) correspond to different sets
of strong nucleon form factors as explained in the text.  Dashed
line: ia+pc with $F_1^{\rm (V)} (q^2)$, dash-dotted line: ia+pc
with $H (q^2)$, long-dashed line: ia+pc with $G_{\rm E}^{\rm (V)}
(q^2)$. (d) Calculations using $F_1^{\rm (V)} (q^2)$ only but
different strong nucleon form factor
 sets as given in the text. Dashed line: ia+pc with set
a, dash-dotted line: ia+pc with set b, and long-dashed line: ia+pc
with set c.}
\end{minipage}
\end{tabular}
\end{figure}

It is well known that some uncertainty is related to the strong
nucleon form factors. Without dwelling too much on that point we
would like to compare three different form factors to see how this
uncertainty propagates. The introduction of  the strong nucleon
form factors changes the expressions for the pair-current
contribution~\cite{pair}. Three sets of strong nucleon form
factors (for $\pi NN$-vertex) have been employed in calculations.
The results are displayed in Figs.~\ref{fig:iapc} a-c,
respectively, a monopole vertex~\cite{machl87} with a cut-off mass
of $\Lambda = 1.25$ GeV (set a) and $\Lambda = 0.85$ GeV (set b).
In addition, a vertex inspired by a QCD analysis has been used
with two parameters chosen to be $\Lambda_1 = 0.99$ GeV and
$\Lambda_2 = 2.58$ GeV ~\cite{gari83} (set c).

It is seen from Fig.~\ref{fig:iapc} that there is a strong
dependence of the differential cross section on the nucleon
electromagnetic form factors. It is also seen that the minimum at
$t=12$ fm$^{-2}$ in the impulse approximation contribution is
shifted by pionic pair current to the region $t > 16$ fm$^{-2}$
using $G_{\rm E}(q^2)$, to $t > 18$ fm$^{-2}$ using the function
$H(q^2)$ and to $t > 22$ fm$^{-2}$ using $F_1^{\rm (V)}(q^2)$ as a
form factor. The largest shift in the cross section is obtained
where $F_1^{\rm (V)}(q^2)$ is used in the calculations. Both the
size of the shift and the behavior of the cross section
considerably depend on the set of parameters as well as the type
of the $\pi NN$-vertex used. This is illustrated in
Fig.~\ref{fig:iapc}d for calculations using $F_1^{\rm (V)}(q^2)$
alone, but for different
parameterizations of the strong vertex.
Thus we conclude that to compare the RIA calculations with the
experimental data one must take into account the contribution of
the $P$ waves for the deuteron and include the contribution of the
final state interaction within the BS approach. It should be done
in future, in the meantime we shoewd here the important points of
this calculations.

\section{Deep Inelastic Scattering\label{DIS}}
Studies of the electromagnetic processes with finite momentum
transfer to a bound state within the BS formalism allowed to
extract important features of the relativistic bound states. It
has provided the general relations between the relativistic
structure of the bound state and its dynamical properties.
However, such a study triggers many questions. At present,
numerical calculations are restricted to the region of small
relative momenta of nucleons which excludes from the analysis of
the BS formalism a very important range of high relative momenta
in Minkowski space. The role of the non-nucleon degrees of freedom
should be studied and the BSA should be reformulated in order to
take into account these states. In this section, we discuss a
process which is free from such uncertainties. This is deep
inelastic scattering in Bjorken limit. Due to the unitarity
condition, the amplitude of this process is defined by the forward
amplitude of the Compton scattering, where momentum transfer to
the bound state is zero. In analysis of this process we can
concentrate on a specific feature of the bound state --- relative
time of the constituents.

The problem of the relative time of the constituents was widely
discussed for a long time, and it was accepted that this feature
is an unphysical property of the relativistic theory and the
problem can be solved by fixing the relative time in a consistent
way. This solution produced large number of similar quasipotential
approaches. However, the analysis of different methods of fixing
the relative time performed in recent papers~\cite{TjonPasc} has
shown that these different methods are not equivalent. The
analysis of the DIS off the deuteron and light nuclei has shown
that this is a property of a bound state, and it allows to solve
the long-standing problem of the EMC effect. In this section we
will consider it in detail.

\subsection{\em Basic Definitions}
\begin{figure}[ht]
\begin{tabular}{cc}
\begin{minipage}[t]{0.3\linewidth}
\hskip 10mm
\includegraphics[width=1.\textwidth]{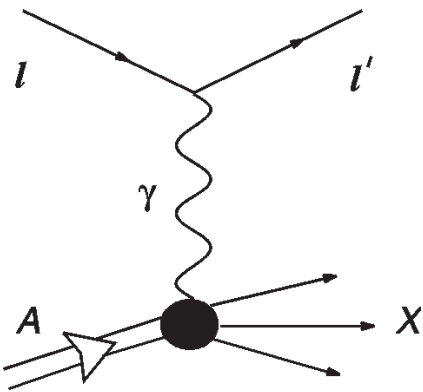}
\end{minipage}
& \hskip 20mm
\begin{minipage}{0.4\linewidth}
\vskip -60mm \caption{\label{dis:OPA} Deep inelastic scattering
off a nucleus $A$  in the one-photon-exchange approximation. }
\end{minipage}
\end{tabular}
\end{figure}

In the process of the deep inelastic scattering (DIS)  of  leptons
from nuclei
\begin{equation} l + A \to l' + X ~,
\label{one} \end{equation} a lepton $l$ with momentum $k$ is
scattered off a nucleus $A$ with initial four-momentum $P$
transferring momentum, $q=l-l^\prime$. Inclusive experiments on
DIS record only the final lepton with momentum $l^\prime$, while
the state $X$ is unobserved final hadronic states of the reaction.
In the lowest order of the electromagnetic constant
$\alpha=e^2/4\pi$ this process can be depicted schematically by
the one-photon exchange graph, as shown in Fig.~\ref{dis:OPA}.\\

\vspace{0.8cm}
In this approximation, the cross section for the
reaction~(\ref{one}) can be written as a contraction of hadronic
and leptonic tensors:
\begin{eqnarray}  &&d\bar\sigma\propto
\frac{\alpha^2}{q^4} L^{\mu \nu}(k, k^\prime) W_{\mu \nu}(P,q)
.\label{cross-sec} \end{eqnarray} The leptonic tensor describes
the hard photon emission by a lepton. Since the lepton is assumed
to be a point particle, the expression for $L^{\mu\nu}$ takes the
simple form
\begin{equation}
L_{\mu\nu}(l, l^\prime) =\frac 12 \sum\limits_{s^\prime} \bar
u^{s^\prime}(l^\prime) \gamma_\nu u^{s}(l) \bar u^{s}(l)
\gamma_\mu u^{s^\prime}(l^\prime). \end{equation}

All the information about the target and its structure is
contained in the hadronic tensor, which has the form:
\begin{equation} W_{\mu\nu}(P,q)=\frac 12 \sum\limits_{n} \langle
P|J_\mu^+|n\rangle \langle n|J_\nu|P\rangle
(2\pi)^4\delta^4(P+q-p_n). \end{equation} This definition allows
the hadronic tensor of the nucleus to be related to the amplitude
for forward Compton scattering by means of the unitarity
condition:
\begin{equation} W_{\mu\nu}(P,q)=\frac{1}{2\pi}{\rm
Im}T_{\mu\nu}(P,q). \label{unit} \end{equation} In the case of
electron (muon) scattering on an unpolarized target, the tensor
$W_{\mu\nu}$  can be written most generally as
\begin{equation}
W^{\mu\nu}(P,q)=\\ W_1(\nu, q^2)g^{\mu\nu}+\frac{W_2(\nu,
q^2)}{M^2}P^\mu P^\nu+ \frac{W_4(\nu, q^2)}{M^2}q^\mu q^\nu +
\frac{W_5(\nu, q^2)}{M^2}(P^\mu q^\nu + q^\mu P^\nu),
\end{equation}
where  $\nu=q_0$ is the photon energy, $W_i$ are the target
structure functions. Due to the gauge invariance condition,
\begin{equation}
q_\mu W^{\mu\nu}(P,q)=0, \label{calib}\end{equation} the hadronic
tensor depends only on two structure functions,
\begin{eqnarray} W_{\mu \nu}(P,q)=
W_1(\nu, q^2) \left( -g_{\mu\nu} + \frac{q_\mu q_\nu}{q^2}\right)
+ \frac{W_2(\nu, q^2)}{M^2}\left(P_\mu - \frac{P\cdot
q}{q^2}q_\mu\right) \left(P_\nu - \frac{P\cdot
q}{q^2}q_\nu\right).\nonumber \end{eqnarray}
 In the Bjorken limit
($-q^2=Q^2\rightarrow \infty$, $\nu\rightarrow \infty$) the
condition for the scale invariance is realized, and it is possible
to change over to the structure functions independent of $q^2$,
\begin{eqnarray} M W_1(\nu, q^2)\rightarrow F_1(x)\nonumber \\ \nu
W_2(\nu, q^2)\rightarrow F_2(x), \label{invsf}\end{eqnarray}
 where
$F_1$ and $F_2$ are the scale-invariant structure functions (SF),
and $x=-q^2/(P\cdot q)$ is a new scale-invariant variable, called
$x$-Bjorken variable. Using~(\ref{invsf}), $W_{\mu\nu}$ can be
written as
\begin{equation} W_{\mu \nu}(P,q)=\left( -g_{\mu\nu} +
\frac{q_\mu q_\nu}{q^2}\right)F_1(x) + \frac{1}{P\cdot
q}\left(P_\mu - \frac{P\cdot q}{q^2}q_\mu\right) \left(P_\nu -
\frac{P\cdot q}{q^2}q_\nu\right)F_2(x). \label{inlorentz}
\end{equation} \noindent

The experimental study of the deep-inelastic scattering of muons
on the deuteron and the iron nucleus has led to the discovery of
the EMC effect, which is the manifestation of the nontrivial
nucleon structure changes in a bound state. Strictly speaking, the
EMC effect, which is interpreted by most authors as a decrease of
the value of the SF of the {\em free nucleon} in the iron nucleus
in the range $0.3<x<0.7$, most likely reflects the difference in
the structure of the {\em deuteron} and the helium nucleus. In
fact, if we restrict ourselves to the range $10^{-3}<x<0.7$, it is
easy to verify that the form of the ratio $r(x)=F_2^{\rm
He^4}/F_2^{\rm D}$ is mimicked in heavier nuclei. The universality
of the $x$ dependence of the modification of the nucleon structure
in nuclei with mass $A\le 4$ was established in Refs.
\cite{sm94,sm95}, where the world data on DIS of electrons and
muons on nuclei were analyzed. These results obviously indicate
that the saturation of the modification of the structure function
$F_2(x)$ already occurs in the helium nucleus.
The
evolution of the modifications from $A=4$ to $A\sim 200$ is
manifested as an increase of the oscillation amplitude $a_{_{\rm
EMC}}=1-r^A_{min}$ by a factor of $\sim 3$ and is well described
as the effect of evolution of the nuclear density as a function of
$A$.

The cessation of the modifications of $F_2(x)$ for $A \ge 4$ is
demonstrated most clearly by the unchanging form of $r^A(x)$,
fixed by the location of the three points $x_1=0.0615$,
$x_2=0.287$, and $x_3=0.84$ at which $r^A(x)=1$ independently of
$A$~\cite{sm99}. Thus, the detailed study of the structure
function of the light nuclei helps to resolve not only the problem
of the EMC effect but to understand the nature of the short range
interactions as well.


\subsection{\em Basic Approximations}\indent
Lets us consider the standard assumptions used in the  analysis of
DIS:
\begin{itemize} \item  the one-boson approximation in the bound
state equation;\\[-0.8cm]
\item  treatment of the DIS amplitude for a nucleus as an
incoherent sum of amplitudes for individual constituents;\\[-0.8cm]
\item representation of the hadronic tensor of the bound nucleon
in terms of scalar functions in the same form as the free nucleon.
\end{itemize}

The first assumption allows us to solve the corresponding equation
for a bound state wave function. In case of the Bethe--Salpeter
formalism within the meson-nucleon field theory this assumption
leads to the kernel of the BS equation in the form of one-meson
exchange. As it was discussed in the previous sections, it results
in a certain success in describing the low-energy properties of
the deuteron and its elastic form factors. However, for a
successful description of high energy behavior of bound state we
need nonperturbative methods for deriving the kernel of the BS
equation. As one of the methods, the separable form of the kernel
can be used, which is beyond this approximation. As we noted
above, the DIS in the Bjorken limit does not depend on high energy
behavior of the BSA, so we may rely on this approximation. Below
we will use Graz II potential for numerical calculations which
also successfully describes low-energy properties of the deuteron.

The second assumption allows to treat the squared amplitude
$W^{A}_{\mu\nu}$ of DIS on the nucleus as the sum of the squared
amplitudes for scattering on individual constituents, while the
interference terms are neglected. As it will be shown below, the
justification for this is the suppression of the interference
terms in $W^{A}_{\mu\nu}$ as inverse powers of $Q^2$. Therefore,
these terms are important at small $Q^2$ where they can affect the
$Q^2$ dependence of $W^{A}_{\mu\nu}$, while in the Bjorken limit
they can be neglected.

The available experimental data for DIS on nuclei is mainly in the
region $x>10^{-3}$ and $Q^2>1~GeV^2$, and shows that the ratio
$F^{\rm A}_2/F^{\rm D}_2$ is independent of $Q^2$. In the
calculations we shall restrict ourselves to the Bjorken limit,
where the first and second approximations are well justified.

The third assumption allows the hadronic tensor of a virtual
nucleon to be represented in the form ~(\ref{inlorentz}). But this
representation is valid when the nontrivial differences between
scattering on free and bound nucleon are small. There are three
such differences which result in the so-called off-shell effects:
\begin{itemize}
  \item the impossibility of using the condition of gauge invariance
  for the bound nucleon in the form~(\ref{calib});\\[-0.8cm]
  \item the contribution of antinucleon degrees of freedom;\\[-0.8cm]
  \item the relative time separating the bound nucleons.
\end{itemize}

Since it is impossible to use the condition~(\ref{calib}), which
has been formulated systematically only for physical particles,
the expression for $W^{N}_{\mu\nu}$ for a bound nucleon turns out
to be more complicated than~(\ref{inlorentz}). In general, as
analysis in the quark-parton model has shown~\cite{offshell}, the
amplitude for DIS on a bound nucleon can be constructed in terms
of $14$ structure functions, of which only three are important in
the high $Q^2$ limit. From this point of view, the choice of the
actual number of the SFs parameterizing the Lorentz structure of
the hadronic tensor depends strongly on the model assumptions. The
model independent solution of the problem has been found  within
the Bethe-Salpeter formalism, which gives in the Bjorken limit a
rigorous relation between nuclear structure functions and on-shell
nuclear constituent structure functions $F_1(x)$, $F_2(x)$ and
their derivatives~\cite{bma3,bma4}.

The role of the contribution of the antinucleon degrees of freedom
in the structure of relativistic nucleus is not yet completely
clear. Recent studies in the framework of the Bethe-Salpeter
formalism discussed in the previous sections have shown that in
the electron elastic scattering and electro-disintegration of the
deuteron the effects of the antinucleon degrees of freedom can be
connected with mesonic-pair currents. They are
also important for describing deuteron static
 properties~(see section \ref{em-moms}).
However, as it will be shown below, their contribution to
$W^{N}_{\mu\nu}$ in the Bjorken limit is negligible.

The third off-shell effect was dropped out from all
semi-relativistic approaches. The different ways of doing it
consistently leads to different quasi-potential approaches. The
disregard of the effect was justified by the assumption that the
relative time of the bound nucleons is an unphysical feature of
the relativistic bound state. Thus all observables should not
depend on this property and different  quasi-potential approaches
should produce physically equivalent results.
 However, simple analysis of the analytic properties of the
off-shell nucleon hadronic tensor
has shown that the relative time can affect
observables in the DIS. Another indirect note on such possible
effects was made by the analysis~\cite{TjonPasc}, where it was
shown that different quasi-potential approaches are not equivalent
in sense of relativistic covariance. Further analysis of the
amplitude of the DIS has shown that the relative time effect is
responsible for the deviations of nuclear to deuteron SF's ratio
and can provide universal understanding of the EMC effect for all
nuclei. In the succeeding sections we will consider in details
these results.

\subsection{\em DIS on the Deuteron}~\label{deuteron}

\subsubsection{\em BSA for  Compton Scattering on the Deuteron} \label{deuteron:Compton}
Due to the unitarity relation~(\ref{unit}), the calculation of the
hadronic part of the amplitude reduces to the calculation of the
amplitude for forward Compton scattering on the deuteron, which by
the definition is the expectation value of the $\rm T$-product of
nucleon electromagnetic currents in deuteron states:
\begin{eqnarray}
T^{\rm D}_{\mu \nu}(P,q) = i \int d^4x e^{ i q x } \langle {\rm
D}|  {\rm T} \left(J_\mu \left(x \right) J_\nu \left(0\right)
\right) |{\rm D} \rangle. \label{imp33}\end{eqnarray}
Using~(\ref{matrres}), this definition can be rewritten in terms
of the solutions of the BS equation for the deuteron $\Gamma^{\rm
D}(P,k)$ and two-nucleon Green's functions ${\overline
G_6}_{\mu\nu}$:
\begin{equation} T^{\rm D}_{\mu\nu}(P,q) = \int \frac{d^4k_1}{(2\pi )^4} \frac{d^4k_2}{(2\pi
)^4}\overline{\Gamma}^{\rm D}(P,k_1)S_{(2)}(P,k_1) {\overline
G_6}_{\mu \nu}(q,P,k_1,k_2)S_{(2)}(P,k_2)\Gamma^{\rm D}(P,k_2).
\label{tvert}\end{equation} According to~(\ref{barg2nk}), the
function ${\overline G_6}_{\mu\nu}$ is related to the exact
two-nucleon Green's function with an insertion describing the
Compton scattering of virtual photons on a system of two
interacting nucleons: \begin{eqnarray} {\overline
G_6}_{\mu\nu}(q,P,k,k^\prime) =\int \frac{d^4k_1}{(2\pi)^4}
\frac{d^4 k_2}{(2\pi)^4} G^{-1}_4(P,k,k_1) {G_{6}}_{\mu \nu
}(q,P,k_1,k_2)G^{-1}_4(P,k_2,k^\prime),
\label{mandv}\end{eqnarray} where
\begin{eqnarray}
&&{G_6}_{\mu\nu}(q,P,k^\prime,k)= i\int d^4x d^4y d^4y^\prime d^4Y
d^4Y^\prime e ^{-iky+ik^\prime y^\prime} e^{-iqx}
e^{-iP(Y-Y^\prime)} \\ \nonumber &&\times <0|{\rm T}({\bar
\psi}(Y+\frac y2){\bar \psi}(Y-\frac y2)J_\mu (x) J_\nu (0) \psi
(Y^\prime+\frac {y^\prime}{2}) \psi (Y ^\prime-\frac
{y^\prime}{2}))|0>. \end{eqnarray} If a specific form for
$\overline G_4$ is assumed, the function ${\overline
G_6}_{\mu\nu}$ can be obtained explicitly. To find the amplitude
for Compton scattering on the deuteron in general, it is
sufficient to determine the relation between ${\overline
G_6}_{\mu\nu}$ and the expansion of ${G_6}_{\mu\nu}$ in terms of
the functions $\overline G_4$.

Expressing the functions $G_4$ using~(\ref{expansion}), we obtain
the following expansion of $G_4$ in terms of $\overline G_4$:
\begin{eqnarray}
&&G_4(P;k,k^\prime)=S_{(2)}(P,k)\left((2\pi)^4\delta(k-k^\prime)+
\phantom{\frac{dx}{(2\pi)^4}}\right.\label{g4}\\ &&\left. +
\sum\limits_{n\ge 1}\frac{1}{n!}\int\frac{d^4k_1}{(2\pi)^4}...
\frac{d^4k_n}{(2\pi)^4}\overline{G}_4(P;k,k_1)S_{(2)}(P,k_1)...
\overline{G}_4(P;k_n,k^\prime)S_{(2)}(P,k^\prime)\right).
\nonumber\end{eqnarray} Furthermore, expanding ${G_6}_{\mu\nu}$
and substituting this expression into~(\ref{mandv}), we obtain a
series whose $n$-th term has the form
\begin{eqnarray} &&{G_6}_{\mu\nu}^{(n)}(q,P,k,k^\prime)=\label{g6n}\\
&&\sum\limits_{n_1+n_2+n_3= n} \int
\frac{d^4k_2}{(2\pi)^4}\frac{d^4k_1}{(2\pi)^4}
G_4^{(n_1)}(P,k,k_1) {\overline
G_6}^{(n_2)}_{\mu\nu}(q,P,k_1,k_2)G_4^{(n_3)}(P,k_2,k^\prime).
\nonumber\end{eqnarray} Choosing the term of zeroth order in
$\overline G_4$, we immediately obtain the corresponding
contribution to ${\overline G_6}_{\mu\nu}$:
\begin{eqnarray} &&{\overline G_6}^{(0)}_{\mu\nu}(q,P,k,k^\prime)=\label{lam0}\\
&&={S_{(2)}}^{-1}(P,k)\left[{G^{(0)}_6}^{\bf
a}_{\mu\nu}(q,P,k)(2\pi)^4\delta^4( k-k^\prime) +{G^{(0)}_6}^{\bf
b}_{\mu\nu}(q,P,k)(2\pi)^4\delta^4(k-k^\prime-q)
\right]{S_{(2)}}^{-1}\left(P,k^\prime \right). \nonumber
\end{eqnarray}
Thus, in the zeroth order in $\overline G_4$ the function
${\overline G_6}_{\mu\nu}$ contains both the one-nucleon
contribution ${\small\bf a})$ \begin{equation} {G^{(0)}_6}^{\bf
a}_{\mu\nu}(q,P,k)= {G_4}_{\mu\nu}\left(q,\frac{P}{2}+k\right)
\otimes S\left(\frac{P}{2}-k\right)+
{G_4}_{\mu\nu}\left(q,\frac{P}{2}-k\right)\otimes
S\left(\frac{P}{2}+k\right),
\end{equation}
and the contribution corresponding to scattering on various
nucleons ${\small\bf b})$: \begin{equation} {G^{(0)}_6}^{\bf
b}_{\mu\nu} (q,P,k)={G_3}_{\mu}\left(q,\frac{P}{2}+k\right)
\otimes {G_3}_{\nu}\left(q,\frac{P}{2}-k\right)+
{G_3}_{\mu}\left(q,\frac{P}{2}-k\right) \otimes
{G_3}_{\nu}\left(q,\frac{P}{2}+k\right).
\end{equation}

The Green's functions  ${G_4}_{\mu\nu}$ and ${G_3}_{\mu}$
respectively describe the Compton and elastic scattering of a
virtual photon on a virtual nucleon.

The first order contribution to ${G_6}_{\mu\nu}$ depends on
${\overline G_6}^{(0)}_{\mu\nu}$ and ${\overline
G_6}^{(1)}_{\mu\nu}$, and this leads to the expression
\begin{eqnarray} &&{\overline G_6}^{(1)}_{\mu\nu}(q,P,k,k^\prime)=
{S_{(2)}}^{-1}(P,k){G^{(1)}_6}_{\mu\nu}(q,P,k,k^\prime)
{S_{(2)}}^{-1}(P,k^\prime)- \label{lam2}\\
&&\!\!\!\!\!\!\!\!\!\!\!\! - \int
\frac{d^4k^{\prime\prime}}{(2\pi)^4}\left\{{S_ {(2)}}^{-1}(P,k)
{G_4^{(1)}}(P,k,k^{\prime\prime}) {\overline
G_6}^{(0)}_{\mu\nu}(q,P,k^{\prime\prime},k^\prime) +{\overline
G_6}^{(0)}_{\mu\nu}(q,P,k,k^{\prime\prime})
+{G_4^{(1)}}(P,k^{\prime\prime},k^\prime)
{S_{(2)}}^{-1}(P,k^\prime)\right\}, \nonumber\end{eqnarray} where
the function ${G^{(1)}_6}_{\mu\nu}$ is expressed in terms of the
Green's functions  ${\overline G_5}_{\mu}$ and the zero-order term
of the function ${G_6}_{\mu\nu}$: \begin{eqnarray}
&&{G_6}^{(1)}_{\mu\nu}(q,P,k,k^\prime)= \nonumber\\
&&\int\frac{d^4k^{\prime\prime}}{(2\pi)^4}\frac{d^4k^{\prime\prime\prime}}
{(2\p i)^4}S_{(2)}(P,k)
\overline{{G_5}_{\mu}}\left(q,k,k^{\prime\prime}+q\right)
\overline G_4(P,k^{\prime\prime}+q,k^{\prime\prime\prime}+q)
\overline{{G_5}_{\nu}}\left(q,k^{\prime\prime\prime}+q,k^\prime\right)
S_{(2)}(P ,k^\prime)+ \label{lam3}\\ && + \int
\frac{d^4k^{\prime\prime}}{(2\pi)^4}\left\{
{G_4^{(1)}}(P,k,k^{\prime\prime }) {\overline
G_6}^{(0)}_{\mu\nu}(q,P,k^{\prime\prime},k^\prime)
S_{(2)}(P,k^\prime) +S_{(2)}( P,k){\overline
G_6}^{(0)}_{\mu\nu}(q,P,k,k^{\prime\prime})
{G_4^{(1)}}(P,k^{\prime\prime},k^\prime)\right\}.
\nonumber\end{eqnarray}

According to Eq.~(\ref{barg2nk}), the function  ${\overline
G_5}_{\mu}$ is determined by the Green's function ${G_5}_{\mu}$
describing the absorption of a virtual photon by a system of two
virtual nucleons.

Following this procedure, we can obtain ${\overline G_6}_{\mu\nu}$
in any order in $\overline G_4$. However, the general structure of
Eq.~(\ref{tvert}) is such that all the higher contributions reduce
to the leading term already studied. This can be easily checked by
using~(\ref{BSgam}) to go to higher order of $\overline G_4$
in~(\ref{tvert}).

Substituting the expressions obtained for ${\overline
G_6}_{\mu\nu}$ into (\ref{tvert}) and taking into account the
definition (\ref{mandv}), we obtain the amplitude for Compton
scattering on the deuteron in general form: \begin{eqnarray}
&&T_{\mu \nu}^{\rm D}(P,q)=\int
\frac{d^4k}{(2\pi)^4}\overline{\Gamma^{\rm D}}(P,k)
{G_6}^{(0)}_{\mu\nu}(q,P,k)\Gamma^{\rm D}(P,k)+ \nonumber
\\ &&+\int
\frac{d^4k}{(2\pi)^4}\frac{d^4k^\prime}{(2\pi)^4}
\frac{d^4k^{\prime\prime}}{(2\pi)^4}\frac{d^4k^{\prime\prime\prime}}{(2\pi)^4}
\overline{\Gamma^{\rm D}}(P,k)S^{(2)}(P,k) {\overline
G_5}_{\mu}(q,P,k,k^\prime)G_4(P,k^\prime,k^{\prime\prime})\times
\nonumber\\ &&\times{\overline
G_5}_{\nu}(q,P,k^{\prime\prime},k^{\prime\prime\prime})
S^{(2)}(P,k^{\prime\prime\prime})\Gamma^{\rm
D}(P,k^{\prime\prime\prime}). \nonumber\end{eqnarray}

\begin{figure}[h] \epsfxsize=10cm \hspace*{4cm}\epsfbox{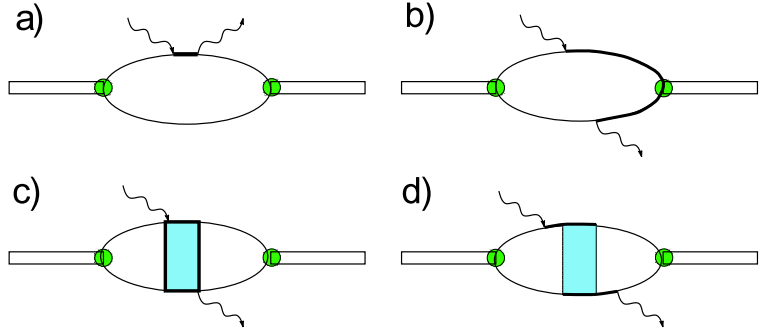}
\caption{\label{diags} Diagrams of forward Compton scattering on
the deuteron }\end{figure}

In Fig.~\ref{diags} we show schematically the various
contributions to the amplitude of forward Compton scattering on
the deuteron. Transpositions of virtual nucleon lines are implied
for all diagrams.  The explicit form of the expressions
represented by these graphs is given by the terms ({\bf a}) and
({\bf b}) in~(\ref{lam0}).
 Here the heavy and light lines denote the nucleon
propagators with high and low momenta, respectively. The graph a)
represents the relativistic impulse approximation in which only
scattering off single nucleon is taken into account. It
corresponds to contribution of the first term in Eq.~(\ref{lam0}).
The diagram b) represents contribution of interference terms in
the impulse approximation (the second term in Eq.~(\ref{lam0}).
The contribution of the terms contain the BS vertex functions with
high momenta to imaginary part of the Compton amplitude is
suppressed as $(1/Q^2)^l$, $l \geq 2$. The diagrams c) and d)
represent contribution of interaction corrections to
${\overline{G}_{6}}_{\mu\nu}$ (see Eq.~(\ref{lam2}). These terms
contain the contributions of two or more nucleon propagators with
high momenta and, therefore, are suppressed as $(1/Q^2)^l$.

Thus the only $Q^2$ independent term comes from the relativistic
 impulse approximation,
while irreducible interaction corrections to the imaginary part of
$T_{\mu \nu}^{\rm D}$ are suppressed by powers of
$1/Q^2$~\cite{deut}. This justifies consideration of the
 zeroth order term of
${{G}_{6}}_{\mu\nu}$ presented by the first term in~(\ref{lam0}).

\subsubsection{\em Hadronic Tensor of the Deuteron}
Let us consider now approximation  for the hadronic tensor of the
 deuteron which neglects  terms of the order  $1/Q^2$.
Substituting ~(\ref{lam0}) into ~(\ref{tvert}) and discarding the
$1/Q^2$ terms, we can write the amplitude for unpolarised
scattering on the deuteron as: \begin{eqnarray}\label{imp17} & &
T_{\mu \nu}^{\mathrm{D}}(P,q)=\\ \nonumber
&=&\int\frac{d^4k}{(2\pi)^4} \overline{\Gamma}^D(P,k)S\left(\frac
P2-k\right)\left(S\left(\frac P2+k\right)
{\overline{G}_4}_{\mu\nu}\left(q,\frac P2+k\right) S\left(\frac
P2+k\right)\right) \Gamma^D(P,k).
\end{eqnarray}
The nucleon propagator $S$ can be expanded in terms of Dirac
spinors: \begin{eqnarray} S^{ss^\prime}(p)=\frac{m}{\widetilde
E}\frac{u^s({\bf p}){\bar u}^{s^\prime}({\bf p})} {(p_0-\widetilde
E+\mathrm{i}\delta)} -\frac{m}{\widetilde E}\frac{v^s({\bf
p}){\bar v}^{s^\prime}({\bf p})} {(p_0+\widetilde
E-\mathrm{i}\delta)}. \label{prop}\end{eqnarray} Here  $\widetilde
E=\sqrt{{\bf p}^2+m^2+(\hat p + m)\overline G_2(p)}$ is the
nucleon energy, which becomes the nucleon energy on the mass shell
($E=\sqrt{{\bf p}^2+m^2}$) for $p^2=m^2$.

The Green's function ${\overline{G}_4}_{\mu\nu}$ is directly
related to the amplitude of Compton scattering on the nucleon:
\begin{eqnarray} T_{\mu \nu}^{\rm\tilde N} \left(\frac
P2+k,q\right)=2m\sum_s {\bar u}^{s}\left({\frac \bP2+\bk }\right)
{\overline{G}_4}_{\mu\nu}\left(q,\frac P2+k\right) u^{s}\left({
\frac \bP2+\bk }\right)\nonumber \\ T_{\mu \nu}^{\overline{\rm
\tilde N}} \left(\frac P2+k,q\right)=2m\sum_s {\bar v}^{s}\left({
\frac \bP2+\bk }\right) {\overline{G}_4}_{\mu\nu}\left(q,\frac
P2+k\right) v^{s}\left({ \frac \bP2+\bk }\right),
\label{TtoG}\end{eqnarray} where  ${\rm\tilde N} (\overline{\rm
\tilde N})$ denotes a bound nucleon (antinucleon). Using the
representation~(\ref{prop}) and Eq.~(\ref{TtoG}) and taking into
account the azimuthal symmetry of the Bethe--Salpeter vertex
function, we rewrite the Compton scattering amplitude in terms of
the nucleon and antinucleon amplitudes: \begin{eqnarray}
\label{factor}\frac 16\sum\limits_{s,s^{\prime }}\widetilde T_{\mu
\nu }^{{\rm\tilde N}^{s,s^{\prime }}}(\frac P2 +k,q)
f^{s,s^{\prime
};S}(P,k)&=&T_{\mu \nu }^{\rm\tilde N}(\frac P2 +k,q)f(P,k), \\
\nonumber f(P,k)&=&\frac 13\sum\limits_{s,S}f^{ss;S}(P,k).
\end{eqnarray} This leads to the analog of the convolution formula
for the Compton amplitude:
\begin{eqnarray} \label{impcomp21} \!\!\! T_{\mu \nu
}^{\mathrm{D}}(P,q)=\int \frac{d^4k}{(2\pi )^4}T_{\mu \nu
}^{\rm\tilde N}\left(\frac P2+k,q\right)f^{\rm\tilde N}(P,k) +
\int \frac{d^4k}{(2\pi )^4}T_{\mu \nu}^{\overline{\rm\tilde N}}
\left(\frac P2+k,q\right)f^{\overline{\rm\tilde N}}(P,k).
\nonumber \end{eqnarray} The total averaged nucleon Compton
scattering amplitude ${\overline{G}_4}_{\mu\nu}\left(q,\frac
P2+k\right)$ has singularity associated with the continuum in the
intermediate state. Therefore, at large $Q^2$, $T_{\mu
\nu}^{{\rm\tilde N}({\rm\overline{\tilde N}})} \left(\frac
P2+k,q\right)$ can be related to the hadronic tensor of the
nucleon
 by the unitarity condition:
 { \begin{eqnarray} W_{\mu \nu}^{\rm D}(P,q)=
\int \frac{d^4k}{(2\pi )^4}W_{\mu \nu }^{\rm \tilde N}\left(\frac
P2+k,q \right)f^{\rm\tilde N}(P,k) + \int \frac{d^4k}
{(2\pi)^4}W_{\mu \nu}^{\rm\overline{\tilde N}} \left( \frac
P2+k,q\right)f^{\rm\overline{\tilde N}}(P,k), \label{hadron}
\end{eqnarray} }
 where distribution functions have the form:
\begin{eqnarray} \!\!\!\!\!f^{\rm\tilde N}(P,k)= \frac{{\mathrm{i}}m^2}{2E^3}\frac{1}
{\left(\frac{M_{\mathrm{D}}}{2}+k_0-E+\mathrm{i}\delta\right)^2}
\left[\frac{\Phi^2_{++}(P,k)} {-k_0-(E-\frac{M_{\mathrm{D}}}{2})+
\mathrm{i}\delta}+ \frac{\Phi^2_{+-}(P,k)}
{-k_0+(E+\frac{M_{\mathrm{D}}}{2})-
\mathrm{i}\delta}\right],\nonumber\\
\!\!\!\!\!f^{\overline{\rm\tilde N}}(P,k)=
\frac{{\mathrm{i}}m^2}{2E^3}\frac{1}
{\left(\frac{M_{\mathrm{D}}}{2}+k_0+E-\mathrm{i}\delta\right)^2}
\left[\frac{\Phi^2_{--}(P,k)}
{-k_0+(E+\frac{M_{\mathrm{D}}}{2})-\mathrm{i}\delta}+
\frac{\Phi^2_{-+}(P,k)}
{-k_0-(E-\frac{M_{\mathrm{D}}}{2})+\mathrm{i}\delta}\right].\nonumber
\end{eqnarray}
The functions  $\Phi$ are related to the BS vertex functions as
follows:
\begin{eqnarray} \Phi^2_{++}(M_D,k)&=&\overline{\Gamma }^{\rm D}_{\alpha\beta}(M_D,k)
\sum_s u^s_\alpha({\bf k})\overline{u}^s_\delta({\bf k}) \sum_s
u^s_\beta(-{\bf k})\overline{u}^s_\gamma(-{\bf k})
\Gamma^{\rm D}_{\delta\gamma}(M_D,k),\nonumber \\
\Phi^2_{+-}(M_D,k)&=&-\overline{\Gamma}^{\rm
D}_{\alpha\beta}(M_D,k) \sum_s u^s_\alpha({\bf
k})\overline{u}^s_\delta({\bf k}) \sum_s v^s_\beta({\bf
k})\overline{v}^s_\gamma({\bf k})
\Gamma^{\rm D}_{\delta\gamma}(M_D,k),\nonumber \\
\Phi^2_{-+}(M_D,k)&=&-\overline{\Gamma}^{\rm
D}_{\alpha\beta}(M_D,k) \sum_s v^s_\alpha(-{\bf
k})\overline{v}^s_\delta(-{\bf k}) \sum_s u^s_\beta(-{\bf
k})\overline{u}^s_\gamma(-{\bf k})
\Gamma^{\rm D}_{\delta\gamma}(M_D,k),\nonumber \\
\Phi^2_{--}(M_D,k)&=&\overline{\Gamma}^{\rm
D}_{\alpha\beta}(M_D,k) \sum_s v^s_\alpha(-{\bf
k})\overline{v}^s_\delta(-{\bf k}) \sum_s v^s_\beta({\bf
k})\overline{v}^s_\gamma({\bf k}) \Gamma^{\rm
D}_{\delta\gamma}(M_D,k). \end{eqnarray} Thus, we have obtained an
expression relating the hadronic tensor of the relativistic
deuteron to the hadronic tensors of the off-shell nucleon and
antinucleon bound in the nucleus.

\paragraph{Structure Function of the Deuteron \boldmath $F_2^{\rm D}$\unboldmath}
In order to calculate the deuteron structure function $F_2^{\rm
D}(x)$, it is necessary to express the corresponding hadronic
tensor in terms of the scalar structure function. This procedure
can be performed by using the representation~(\ref{inlorentz}),
which is valid only for free particles. This makes it inapplicable
for a nucleon bound in the deuteron.

The  problem can be overcome by integrating~(\ref{hadron}) with
respect to $k_0$, taking into account the analytic properties of
the integrand.  The integrand contains singularities of the
propagator, and the BS vertex functions. The nucleon propagators
contain nucleon and antinucleon poles and cuts connected with the
self energy $\overline G_2(p)$. The latter contributes only at
large nucleon energy and can be neglected in our basic
approximation. The singularities of the Bethe--Salpeter vertex
functions can be fixed by means of the relation between these
vertices and the two-nucleon Green's function: \[ \Gamma^{\rm
D}_{\alpha\beta}(P,k)\overline{\Gamma}^{\rm
D}_{\delta\gamma}(P,k')= \lim_{P^2\rightarrow M_{\mathrm{D}}^2}
(P^2-M_{\mathrm{D}}^2)
S^{-1}_{(2)}(P,k){G_4}_{\alpha\beta\delta\gamma}(P,k,k')S^{-1}_{(2)}(P,k^\prime).
\]
Thus the BS vertex functions for the deuteron has the same
singularities in the relative momentum as the two-nucleon Green's
function. Since the singularities in this function closest in
energy are determined by the cut beginning at $k^2=m_{\pi}^2$,
they can be neglected in integrating~(\ref{hadron}) with respect
to $k_0$, following our assumptions. This makes it possible to
approximate the integral with respect to $k_0$ by the residues at
the nucleon and antinucleon poles of the corresponding
propagators. As a result, we obtain the following expression for
the hadronic tensor: {
\begin{eqnarray}
W_{\mu\nu}^D(M_D,q)=\int\frac{d^3k}{(2\pi)^3}\frac{m^2}{2E^3(M_D-2E)^2}
\left\{\phantom{\frac{a^2}{b^2}}
\Phi^2_{++}(M_D,k)W_{\mu\nu}^N({\bk},q)+ \right.
\label{hadr}\nonumber \\
+\left. (M_D-2E)\frac{\partial}{\partial
k_0}\left(W_{\mu\nu}^N(k,q) \Phi^2_{++}(M_D,k)\right)_{k_0=k^N_0}+
\right.  \\ \left.
+\frac{(M_D-2E)^2}{M_D^2}\left[\phantom{\frac{a^2}{b^2}}
+\Phi^2_{+-}(M_D,k)W_{\mu\nu}^N({\bk},q)+
\Phi^2_{-+}(M_D,k)W_{\mu\nu}^{\overline{N}}({\bk},q)+
\right.\right.\nonumber\\  \left. \left.
+M_D\frac{\partial}{\partial k_0}\left(W_{\mu\nu}^N(k,q)
\Phi^2_{+-}(M_D,k)\right)_{k_0=k^N_0}+ M_D\frac{\partial}{\partial
k_0}\left(W_{\mu\nu}^{\overline{N}}(k,q)
\Phi^2_{-+}(M_D,k)\right)_{k_0=k^N_0}+\right. \right. \nonumber \\
\left. \left. +\frac{M_D^2}{(M_D+2E)} \frac{\partial}{\partial
k_0}\left(W_{\mu\nu}^{\overline{N}}(k,q)
\Phi^2_{--}(M_D,k)\right)_{k_0=k^N_0}+
\frac{M_D^2}{(M_D+2E)^2}\Phi^2_{--}(M_D,k)W_{\mu\nu}^{\overline{N}}
({\bk},q)\right] \right\}.\nonumber \end{eqnarray} } Thus we have
expression relating the hadronic tensor of the deuteron to the
hadronic tensor of the on-shell nucleons and their derivatives
near the mass shell.
 Now we can use Eq.~(\ref{inlorentz}) and obtain $F_2^{\rm
D}$ by means of a projection operator:
$$W_j^{\rm \widetilde
N}(q,k_i)= P_{j}^{\mu\nu} W_{\mu\nu}^{\rm \widetilde N}(k_i\cdot
q,q^2,k_i^2).$$ In the Bjorken limit we can use the metric tensor
$g_{\mu\nu}$ as this operator:
$$\lim _{Q^2\rightarrow \infty }g^{\mu \nu }W^{\rm
N(A)}_{\mu \nu }(P,q) =-\frac 1x F^{\rm N(A)}_2(x)~.$$ This
operator is independent of the relative momentum, and so the
derivative of the hadronic tensor has the following form:
\begin{eqnarray} &&g^{\mu\nu}\frac {d}{d{k_i}_0} W_{\mu\nu}^{\rm
\widetilde N}(k_i,q)= \frac {d}{d(k_i\cdot q)}W^{\rm \widetilde
N}(k_i\cdot q,q^2,k_i^2) \frac{d(k_i\cdot q)}{d{k_i}_0} +
2{k_i}_0\frac {d}{dk_i^2}W^{\rm \widetilde N}(k_i\cdot
q,q^2,k_i^2), \nonumber\\ &&W^{\rm \widetilde N}(k_i\cdot
q,q^2,k_i^2)= g^{\mu\nu} W_{\mu\nu}^{\rm \widetilde
N}(k_i,q).\label{HadrtoSF} \end{eqnarray} Here the first term
reflects the modification of the structure of the bound nucleon.
The second term reflects the variations of the nucleon hadronic
tensor with nucleon energy and its value is proportional to
$(M_{\rm D}-2E)/M_{\rm D}$. This allows us to neglect the
dependence of $W_{\mu\nu}^{\widetilde N}$ on $k^2_i$:
\begin{equation}
\frac{d}{d k_0} \lim _{Q^2\rightarrow \infty }g^{\mu \nu } W^{\rm
\widetilde N}_{\mu \nu }(P,q)|_{k_0=k^{\mathrm{N}}_0}
=\left[\frac{1}{x^2}F_2(x)-\frac{1}{x}\frac{d}{d x}F_2(x)\right]
\left(\frac{d x}{dk_0}\right)_{k_0=k^{\mathrm{N}}_0}.
\label{dhadr}\end{equation}

Neglecting terms of order  $(M_D-2E)^2$ we can write the deuteron
structure function in the following form: {
\begin{eqnarray}
F_2^{\rm D}(x_{\rm D})=\int \frac{d^3k}{(2\pi)^3}
\frac{m^2}{4E^3(M_{\rm D}-2E)^2}\left\{F_2^N(x_N)
\left(\frac{E-k_3}{M_{\rm D}}+\frac{M_{\rm D}-2E}{M_{\rm
D}}\right) \Phi^2(M_{\rm D},k) -  \right.\nonumber\\ \label{f2}\\
\left. -\frac{M_{\rm D}-2E}{M_{rm D}} x_{\rm N}\frac{dF_2^{\rm
N}(x_{\rm N})}{dx_{\rm N}}\Phi^2(M_{\rm D},k) + F_2^{\rm N}(x_{\rm
N})\frac{E-k_3}{M_{\rm D}}(M_{\rm D}-2E) \frac{\partial}{\partial
k_0}\Phi^2(M_{\rm D},k)\right\}_{k_0=E-{\rm M_{\rm D}}/2}.
\nonumber\end{eqnarray} }

The normalization condition for the Bethe-Salpeter vertex
function~(\ref{norm_chi}) gives two normalization conditions for
the distribution function in this expression. They are the
momentum sum rule:
\begin{eqnarray} \!\!\!\!\!\!\int\frac{d^3k}{(2\pi)^3} \frac{m^2
(2E)}{2E^3(M_{\mathrm{D}}-2E)^2}\left\{
\left(\frac{M_{\mathrm{D}}-E}{M_{\mathrm{D}}}\right)
\Phi^2(M_{\mathrm{D}},k)+\right.\label{esr}\\ \left.
+\frac{M_{\mathrm{D}}-2E}{M_{\mathrm{D}}} \frac{\partial}{\partial
k_0} \Phi^2(M_{\mathrm{D}},k)\right\}_{k_0=k^{\mathrm{N}}_0}=
M_{\mathrm{D}} \nonumber\end{eqnarray}
 and the baryon sum rule
\begin{eqnarray}
\!\!\!\!\!\!\!\int\frac{d^3k}{(2\pi)^3}
\frac{m^2}{2E^3(M_{\mathrm{D}}-2E)^2}\left\{
\frac{M_{\mathrm{D}}-E}{M_{\mathrm{D}}}\Phi^2(M_{\mathrm{D}},k)+
(M_{\mathrm{D}}-2E) \frac{\partial}{\partial k_0}
\Phi^2(M_{\mathrm{D}},k)\right\}_{k_0=k^{\mathrm{N}}_0}=2.
\nonumber\end{eqnarray} That means that in frame of the
Bethe-Salpeter formalism the baryon and momentum sum rules are
different form of the same normalization
condition~(\ref{norm_chi}). What gives solution of the
longstanding problem of the simultaneous satisfaction of these sum
rules in the framework of the phenomenological models of the
EMC-effect.

\paragraph{The Nonrelativistic Limit.}
Let us expand the energy of the bound nucleon in (\ref{f2}) in
powers of  ${{\bf k}^2}/{m^2}$. This leads to
\begin{eqnarray}\label{f2nr}  F_2^{\rm D}(x_{\rm D})=\int
\frac{d^3k}{(2\pi)^3} \left\{F_2^{\rm N}(x_{\rm N})
\left(1-\frac{k_3}{m}\right) \Psi^2({\bf k})-
\frac{-T+\varepsilon}{m} x_{\rm N}\frac{dF_2^{\rm N}(x_{\rm
N})}{dx_{\rm N}}\Psi^2({\bf k})\right\},
\end{eqnarray} where $T=2E-2m$ is the nucleon kinetic energy and
$\varepsilon=M_{\rm D}-2m$ is the binding energy.

We introduce the analog of the nonrelativistic wave function
$\Psi^2({\bf k})$ (see also Eq.(\ref{eqn:psi1})), related to
$\Phi^2(M_{\mathrm{D}},k)$ as
$$ \Psi^2({\bf k})=\frac{m^2(M_{\rm D}-E)}{4E^3M_D(M_D-2E)^2}
\left\{\Phi^2(M_D,k)\right\}_{k_0=E-{\rm M_D}/2}. $$ The
normalization condition for $\Psi^2({\bf k})$ has the form:
$$\int
\frac{d^3k}{(2\pi)^3}\Psi^2({\bf k})=1.$$ We compare (\ref{f2nr})
with the calculations in the nonrelativistic limit of the
meson-nucleon field theory with synchronous nucleons:
\begin{eqnarray}\label{f2nkap}
F_2^{\rm D}(x_{\rm D})=\int \frac{d^3k}{(2\pi)^3} F_2^{\rm
N}(x_{\rm N}) \left(1-\frac{k_3}{m}\right) \Psi^2({\bf k})-
\frac{-<T>+\varepsilon}{m} x_{\rm D}\frac{dF_2^N(x_{\rm
D})}{dx_{\rm D}}.
\end{eqnarray}
Here $\Psi^2({\bf k})$ is the solution of the Shr\"odinger
equation~\cite{kapnonrel}.

Equations (\ref{f2nr}) and (\ref{f2nkap}) obviously have the same
structure. In the nonrelativistic calculation the term containing
the derivative of the nucleon structure function arose as a result
of the inclusion of the meson corrections associated with the
nucleon potential, while the analogous term in (\ref{f2nr}) is
ensured by the relative time of a bound nucleon. It is this
contribution which makes the deuteron to nucleon structure
functions ratio to differ from  unity in the range $0.3<x<0.6$.

From the obtained results the $x$-rescailing model can be derived.
Introducing the following variables,
$$\epsilon=<T>-\varepsilon = 2E-M_{\rm D}\hspace*{.5cm} {\rm and
}\hspace*{.5cm} y=\frac{x_{\rm D}}{x_{\rm N}}\frac{M_{\rm D}}{m}=
\frac{(E-k_3)}{m},
$$ we rewrite the expression~(\ref{f2nr}) in the form:
\begin{eqnarray}
F_2^{\rm D}(x_{\rm D})=\int\limits^{\infty}_{-\varepsilon}
d\epsilon&&\hspace*{-0.8cm}\int\limits^{M_{\rm D}/m}_0dy \left[
F_2^{\rm N}\left(\frac{x_{\rm D}}{y}\frac{M_{\rm
D}}{m}\right)+\frac{\epsilon}{m}\frac{x_{\rm
D}}{y^2}\frac{dF_2^{\rm N}\left(\frac{x_{\rm D}}{y}\frac{M_{\rm
D}}{m}\right)}{d(x_{\rm
D}/y)}\right]\times\label{xresc1}\\
&&\int\frac{d^3{\bf k}}{(2\pi)^3}y\frac{m}{E}\Psi^2({\bf
k})\delta\left(y-\frac{E-k_3}{m}\right)\delta(\epsilon-(M_{\rm
D}-2E))\nonumber
\end{eqnarray}
The variable $y$ has the meaning of the fraction of the deuteron
4-momentum projection onto direction of the photon momentum
transfer carried by the struck nucleon. It is normalized on the
nucleon mass in our notation. The variable $\epsilon$
characterizes nucleon separation energy. If we assume that the
mean value of this variable is small, then the expression in the
square brackets can be considered as an expansion with respect to
the parameter $\epsilon$. Thus we can convert it to the following
expression
$$F_2^{\rm N}\left(\frac{x_{\rm D}}{y}\frac{M_{\rm D}}{m}\right)+\frac{\epsilon}{m}\frac{
x_{\rm D}}{y^2}\frac{dF_2^{\rm N}\left(\frac{x_{\rm
D}}{y}\frac{M_{\rm D}}{m}\right)}{d(x_{\rm D}/y)}\simeq F_2^{\rm
N}\left(\frac{x_{\rm D}}{y-\frac{\epsilon}{m}}\frac{M_{\rm
D}}{m}\right).$$ To put this another way, the
 assumption is correct only when the states with
high separation energy $\epsilon$ give negligibly small
contribution to the integral in Eq.(\ref{xresc1}). Then the
expression~(\ref{f2nr}) takes the following form:
\begin{equation}
F_2^{\rm D}\left(\frac{m}{M_{\rm
D}}x\right)=\int\limits^{\infty}_{-\varepsilon} d\epsilon
\int\limits_{0}^{M_{\rm D}/m} \left\{F_2^{\rm
N}\left(\frac{x}{y-\epsilon/m}\right) f^{\rm
N/{D}}(y,\epsilon)\right\}~. \label{xresc2}\end{equation} Here
$x=x_{\rm D}M_{\rm D}/m$ is the Bjorken $x$ normalized to the
nucleon mass, $f^{\rm N/D}(y, \epsilon)$ is the deuteron spectral
functions for a bound proton:
\begin{eqnarray} f^{\rm N/D}(y,
\epsilon)=\int \frac{d^3k}{(2\pi)^3} \Psi^2({\bf k})
\frac{m}{E_{\rm N}} y\delta\left(y-\frac{E_{\rm N}-k_3}{m}\right)
\delta\left(\epsilon-(2E_{\rm N}-M_{\rm D})\right).\nonumber\\
\nonumber\end{eqnarray} These formulas precisely reproduce
expressions obtained in the $x$-rescailing model~\cite{xresc}. It
is clear that this expression is incorrect if the nucleon
separation energy becomes large, which seriously constrains this
model to small values of the parameter $\epsilon$. This fact can
be important in calculations of the structure functions of heavy
nuclei at large values of $x$ and can explain why the
$x$-rescailing model fails to reproduce experimental data at large
$x$. In next section we will consider extension of our formalism
for nuclei heavier then deuterium.

 Thus, the
Bethe-Salpeter formalism allows the deuteron structure function to
be expressed in terms of the structure functions of the bound
proton and neutron. The antinucleon contributions are suppressed
as the square of the mass defect. The inclusion of the dependence
on the relative time in the amplitudes for DIS on bound nucleons
leads to a modification of the nucleon structure reminiscent of
the EMC effect in heavy nuclei. This allows us to conjecture that
the nature of the effect can be attributed to the evolution of the
bound nucleons relative time effect from nuclei with $A=2$ to the
nuclei with values of $A$ at which the saturation of binding
effects sets in. In the next section we will consider in detail
the application of the Bethe-Salpeter formalism for light nuclei
and provide an analysis of the relativistic effects in the
evolution with $A$.


\subsection{\em Structure Functions of Light Nuclei and the EMC effect\label{DIS:Nuclei}}
Here we shall study the derivation of the relative changes of the
structure function, $F_2^A$, in relation to the structure function
of the isoscalar nucleon, $F_2^{\rm
N}=\frac{1}{2}[F_2^p(x)+F_2^n(x)]$, where $p$ and $n$ denote the
free proton and free neutron obtained from the recent world data
fit. On the other hand, comparison with the experimental data can
be made only for ratios of the structure functions of a nucleus
$A$ and the deuteron --- $A/$D. This is why we also calculate  the
ratios $A/{\rm D}$ using the results of the
section~\ref{deuteron}. In this way we will analyze also the
evolution of the effect of the relative time in the bound
nucleons, discussed in section~\ref{deuteron}, from the deuteron
to light nuclei and its manifestation in the EMC effect.

\subsubsection{\em Amplitude of the DIS for light nuclei}
We consider the generalization of the formalism developed in the
preceding section
 for the analysis of DIS off $n$-nucleon bound system, $n =$ 2$\div$4.
According to Eq.~(\ref{matrres}) the nuclear Compton amplitude can
be written in the form:
\begin{eqnarray}
\label{compt}T_{\mu \nu}^{A}(P,q)=\int d{\cal K} d{\cal K}^\prime
\overline{\Gamma}^{A}(P,{\cal K}){S_{(n)}}(P,{\cal K})
{\overline{G}_{2(n+1)}}_{\mu\nu}(q;P,{\cal K},{\cal K}^\prime)
{S_{(n)}}(P,{\cal K}^\prime) \Gamma^{A}(P,{\cal K}^\prime),
\label{Compton}\end{eqnarray} where ${\cal K}$ denotes a set of
momenta which  describes the relative motion of nucleons, ${\cal
K}=k_1,\dots, k_{n-1}$, $d{\cal K}={d^4k_1}/(2\pi)^4\dots
{d^4k_{n-1}}/(2\pi)^4$, and $P$ is the total momentum of the
nucleus. The function $\Gamma^A(P,{\cal K})$ is the BS vertex
function in momentum space:
\begin{eqnarray} S_{(n)}(P,{\cal
K})\Gamma^A_\alpha(P,{\cal K})=\int d^4x_1\dots
d^4x_{n}e^{-i\sum\limits_{j=1}^{n} k_j x_j} {\Phi}_{\alpha,
P}(x_1\dots x_{n}). \end{eqnarray} Here the Green's function
${\overline{G}_{2(n+1)}}_{\mu\nu}$ represents Compton scattering
of a virtual photon on a system of $n$-virtual nucleons. As in the
case of the deuteron~(see section~\ref{deuteron}), the only $Q^2$
independent term comes from the relativistic impulse
approximation, while irreducible interaction corrections to the
imaginary part of $T_{\mu \nu}^{A}$ are suppressed by powers of
$1/Q^2$~\cite{deut}. This justifies the consideration of the
zeroth order term of ${\overline{G}_{2(n+1)}}_{\mu\nu}$:
\begin{equation}
{\overline{G}_{2(n+1)}}_{\mu\nu}(q;P,{\cal K})=
\sum\limits_{i}{\overline G_4}_{\mu\nu}(q;P,k_i) \otimes
S^{-1}_{2n-1} (k_1,\dots k_{i-1},k_{i+1}, \dots
k_{n-1})\delta({\cal K}-{\cal K}^\prime)  + O(1/Q^2).
\end{equation} Then  $T_{\mu \nu}^A$ can be rewritten in terms of
the off-mass-shell nucleon Compton amplitude $T^{\rm \widetilde
N}_{\mu\nu}(k_i,q)$:
\begin{eqnarray}
&&T_{\mu\nu}^A(P,q)=\int d{\cal K} \sum\limits_{i}T^{\rm
\widetilde N}_{\mu\nu}(k_i,q) \overline u({\bf k}_i)S_{(n)}(P,
k_i)u({\bf k}_i) \overline{\Gamma}^A(P,{\cal K})S_{(n)}(P, {\cal
K})
\Gamma^A(P,{\cal K})\label{compt0}\\
&&T^{\rm \widetilde N}_{\mu\nu}(k_i,q)=\overline u({\bf
k}_i){\overline G_4}_{\mu\nu}(q;P,k_i)u({\bf k}_i)\end{eqnarray}

Following the procedure described in the previous section, we can
make integration over ${k_i}_0$ in this expression, relating
$T^A_{\mu\nu}$ with on-shell nucleon Compton amplitude, $$ T^{\rm
N}_{\mu \nu}(p,q) = i \int d^4x e^{ i q x } \langle {\rm N}|  {\rm
T} \left(J_\mu \left(x \right) J_\nu \left(0\right) \right) |{\rm
N} \rangle .$$ However, this can be realized only after the
singularities in nucleon propagators and the BS vertex functions
are taken into  account~\cite{deut}. Unlike the deuteron case,
where singularities in the BS vertex function can be neglected, in
case of $A>2$ there are additional singularities in  $\Gamma^A$
related with nucleon-nucleon bound states. These are poles in the
range of low relative momenta, and we have to take them into
account in the integration over ${k_i}_0$. To this end, the
``bare'' BS vertex function ${\cal G}^A$ can be introduced, which
is regular with respect to the relative nucleon
momenta~\cite{bma4}:
\begin{eqnarray} \Gamma^A(P,{\cal K})=-\int d{\cal K} g_{2n}(P,{\cal K}, {\cal
K}^\prime) {S_{(n)}}(P,{\cal K}^\prime) {\cal G}(P, {\cal
K}^\prime)\label{bare},
\end{eqnarray}
where $g_{2n}$ denotes the regular part of $n$-nucleon Green's
function at $P^2 \rightarrow$ $M_A^2$:
\begin{eqnarray}
g_{2n}(P,{\cal K}, {\cal K}^\prime)=\sum\limits_{m=1,\{1\dots
n\}}^{n-1} G_{2m}(P,k_1\dots k_{m};k^\prime_1\dots
k^\prime_{m})\otimes G_{2(n-m)}(P,k_{m+1}\dots
k_{n};k^\prime_{m+1}\dots k^\prime_{n}).
\end{eqnarray}
The sum implies all the possible transpositions of bound
particles. The function $g_{2n}$, however, contains singularities
of $m$-nucleon Green's functions ($m<n$).  For example, in the
case of $^3{\rm He}$ the function  $g_6$ depends on the exact
two-nucleon propagator $G_4$, which contains the deuteron pole and
the nucleon-nucleon continuous spectrum $\tilde g_4$:
\begin{equation} G_4\left(\frac{2P}{3}+k,k_1,k^\prime_1\right)=
\frac{\Gamma^{\rm D}(2P/3+k,k_1) \overline{\Gamma}^{\rm
D}(2P/3+k,k^\prime_1)} {\left(2P/3+k\right)^2 -M_D^2} + \tilde
g_4\left(\frac{2P}{3}+k,k_1,k^\prime_1\right)~.
\label{pole}\end{equation} For ${\rm ^4He}$ one has, additionally,
the ${\rm ^3He}$ and ${\rm ^3H}$ poles. Thus, unlike the deuteron
case, the nuclear amplitude of DIS is determined not by the
nucleon
 amplitude alone but also
by amplitudes of all possible bound fragments of the nucleus.

This can be demonstrated by the example of the ${\rm He^3}$
nucleus. Substituting expression~(\ref{bare}) into
Eq.~(\ref{compt0}) and using the relation~(\ref{unit}) we obtain
the hadronic tensor of ${\rm ^3He}$ in the form:
\begin{eqnarray}
&&W_{\mu\nu}^{\rm ^3He}(P,q)=
\int\frac{d^4k}{(2\pi)^4}\frac{d^4k^\prime}{(2\pi)^4}\frac{d^4K}{(2\pi)^4}
W^{\rm D}_{\mu\nu}\left(2P/3+K,q\right)\overline{\cal G}^{\rm
^3He}(P,K,k)
\nonumber\\
&&\frac{\Gamma^{\rm D}(2P/3+K,k)\overline\Gamma{\rm
D}(2P/3+K,k^\prime)} {\left((2P/3+K)^2-M_{\rm D}^2\right)^2}
\otimes S(P/3-K){\cal G}^{\rm ^3He}(P,K,k)+\nonumber\\
&&\overline{\cal G}^{\rm ^3He}(P,K,k)
\left[\int\frac{d^4k_1}{(2\pi)^4}G_4(2P/3+K,k,k_1)
S(P/3+K/2+k_1)\otimes S(P/3+K/2-k_1)\right.\nonumber\\&&\left.
G_4(2P/3+K,k,k_1)\right]\otimes S(P/3-K){\cal G}^{\rm ^3He}(P,K,k)
\frac{W^{\rm N}_{\mu\nu}(P/3-K)}{(P/3-K)^2-m^2}.
\end{eqnarray}
Integrating over the zeroth component of the relative momentum of
the fragments we obtain the hadronic tensor for $\rm ^3He$
expressed in terms of physical amplitudes of the fragments and its
derivatives over $k_0$ at the mass-shell.

The scalar structure functions can be extracted from the hadronic
tensors with the help of the projection operators~(\ref{dhadr}).
Introducing now Bjorken variables for a nucleus $x_{\mbox{\tiny\em
A}}=Q^2/(2P_A\cdot q)$ and for a nucleon $x_{\mbox{\tiny
N}}=Q^2/(2P_{\rm N}\cdot q)$, where N is either p or n,
 we find $F^A_2$ for ${\rm ^3He}$ and
${\rm ^3H}$ in the form: \begin{eqnarray} &&F_2^{{\rm
^3He}}(x_{{\rm ^3He}})=\nonumber\\ &&\hspace*{-0.7cm}
\int\frac{d^3k}{(2\pi)^3} \left[\frac{E_{\rm p}-{k_3}}{E_{\rm
p}}F_2^{\rm p}(x_{\rm p}) +\frac{E_{\rm D}-{k_3}}{E_{\rm D}}
F_2^{\rm D}(x_{\mbox{\tiny D}}) +\frac{\Delta^{{\rm ^3He}}_{\rm
p}}{E_{\rm p}} x_{\rm p}\frac{dF_2^{\rm p}(x_{\rm p})} {dx_{\rm
p}}+\frac{\Delta^{{\rm ^3He}}_{\rm p}}{E_{\rm D}}x_{\mbox{\tiny
D}} \frac{dF_2^{\rm D}(x_{\mbox{\tiny D}})}{dx_{\mbox{\tiny
D}}}\right]
{\Phi^2_{{\rm^3He}}}({\bf k}),\nonumber\\
&&F_2^{{\rm ^3H}}(x_{{\rm ^3H}})=F_2^{{\rm ^3He}}(x_{{\rm
^3He}})|_{ p\leftrightarrow n} \label{he3} \end{eqnarray} and for
${\rm ^4He}$ in the form: \begin{eqnarray} &&F_2^{{\rm
^4He}}(x_{{\rm ^4He}})=\label{he4}\\ &&\hspace*{-0.7cm}
\int\frac{d^3k}{(2\pi)^3} \left[\frac{E_{\rm p}-{k_3}}{E_{\rm
p}}F_2^{\rm p}(x_{\rm p}) +\frac{E_{\rm ^3H}-{k_3}}{E_{\rm
^3H}}F_2^{\rm ^3H}(x_{\rm ^3H}) +\frac{\Delta^{{\rm ^4He}}_{\rm
p}}{E_{\rm p}} x_{\rm p}\frac{dF_2^{\rm p}(x_{\rm p})} {dx_{\rm
p}}+\frac{\Delta^{{\rm ^4He}}_{\rm ^p}}{E_{\rm ^3H}}x_{\rm ^3H}
\frac{dF_2^{\rm ^3H}(x_{\rm ^3H})}{dx_{\rm ^3H}}\right.\nonumber\\
&&\hspace*{-0.7cm} +\left.\frac{E_{\rm n}-{k_3}}{E_{\rm
n}}F_2^{\rm n}(x_{\rm n}) +\frac{E_{\rm ^3He}-{k_3}}{E_{\rm
^3He}}F_2^{\rm ^3He}(x_{\rm ^3He}) +\frac{\Delta^{{\rm ^4He}}_{\rm
n}}{E_{\rm n}} x_{\rm n}\frac{dF_2^{\rm n}(x_{\rm n})} {dx_{\rm
n}}+\frac{\Delta^{{\rm ^4He}}_{\rm ^n}}{E_{\rm ^3He}}x_{\rm ^3He}
\frac{dF_2^{\rm ^3He}(x_{\rm ^3He})}{dx_{\rm ^3He}}\right]
{\Phi^2_{{\rm ^4He}}}({\bf k}), \nonumber\end{eqnarray} where
$\Delta^A_{\rm N}=-M_A+E_{\rm N}+E_{A-1}$ can be interpreted as
the removal energy of the corresponding nuclear fragment. The
three-dimensional momentum distributions $\Phi^2_{A}({\bf k})$ are
defined via the bare Bethe--Salpeter  vertex functions. For
example for $^3{\rm He}$ one has: \begin{eqnarray}
&&\Phi^2_{^3{\rm He}}({\bf k})=\frac{m M_{\rm D}}{4E_{\rm p}E_{\rm
D} M_{{\rm ^3He}}(M_{{\rm D}}-E_{\rm p}-E_{\rm D})^2} \left\{\int
\frac{d^4k_1}{(2\pi)^4}\frac{d^4k^\prime_1}{(2\pi)^4}
\overline{{\cal G}}^{^3{\rm He}}(P,k,k_1)
S_2\left(\frac{2P}{3}+k,k_1\right)\right. \label{phi}\\
&&\hspace*{-0.7cm}\times\left. \Gamma^{\rm
D}\left(\frac{2P}{3}+k,k_1\right) \overline{\Gamma}^{\rm
D}\left(\frac{2P}{3}+k,k^\prime_1\right)
S_2\left(\frac{2P}{3}+k,k^{\prime}_1\right) \otimes \left(\sum_s
u^{s}_\alpha({\bf k})\overline{u}^{s}_\delta({\bf k})\right) {\cal
G}^{^3{\rm He}}(P,k,k^\prime_1)\right\}_{k_0={k_0}_{{\rm p}}}~,
\nonumber\end{eqnarray} where ${k_0}_{{\rm p}}=M_{\rm
^3H}/3-E_{{\rm p}}$. Since at present there are no realistic
solutions of the BS equation for a bound system of three or more
nucleons, we have to use the phenomenological momentum
distributions for numerical evaluations.

The momentum distribution~(\ref{phi}) describes the motion of a
nuclear constituent (N, D, ...) in the field of the off-mass-shell
spectator system. It is directly related with the nuclear momentum
distribution measured in the $e$--$A$ scattering when only a
struck nuclear constituent is detected. It is reasonable, thus, to
assume that the momentum distributions in Eqs.~(\ref{he3}) and
(\ref{he4}) can be related with those extracted from the
experimental data. In the calculations we make use of the
distributions available from~\cite{he3} and~\cite{new}. The
contribution arising from the continuous spectra ($ppn$ for $\rm
^3He$ and $ppnn$ for $\rm ^4He$) is small in the considered
kinematic range and does not change comparison of the final result
with the data. This justifies some simplifications which results
in a rather transparent form of Eqs.~(\ref{he3}) and~(\ref{he4}).
The contributions neglected in the derivations have been
consistently taken into account in the normalization of the
momentum distributions  $\Phi^2_{\rm ^3He}$ and $\Phi^2_{\rm
^4He}$.

It should be stressed that both the modification of $F_2^{\rm N}$
and its evolution from $A=1$ to $4$ found in the
approach~\cite{myadep,bma4,deut} are the consequences of the
relativistic nature of the nuclear structure. In the analytical
calculations, it is essential to use the fact that the nucleons in
the nucleus are separated by the relative time $\tau_i$. It is
this feature which is the cause of the nuclear effect in $F_2^A$,
which appears as a result of the dependence of the hadronic tensor
of the bound nucleon on $\tau_i$.

\subsection{\em Results}
We emphasize that both the modification of the $F_2^{\rm N}$ and
its evolution from $A$ = 1 to 4 obtained in the framework of our
method result from the relativistic consideration of the nuclear
structure.  In the derivations we essentially exploited the fact
that the nucleons behave in a nucleus as asynchronous objects.
This particular feature is responsible for the binding effects in
$F_2^A(x)$ which arise from the dependence of the bound nucleon
hadronic tensor on $\tau_i$. The developed approach has a twofold
merit. First, we can naturally reproduce the results of
nonrelativistic models (e.g.~\cite{xresc}) which offer the
parametrization of the relativistic binding effects. Second, the
outcome of the present study is particularly easy to understand
when compared with the results of the $x$-rescaling
model~\cite{xresc}.

As an example let us consider the structure function of ${\rm
^3He}$. If we regard the integrand in Eq.~(\ref{he3}) as the terms
of the expansion in the binding energy and take into account the
fact that terms above the first order in this quantity are
negligible, we can add higher-order terms in such a way that the
resulting series can be represented by the expression,
\begin{equation}
F_2^{\rm ^3He}\left(\frac{m}{M_{\rm ^3He}}x\right)=\int dy
d\epsilon \left\{F_2^{\rm p}\left(\frac{x} {y-\epsilon/m}\right)
f^{\rm p/{^3He}}(y,\epsilon) + F_2^{\rm D}\left(\frac{x}
{y-\epsilon/M_{\rm D}}\right)f^{\rm
D/{^3He}}(y,\epsilon)\right\}~. \label{xresc3}\end{equation} Here
$\epsilon=\Delta^{{\rm ^3He}}_{\rm p}$ has the meaning of a
nucleon (deuteron) separation energy, $x=x_{\rm ^3He}M_{\rm
^3He}/m$ is the Bjorken $x$ normalized to the nucleon mass, and
$f^{\rm p(D)/^3He}(y, \epsilon)$ are the $\rm ^3He$ spectral
functions for a bound proton (deuteron):
\begin{eqnarray} f^{\rm p(D)/^3He}(y,
\epsilon)=\int \frac{d^3k}{(2\pi)^3} \Phi^2_{^3{\rm He}}({\bf k})
\frac{m}{E_{\rm p(D)}} y\delta\left(y-\frac{E_{\rm
p(D)}-k_3}{M_{\rm p(D)}}\right) \delta\left(\epsilon-(E_{\rm
p}+E_{\rm D}-M_{^3He})\right).\nonumber\\ \nonumber\end{eqnarray}

Indeed, from the comparison of Eqs.~(\ref{dhadr}) and (\ref{he3})
with Eq.~(\ref{xresc3}), we find that the relative time dependence
in the off-shell nucleon Compton amplitude results in the
rescaling of the nucleon Bjorken~$x$. We notice also that due to
the relation between the nucleon mass and the four-dimensional
radius of its localization region, $r^2\sim 1/m^2$, the dependence
on $\tau_i$  has to lead to the increase of the localization
region of the nucleon. In a way this result resembles the model
considerations of the effect of the increase of the deconfinement
radius or nucleon swelling~\cite{rev3}.

The binding effects are expressed in Eq.~(\ref{he3}) and
Eq.~(\ref{he4}) as the first-order derivatives of the structure
functions of nuclear fragments. Thus the input structure
functions, $F_2^{\rm p(n)}(x)$, are responsible here not only for
the internal nucleon structure but also for the dynamics of the
two-nucleon interactions as well. Similarly,
 $F_2^{\rm D}$ is responsible for the structure
of the two-nucleon bound state and for the dynamics of
three-nucleon interactions. The derivative of $F_2^{\rm D}$ is
expressed in terms of the first- and second-order derivatives of
$F_2^{\rm N}$ with the corresponding coefficients as shown in
Eq.~(\ref{f2nr}). Since the off-shell deformation of the bound
deuteron structure is determined by the second derivative of
$F_2^{\rm N}$, this very term accounts for the three-nucleon
dynamics. However, the second derivative of $F_2^{\rm N}$
contributes to $F_2^{\rm ^3He}$ with a very small coefficient,
$\Delta^{{\rm ^3He}}_{\rm D}\Delta^{{\rm D}}_{\rm p}$, and the
three-nucleon dynamics can thus be neglected in the consideration
of the binding effects in DIS.

The nucleon structure functions are introduced by
parameterizations based on the measurements of the proton and the
deuteron structure functions in DIS experiments. We have used the
most recent parametrization of  $F_2^{\rm p}(x,Q^2)$ found
in~\cite{smc} and fixed the value of $Q^2$ to 10 GeV$^2$. The
structure function $F_2^{\rm n}(x)$ is evaluated from $F_2^{\rm
p}(x)$ and from the ratio  $F_2^{\rm n}(x)/F_2^{\rm p}(x)$
determined in~\cite{bcdms}. We have verified that the
uncertainties in $F_2^{\rm p(n)}(x)$ are suppressed in the
obtained ratio $r^A(x)$ and, thus, can be neglected in the
considered kinematic range. On the other hand we have checked that
an unrealistic input $F_2^{\rm N}(x)$ would have completely
destroyed the evolution of the modifications we find in the
lightest nuclei.

\begin{figure}
\begin{center}
\epsfxsize=120mm \epsfbox{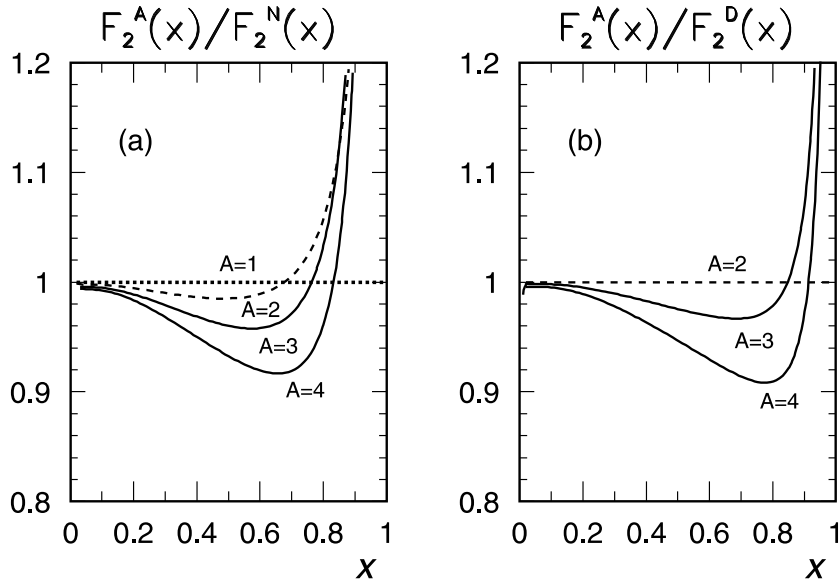} \epsfxsize=120mm
\caption{\label{Adep1}(a) The ratio of structure functions
$F_2^{A}/F_2^{\rm N}$. (b) The ratio of the structure functions
$F_2^{A}/F_2^{\rm D}$. The dashed curve in the left panel
 shows the result of the calculation for the deuteron ($A=2$) made in
section~\protect\ref{deuteron}. The results for $A=3,4$ are shown
by the solid lines. For $A=3$ the isoscalar combination $[F_2^{\rm
^3He}+ F_2^{\rm ^3H}]/2$ is used.}
\end{center}
\end{figure}

The results of the numerical calculations are presented in
Fig.~\ref{Adep1}(a). Here, we show how the free nucleon structure
function $F_2^{\rm N}(x)$ ($A$ = 1) evolves to the deuteron ($A$ =
2) and helium ($A$ = 3 and 4) structure functions. The evolution,
which starts from $F_2^{\rm D}(x)$, is shown in
Fig.~\ref{Adep1}(b). In contrast to the modifications observed for
nuclei with masses $A >$ 4, the pattern of the oscillation of
$r^A(x)$ changes in the range of $A \leq$ 4, causing the
coordinate of the cross-over point, $x_3$, to move toward a larger
value of~$x$.

The modifications with respect to $F_2^{\rm N}(x)$
(Fig.~\ref{Adep1}(a)) are not only of academic interest. We use
them to demonstrate that the change of the nucleon structure
function in the deuteron cannot be regarded as negligible, and,
therefore, the relation $F_2^A(x)/ F_2^{\rm D}(x)$ $\approx
F_2^A(x)/ F_2^{\rm N}(x)$ cannot be justified. The position of
$x_3$ is displaced by 0.08 when $F_2^{\rm N}(x)$ is replaced by
 $F_2^{\rm D}(x)$ for the case $A$ = 3 (Fig.~\ref{Adep1}(b)).
The displacement is eight times larger than the experimental error
for $\overline {x_3}$ found from the analysis of the measurements
of the ratios $F_2^A(x)/ F_2^{\rm D}(x)$~\cite{sm99}. According
to~\cite{sm99}, $\overline{x_3}$ = 0.84 $\pm$ 0.01 independently
of $A$ if $A >$~4.
 This accuracy allows one to reliably discriminate the effect
of modification of the {\em deuteron} structure from that of the
structure of the {\em free nucleon}.

It is remarkable that the value of $(1-x_3)$, which is found for
$F_2^{\rm D}(x)/F_2^{\rm N}(x)$ to be $\sim 0.32$, decreases for
the ratios $F_2^{A=3}(x)/F_2^{\rm D}(x)$ and
$F_2^{\rm^4He}(x)/F_2^{\rm D}(x)$ to $\sim 0.16$ and $\sim 0.08$
respectively. Further evolution of the modifications of $F_2^{\rm
N}(x)$ beyond $A$ = 4 is forbidden by Pauli exclusion principle.
As it follows from the pattern displayed in Fig.~\ref{Adep1}(a)
and from the relation between the cross-over points $x_3$, the
modifications of the nucleon structure resemble a saturation-like
process which is fully consistent with the rapid saturation of the
nuclear binding forces. This phenomenon allows us to introduce a
class of $x$-dependent modifications caused by the binding
effects. Within this class there are no mechanisms which could
lead to further changes in the pattern of $r^A(x)$ formed at the
first stage of the evolution, $A \leq 4$. The evolution of the
modifications to heavier nuclei, where the EMC effect was
discovered, has to proceed independently of $x$ and should be
viewed  as the second stage~\cite{sm95}. The two-stage concept of
the evolution of the free nucleon structure in nuclear environment
is crucial for understanding of the long-standing problem of the
EMC effect.

\begin{figure}
\begin{center}
\epsfxsize=90mm \epsfbox{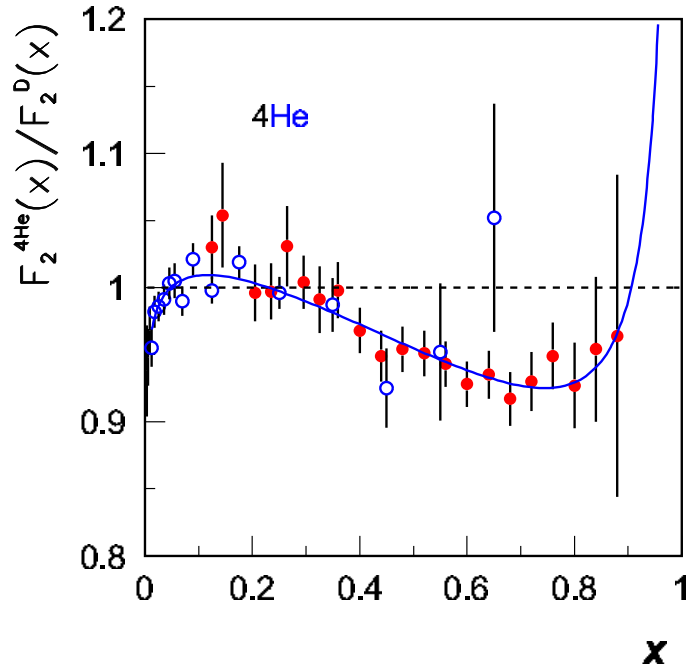} \epsfxsize=90mm
\caption{\label{He4} Results of the calculations of $F_2^{\rm
^4He}(x) / F_2^{\rm D}(x)$ performed in
Refs~\protect\cite{bma3,bma4} (solid line). The experimental
values are shown by the dark~\protect\cite{gomez} and
light~\protect\cite{ama95} circles.}
\end{center}
\end{figure}

As long as the experimental data for $A$ = 2 and 3 are not
available, our predictions can be only confronted with the results
on $F_2^{\rm ^4He}(x) / F_2^{\rm D}(x)$ reported in
Refs.~\cite{gomez,ama95}. This comparison is shown in
Fig.~\ref{He4}. The position of the cross-over point, obtained
from our calculations is $x_3$ = 0.913, which is in good agreement
with the extrapolated data. It is of course of high importance to
improve the accuracy of the data.

\begin{figure}
\begin{center}
\epsfxsize=90mm \epsfbox{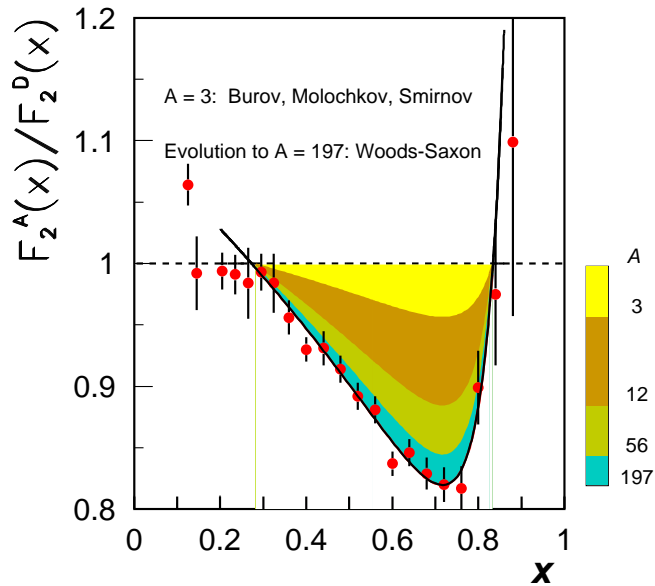} \epsfxsize=90mm
\caption{\label{Adep2} Evolution of the nucleon structure in
nuclei from $A=3$ to $A=197$, obtained by multiplying of
$F_2^{A=3}(x) / F_2^{\rm D}(x)$ by a scale parameter. The $A$
dependence of the parameter is determined by the Wood-Saxon
potential and is shown by different degrees of shading. The
results of measurement on a gold target were obtained in
ref.~\protect\cite{gomez}.}
\end{center}
\end{figure}

On the other hand, we note particularly a good agreement between
the corresponding point for $A$  = 3, $x_3$ = 0.845, and the
average
 $x_3$-value for
nuclei in the range $A = 9 \div 197$ found in  ref.~\cite{sm99}.
  Such an agreement naturally follows from the two-stage concept
 of the $F_2^{\rm N}(x)$ evolution which is
$x$ and $A$ dependent for $A \leq$ 4 and $A$ dependent only for
higher masses. The remarkable feature of our result is that the
$x$ dependent pattern of the EMC effect found experimentally in Fe
develops already at $A$ = 3 and therefore can be viewed as a
particular case of the class of the modifications. This is
demonstrated by comparison of our calculation with data for
 $A=197$  presented in Fig.~\ref{Adep2}.
Thus the EMC effect in heavy nuclei can be considered as the
reflection of the relative time effect of bound nucleons scaled by
the effect of the mean nuclear density.

A fundamental relation follows from the obtained results. Since
binding corrections for $\rm ^3He$ and $\rm ^3H$ have the same
form (see Eq.~(\ref{he3})) we can write
\begin{eqnarray}
I ~= ~ \int\limits_0^1 \frac{{\d} x}{x} \left(F_2^{{\rm
^3He}}(x)-F_2^{{\rm ^3H}}(x)\right)~ = \int\limits_0^1 \frac{{\d}
x}{x} \left( F_2^{\rm p}(x)-F_2^{{\rm n}}(x)\right).
\label{A3A1}\end{eqnarray} The result represents the Gottfried sum
$I$, which has often been studied experimentally  from the
combination of $F_2^{{\rm p}}(x)$ and $F_2^{{\rm D}}(x)$
(cf.~Ref.~\cite{nmcGSR}). Such a combination is equal to $I$ to
within a correction  proportional to $F_2^{\rm N}(x=0)$. Indeed,
as follows from Eq.~(\ref{f2nr}),
\[
I_{\rm D}=\int\limits_0^1 \frac{{\d} x}{x} \left(2F_2^{{\rm p}}(x)
-2F_2^{{\rm D}}(x)\right) =~I~ - 2\frac{\langle M_{\rm D}-2E_{\rm
N}\rangle_{\rm D}}{m} F_2^{\rm N}(x=0)~.
\]
Apparently, such tests cannot be performed rigorously because
$F_2^{\rm N}(x)$ is unknown at $x=0$. On the other hand, if the
difference of the $^3$He and $^3$H binding energy is considered
negligibly small, an experiment, which used these targets, would
be able to measure the nucleon isospin asymmetry independently of
the model uncertainties in the binding corrections.

 \subsection{\em Conclusions}

We have shown in this section that the BS formalism can be applied
to solve the topical problems of DIS of leptons off the lightest
nuclei. In particular, the method for calculations of the
evolution of the nucleon structure function as a function of $A$
has been developed. The method allows us to express the structure
functions $F_2^A(x)$ of the bound nucleon in terms of the
structure functions of nuclear fragments and three-dimensional
momentum distributions.

The pattern of the evolution
 of the modification of the nucleon structure function
in the lightest nuclei, {\rm D}, ${\rm ^3H}$, ${\rm ^3He}$ and
${\rm ^4He}$, is consistent with the saturation property of the
short range nuclear binding forces. The evolution is totally
different from that observed previously for heavy nuclei, in which
only the amplitude of deviations of $F_2^A$/$F_2^{\rm D}$ from
unity increased with $A$. The quantitative predictions for ${\rm
^3He}$ and ${\rm ^4He}$ nuclei, which have to be verified in
future experiments at HERA or TJNAF, imply that the  EMC effect in
heavy nuclei can  be naturally understood as distortions of the
parton distributions in $^3$He or $^3$H  which are modified by the
 nuclear density effects.\\

\section{Summary}

In the present review we have considered the formulation of the
Bethe-Salpeter equation. It is realized for the two-nucleon system
by using the multipole expansion with the spinor structure of the
two nucleons. The separable ansatz for the interaction kernel for
each partial wave has provided a manageable system of the linear
homogeneous equations for the BS amplitude. We have demonstrated
then the construction of the separable interaction by taking only
one term in the Yamaguchi form. Even with the two parameters for
the $^1S_0$ and $^3S_1$ channels, we have found good reproduction
of the phase shifts up to about 100 MeV in addition to the
deuteron binding energy. This part has demonstrated the details of
the formulation of the BS equation with the separable interaction.

We have switched then to the case with the use of the covariant
revision of the Graz II separable potential with the summation of
several separable functions. The calculated results have been
compared, first, to the static deuteron properties. The comparison
shows very good agreement. We have applied then the BS amplitude
for the calculation of the nucleon form factors that determine
$e{\rm D}$ elastic scattering cross sections. Comparison of the
obtained results with the experimental data is good in general.
But there exist definitely some defects in the comparison with the
data as the charge form factor and the tensor polarizations, which
indicate the necessity of improvement.

The comparison with the elastic scattering has brought us to
discuss the ingredient of the BS formalism by taking the simple
cases as the deuteron magnetic moment, deuteron quadruple moment
and further the electro-disintegration of the deuteron. In this
discussion, we have taken all the possible channels in the BS
amplitude. We have found that the relativistic covariant
description automatically include the meson exchange currents, in
particular, the pair current through $P$-wave (negative energy
state) component in the BS amplitude. We find it necessary to
extend the partial waves to include the $P$-wave in order to
construct the relativistic deuteron state. The $P$-wave components
were not included in considered separable interaction kernels.

 Reactions of the elastic  lepton scattering off the deuteron
and the deuteron electro-disintegration served as a testing ground
for the method under investigation and helped to outline both
strong and weak points of the approach. The analysis has proved
the technique to be very promising, even if we find a few evident
discrepancies with data at this stage of development. Several
items can be suggested for the  program of further theoretical
studies. 1) Construction of the separable potential of higher rank
in order to reach better understanding of the properties of the
deuteron, phases of $NN$- scattering, and to study hadron-deuteron
processes (for example, the reaction  $p+D \to p+X$); 2) Research
into the  relativistic two-body currents; 3) Studies of the
off-shell effects in lepton-deuteron scattering.

One specific feature of the BS formalism deserves a special
comment.
 The BS amplitude depends on the zeroth component of the relative coordinate
(relative time)  of the bound nucleons, which is reflected  in the
dynamical  observables of  the $n$-nucleon bound state. In
momentum space, this  leads to the dependence on the zeroth
component of the nucleon relative momentum  (relative energy). The
dependence is manifested as observable effects in DIS of leptons
off the lightest nuclei.

The extension of the  approach to the process of deep inelastic
lepton scattering off the deuteron has been realized in a
model-independent way. This quality is of particular importance
for the consideration of the relatively small effects of the
modification of the nucleon structure function $F_2^{\rm N}(x)$ by
the $NN$ binding forces. Based on the  the model-free technique,
the method for calculations of the evolution of the nucleon
structure in the lightest nuclei as a function of $A$ has been
developed.

We have found that the effects from asynchronous nucleons which
naturally follow from the relativistic treatment of the
two-nucleon bound state are decisive in obtaining differences
between the structure functions of bound and free nucleons. The
characteristic modification of the nucleon structure functions
found for $A$ = 2  serves as a priming for the modifications in
the three- and four-nucleon systems and plays, therefore, a
fundamental role in the evolution of the bound nucleon structure.
The EMC effect, which was essentially the observation that
partonic structures of $A$ = 2 and $A$ = 56 nuclei were different,
can be now regarded as a particular case of the whole class of
modifications of the free nucleon structure in nuclear
environment.

When translated to a nonrelativistic language, the event of
asynchronous nucleons can be associated  with the increase of the
localization region for the bound nucleon and is observed as the
modification of  $F_2^{\rm N}(x)$. The developed approach does not
require consideration of the three-nucleon forces to describe
correctly the data available for the ratio $F_2^{\rm
^4He}$/$F_2^{\rm D}$. The two-nucleon interactions can be,
therefore, considered as the dominant mechanism for the evaluation
of the nuclear binding effects in the kinematic range 0.3 $<x<$
0.9.

The results obtained for the lightest nuclei,
 {\rm D}, ${\rm ^3H}$, ${\rm ^3He}$ and ${\rm ^4He}$,
demonstrate that the BS formalism can have interesting practical
applications for a certain class of electromagnetic reactions,
namely DIS of leptons in asymptotic regime. We see future
prospects of the covariant BS formalism in the confrontation of
forthcoming high precision data with quantitative theoretical
predictions.

We have reviewed the covariant description of few-nucleon systems
in the framework of the Bethe-Salpeter approach. This trial is
still at primitive stage as compared to the widely used
quasi-potential description of the BS equation. We thought it
important though to write up the present status of this new
development in the review form. Hence, the purpose of this article
is not to tell all the consequences of the present approach but to
show what is included in the covariant description of the BS
approach and what has to be done further in this framework. In
this sense, we would like to mention also the connection of the
covariant BS formalism to the one of the light front formalism
discussed in the article. Even at this simple stage it is clear
that covariant four-dimensional approach give qualitatively new
point of view on the present problems in the nuclear physics.


\paragraph{Acknowledgment}
 We are indebted to the late academician A.M.~Baldin for his
continuous interest in the covariant description of a composite
system, which has stimulated strongly our research project.

We wish to thank our collaborators M.~Beyer, A.A.~Goy,
S.M.~Dorkin, K.Yu.~Kazakov, A.V.~Shebeko and S.Eh.~Shirmovsky for
their contribution to the present subject. We would like to thank
Professor Y.~Yamaguchi and Professor Y.~Nambu for interest in the
present work and fruitful discussions. V.V.B. and A.V.M. expresses
their deep gratitude to the Director of the Research Center for
Nuclear Physics of Osaka University (Japan) for warm hospitality
and support in preparing the manuscript. S.G.B., V.V.B., A.V.M.
thank Professor G.~Roepke, Professor D.~Blaschke for the
hospitality and interest in the work and acknowledge the support
of the Rostock University (Germany). A.V.M thank to the INFN
section Roma II for the warm hospitality. G.I.S. gratefully
acknowledges the support of the University of Gent (Belgium), of
the University Blaise Pascal (Clermont-Ferrand) and Laboratoire de
Physique Corpusculaire IN2P3--CNRS (France).

The work was supported in part by RFBR grant No. 00-15-96737.

\section*{Appendix A}
\addcontentsline{toc}{section}{Appendix A}
\setcounter{equation}{0}
\def\theequation{A.\arabic{equation}}
\def\theqnarray{A.\arabic{eqnarray}}
\def\thetable{A.\arabic{table}}
In this appendix we give the matrix elements used for calculation of the
deuteron quadrupole moment.
\begin{eqnarray}\label{upwp}
&&Q_{\bk}^{(+,+)}=- \frac{e}{2M} \int\!\frac{ d k_0\bk^2 d|\bk|}{
i (2\pi)^4}(E_{\bk}-{M\over 2}+k_0) \left\{ (1-{2k_0\over
M})^2\left[-{1\over 12}
\frac{(E_{\bk}-m)^2}{\bk^2E_{\bk}^2}\Upp^2\right.\right.
\\
&& -{1\over 120}\frac{14E_{\bk}^4+5E_{\bk}^2m^2-3m^4+
20E_{\bk}^3m}{\bk^2E_{\bk}^4}\Wpp^2 +{1\over
10}\Wpp\frac{1}{|\bk|} \frac{\partial}{\partial |\bk|}{\Wpp}
+{1\over 20}\Wpp\frac{\partial^2}{\partial \bk^2}{\Wpp}
\nonumber\\
&&+{\sqrt{2}\over 60}
\frac{3m^4-4E_{\bk}^4+5E_{\bk}^2m^2+5E_{\bk}^3m}
{\bk^2E^4_{\bk}}\Upp\Wpp +{\sqrt{2}\over
20}\frac{2E_{\bk}+3m}{E_{\bk}}\Upp
\frac{1}{|\bk|}\frac{\partial}{\partial |\bk|}{\Wpp}
\nonumber\\
&&+{\sqrt{2}\over 20}\frac{2E_{\bk}-3m}{E_{\bk}}\Wpp
\frac{1}{|\bk|}\frac{\partial}{\partial |\bk|}{\Upp}
\left.+{\sqrt{2}\over 20}\Wpp \frac{\partial^2}{\partial
\bk^2}{\Upp} +{\sqrt{2}\over 20}\Upp \frac{\partial^2}{\partial
\bk^2}{\Wpp}\right]
\nonumber\\
&&+{1\over 5}\frac{\bk^2}{M^2} \left[{3\over
2}\frac{1}{\bk^2}\Wpp^2+\Wpp
\frac{1}{|\bk|}\frac{\partial}{\partial |\bk|}{\Wpp}
+\frac{3\sqrt{2}}{\bk^2}\Upp\Wpp\right.\nonumber\\[2mm]
&&\left.\left.+\sqrt{2}\Wpp
\frac{1}{|\bk|}\frac{\partial}{\partial |\bk|}{\Upp} +\sqrt{2}\Upp
\frac{1}{|\bk|} \frac{\partial}{\partial
|\bk|}{\Wpp}\right]\right\} \nonumber\end{eqnarray} and
\begin{eqnarray}
\label{Qeps} &&Q_{k_0}^{(+,+)}= \frac{e}{2M} \int\!\frac{ d
k_0\bk^2 d|\bk|} { i (2\pi)^4}{1\over 5}
\frac{\bk^2}{M^2}(E_{\bk}-{M\over 2}+k_0) \left\{ -\sqrt{2}
\left[\Upp\frac{\partial^2}{\partial k_0^2}{\Wpp} \right.\right.
\\
&& \left.\left. +\Wpp\frac{\partial^2}{\partial
k_0^2}{\Upp}\right] -\Wpp\frac{\partial^2}{\partial k_0^2}{\Wpp}
\right\} +\frac{e}{2M} \int\!\frac{ d k_0\bk^2 d|\bk|} { i
(2\pi)^4} {3\over 10M}(1-{2k_0\over M})^2(E_{\bk}-{M\over 2}+k_0)
\nonumber\\
&&\left\{ \sqrt{2}\left[\left(1+{m\over E_{\bk}}\right)
\Upp\frac{\partial}{\partial k_0}{\Wpp} +\left(1+{m\over
E_{\bk}}\right) \Wpp\frac{\partial}{\partial k_0}{\Upp}\right]
+\Wpp\frac{\partial}{\partial k_0}{\Wpp}\right]
\nonumber\\
&&+\frac{e}{2M} \int\!\frac{ d k_0\bk^2 d|\bk|} { i (2\pi)^4}
{|\bk|\over 5M}(1-{2k_0\over M})(E_{\bk}-{M\over 2}+k_0)
\left\{ \sqrt{2}\left[\Upp\frac{\partial^2}{\partial k_0\partial
|\bk|}{\Wpp} \right.\right.
\nonumber\\
&& \left.\left. +\Wpp\frac{\partial^2}{\partial k_0\partial
|\bk|}{\Upp}\right] +\Wpp\frac{\partial^2}{\partial k_0\partial
|\bk|}{\Wpp}\right]. \nonumber
\end{eqnarray}

\begin{eqnarray}
\label{LB1} &&Q^{(++)}_{LB}=\frac{e}{2M} \int\!\frac{ d k_0 \bk^2
d |\bk|}{ i (2\pi)^4}
(E_{\bk}-\frac{M}{2}+k_0)(1-\frac{2k_0}{M})\frac{1}{5M}
\frac{1}{E_{\bk}} \left\{
\frac{6E_{\bk}^2-2mE_{\bk}-m^2}{E^2_{\bk}} \left[
\frac{1}{2}\Wpp^2 \right.\right.
\nonumber\\
&&\left.\left. +\sqrt{2}\Upp\Wpp\right] +\sqrt{2}|\bk|
\left[\Upp\frac{\partial}{\partial |\bk|}\Wpp
+\Wpp\frac{\partial}{\partial |\bk|}\Upp\right] \right.
\nonumber\\
&&+\left.|\bk|\Wpp\frac{\partial}{\partial |\bk|} \Wpp\right\}
-\frac{e}{2M} \int\!\frac{ d k_0\bk^2 d|\bk|}{ i (2\pi)^4}
(E_{\bk}-\frac{M}{2}+k_0)\frac{1}{5} \frac{\bk^2}{M^2E_{\bk}}
\left\{ \sqrt{2}\left[\Upp\frac{\partial}{\partial k_0}\Wpp+
\right.\right.
\nonumber\\
&&+\left.\left. \Wpp\frac{\partial}{\partial k_0}\Upp\right]
\Wpp\frac{\partial}{\partial k_0}\Wpp\right\}.
\nonumber\end{eqnarray}

Here we omit arguments $(k_0,|\bk|)$ in the BS amplitudes
$\phi_{\sp}$ and $\phi_{\Dp}$.

\section*{Appendix B}
\addcontentsline{toc}{section}{Appendix B}
\setcounter{equation}{0}
\def\theequation{B.\arabic{equation}}
\def\theqnarray{B.\arabic{eqnarray}}
\def\thetable{B.\arabic{table}}
In this appendix we give the functions $V_i^{(1,2)}$ which defines
the Lorenz invariant part $V(s,q^2)$ of the matrix element for the
deuteron electrodisintegration. This part is defined by the
Eqs.(\ref{eqn:V}) and (\ref{v1v2}). The function used in the
Eq.(\ref{v1v2}) are follows:
\begin{eqnarray}
V_1^{(1)} =&& 4\omega_1 c_1b_1 \Bigl[ h_3+h_5+2h_7 \Bigr],
\\
V_2^{(1)} =&&\omega_1 b_2 \Bigl[ -2\Bigl(2(c_1-c_2)+1\Bigr)h_1 +
8c_2h_3 - 4\omega_2 dh_4 +8\Bigl(2c_1-c_2+1\Bigr)h_5 -4\omega_2
dh_6
\nonumber\\
&& + \frac{\omega_3}{2}
\Bigl(2(4k^2+12m^2+M^2-4(Pk))c_1+2(4k^2-4m^2-4(Pk)+M^2)c_2
\nonumber\\
&& + 4k^2+12m^2+M^2-4(Pk)\Bigr)h_7 - 8\omega_2 dh_8 \Bigr],
\\
V_3^{(1)} =&&\omega_1 b_3 \Bigl[ -2\Bigl(2(c_1-c_2)+1\Bigr)h_1
-8\Bigl(2c_1-c_2\Bigr)h_3 -4\omega_2 dh_4 +8\Bigl(1-c_2\Bigr)h_5 -
4\omega_2 dh_6
\nonumber\\
&& - \frac{\omega_3}{2} \Bigl(
2(M^2+12m^2+4k^2-4(Pk))c_1+2(4m^2-4k^2-M^2+4(Pk))c_2
\nonumber\\
&& - M^2-12m^2-4k^2+4(Pk)\Bigr)h_7 - 8\omega_2 dh_8 \Bigr],
\\
V_4^{(1)} =&&\omega_1 b_4 \Bigl[ -4\Bigl(2(c_1-c_2)+1\Bigr)h_1-
\omega_3 \Bigl((12m^2+M^2-8(kq)+4k^2-4(Pk)+4(Pq))c_1
\nonumber\\
&& + 2(M^2-4m^2-4k^2)c_2-4(Pk)+8k^2-16dM^2)\Bigr)h_3 -8\omega_2
dh_4
\nonumber\\
&& + \omega_3 \Bigl((12m^2-4(Pq)+M^2+4k^2+8(kq)-4(Pk))c_1+
4(2(Pk)-4m^2-M^2)c_2
\nonumber\\
&& + M^2+12m^2-4k^2+16dM^2\Bigr)h_5 -8\omega_2 dh_6
+4\Bigl(2(kq)-(Pq)\Bigr)c_1
\nonumber\\
&& + 2\omega_3 \Bigl(
(4(Pk)-4m^2+4k^2-3M^2)c_2-4k^2+M^2+4m^2+16dM^2\Bigr)h_7
\nonumber\\
&& - \omega_2 \Bigl(M^2+12m^2+4k^2-4(Pk)\Bigr)dh_8 \Bigr],
\end{eqnarray}

\begin{eqnarray}
V_1^{(2)}=&&\omega_1 b_1 \Bigl[
-\Bigl(1-2c_2\Bigr)h_1+4c_2h_3-2\omega_2
dh_4+4\Bigl(1-c_2\Bigr)h_5 -2\omega_2 dh_6
\nonumber\\
&& +\frac{\omega_3}{4}\Bigl(2(4k^2-4m^2-4(Pk)+M^2)c_2
+4k^2+12m^2+M^2-4(Pk)\Bigr)h_7 -4\omega_2 dh_8 \Bigr],
\\
V_2^{(2)}=&&\omega_1 b_2 \Bigl[ 4c_2h_1-2\omega_2 dh_2 +2\omega_3
\Bigl((4(kq)+q^2)c_1+4c_2k^2-2k^2+6dM^2\Bigr)h_3 -4\omega_2 dh_4
\nonumber\\
&& + 2\omega_3
\Bigl((4(kq)+q^2)c_1+(4(Pk)-M^2-4m^2)c_2-2k^2+6dM^2\Bigr)h_5
+4\omega_2 dh_6
\nonumber\\
&& +\omega_3 \Bigl(16c_1(kq)+4c_1q^2+(12(Pk)-4m^2+4k^2-3M^2)c_2
-8k^2+24dM^2\Bigr)h_7
\nonumber\\
&& -\frac{\omega_2}{2}\Bigl(M^2-4m^2+4k^2-4(Pk)\Bigr)dh_8 \Bigr],
\\
V_3^{(2)}=&&\omega_1 b_3 \Bigl[ 4\Bigl(1-c_2\Bigr)h_1-2\omega_2
dh_2 +2\omega_3 \Bigl((2(Pq)+q^2)c_1+(M^2-4m^2)c_2
\nonumber\\
&& -2(Pk)+2k^2-2dM^2\Bigr)h_3 +4\omega_2 dh_4 +\omega_3 \Bigl(
2(2(Pq)+q^2)c_1+2(4m^2+M^2)c_2
\nonumber\\
&&-12m^2-M^2-4dM^2\Bigr)h_5+12\omega_2 dh_6+ \Bigl(4(q^2+2(Pq))c_1
\nonumber\\
&& +(3M^2+4m^2-4k^2+4(Pk))c_2 -8m^2-2M^2-8dM^2\Bigr)h_7
\nonumber\\
&&+\frac{\omega_2}{2}\Bigl(28m^2+M^2+4k^2-4(Pk)\Bigr)dh_8 \Bigr],
\\
V_4^{(2)}=&&\omega_1 b_4 \Bigl[ \frac{\omega_3}{4}
\Bigl(8(2(kq)-(Pq))c_1+2(4k^2-4m^2+4(Pk)-3M^2)c_2
\nonumber\\
&& -12k^2+M^2+12m^2+4(Pk) + 32dM^2\Bigr)h_1 -4\omega_2 dh_2 +
\Bigl(4(q^2+(Pq)+2(kq))c_1
\nonumber\\
&&+(M^2+4k^2-4m^2+4(Pk))c_2 - 4(Pk)+8dM^2\Bigr)h_3
\nonumber\\
&& -\frac{\omega_2}{2}\Bigl(M^2-4m^2+4k^2-4(Pk)\Bigr)dh_4
-\omega_3 \Bigl(4(2(kq)-3(Pq)+4(kq)-q^2)c_1
\nonumber\\
&&+(4k^2+4(Pk)-4m^2-7M^2)c_2 -8k^2+2M^2+8m^2+4(Pk)+24dM^2\Bigr)h_5
\nonumber\\
&&+\frac{\omega_2}{2}\Bigl(28m^2+M^2+4k^2-4(Pk)\Bigr)dh_6
+\frac{\omega_4}{2} \Bigl( (M^2(Pq)+12q^2m^2+2M^2(kq)
\nonumber\\
&& -8(kq)(Pk)+q^2M^2-8(kq)m^2+28(Pq)m^2+4q^2k^2+4(Pq)k^2+8k^2(kq)
\nonumber\\
&& -4(Pq)(Pk)-4q^2(Pk))c_1 +2(16m^4+16k^4+M^4-16(Pk)^2-32k^2m^2
\nonumber\\
&& +8M^2k^2+56M^2m^2)c_2 +96k^2m^2-8M^2k^2 -64(Pk)m^2-40M^2m^2-M^4
\nonumber\\
&& +16(Pk)^2-16k^4-80m^4 +16(4k^2+M^2-4(Pk)-20m^2)M^2d\Bigr)h_7
\nonumber\\
&& +\omega_5\Bigl(M^2-4(Pk)+4k^2+12m^2\Bigr)dh_8 \Bigr],
\end{eqnarray}
with the notations
$$
\omega_1=1/m,\quad \omega_2=M^2/m^2,\quad \omega_3=1/m^2,\quad
\omega_4=1/m^4,\quad \omega_5=M^2/m^4.
$$

\end{document}